\documentclass[longauth]{aa}  

\usepackage{natbib}
\bibpunct{(}{)}{;}{a}{}{,} 

\usepackage{graphicx}

\usepackage{hyperref}
\usepackage{orcidlink}

\usepackage{txfonts}

\begin{document} 

\title{ALMAGAL} 
\subtitle{VI. The spatial distribution of dense cores during the evolution of cluster-forming massive clumps}

\author{E.~Schisano\inst{\ref{iaps}}\orcidlink{0000-0003-1560-3958}
\and
S.~Molinari\inst{\ref{iaps}}\orcidlink{0000-0002-9826-7525}
A.~Coletta\inst{\ref{iaps},\ref{sapienza}}\orcidlink{0000-0001-8239-8304}\and
D.~Elia\inst{\ref{iaps}}\orcidlink{0000-0002-9120-5890}
P.~Schilke\inst{\ref{u-koln}}\orcidlink{0000-0003-2141-5689}\and
A.~Traficante\inst{\ref{iaps}}\orcidlink{0000-0003-1665-6402}\and
\'A.~S\'anchez-Monge\inst{\ref{icecsic}, \ref{ieec}}\orcidlink{0000-0002-3078-9482} \and
H.~Beuther\inst{\ref{mpia}}\orcidlink{0000-0002-1700-090X} \and
M.~Benedettini\inst{\ref{iaps}}\orcidlink{0000-0002-3597-7263} \and
C.~Mininni\inst{\ref{iaps}}\orcidlink{0000-0002-2974-4703}\and
R.~S.~Klessen\inst{\ref{u-hei},\ref{ralf-2},\ref{cfa},\ref{cfa-rad}}\orcidlink{0000-0002-0560-3172}\and
J.~D.~Soler\inst{\ref{iaps}}\orcidlink{0000-0002-0294-4465} \and
A.~Nucara\inst{\ref{iaps},\ref{torverg}}\orcidlink{0009-0005-9192-5491}\and
S.~Pezzuto\inst{\ref{iaps}}\orcidlink{0000-0001-7852-1971}\and
F.~van der Tak\inst{\ref{sron}, \ref{u-gron}}\orcidlink{0000-0002-8942-1594}\and
P.~Hennebelle\inst{\ref{saclay}}\orcidlink{0000-0002-0472-7202}\and
M.~T.~Beltr\'an\inst{\ref{arcetri}}\orcidlink{0000-0003-3315-5626} \and
L.~Moscadelli\inst{\ref{arcetri}}\orcidlink{0000-0002-8517-8881}\and
K.~L.~J.~Rygl\inst{\ref{ira}}\orcidlink{0000-0003-4146-9043}\and
P.~Sanhueza\inst{\ref{bunkyo}},\orcidlink{0000-0002-7125-7685}\and
P.~M.~Koch\inst{\ref{asiaa}}\orcidlink{0000-0003-2777-5861}\and
D.~C.~Lis\inst{\ref{caltech}}\orcidlink{0000-0002-0500-4700}\and
R.~Kuiper\inst{\ref{u-duisb}}\orcidlink{0000-0003-2309-8963}\and
G.~A.~Fuller\inst{\ref{u-koln},\ref{u-man}}\orcidlink{0000-0001-8509-1818} \and
A.~Avison\inst{\ref{skao}}\orcidlink{0000-0002-2562-8609}\and
L.~Bronfman\inst{\ref{u-chile}}\orcidlink{0000-0002-9574-8454}\and
U.~Lebreuilly\inst{\ref{saclay}}\orcidlink{0000-0001-8060-1890}\and
T.~M\"{o}ller \inst{\ref{u-koln}}\orcidlink{0000-0002-9277-8025}\and
T.~Liu\inst{\ref{shanghai}}\orcidlink{0000-0002-5286-2564}\and
V.~-M.~Pelkonen\inst{\ref{iaps}}\orcidlink{0000-0002-8898-1047}\and
L. Testi\inst{\ref{u-bo}}\orcidlink{0000-0003-1859-3070}\and
Q.~Zhang\inst{\ref{cfa}}\orcidlink{0000-0003-2384-6589}\and
T.~Zhang\inst{\ref{zhejiang},\ref{u-koln}}\orcidlink{0000-0002-1466-3484}\and
A.~Ahmadi\inst{\ref{leiden}}\orcidlink{0000-0003-4037-5248}\and
J.~Allande\inst{\ref{arcetri},\ref{unifi}}\orcidlink{0009-0007-4060-0560}\and
C.~Battersby\inst{\ref{u-conn}}\orcidlink{0000-0002-6073-9320}\and
J.~Wallace\inst{\ref{u-conn}}\orcidlink{0009-0002-7459-4174}\and
C.~L.~Brogan\inst{\ref{nraoCH}}\orcidlink{0000-0002-6558-7653}\and
S.~Clarke\inst{\ref{u-koln},\ref{asiaa}}\orcidlink{0000-0001-9751-4603} \and
F.~De~Angelis\inst{\ref{iaps}}\orcidlink{0009-0002-6765-7413}\and
F.~Fontani\inst{\ref{arcetri},\ref{mpe},\ref{lux}}\orcidlink{0000-0003-0348-3418}\and 
P.~T.~P.~Ho\inst{\ref{asiaa},\ref{hawaii}}\orcidlink{0000-0002-3412-4306}\and
T.~R.~Hunter\inst{\ref{nraoCH}}\orcidlink{0000-0001-6492-0090}\and
B.~Jones\inst{\ref{u-koln}}\orcidlink{0000-0002-0675-0078}\and
K.~G. Johnston\inst{\ref{lincoln}}\orcidlink{0000-0003-4509-1180}\and
P.~D.~Klaassen\inst{\ref{roe}}\orcidlink{0000-0001-9443-0463}\and
S.~J.~Liu\inst{\ref{iaps}}\orcidlink{0000-0001-7680-2139}\and
S.-Y.~Liu\inst{\ref{asiaa}}\orcidlink{0000-0003-4603-7119}\and
Y.~Maruccia\inst{\ref{napoli}}\orcidlink{0000-0003-1975-6310}\and
A.~J.~Rigby\inst{\ref{leeds}}\orcidlink{0000-0002-3351-2200}\and
Y.-N.~Su\inst{\ref{asiaa}}\orcidlink{0000-0002-0675-276X}\and
Y. Tang\inst{\ref{asiaa}}\orcidlink{0000-0002-0675-276X}\and
S.~Walch\inst{\ref{u-koln},\ref{datacologne}}\orcidlink{0000-0001-6941-7638}\and
H. Zinnecker\inst{\ref{u-auton-chile}}\orcidlink{0000-0003-0504-3539}
    }

   \institute{INAF - IAPS, via Fosso del Cavaliere, 100, I-00133 Roma, Italy\\
              \email{eugenio.schisano@inaf.it}             
Istituto Nazionale di Astrofisica (INAF)-Istituto di Astrofisica e Planetologia Spaziali, Via Fosso del Cavaliere 100, I-00133 Roma, Italy \label{iaps}
\and
Dipartimento di Fisica, Sapienza Universit\`a di Roma, Piazzale Aldo Moro 2, I-00185, Roma, Italy \label{sapienza}
\and
Physikalisches Institut der Universit\"at zu K\"oln, Z\"ulpicher Str. 77, D-50937 K\"oln, Germany\label{u-koln}
\and
Institut de Ci\`encies de l'Espai (ICE, CSIC), Campus UAB, Carrer de Can Magrans s/n, E-08193, Bellaterra (Barcelona), Spain\label{icecsic} 
\and
Institut d'Estudis Espacials de Catalunya (IEEC), E-08860, Castelldefels (Barcelona), Spain\label{ieec} 
\and
Max Planck Institute for Astronomy, K\"onigstuhl 17, 69117 Heidelberg, Germany\label{mpia}
\and
Universit\"{a}t Heidelberg, Zentrum f\"{u}r Astronomie, Institut f|"{u}r Theoretische Astrophysik, Albert-Ueberle-Straße 2, D-69120 Heidelberg, Germany \label{u-hei} 
\and
Universit\"{a}t Heidelberg, Interdisziplin\"{a}res Zentrum f\"{u}r Wissenschaftliches Rechnen, Im Neuenheimer Feld 205, D-69120 Heidelberg, Germany \label{ralf-2} 
\and
Center for Astrophysics,Harvard \& Smithsonian, 60 Garden Street, Cambridge, MA 02138, USA\label{cfa} 
\and
Radcliffe Institute for Advanced Studies at Harvard University, 10 Garden Street, Cambridge, MA 02138, U.S.A. \label{cfa-rad}
\and
Dipartimento di Fisica, Università di Roma Tor Vergata, Via della Ricerca Scientifica 1, I-00133 Roma, Italy\label{torverg}
\and
SRON Netherlands Institute for Space Research\label{sron}
\and
Kapteyn Astronomical Institute, University of Groningen, Landleven 12, 9747 AD Groningen, The Netherlands\label{u-gron}
\and 
Universit\'e Paris-Saclay, Universit\'e Paris-Cit\'e, CEA, CNRS, AIM, 91191 Gif-sur-Yvette, France\label{saclay}
\and
Istituto Nazionale di Astrofisica (INAF), Osservatorio Astrofisico di Arcetri, Largo E. Fermi 5, Firenze, Italy \label{arcetri} 
\and
INAF-Istituto di Radioastronomia \& Italian ALMA Regional center, Via P. Gobetti 101, I-40129 Bologna, Italy \label{ira}
\and
Department of Astronomy, School of Science, The University of Tokyo, 7-3-1 Hongo, Bunkyo, Tokyo 113-0033, Japan\label{bunkyo}
\and
Institute of Astronomy and Astrophysics, Academia Sinica, 11F of ASMAB, AS/NTU No.\ 1, Sec.\ 4, Roosevelt Road, Taipei 10617, Taiwan \label{asiaa}
\and
Jet Propulsion Laboratory, California Institute of Technology, 4800 Oak Grove Drive, Pasadena, CA 91109, USA\label{caltech}
\and
Faculty of Physics, University of Duisburg-Essen, Lotharstra{\ss}e 1, D-47057 Duisburg, Germany\label{u-duisb}
\and
Jodrell Bank center for Astrophysics, University of Manchester\label{u-man}
\and
SKA Observatory, Jodrell Bank, Lower Withington, Macclesfield, SK11 9FT, UK\label{skao}
\and
Departamento de Astronomía, Universidad de Chile, Casilla 36-D, Santiago, Chile\label{u-chile}
\and
Shanghai Astronomical Observatory, Chinese Academy of Sciences, 80 Nandan Road, Shanghai 200030, China\label{shanghai}
\and
Dipartimento di Fisica e Astronomia, Alma Mater Studiorum - Universit\`a di Bologna\label{u-bo}
\and
Research Center for Astronomical computing, Zhejiang Laboratory, Hangzhou, China\label{zhejiang}
\and
Leiden Observatory, Leiden University, PO Box 9513, 2300 RA Leiden, The Netherlands\label{leiden}
\and
Dipartimento di Fisica e Astronomia, Università degli Studi di Firenze, Via G. Sansone 1, I-50019 Sesto Fiorentino, Firenze, Italy\label{unifi}
\and
University of Connecticut, Department of Physics, 2152 Hillside Road, Unit 3046 Storrs, CT 06269, USA\label{u-conn}
\and 
National Radio Astronomy Observatory, 520 Edgemont Road, Charlottesville VA 22903, USA\label{nraoCH}
\and
Max-Planck-Institute for Extraterrestrial Physics (MPE), Garching bei M\"unchen, Germany \label{mpe} 
\and
LUX, Observatoire de Paris, Université PSL, Sorbonne Université, CNRS, 75014, Paris, France \label{lux}
\and
East Asian Observatory, 660 N.\ A'ohoku, Hilo, Hawaii, HI 96720, USA \label{hawaii}
\and
School of Engineering and Physical Sciences, The University of Lincoln, Brayford Way, Lincoln, LN6 7TS, United Kingdom\label{lincoln}
\and
UK Astronomy Technology center, Royal Observatory Edinburgh, Blackford Hill, Edinburgh EH9 3HJ, UK\label{roe}
\and
INAF - Astronomical Observatory of Capodimonte, Via Moiariello 16, I-80131 Napoli, Italy \label{napoli}
\and
School of Physics and Astronomy, University of Leeds, Leeds LS2 9JT, UK\label{leeds}
\and
Center for Data and Simulation Science, University of Cologne, Germany\label{datacologne}
\and Universidad Aut\'onoma de Chile, Av. Pedro de Valdivia 425, Providencia, Santiago de Chile, Chile\label{u-auton-chile}
}

\date{Received XXX, xxx, 2025; accepted XXX, XXX }

  \abstract
  {High-mass stars and star clusters form from the fragmentation of massive dense clumps driven by gravity, turbulence, and magnetic fields. The extent to which each of these agents impacts the fragmentation depending on the clump mass, density, and evolutionary stage is still largely unknown.
  }
  {The ALMA evolutionary study of high-mass protocluster formation in the GALaxy (ALMAGAL) project, with $\sim1000$ clumps observed at $\sim$1000\,au resolution, allows a statistically significant characterization of the fragmentation process over a large range of clump physical parameters and evolutionary stages.
  Our goal is to characterize where and how the dense cores revealed by ALMA are distributed in massive potentially cluster-forming clumps to trace how fragmentation is initially set and how it proceeds before gas dispersal due to stellar feedback.
  }
  {We characterized the spatial distribution of dense cores in the 514 ALMAGAL clumps that host at least four cores, using a set of quantitative descriptors that we evaluated against the clump bolometric luminosity-to-mass ratio, which we adopted as an indicator of the evolution of the system. We measured the separations between cores with the minimum spanning tree (MST) method, which we compared with the predictions of gravitational fragmentation from Jeans theory. We investigated whether cores have specific arrangements using the $Q$ parameter or variations due to their masses with the mass segregation ratio, $\Lambda_{MSR}$.
  }
  {ALMAGAL cores are distributed throughout the entire area of the clump, usually arranged in elliptical groups with an axis ratio $e\sim2.2$, although high values with $e\,\geq$\,5 are also observed. We found a single characteristic core separation per clump in $\sim76$\% of cases, suggesting that multiple fragmentation lengths may be frequently present. Typical core separations are compatible with the clump-averaged thermal Jeans length, $\lambda^{th}_{J}$. However, we found an additional  population of cores, typical of low-fragmented and young clumps, which are on average more widely separated with $l\,\approx\,3\times\lambda^{th}_{J}$. By stacking the distributions of the core separations in clumps of similar evolutionary stage, we also found that the separation decreases on average from $l\sim22000$\,au in younger systems to $l\sim7000$\,au in more evolved ones. The ALMAGAL cores are typically distributed in fractal-type subclusters, while centrally concentrated patterns appear only at later stages, but we do not observe a progressive transition between these configurations with evolution.
  Finally, we also found 110 ALMAGAL systems with a signature of mass segregation, with an occurrence that increases with evolution.
  }
  {}
   
   \keywords{Stars: formation -- Stars: protostars --
                ISM: clouds --
                Submillimeter: ISM
            }  
   \titlerunning{The spatial distribution of ALMAGAL dense cores}
   \maketitle

\section{Introduction}

At least half of all stars, including our Sun and almost all massive stars, formed in clustered systems of hundreds to thousands of objects \citep{Lada2003, Zinnecker2007, Bressert2010, Adams2010, Megeath2016, Megeath2022, Adamo2020}. The prevalence of clustered star formation is a direct consequence of the  hierarchical nature of the interstellar medium (ISM) from which stars form \citep{Elmegreen2010}. The ISM shows substructures on all spatial scales \citep{Scalo1985, Elmegreen1996, Williams2000, Molinari2010}: molecular clouds ($\geq1$\,pc), filaments ($\sim0.1-1$\,pc), clumps ($\sim0.3-1$\,pc), and cores ($\lesssim0.1$\,pc). The fragmentation of these structures leads to the observed distribution of young stars, which are typically arranged in groups or clumpy subclusters \citep{Kuhn2014, Zhou2024}. Only a fraction of these young  systems ultimately remain as stellar bounded clusters after they merge into smoother distributions \citep{Kruijssen2012,Grudic2018, Li2019, Krumholz2020,Krause2020, Zhou2025}. Star clusters form in dense clumps that are typically located at the intersection of multiple filaments \citep{Myers2009,Schneider2012, Motte2018, Zhou2022}.  These dense clusters contain one or more high-mass stars (M$\,\geq\,8$\,M$_{\odot}$), whose stellar winds, ionization, heating, radiation pressure, and ultimately supernova explosions play a fundamental role in shaping the evolution of the interstellar medium, from planetary systems to the entire galactic ecosystem \citep{Portegies2010}.

The formation of dense clusters is a complex theoretical problem. It is governed by the interplay of gravitational collapse, turbulence, and magnetic fields in an environment that is continuously altered by radiative and mechanical feedback from the embedded high-mass stars still in formation \citep{MacLow2004,Dale2011,Krumholz2014,Krumholz2020,Li2014,Girichidis2020,Padoan2020,Grudic2021,Guszejnov2021,VazquesSemadeni2019,VazquesSemadeni2024}. In addition to these processes, the final structure and properties of clusters are influenced by other factors. These include the initial cloud conditions and morphology \citep{Girichidis2011}, which may also be related to the local galactic environment \citep{Smith2020}; the possible coalescence of protostellar cores in the early evolutionary phases \citep{Dib2023}; and the hierarchical merging where distinct groups of pre-existing star-forming cores or young subclusters assemble \citep{Maschberger2010}.

Current models of high-mass stars and cluster formation are divided into categories that mostly differ by the dynamical state of the parent structure that fragments and where the material that ends up in the protostars comes from. One category is composed of static models that assume that high-mass star formation starts in substructures created by the supersonic turbulence present in molecular clouds \citep{McKee2003}, with possible contributions from the magnetic field. The material is initially gathered in these substructures and builds up the entire mass reservoir for the star formation process. In this category, there are different scenarios distinguished by the structure type that is initially assembled: cores for the {core-fed} models, for example the turbulent model of \citet{McKee2003}, or the entire clump for the in situ formation scenario introduced by \citet{Longmore2014}. In contrast, another category posits a more dynamical scenario where star formation is coeval to a further accumulation of gas in the high-density structures: cores and clumps. This additional contribution increases the reservoir of material available for accretion onto young protostars. \citet{Longmore2014} refer to this \text{category as conveyor belt mode} to recall that the mass of star-forming material is never assembled at a single time (see also the discussion in \citealt{Krumholz2020}), but is continuously transferred. The competitive accretion model introduced by \citet{Bonnell2001} falls into this category since it prescribes that cores gather their masses from the clump local environment, with a mass growth that is more efficient for objects close to the bottom of the gravitational well \citep{Bonnell2006}. While many models agree that flows funnel material from low to high densities to assemble massive stars, they differ on what drives the local accretion flow. For example, it is driven by gravity in the global hierarchical collapse (GHC) model  \citep{Ballesteros2018,VazquesSemadeni2019,VazquesSemadeni2024}), while by the large-scale random (turbulent) velocity field in the inertial flow scenario of \citet{Padoan2020}.\\

Over the years, several studies have been dedicated to characterizing the initial conditions of star formation by observing young stellar clusters \citep{Gutermuth2009, Kuhn2014,Parker2017, Parker2018, Dib2018a}. Although they are considerably young, these systems are already influenced by their dynamical evolution, which does not allow one to trace back where and under which initial conditions the stars originally formed. This leaves severe uncertainties about how the time-dependent process of fragmentation started and how it proceeds during the early evolution of the system, before most of the gas is dispersed, questions that can be answered only by  observing the distribution of star-forming cores.

Modern millimeter interferometers, such as the Atacama Large Millimeter/submillimeter Array (ALMA), the NOrthern Extended Millimeter Array (NOEMA), and the SubMillimeter Array (SMA) have achieved the angular resolution and sensitivity required to unveil the fragmentation properties in the dense clumps, which are the crowded and heavily extincted sites where high-mass stars and clusters form, typically located a few kiloparsec from the Sun \citep{Zhang2014,Zhang2015,Csengeri2017,Sanhueza2017,Beuther2018,Fontani2018,Svoboda2019,Sadaghiani2020,Moscadelli2021, Palau2021,Liu2020,Liu2022,Morii2021,Morii2023,Traficante2023,Avison2023,Ishihara2024,Liu2024, Zhou2024a}. In recent years, an important observational effort has been carried out to study these massive clumps, identifying and characterizing the embedded population of star-forming fragments in order to provide important observational constraints to the models of high-mass stars and cluster formation. Several surveys focus on clumps in specific evolutionary phases, either very young and quiescent systems embedded in infrared dark clouds \citep{Zhang2014, Sanhueza2017, Svoboda2019, Sanhueza2019, Anderson2021, Morii2023, Rigby2024} or systems in more advanced stages that are associated with infrared bright emission \citep{Beuther2018, Liu2020, Liu2024, Xu2024}. Other surveys selected their targets to cover a wide range of evolutionary phases \citep{Avison2023, Traficante2023, Ishihara2024}, but observed only a limited number of systems. The ALMA-IMF Large Program \citep{Motte2022} surveyed 15 massive Galactic protoclusters near the Sun (d$\,<\,5.5$\,kpc) for a detailed analysis of the core mass function in different environments and evolutionary stages. These observations have built up a rich, but rather heterogeneous dataset of observed properties of the fragments hosted in about two hundred massive clumps. 

Although observed core masses are an important reference point for star formation models, the spatial distribution of cores offers an additional and robust diagnostic with critical information on how clusters assemble and the relevance of various physical mechanisms in their formation process. Fragmentation leaves a different signature imprinted on the separations of cores in a turbulent or gravity-dominated medium, which is assumed in different categories of models, characterized by different values for the Jeans length \citep{MacLow2004, VazquesSemadeni2024}. Several observations have reported separations that are close to or smaller than the average thermal Jeans length of the hosting structure (clump or cloud; \citealt{Palau2015, Kainulainen2017, Beuther2018, Sanhueza2019, Morii2024,  Ishihara2024, Beuther2025}). However, turbulent fragmentation is not completely ruled out as an additional contribution from turbulent support is required to explain observations in other systems \citep{Rebolledo2020, Avison2023} and in nearby clouds \citep{Ishihara2025}. In general, distinct fragmentation patterns with strong core alignments, clustered arrangments with cores randomly distributed, or low fragmentation and rather symmetrical spatial distributions are observed, whether the dominant energetic contribution in the system is the magnetic field, turbulence, or gravity, respectively \citep{Tang2019}. Anisotropic accretion flows, whose presence may be traced by small-scale subfilaments, provide additional fragmentation, increasing the number of cores and their location in the system. In addition, dynamical effects, such as the merging of subclusters \citep{Bonnell2003}, clump global gravitational collapse \citep{VazquesSemadeni2017}, and tidal interactions between the cores \citep{Zhou2024a, Li2024G}, modify the overall structure and spatial distribution of young systems. Consequently, variations in these structural properties across different evolutionary stages can be used to determine the presence and impact of such dynamics. Observations suggest that the typical separation between cores is shorter in more evolved systems \citep{Beuther2021, Traficante2023, Ishihara2024}. Moreover, the core distribution is also found to be more compact in the later stages \citep{Xu2024} and in regions with higher star formation rates \citep{Dib2019}. These two pieces of evidence both support the presence of a strong gravitational pull, a signature that a global collapse, as predicted by the GHC theory \citep{VazquesSemadeni2024}, is active in massive clumps.

Additional information is provided by looking at the location of the massive members of the clustered system. Although young stellar clusters often show the presence of mass segregation (MS), with high-mass objects that are most likely to lie close to the cluster center \citep{Allison2009, Parker2017, Parker2018, Dib2019, Jia2025}, it is unclear whether those patterns are connected to their initial formation or are due to their subsequent dynamical evolution (see also \citealt{Portegies2010}). A similar statistical analysis carried out on the dense core population shows that MS may be present since the early stages \citep{Sanhueza2019, Sadaghiani2020, Nony2021, Morii2024, Xu2024}, which is indicative that competitive accretion may be a dominant path of their formation \citep{Bonnell2006}.

Although several indications were drawn from the available dataset, the diverse selection criteria of the targets, the different spatial resolutions and sensitivity levels achieved by the various surveys, and the low number of observed clumps limit the statistical analysis and direct comparison of the observed fragment population. The ALMA Cycle 7 large program ALMA evolutionary study of high-mass protocluster formation in the GALaxy (ALMAGAL) (2019.1.00195L, P.I.: S.Molinari, P.Schilke, C.Battersby, and P.Ho) was designed to mitigate these problems by observing an unprecedented large sample of 1013 clumps covering a wide range of physical conditions and Galactic locations \citep{Molinari2025}. The ALMAGAL program provides an extended and homogeneous dataset to study the fragmentation process in dense and massive clumps. In this paper we analyzed the spatial distribution of the cores identified in the ALMAGAL fields by \citet{Coletta2025} to provide a robust characterization of where the cores are located in massive and potentially cluster-forming clumps, what their typical separation is, how they are distributed, and how these properties vary across the early evolutionary phases. This work is a complement to the comparison between the physical properties of the ALMAGAL core population, specifically the level of fragmentation, core masses, and efficiency, and the large-scale clump parameters presented by \citet{Elia2025arXiv}. \\

The structure of this paper is as follows. We discuss the details of the ALMAGAL program in Sect.\,\ref{Sect:Data}, summarizing the observational strategy for the survey and the properties of the ALMAGAL clumps and the extracted core population used in the analysis. We present the results of the analysis of the spatial distribution of ALMAGAL cores in Sect.\ref{Sect:Results}. Specifically, in Sect.\,\ref{Sect:SpatialDistributionALMAGAL} we describe the general morphological properties of the groups of cores, defining where they are distributed. 
In Sect.\,\ref{sect:CoreSeparation} we determine the separations between the cores and compare them with the clump averaged properties in Sect.\,\ref{sec:SeparationvsClumpProperties}, focusing on how the separations vary during different evolutionary phases. We discuss how they relate to the expectation of the Jeans fragmentation theory in Sect.\ref{sect:SectJeans}. In Sect.\ref{Sect:DiagnosticsFragmentation} we present additional statistical descriptors to illustrate how cores are distributed and whether mass segregation is present in systems, checking if they show a variation across the different evolutionary phases. Finally, we discuss our results within the framework of the generally adopted models in Sect.\ref{sect:Discussion}. \\

\section{Data}
\label{Sect:Data}

\subsection{The ALMAGAL clump sample} 
\label{Sect:ClumpProperties}

\begin{figure*}
    \centering
    \includegraphics[width=0.95\linewidth]{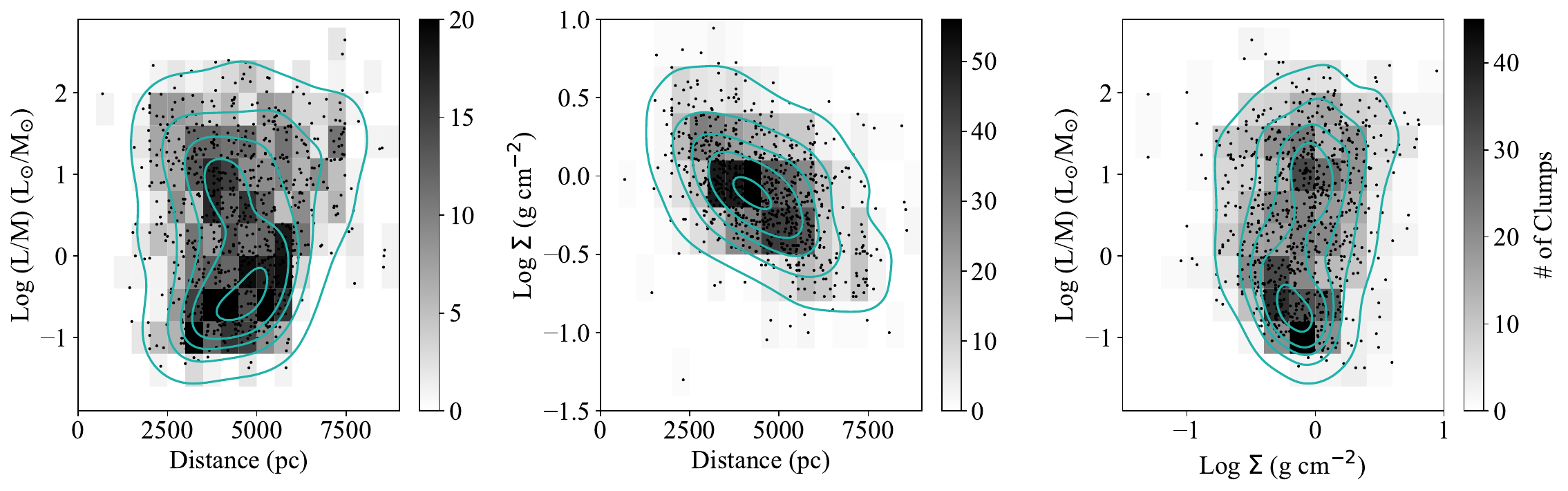}
    \caption{Distribution of the $L/M$ ratio and average surface density, $\Sigma_{cl}$, as function of the clump heliocentric distance for the ALMAGAL sample, and comparison of these two properties. Each panel includes the 2D density distribution computed dividing the intervals of $L/M$, $\Sigma_{cl}$, and  $d_{cl}$ into bins of width 0.3 dex, 0.2 dex, and 500\,pc, respectively. The solid light gray lines indicate the contour levels of the 2D density distribution derived with kernel density estimation (KDE) corresponding to 10, 30, 50, 70, 95 percent of the sample.
    }
    \label{Fig:ClumpProperties}
\end{figure*}

The ALMAGAL large program (2019.1.00195.L, P.I.: Molinari, Schilke, Battersby, Ho) observed with ALMA in Band 6 ($\sim220$\,GHz, $\sim1.4$\,mm) 1013 dense massive clumps located close to the Galactic Plane, and selected from the Hi-GAL \citep{Molinari2010,Molinari2016, Elia2017, Elia2021} and RMS surveys \citep{Lumsden2013}, to investigate their fragmentation properties. The criteria adopted for the target selection are described in \citet{Molinari2025} and are based on thresholds on the {\it Herschel} continuum emission, the clump mass, $M$, and the average surface density, $\Sigma_{cl}$, determined from the {\it Herschel} Hi-GAL data \citep{Molinari2016,Elia2017,Elia2021}. An initial analysis of the ALMA ACA data provided new estimates for the clump heliocentric distances, derived from the radial velocities measured from the spectral lines of multiple molecular tracers (Benedettini et al. in prep.). The ALMAGAL clumps, with physical properties updated thanks to these new measurements, have masses $M$ between $\sim100$ and 12000\,M$_{\odot}$, bolometric luminosity, $L$, between $\sim50$ and $\sim5\times10^{5}$\,L$_{\odot}$, and are located in the Galaxy at heliocentric distances, $d_{cl}$, from $\sim1$ up to 14\,kpc, with only one clump lying closer than 1\,kpc \citep{Molinari2025,Elia2025arXiv}. Notwithstanding these updates, the observed ALMAGAL clumps effectively sample the entire evolutionary path of massive Galactic clumps that are likely to host the formation of high-mass stars and clusters.
In fact, they have clump-averaged surface densities, $\Sigma_{cl}$, ranging from $\sim0.1$ up to $\sim10$\,g\,cm$^{-2}$, which covers the thresholds suggested by theoretical modeling and observational constraints for the formation of high-mass stars \citep[][see also the discussion in \citealt{Beuther2025}]{Krumholz2008,Kauffmann2010,Baldeschi2017}. Moreover, we note that the ALMAGAL sample includes clumps with a ratio of their bolometric luminosity to their mass, $L/M$, which is uniformly distributed between $\sim0.5$ and $\sim50$\,L$_{\odot}$/M$_{\odot}$, and includes cases with a $L/M$ down to $\sim0.05$ and up to $\sim400$\,L$_{\odot}$/M$_{\odot}$, although highly evolved phases are less sampled, as there are fewer systems with $L/M\,>100$\,L$_{\odot}$/M$_{\odot}$ compared to the other evolutionary stages. Since the $L/M$ ratio is a well-established indicator of the evolutionary stage of clumps \citep{Molinari2016b}, the ALMAGAL clumps provide a complete and rather homogeneous sampling of the evolution of massive clumps with active star formation. 
We present in Fig.\,\ref{Fig:ClumpProperties} a thorough comparison between the $L/M$ ratio, the surface density, $\Sigma_{cl}$, and the heliocentric distance, $d_{cl}$, of the ALMAGAL targets. There are 25 clumps located at distances greater than 9\,kpc after updating their distances that are not shown in the left and central panels for clarity reasons, but the distribution of their properties is similar to the ones presented. 
Figure\,\ref{Fig:ClumpProperties} shows that there is no relationship between the ratio $L/M$ and either $\Sigma_{cl}$ or $d_{cl}$, indicating that target selection did not introduce any strong bias due to an eventual correlation between these two quantities. In contrast, the ALMAGAL sample presents a relation between $\Sigma_{cl}$ and $d_{cl}$, introduced as a consequence of the selection criteria on the clump mass.

\subsection{ALMAGAL observation strategy and continuum images}

The ALMAGAL program adopted an observational strategy of single pointings centered on the emission peaks identified in the {\it Herschel} Hi-GAL 250\,$\mu$m images \citep{Molinari2016,Elia2017}. Each ALMAGAL clump has been observed with a single pointing observation with ALMA, both with the ACA 7m and the main 12m antenna arrays. Observations have been carried out using two different configurations for the main array to image all ALMAGAL clumps with a similar physical resolution of $\sim1000$\,au \citep{SanchezMonge2025}, although they are located over a wide range of heliocentric distances \citep{Molinari2025}. The C2+C5 configuration has been adopted to observe clumps with $d_{cl}\,\lesssim\,4.7$\,kpc, achieving a resolution of $\sim0.3$\arcsec, while the C3+C6 configuration for clumps with $d_{cl}\,\gtrsim\,4.7$\,kpc for a resolution of $\sim0.15$\arcsec. These configurations cover the long and short baselines observed by the main array which, combined with the ACA compact array, allowed the recovery of the flux over all spatial scales up to $\sim29$\arcsec, corresponding to the maximum recoverable scale of the ACA array in band 6. In the following discussion, we refer to the group of clumps observed with the two configurations as near and far groups for the targets observed with C2+C5 and C3+C6 configurations, respectively. Although the two groups were well separated in $d_{cl}$ when the initial target selection was made, this distinction has been lost  with the update of the kinematic distances (see Benedettini et al. in prep.). However, this revision has a marginal impact on the statistical results, as most of the targets in each group are still located in a limited interval of $d_{cl}$ equal to $2\,\leq\,d_{cl}\,\leq\,4.7$\,kpc and $4.7\,\leq\,d_{cl}\,\leq\,7.5$\,kpc, for near and far, respectively (see also Sect.\,\ref{sect:distributionSeparations}).

The detailed description of data reduction, calibration, and imaging is presented in \citet{SanchezMonge2025}. The continuum images obtained from the combination of the data from all the arrays have a circular field of view (FoV) with a diameter of $\sim36$\arcsec, limited by the primary beam size of the 12m antenna and the cut to an attenuation of 0.3 applied during the image processing \citep{SanchezMonge2025}. In each image, the noise level is rather homogeneous in the central region of the FoV with an average value of $\sim0.05$\,mJy/beam \citep{SanchezMonge2025}, but naturally increases towards the edges due to the antenna primary beam attenuation that causes a decrease in sensitivity by a factor of two at angular distances greater than $\sim13.8$\arcsec\,from the phase center.

\subsection{ALMAGAL core catalog}

\citet{Coletta2025} extracted compact sources in ALMAGAL continuum images using a modified version of the CuTEx photometric package \citep{Molinari2011}. The entire catalog of identified compact sources is composed of 6348 objects detected at 5$\sigma$ in 844 clumps. The nature of these sources can be assessed by their photometric and physical properties. All of these sources have a centrally peaked intensity profile, reflecting the high detection efficiency of the CuTEx algorithm for this morphology and its low sensitivity to extended or flat-profile emission. These sources span sizes between $\sim500$ and $\sim5000$\,au and have mean densities greater than 10$^{6}$\,part\,cm$^{-3}$ \citep{Coletta2025}. However, most of them have sizes between 1000 and 2500\,au  (see Fig. 13 in \citealt{Coletta2025}) and quite high densities between 1.4 to 8$\cdot10^{7}$\,part\,cm$^{-3}$.
These properties are characteristic of centrally condensed star-forming cores whose innermost regions have undergone collapse. 
Hence, it is probable that a large fraction of the ALMAGAL sources already host at least one protostellar object. Following this interpretation, we refer to all sources in the ALMAGAL catalog as cores throughout this paper, with the understanding that they likely represent the inner, high-density regions of larger, more extended structures.

We adopted the positions provided by \citet{Coletta2025} to analyze how the cores are distributed in the fields and to measure their relative separations, with the aim of characterizing the fragmentation process in high-mass clumps. Of the 844 systems, 130 clumps host only one core and are not suitable for our analysis. We recognize that the reliability of our analysis and the diagnostics adopted for the spatial distributions are highly dependent on the number of cores. \text{We consider as our full sample} all the systems with $N_{core}\,\geq\,4$, excluding the 200 clumps with two or three hosted cores, which are 110 and 90 clumps, respectively \citep{Coletta2025}. The inclusion of these systems has a negligible impact on the results discussed in the paper, and the properties of their separations are consistent with the systems with $4\,\leq\,N_{cores}\,\lesssim\,6$. 
We adopted the full sample to discuss the distribution of core separation and the comparison with the Jeans fragmentation theory, as systems with a low number of cores may correspond to phases when fragmentation is at the beginning.
In contrast, to study the global structural properties of the spatial distribution of the cores, i.e., the type of distribution, mass segregation, we restricted our analysis only to systems with a high degree of fragmentation ($N_{core}\,\geq\,8$ or higher). 

In summary, the analyzed dataset comprises 514 clumps ($N_{core}\,\geq\,4$), containing a total of 5728 cores, greatly increasing the statistics available in the literature. However, we identify subsamples with different cuts on the minimum fragmentation level for the various aspects of the analysis of the core spatial distribution. Table \ref{Tab:ALMAGALcores} summarizes the sample sizes of these subsamples.

\begin{table}
    \centering
     \caption{Size of statistical subsamples (number of clumps, $N_{clump}$, and cores, $N_{cores}$) studied in this work, selected with different cuts on the degree of fragmentation.   
     }
     \begin{tabular}{cccccc}
     \hline  
         &  Total & N$_{clump}$ & N$_{clump}$  &  N$_{cores}$ & N$_{cores}$ \\
         &  & near & far  & near & far \\
      \hline  
      N$_{cores}\geq$4 & 514 & 330 &  184 &  4012&  1716 \\
      N$_{cores}\geq$8 & 296   & 218 &  78&  3429  & 1158 \\
      N$_{cores}\geq$10 & 226  & 172 & 54 &  3033 & 956 \\
      \hline  
    \end{tabular}
    \label{Tab:ALMAGALcores}
    \tablefoot{
    Statistics from the ALMAGAL catalog by \citet{Coletta2025}. The subsamples are divided into near and far group, referring to systems with different heliocentric distances, which are observed with different array configurations (see text and \citealt{Molinari2025}).
    }
\end{table}

\section{Results}
\label{Sect:Results}

\subsection{The spatial distribution of ALMAGAL cores}
\label{Sect:SpatialDistributionALMAGAL}

\begin{figure*}
    \centering
    \includegraphics[width=0.45\linewidth]{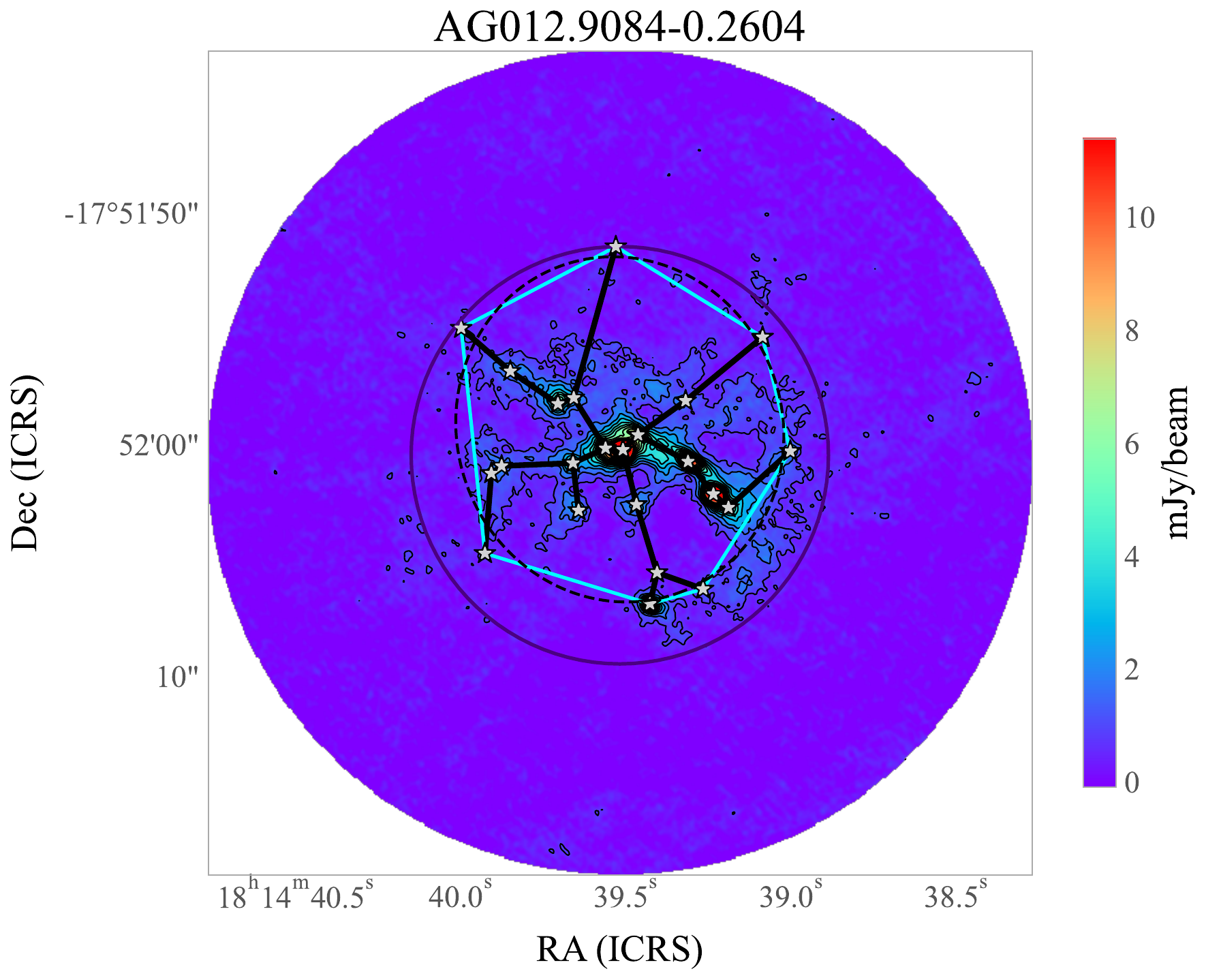}
    \includegraphics[width=0.45\linewidth]{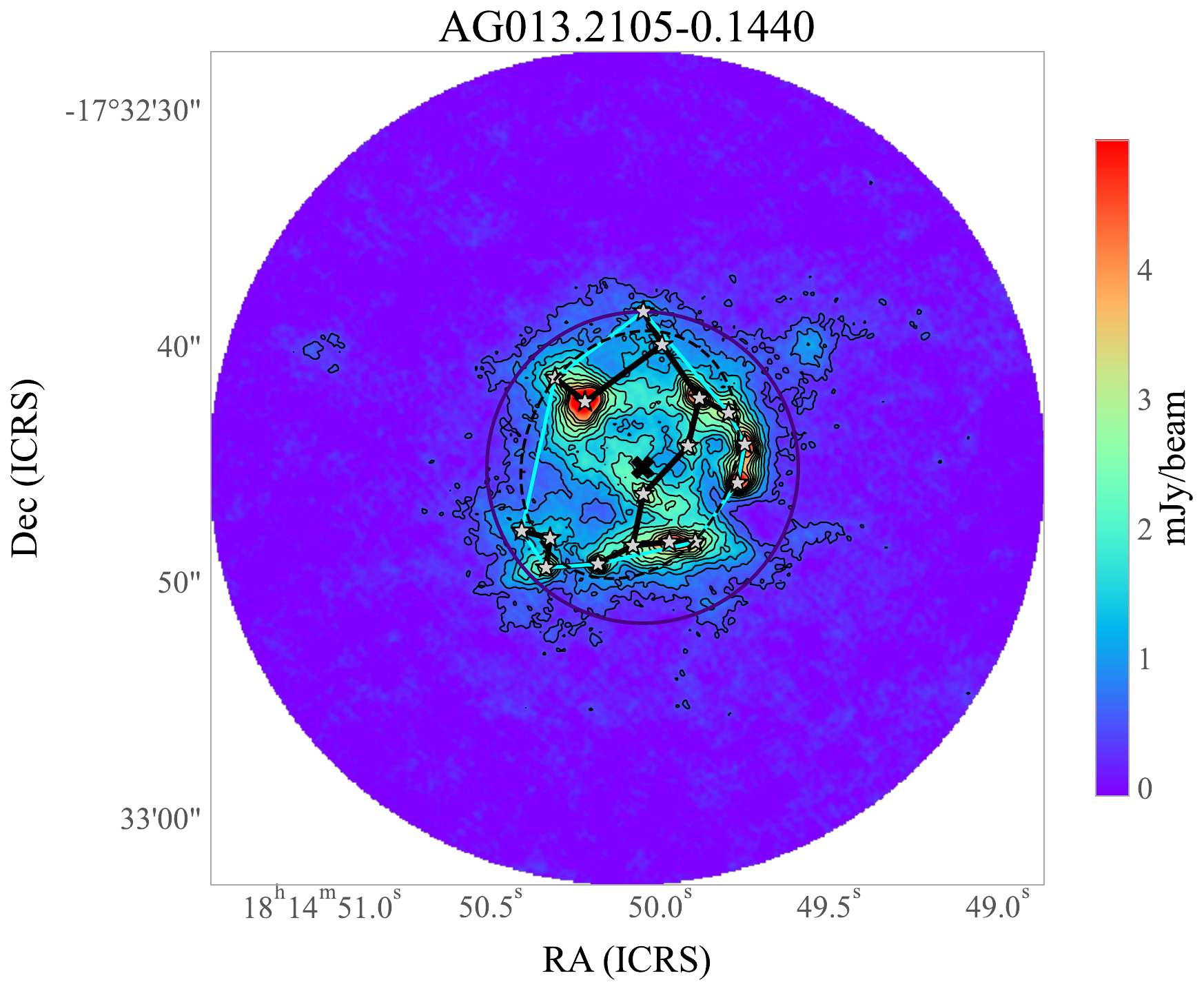}
    \includegraphics[width=0.45\linewidth]{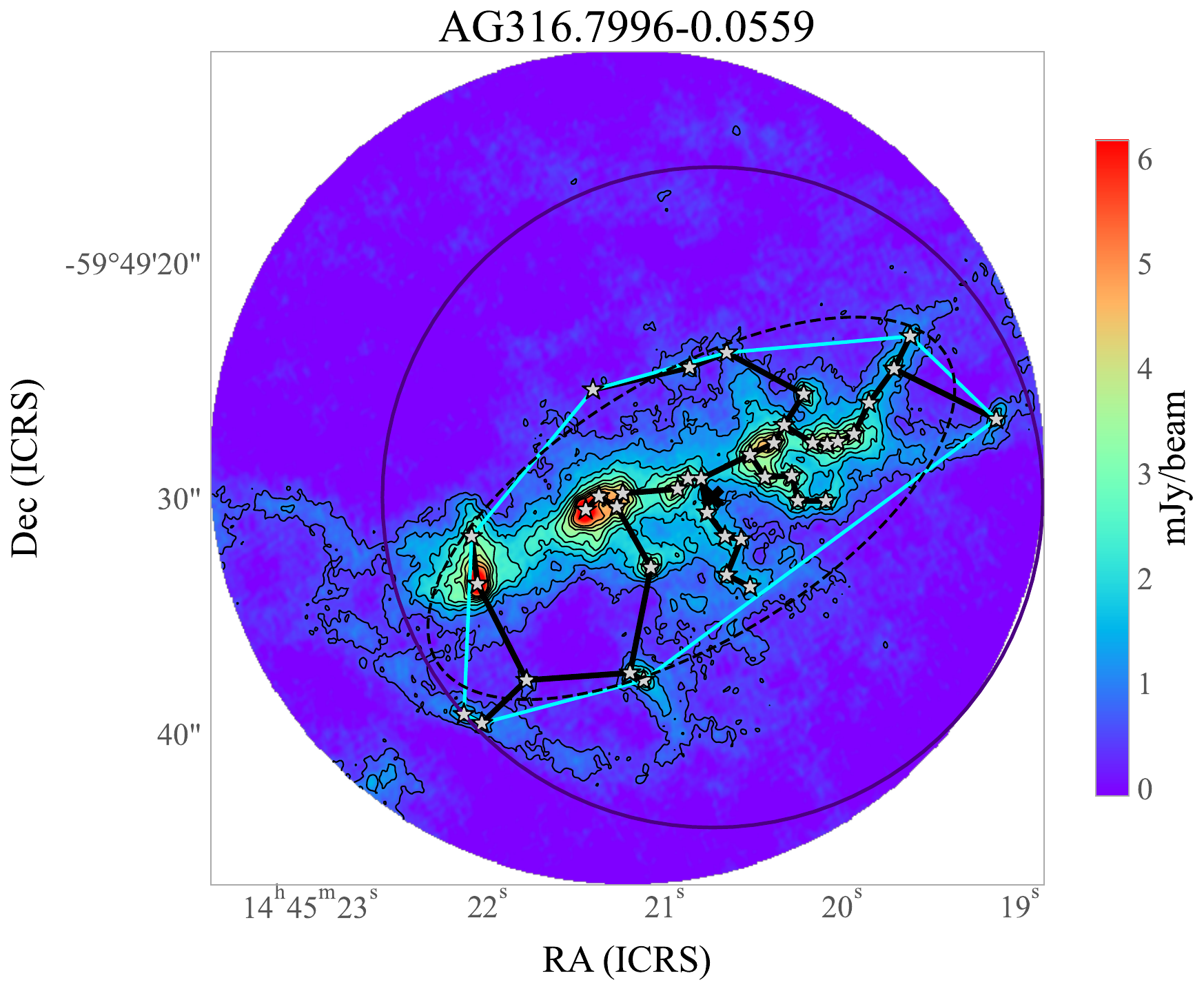}
    \includegraphics[width=0.45\linewidth]{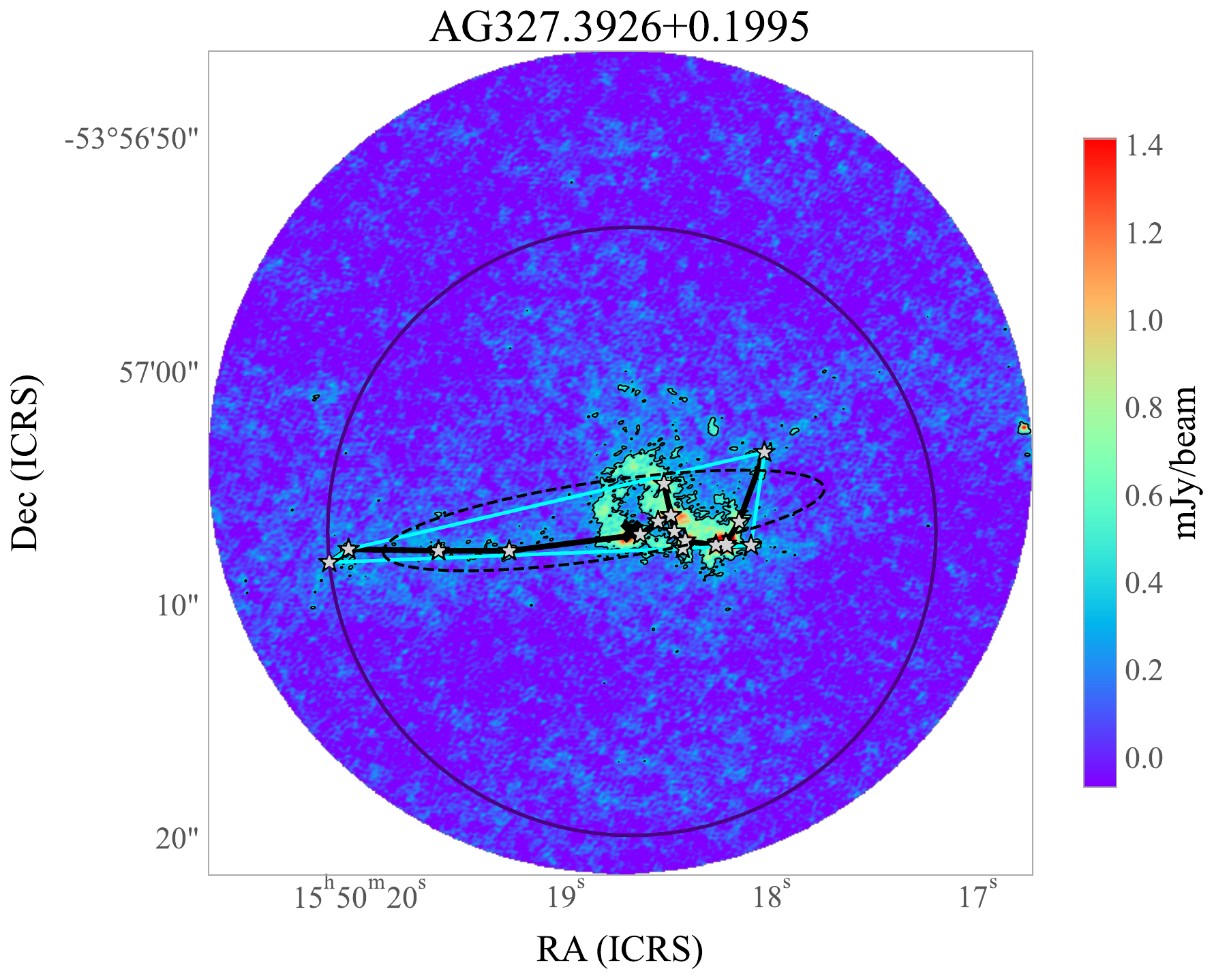}
    \caption{Examples of the distribution of cores found in four ALMAGAL clumps overlapping on the dust continuum map at 1.4\,mm from which they were extracted. These fields show examples of  clustered systems with circular and elliptical patterns and cases of aligned cores. The contour levels correspond to $4n\times\sigma$ level and $\sigma$ equal to the noise level measured on the image. The positions of the cores from the ALMAGAL catalog extracted by \citet{Coletta2025} are shown with gray star markers. In each field we indicate the cluster geometrical center and its radius (see the definition in the text) with a black cross and a dark blue circle, respectively. We also show the convex hull polygon (cyan segments) and the corresponding best-fitting ellipse (black dashed line) adopted for the morphological characterization of the system. The solid thick black segments are the MST edges determined with Prim's algorithm \citep{Prim1957}, connecting all the cores in the field.}
    \label{Fig:ExampleMST_1}
\end{figure*}

\begin{figure*}
    \centering
    \includegraphics[width=0.45\linewidth]{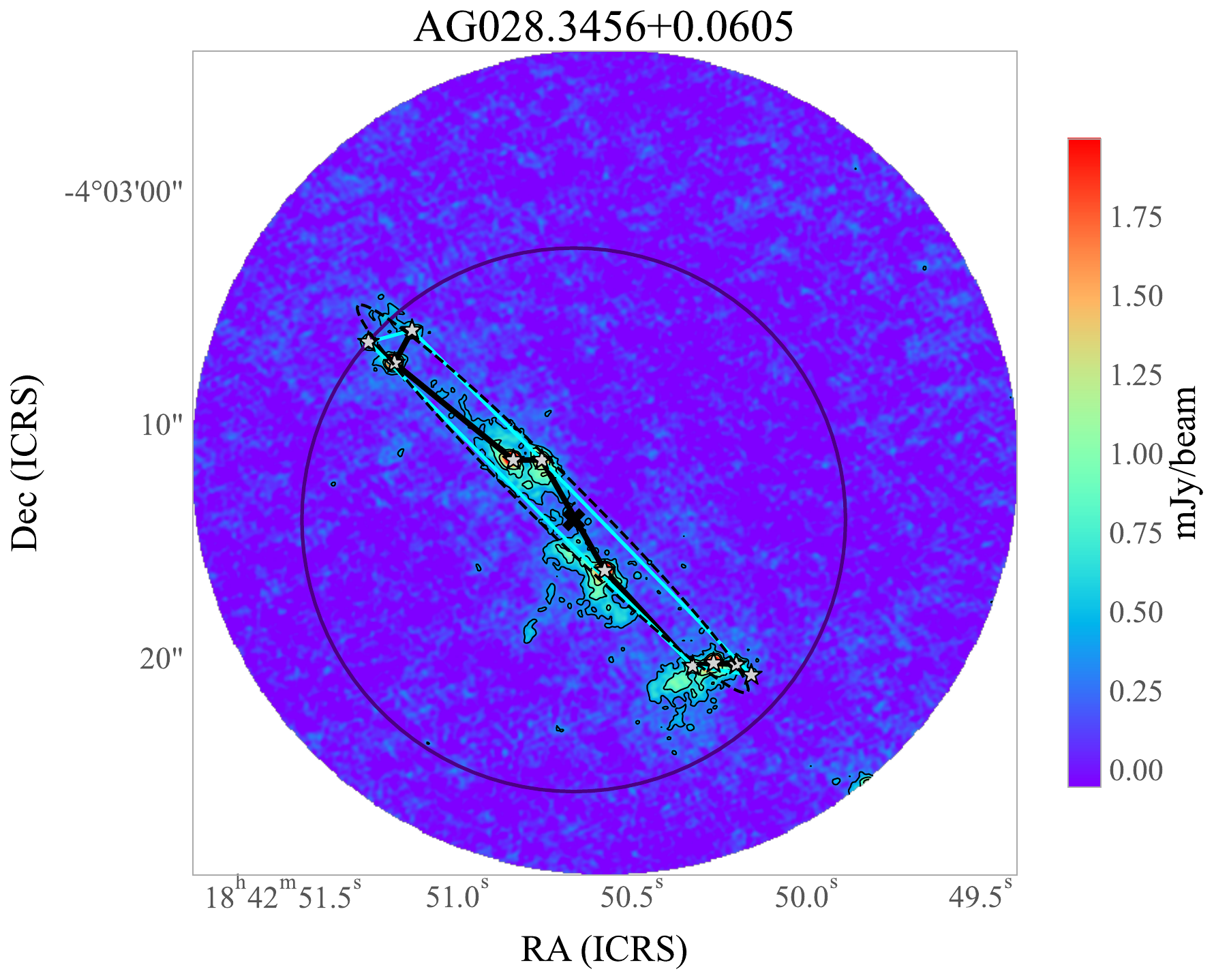}
    \includegraphics[width=0.45\linewidth]{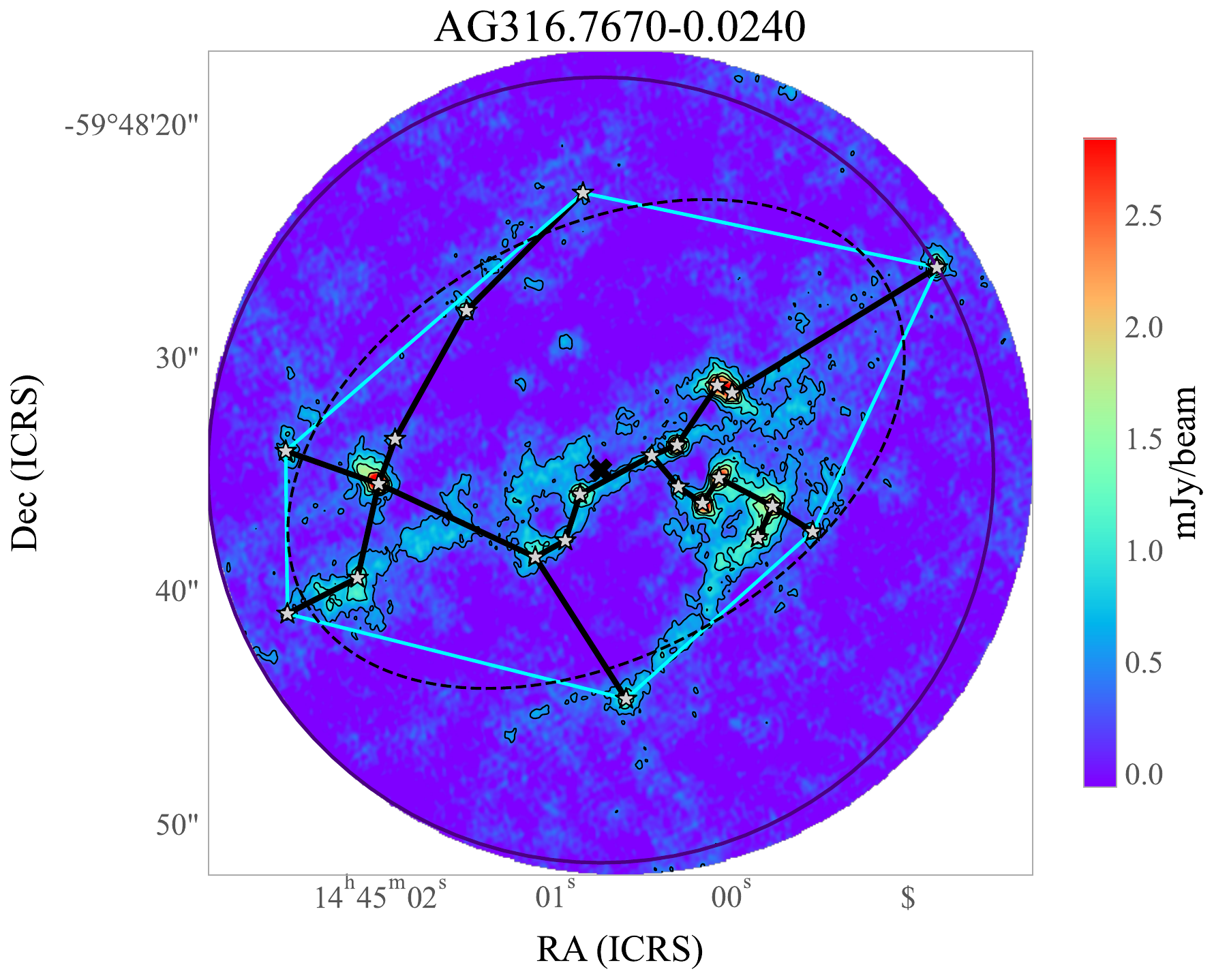}
    \includegraphics[width=0.45\linewidth]{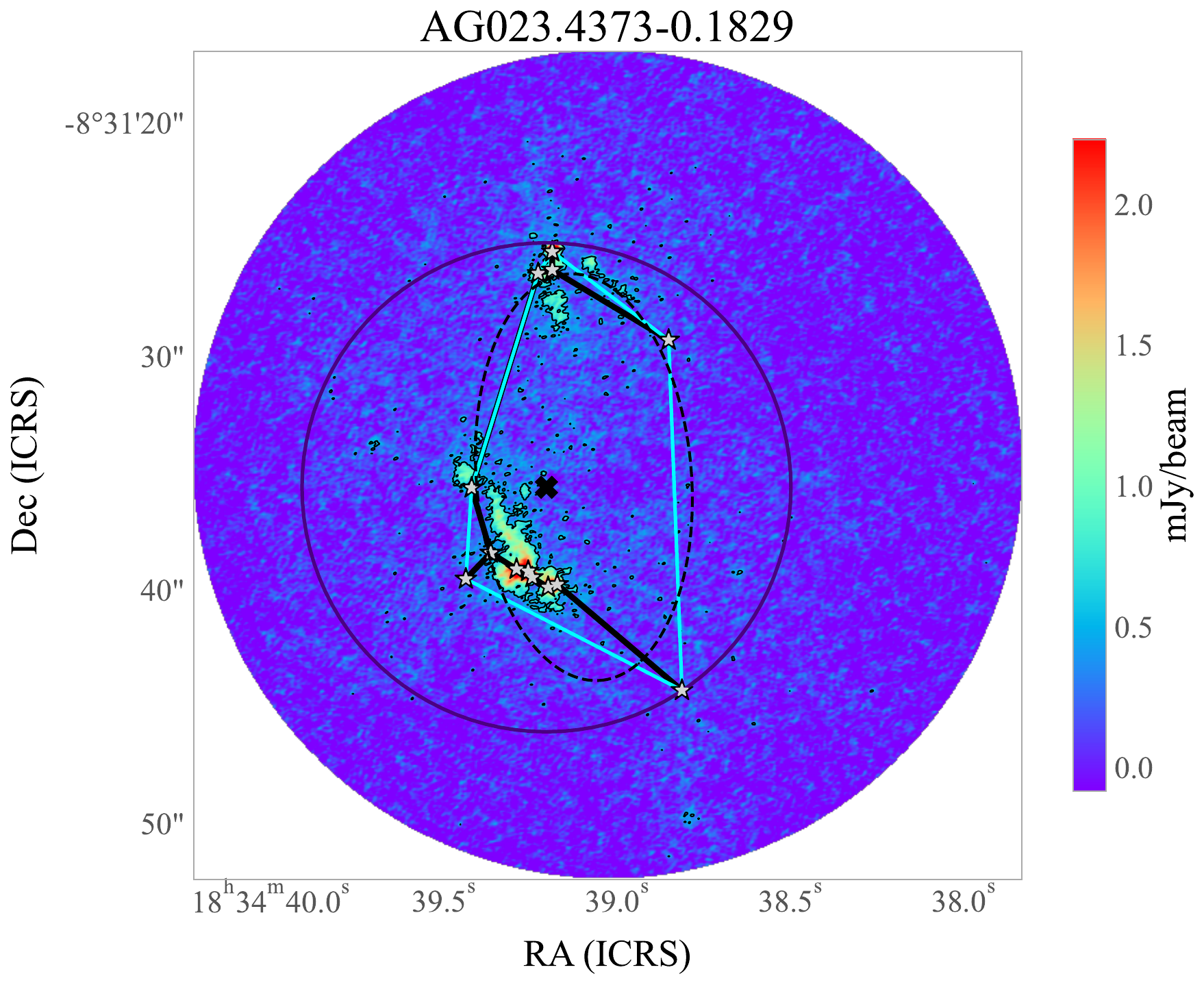}
    \includegraphics[width=0.45\linewidth]{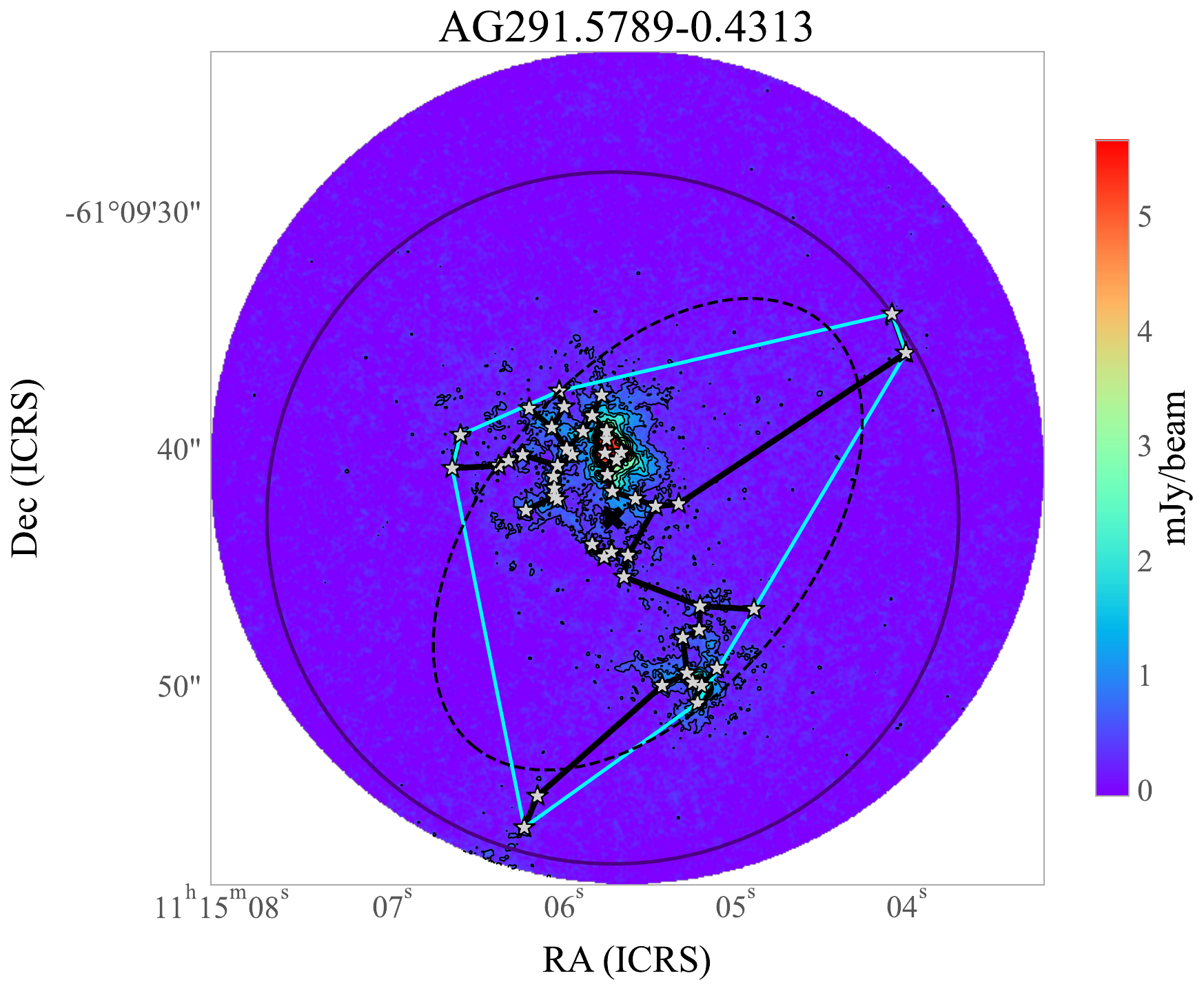}
    
    \caption{ As in Fig.\,\ref{Fig:ExampleMST_1}. These fields show examples of cores distributed over filamentary features and in well-separated local substructures, both patterns that are sometimes found in ALMAGAL continuum images.}
    \label{Fig:ExampleMST_2}
\end{figure*}

The FoV of the ALMAGAL continuum images allows us to probe the fragmentation of massive clumps over circular regions with linear diameters of $\sim0.3 - 1.4$\,pc in clumps located between 2 and 8\,kpc, which are the bulk of the ALMAGAL sample. Visual inspection reveals that the identified cores are arranged in highly diverse configurations. The observed patterns depend on three main factors: the degree of fragmentation (number of  cores), the presence (or absence) of extended continuum emission, and the morphology of that emission. The variety in morphologies disclosed by the ALMAGAL data is partially illustrated by the examples presented in Fig.\,\ref{Fig:ExampleMST_1} and Fig.\,\ref{Fig:ExampleMST_2}.

We observe a difference between systems where multiple cores are detected and those that are poorly fragmented. In the first case, cores are often grouped together and located in a restricted area of the observed field, typically found on a major patch of extended continuum emission with an irregular morphological appearance. Although there are cases where the emission has a circular shape or modest elongation, it is not uncommon to find cases in which it appears quite elongated \citep[see discussion in][]{Molinari2025}. The distribution of cores typically follows the most prominent continuum emission features, but they are also found outside those boundaries. Cores are also found lying in substructures with a highly filamentary appearance, typical of molecular clouds and present at all spatial scales, as highlighted by {\it Herschel} observations \citep{Andre2014, Schisano2014,Schisano2020,Konyves2015, Dib2020,Schneider2022,Hacar2023}. Occasionally, these subparsec filaments are found converging toward the main emission patch as in hub-filament systems \citep{Myers2009, Hacar2018, Motte2018} (see the bottom left panel of Fig.\,\ref{Fig:ExampleMST_1}). In other fields, the extended continuum emission splits into several well-separated regular patches, each of which hosts a group of associated cores. This is indicative that massive clumps can have an additional higher hierarchical level of substructuring (see the lower panels of Fig.\,\ref{Fig:ExampleMST_2}), whose presence introduces a subclustering of the core population.

In contrast, systems characterized by a low level of fragmentation do not show a well-defined pattern in the core distribution. These cores are typically sparse in the observed field and do not present a strong signature of grouping or clustering. However, we identified cases where the cores are aligned along specific directions, which extend over several arcseconds. These aligned patterns sometimes match one or a few extended filaments crossing a fraction of the FoV. Even in the cases lacking the detection of extended emission, it may be possible that the hosting filament is undetected for sensitivity reasons and that a signature of its presence is only revealed by the alignment of the cores. 

In general, ALMAGAL observations show that the cores are associated with almost every extended region of continuum emission. The high angular resolution reveals that these regions have a wide variety of appearances, as discussed in \citet{Molinari2025}. As a consequence of the association, the spatial distribution of star-forming cores shows a similar diversity, likely due to the interplay of different physical factors, such as gravitational force, turbulence, magnetic fields, and feedback mechanisms \citep{MacLow2004,Krumholz2006,Li2014,Shima2017, Tang2019, Sanhueza2025}. All of these effects contribute to shape the fragmentation in clumps, affecting both the number and the distribution of cores during their evolution. In this paper, we refer to any group of cores identified in the ALMAGAL clumps with $N_{cores}\,\geq\,4$ as a ``cluster'', even when they are limited in number, smaller than the members of typical embedded star clusters.

 In the following sections we introduce quantitative metrics and diagnostic tools to characterize the variety of observed patterns described above. Our aim was to answer the questions of where the cores are formed, how they are distributed, and what clues of the fragmentation process are left imprinted in their spatial distribution. To this end, we analyzed the size and morphology of the observed core distributions, and then focused on the mutual separation between cores. This is an important observable of the fragmentation process, which we related to the clump-averaged properties and compared to the Jeans fragmentation theory. Finally, we evaluated additional diagnostics capable of defining how cores are distributed and how such a distribution changes as a function of the evolutionary phase of the entire system.
In addition, we note that a first rough classification of the observed variety of ALMAGAL clusters can be drawn using the convex hull polygon\footnote{The convex hull polygon is the smallest polygon that encloses all the cores and has all interior angles smaller than $\pi$.}(CH) applied to the cores. A simple classification derives from the comparison of the number of vertices of the CH, $N_{v}^{hull}$, with the number of cores located well inside the polygon boundaries, $N_{in}^{hull}$ (see Figs.\,\ref{Fig:ExampleMST_1} and\,\ref{Fig:ExampleMST_2} for reference). About $\sim37$\% of the ALMAGAL clumps studied in this work (190 out of 514) have more cores within the CH boundaries than at the edges ($N_{in}^{hull}\,>\,N_{v}^{hull}$), a pattern that defines a well-localized group of objects with a surrounding sparse population, which is similar to a star cluster.
On the contrary, there are 64 clumps ($\sim 12\%$) where all detected cores form the CH polygon ($N_{in}^{hull} = 0$). These systems have a small number of cores, from four to six sparse cores, and typically lack substantial associated extended continuum emission that would be expected if a large amount of high-density material is present. However, most ALMAGAL systems ($\sim 51\%$) are between these two extreme patterns, since they show only a few cores inside the CH polygon. These systems can be young precursors of a clustered system where fragmentation is still at the beginning, or they can be a population of sparse objects derived from the local fragmentation of the hosting molecular cloud. Nonetheless, it is worth to evaluate the core distribution even in this latter case, as clusters are also recognized to form through mergers of group of cores initially distributed throughout the entire cloud \citep{Chen2021,Dobbs2022}.

\begin{figure}
    \centering
    \includegraphics[width=1\linewidth]{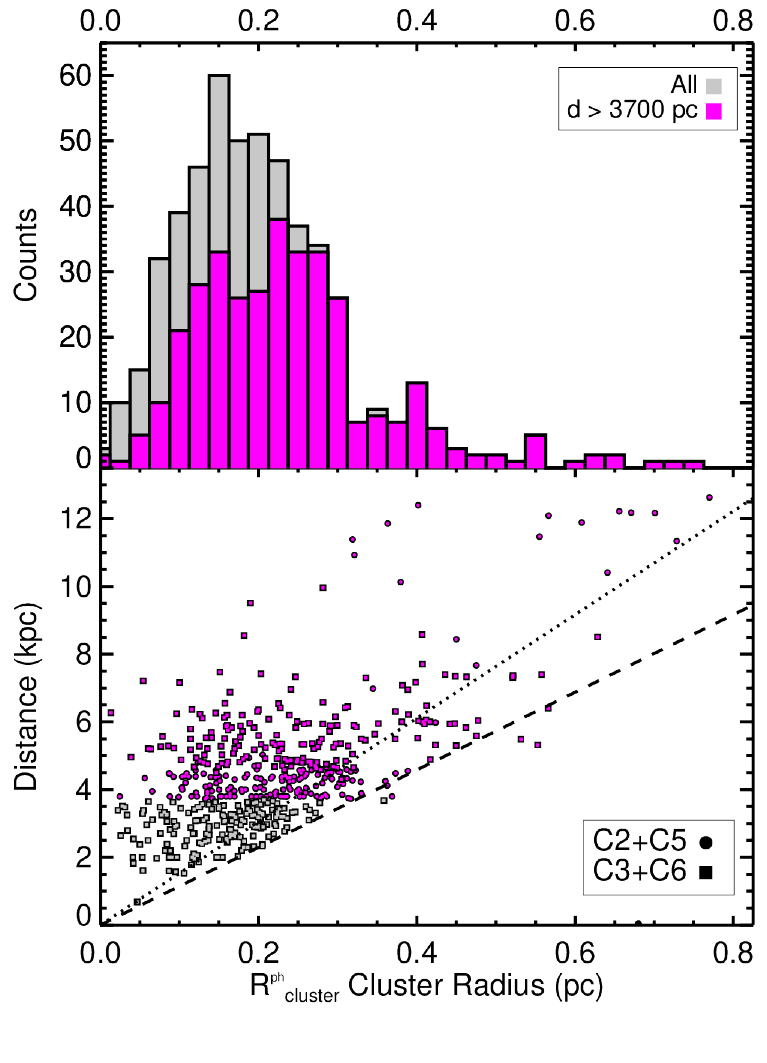}
    \caption{ {\it Top panel}: Distribution of the cluster radius measured from the average core positions for the sample of 514 ALMAGAL clumps with $N_{core}\geq4$ (in gray) and the subsample composed 347 clumps with $d_{cl}\,\geq\,3.7$\,kpc (magenta). {\it Bottom panel}: Clump heliocentric distance vs. measured cluster radius. The dashed line shows the physical size of the ALMAGAL FoV, while the dotted line indicates the area where the sensitivity is constant within a factor of two, i.e., angular sizes $\sim13.5$\arcsec. Cluster radii exceeding the FoV sizes are possible in systems with a substantial offset between the cluster and the image center.
    }
    \label{fig:ClustRadiusDistance_alt}
\end{figure}

\begin{figure}
    \centering
    \includegraphics[width=1\linewidth]{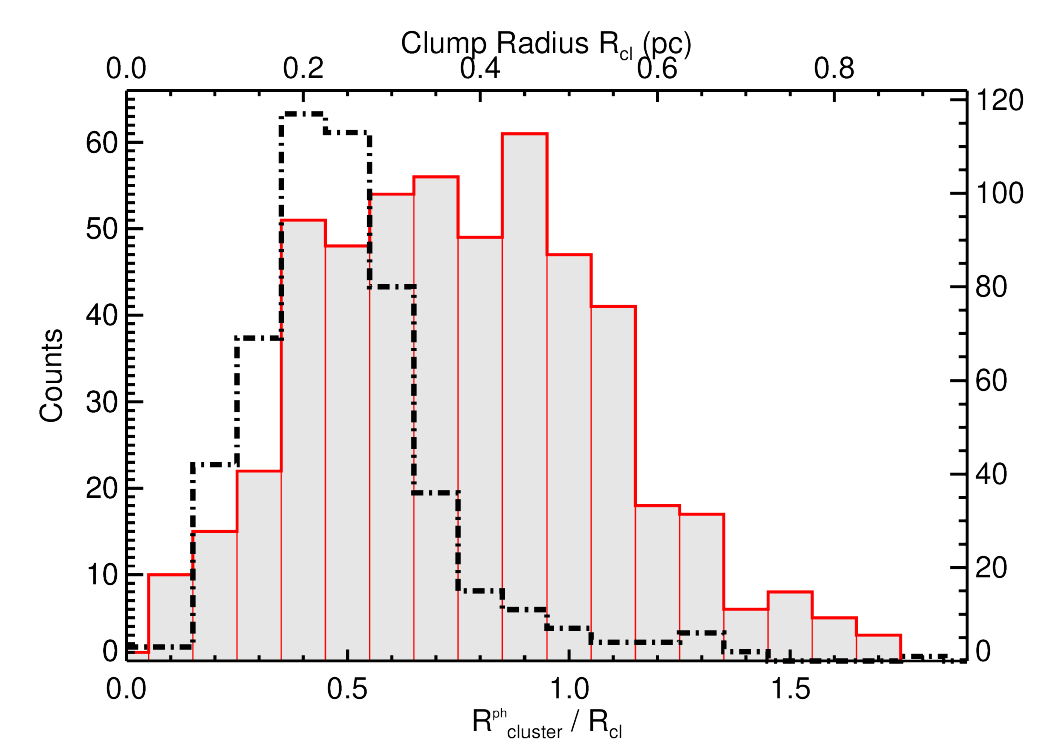}
    \caption{Distribution of the ratio of the radius of the core cluster to the clump \text{radius} (gray histogram). The dot-dashed line, which refers to the top and right axes, shows the distribution of the clump radius derived from {\it Herschel} Hi-GAL data at 250 $\mu$m from \citet{Elia2021}.
    }
    \label{fig:ClumpClusterSize}
\end{figure}

\subsubsection{Center and size of the clusters of ALMAGAL cores}
\label{sect:CenterField}

\begin{figure*}
    \centering
    \includegraphics[width=0.98\linewidth]{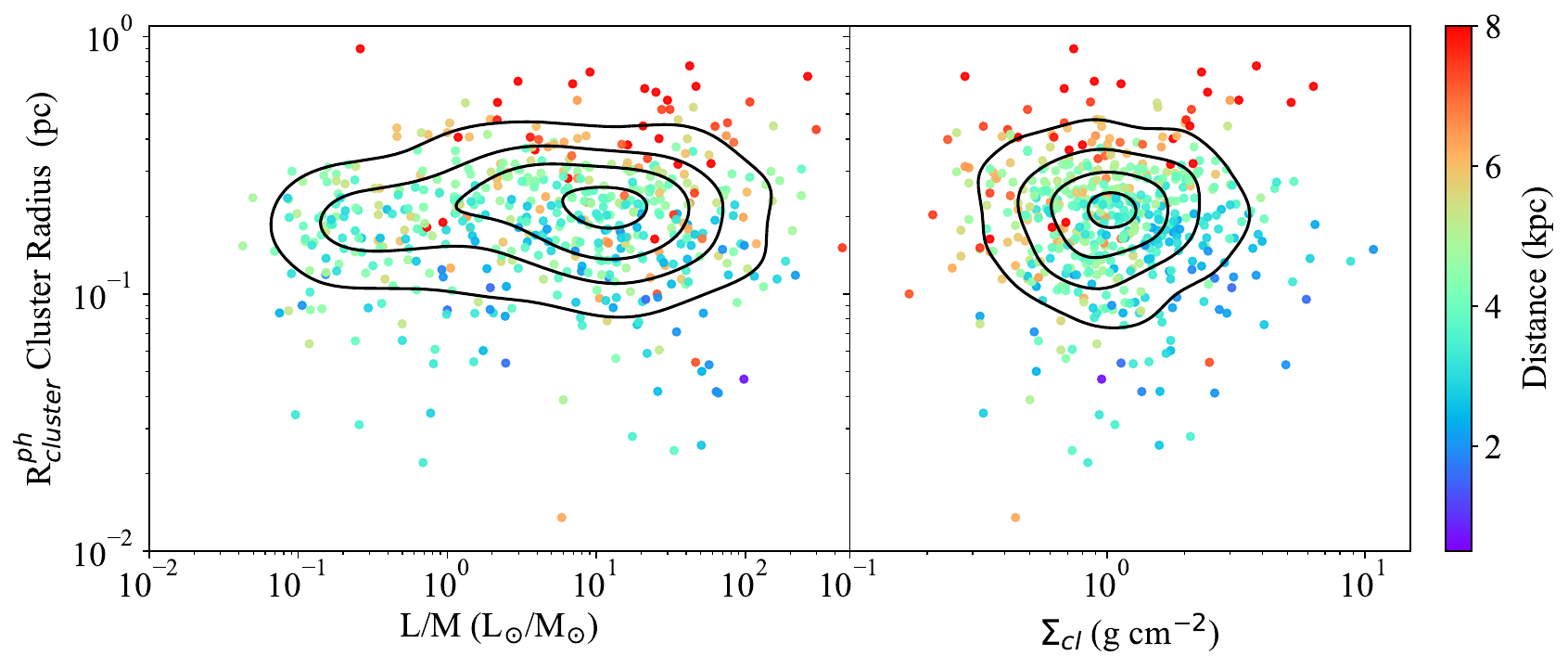}
    \caption{Cluster radius as a function of the clump-averaged properties: $L/M$ (\textit{left panel}) and surface density $\Sigma_{cl}$ (\textit{right panel}), color-coded by the clump distance. The black lines indicate the contours of the 2D probability density distribution corresponding to 10, 25, 50, 75, and 95\% levels.
    }
\label{Fig:DistributionClusterRadius_PhysicalProperties}
\end{figure*}

In general, the ALMAGAL cores are located in a portion of the observed FoV, although there are several cases where they spread over the entire, with cores that are also found near the edges. The centers of the core clusters closely match the center of the observed ALMA fields. We adopted two methods to determine the cluster center: {\it i}) the geometrical and {\it ii}) the mass-weighted average of the core positions. These methods produce similar statistical results, although these positions may differ for individual systems. The ALMAGAL cluster centers are mostly located within 3\arcsec\,from the center of the ALMA observed fields, which corresponds to the photometric center of the Hi-GAL clumps as traced in the {\it Herschel} 250 $\mu$m maps \citep{Molinari2016,Elia2021}. The offset between these positions is significant only in a small number of systems: as a reference, it exceeds 6\arcsec, which is equal to $\sim1/3$ of the FoV, only in 35 fields ($\sim7$\% of the investigated sample).

We estimated the radius of the cluster, $R_{cluster}$, as the distance of the farthest core from the cluster center \citep{Cartwright2004}. These radii are smaller than 13\arcsec\,in 384 systems, i.e., 75\% of the investigated sample. This region, corresponding to $\sim0.72\times$ FoV, is roughly coincident with the fraction of the images where the noise level is more homogeneous \citep{SanchezMonge2025}. However, there are several systems where cores are also detected close to the FoV's edge; most of those clumps are systems with $d_{cl}\,<\,4$\,kpc. It is likely that for these systems, the single pointing mode of the ALMAGAL observations is sampling only the inner region of the clump. This is a critical selection effect that must be taken into account.

The top panel of Fig.\,\ref{fig:ClustRadiusDistance_alt} shows the distribution of the cluster radius measured by converting $R_{cluster}$ into a linear size, $R^{ph}_{cluster}$, using the updated heliocentric distances \citep[ Benedettini et al. in prep.]{Molinari2025}. The distribution for the full sample shows two key features: a peak close to the median value of $\sim0.2$\,pc, and a notable drop at $R^{ph}_{cluster}\approx0.32$\,pc. To determine the impact of the FoV bias on the true physical distribution, we selected the 347 clumps with $d_{cl}\geq3.7$\,kpc. As shown in the bottom panel of Fig.\,\ref{fig:ClustRadiusDistance_alt}, for those systems, the observed area is sufficiently large to identify clusters with radius $R^{ph}_{cluster}\geq$\,0.32\,pc. In this sample, the peak at $\sim0.2$\,pc disappears, revealing a distribution that is rather uniform in the interval $0.1\,\leq\,R^{ph}_{cluster}\,\leq\,0.32$\,pc. However, the cut-off at 0.32 pc remains even in this selected sample, suggesting that it is a true physical upper limit to the cluster size. We find only 36 systems with $R^{ph}_{cluster}\,\geq\,0.32$\,pc ( out of the 302 clumps with $3.7\,\leq\,d_{cl}\,\leq6.5$\,kpc), a number that is sufficiently small to not represent a significant statistic.  Our conclusion is that the clumps with $3.7\,\leq\,d_{cl}\,\leq6.5$\,kpc provide the most robust representation of the true cluster size distribution in a generic massive dense clump. As a consequence, it is quite likely that there would be cores located outside the observed area in the 168 fields with $N_{core}\geq4$ and $d_{cl}\leq3.7$\,kpc, meaning that the measurements of the core separations can be potentially incomplete in these systems, while they offer a better snapshot of the central region of the clump.

Our estimates of $R^{ph}_{cluster}$ can be affected by uncertainties. First, this definition of cluster radius depends on the effective membership of the farthest core, which is not always reliably confirmed. Secondly, there is a decrease in sensitivity towards the image edges, meaning that cores with low fluxes located at the margin of the FoV may remain undetected. It is interesting to compare these estimates with the size of their hosting clump, which defines where there is an overdensity of material that is potentially available for star formation.

Ideally, it would be useful to compare $R^{ph}_{cluster}$ with the extension of the continuum emission measured on the same ALMAGAL data. However, as discussed by \citet{Molinari2025}, ALMAGAL interferometric data are affected by spatial filtering, and in several fields, they may miss a significant flux fraction. This makes ALMAGAL data unsuitable for reliably tracing the full, extended emission of the clump, even if they reveal the complexity of the clump inner structure. For this reason, we adopted the clump sizes, $R_{cl}$, determined from {\it Herschel} 250 $\mu$m data and defined as the circularized half-width at half-maximum of the elliptical Gaussian fit with CuTEx \citep{Molinari2016,Elia2021}. This definition is certainly limited and a simplification of the complexity of the clump structure. However, our goal here is not to define an exact physical boundary, which is also not well defined in molecular clouds. Rather, our aim is to use a consistent and robust proxy for the region containing the bulk of the clump mass. We note that ALMAGAL single pointings are sufficiently wide to probe a wider area than the {\it Herschel} beam at 250 $\mu$m (with a full-width half maximum of $\sim18$\arcsec). 

The size of the clumps ranges from $\sim0.1$ to $\sim0.9$\,pc, and most of them have $R_{cl}\leq0.35$\,pc, as shown in Fig.\,\ref{fig:ClumpClusterSize}, which is comparable to our measurements of the cluster radius. The distribution of the ratio of $R^{ph}_{cluster}$ to $R_{cl}$ reported in Fig.\,\ref{fig:ClumpClusterSize} indicates that the cores are typically located between $\sim0.3$ and $\sim1.2\times R_{cl}$. Hence, the formation of cores takes place in any fraction of the hosting clump, and may even extend to its close outskirts, but rarely beyond the overdensity that has formed in the molecular cloud. Although it is possible that additional cores may reside farther away from the center of the clump, they must have low mass to remain undetected towards the edge of the FoV. As the sensitivity  in this area drops by a factor of $\sim3$, these undetected cores would have masses below $\leq\,0.6$\,M$_{\odot}$, adopting the completeness limit indicated by \citet{Coletta2025}.

\begin{figure}
    \centering
    \includegraphics[width=0.99\linewidth]{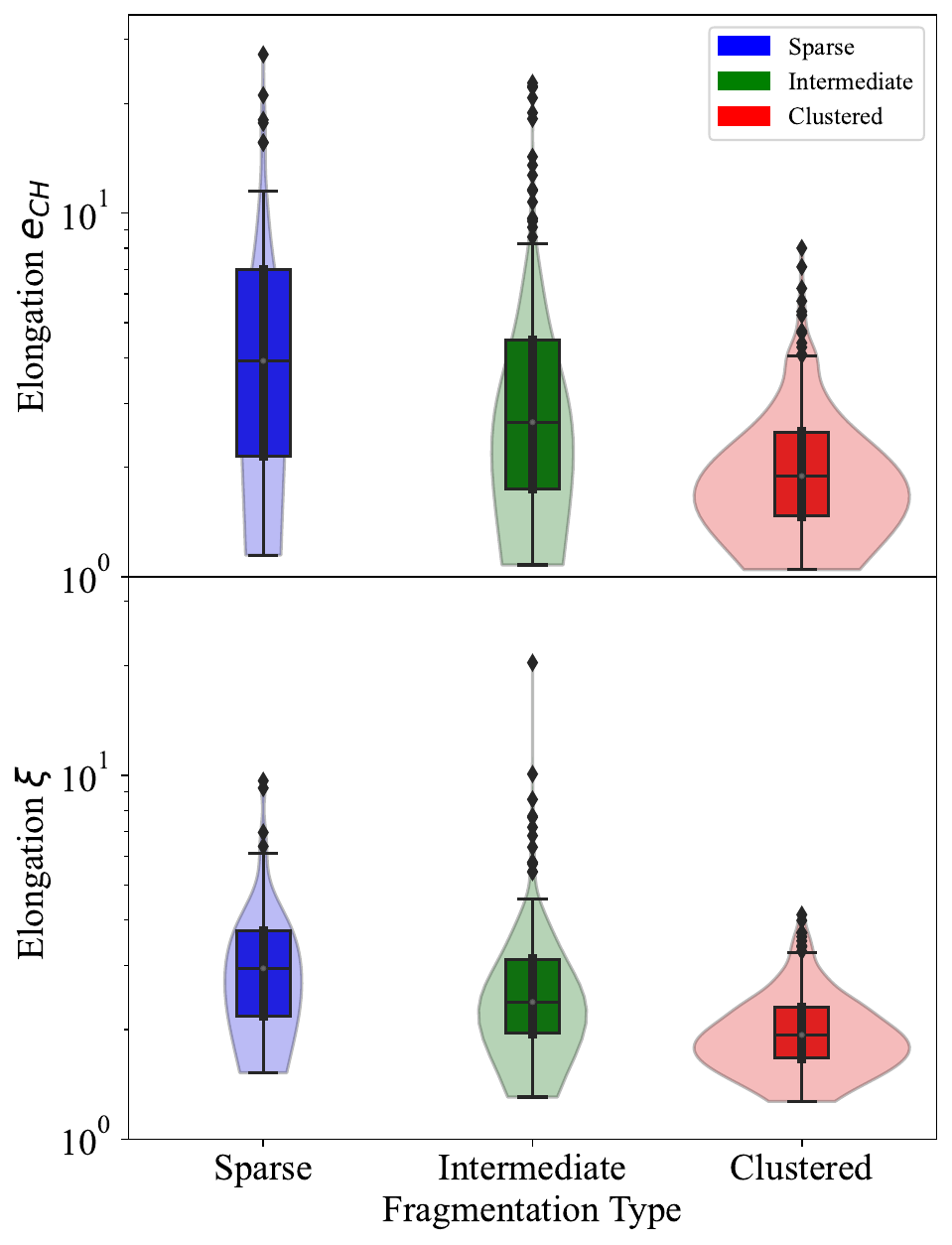}
    \caption{Whisker plots showing the distribution of the two measured estimates of the cluster elongation for the different classes of ALMAGAL clumps identified by the number of cores inside the CH polygon: sparse, where no further core lies inside it; intermediate, where there are fewer cores than the ones defining the polygon itself; clustered, where there are more cores than the ones at polygon vertices. The left panel presents the elongation $e_\mathrm{CH}$, defined as the ratio of the major to the  minor axis of the best fitting ellipse to the CH polygon, while the right panel refers to the ratio $\xi$ of the circular area of the cluster to the area of the CH polygon. The first and third quartiles, and the 10th and 90th percentiles of the distribution are indicated by the boxes and the segment extremes, respectively, while the central segment corresponds to the median. }
    \label{Fig:ElongationClassification}
\end{figure}

Finally, we found that our measurements of $R^{ph}_{cluster}$ do not show any correlation with the clump surface density $\Sigma_{cl}$ or the $L/M$ ratio as shown by Fig.\,\ref{Fig:DistributionClusterRadius_PhysicalProperties}. A similar result is found for the ratio $R^{ph}_{cluster}/R_{cl}$.

\subsubsection{Morphology of the clusters of cores}
\label{sect:MorphologyField}

The definition of cluster radius introduced in the previous section provides a robust measurement for spherical systems, for which the projected circular area $A=\pi R_{cluster}^{2}$ is a valid approximation for the size of the entire cluster. Nonetheless, cores are often distributed over elongated or irregular regions, for which the circular area gives a considerable overestimate of the cluster size (see, e.g., the bottom panels of Fig.\,\ref{Fig:ExampleMST_1}). The CH polygon allows us to better estimate the cluster area and determine the occurrence of spherical, elongated, and irregular clusters. This method is widely adopted in studies of embedded clusters in star-forming regions \citep{Schmeja2006, Gutermuth2009, Dib2019} and it was recently adopted in \citet{Molinari2025} to also characterize the extended continuum emission in the ALMAGAL fields that shows significant morphological diversity.

We note that the observed variety in the spatial distribution of cores is not unexpected, since similar results are also found when observing stellar systems, of which the ALMAGAL cores are the precursors. In fact, \citet{Kuhn2014} observed different appearances in young star systems, with the most recurrent morphologies represented by  chains of subclusters, clumpy ones, and isolated groups. The elongation of the cluster is a simple metric that can identify these possible patterns. To measure the cluster elongation, we fitted an ellipse to the area covered by the CH, measuring its orientation, PA$_{CH}$, and the ratio between the major and minor axes, $e_{CH}$. A different method was proposed by \cite{Schmeja2006}, who introduced the quantity

\begin{equation}
    \xi = \frac{R_{cluster}}{R^{CH}},
\end{equation}

\noindent where $R_{cluster}$ is the circular cluster radius and $R^{CH}$ is the convex hull radius, which is defined as the radius of the circle with an area equivalent to CH, $A^{CH}$. We note that \citet{Schmeja2006} adopted an additional correction factor $1/(1 -N^{hull}_{v} / N_{cores})$ to statistically recover the real area on which the sources are distributed \citep{Hoffman1983} (see also the discussion in Appendix A of \citet{Parker2018}). However, such a factor was found to be reliable only when $\sim200$ points are available \citep{Ripley1977}, while in our case it would introduce large discrepancies since $N^{hull}_{v}$ may be very close to $N_{cores}$. Therefore, we computed $R^{CH}$ without applying this factor, although we also report the CH area normalized by it in Table\,\ref{Tab:FieldDistribution}. 
The distributions of the two estimators $e_{CH}$ and $\xi$ are similar, with similar average values. The distribution of $\xi$ is narrower as $\xi$ has a smaller dynamic despite the irregular shape of the CH. The  most common shape of the cluster is a slightly elongated ellipse, with a median $e_{CH}\sim2.2$ [$\xi\sim2.3$]. While we found systems in which the cluster has a rather circular shape ($e_{CH}\leq1.3$), they represent a minority with only 46 cases ($\sim9$\%) of 514. Other systems have strong elongations as high as $\approx20-30$, depending on the adopted estimator. \\
Approximately circular shapes may result from a distribution where cores are uniformly distributed in the system, but also from patterns where a few filamentary features are converging towards a central hub and are symmetrically distributed over different directions, as in the top left panel of Fig.\,\ref{Fig:ExampleMST_1}. 
Figure\,\ref{Fig:ElongationClassification} shows the distribution of the measured elongations for the clusters classified into the three groups described in Sect.\,\ref{Sect:SpatialDistributionALMAGAL}. Similar results are obtained if the classes are defined in terms of the fragmentation level. High elongations are measured in systems with a limited number of cores and include the cases where they align along specific directions or are distributed over extended filamentary structures. Clustered systems typically have smaller elongation, with a median and interquartile range equal to $\xi\,=\,1.9^{+0.4}_{-0.3}$ ($e_{CH}\,=\,1.9^{+0.6}_{-0.4}$). As they are representative of systems with a high degree of fragmentation, this means that their cores are typically distributed over half of the circular area ascribed to the cluster, in slightly flattened structures like the ones shown in the bottom left panel of Fig.\,\ref{Fig:ExampleMST_1}. However, it is worth noticing that a fraction of $\sim7-16\%$ (depending on the adopted estimator) of the 190 systems classified as clustered have elongations greater than three. While these high values suggest rather elongated morphologies, they also derive from peculiar distributions of the cores with highly asymmetric patterns. An example is shown in the bottom right panel of Fig.\,\ref{Fig:ExampleMST_1}, where a group of cores is aligned along a direction converging towards an extended and rather circular emission causing a large elongation due to their separation from the main system.  

\begin{table*}[]
\caption{Metrics of the spatial distribution of the ALMAGAL cores: position, sizes, and shape of the distribution.} 
\resizebox{\textwidth}{!}{
\begin{tabular}{cccccccccccccc}

\hline
ALMAGAL\_ID       & $N_{cores}$    & RA\tablefootmark{a}  & DEC\tablefootmark{a} & Offset\tablefootmark{a} & $R_{cluster}$\tablefootmark{b}    & $R_{cluster}^{ph}$\tablefootmark{b}    & $A^{CH}$\tablefootmark{c} & $A_{norm}^{CH}$\tablefootmark{c}   & $N_{v}^{Hull}$\tablefootmark{c} & $N_{in}^{Hull}$\tablefootmark{c} & $\xi$\tablefootmark{d}   & $e^{CH}$\tablefootmark{d} & PA\tablefootmark{d}     \\
Field       & -    & deg  & deg & arcsec & arcsec    & pc    & arcsec$^{2}$  & arcsec$^{2}$  & - & -  & -   & -  & degree     \\

\hline

AG010.2283-0.2064 & 9     & 272.25125 & -20.19142 & 3.10      & 13.79 & 1.30E-01 & 51.67    & 93.01       & 4      & 5       & 3.40  & 5.24       & -110   \\
AG010.3226-0.1599 & 11    & 272.25671 & -20.08653 & 3.96      & 13.34 & 2.26E-01 & 130.42   & 204.94      & 4      & 7       & 2.07  & 1.37       & -31    \\
AG010.6185-0.0314 & 6     & 272.28967 & -19.76494 & 2.68      & 5.96  & 1.52E-01 & 7.17     & 21.52       & 4      & 2       & 3.95  & 5.28       & -27    \\
AG010.6691-0.2196 & 5     & 272.49161 & -19.81147 & 4.96      & 9.11  & 1.54E-01 & 35.82    & 179.11      & 4      & 1       & 2.70  & 2.33       & -27    \\
AG010.8851+0.1225 & 4     & 272.28314 & -19.45659 & 0.43      & 2.02  & 2.64E-02 & 2.55     & 10.22       & 3      & 1       & 2.24  & 1.95       & -12    \\
AG011.0829-0.5311 & 6     & 272.99283 & -19.59885 & 3.12      & 7.44  & 1.19E-01 & 34.28    & 102.84      & 4      & 2       & 2.25  & 2.86       & 56     \\
AG011.3614+0.8017 & 5     & 271.89975 & -18.71168 & 3.80      & 13.76 & 3.67E-01 & 112.73   & 563.64      & 4      & 1       & 2.30  & 4.30       & -266   \\

\hline
\end{tabular}}
\tablefoot{
\tablefoottext{a}{Geometric center of the group of cores present in the ALMAGAL core catalog \citep{Coletta2025} and its offset from the image center, which corresponds to the clump center measured in the Hi-GAL 250 $\mu$m maps.}
\tablefoottext{b}{Cluster angular radius, $R_{cluster}$ and corresponding linear size,  $R^{ph}_{cluster}$.}\tablefoottext{c}{Parameters of the convex hull (CH) polygon computed from the core positions: area, $A^{CH}$, normalized area according to  $A_{norm}^{CH}=A^{CH}/(1 -N^{hull}_{v} / N_{cores})$ from  \citet{Schmeja2006}, number of cores at the CH vertices, $N_{\nu}^{hull}$, and inside the polygon, $N_{in}^{hull}$.}\tablefoottext{d}{Elongation of the cluster: ratio of cluster radius to CH equivalent radius, $\xi$, and ellipticity $e$ and position angle PA of the best fitting ellipse to the convex hull.} The complete version of this table is available at the CDS.
}
\label{Tab:FieldDistribution}
\end{table*}

\subsection{Core separations }
\label{sect:CoreSeparation}

\subsubsection{Methods}

Several mathematical tools are available to characterize the spatial distribution of a set of localized points, which are representative of the members of a cluster. In particular, an important metric to measure is the mutual separations between those points, whose distribution may be related both to the fragmentation scales and to the degree of clustering \citep{Cartwright2004,Schmeja2008,Sanchez2009,Allison2009,Parker2018, Dib2019}. Those tools are based on the computation of the distance matrix $\delta(i,j)$, a symmetric matrix whose elements are the angular distances between the points taken pairwise. Given $N$ sources, the distance matrix $\delta(i,j)$ includes all possible $\frac{ (N-1)\times N }{2}$ connections and fully characterizes the relative separations for all elements of the cluster. In this work, we do not require the complete information provided by $\delta(i,j)$, as, to describe the length of fragmentation, it is sufficient to relate each core to its direct neighbor. Those separation are usually computed with the minimum spanning tree (MST) method.

The MST is a graph that connects all cores of the system and corresponds, among all the possible connecting paths, to the set of segments with the shortest total length.
It is composed of $(N-1)$ segments, named MST edges, that are a subset of the $\delta(i,j)$. A simple way to estimate the typical separations between the cores is the statistical analysis of the distribution of the MST edges. \\
We computed the MST with an IDL implementation of the Prim’s algorithm \citep{Prim1957}, identifying in the distance matrix $\delta(i,j)$ the edges $l_{i}$ that compose the MST. Prim's algorithm starts from a single position, adopted as a vertex, and identifies the shortest among all the possible connections with the adjacent sources, then proceeds iteratively until all the sources are finally connected, avoiding to form closed loops in the path. Examples of the computed MSTs are presented in Fig.\,\ref{Fig:ExampleMST_1} and\,\ref{Fig:ExampleMST_2}. We identified a total of 5214 edges, $l_{i}$, measured in all 514 ALMAGAL fields analyzed in this study. 

We followed two different approaches in the analysis of such a large ensemble\footnote{We use the term ensemble as in statistical mechanics to highlight the implicit assumption that the observed systems are statistical realizations of the same fragmentation process active in massive clumps. With this in mind, each observed pattern is a possible state undergone by this process.} of separations. On the one hand, we independently analyzed the distribution of the separations in each clump. We measured statistical quantities, such as the median and mode of distribution, to provide an estimator for the characteristic core separation in each system. On the other hand, we consider the entire ensemble as a diagnostic of the distributions of core separations, relating all the measured separations to the clump physical properties. In such a case, the statistical estimators are determined over subsamples obtained by stacking all the separations measured in clumps with similar values of a specific property, such as the $L/M$ ratio, ignoring that those clumps are distinct structures where possible different external conditions may be present. These two approaches allow for a well-rounded and complete description of the statistical information present in our measurements and to determine the possible link to the clump-averaged properties.

With the first approach, we provided estimates for the core separations that are clump-averaged, similar to other physical properties provided for clumps, such as density or temperature. These estimates also highlight the information provided by poorly fragmented systems, which would be hidden in the entire ensemble, as they are numerically outweighed by the contribution of clumps with several cores. Instead, with the entire ensemble, we provide an overall snapshot of the fragmentation process and relax the assumption of a unique characteristic core separation for clump. This analysis method provides a better statistical characterization thanks to the size of the sample and a more robust handling of outlier cores, whose presence would affect the clump-averaged estimators, especially for systems with a low degree of fragmentation.

\subsubsection{The distribution of core separations} 
\label{sect:distributionSeparations}

We divided the entire ensemble according to the set of array configurations adopted for the ALMAGAL observations, i.e., the near and far groups, respectively. The reason for this choice is twofold. First, the FoV is resolved by these two configurations in a different number of synthesized beams, meaning that these two datasets could potentially provide a substantially different density of point-like sources. In more detail, the FoV of the combined ALMAGAL continuum images is covered by $\sim2000$ and $\sim8100$ synthesized beams with the C2+C5 and C3+C6 configurations, respectively. Secondly, this division allows us to control the impact of the clump distance, as these two groups are composed of clumps located mainly in two different ranges of heliocentric distances. The two subsamples are composed of 4012 and 1716 cores, identified in 330 and 184 clumps, and provide 3682 and 1532 separations, respectively, in the near and far groups. The discrepancy in the number of cores is mainly due to the sensitivity limit of the survey as discussed by \citet{Coletta2025}. 

\begin{figure*}
    \centering
    \includegraphics[width=1\linewidth]{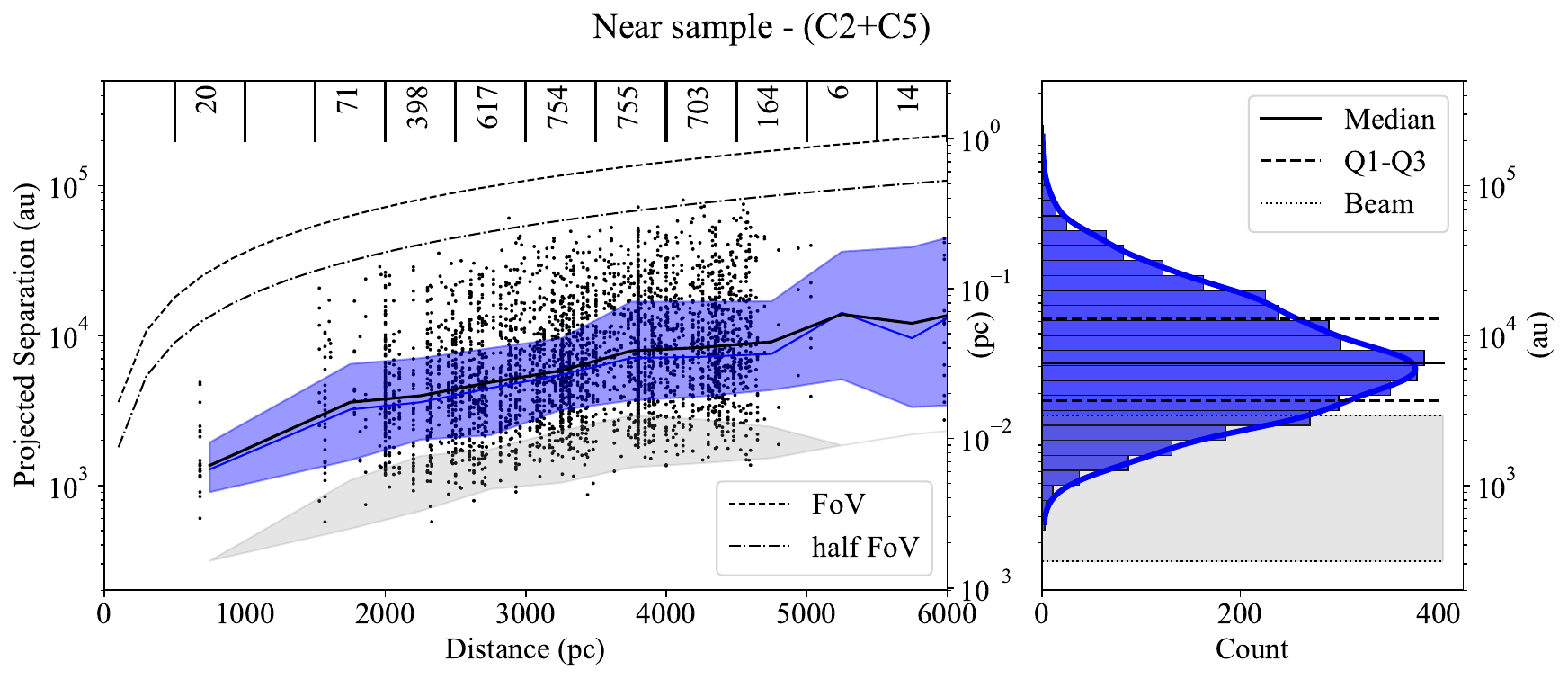}
    \includegraphics[width=1\linewidth]{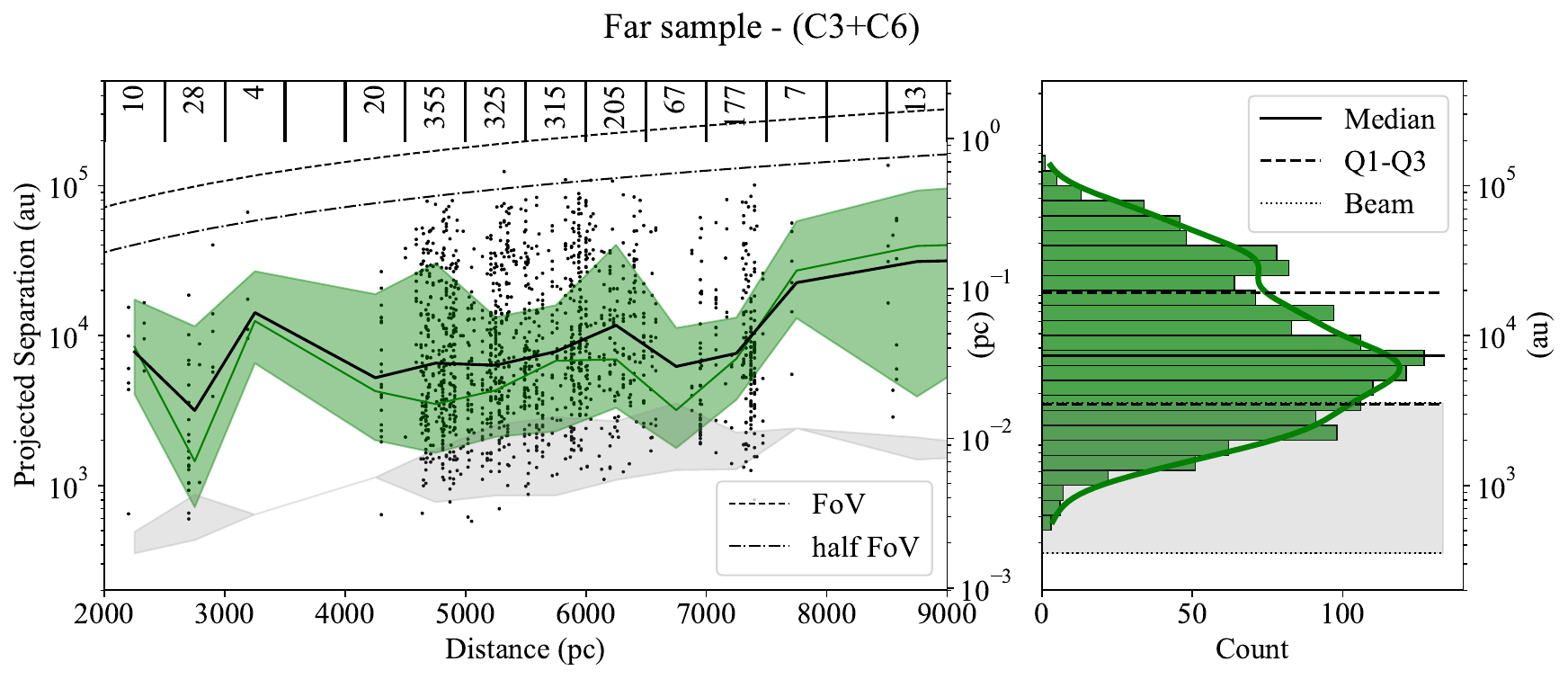}
    \caption{{\rm \textit{Left panels}}: Linear separation between ALMAGAL cores as a function of the distance of the ALMAGAL clump for near (\textit{top panel}) and far configuration sample (\textit{bottom panel}). The gray shaded areas indicate the ranges of the limiting sizes set by the beams of the ALMAGAL images. The black and colored solid lines (blue and green) indicate the median and the mode of the distribution of the separations, respectively, while the colored area traces the distribution width, defined at the 68\% peak intensity. The dashed and dotted lines indicates one-half of the observed field and the full field, respectively. The vertical lines show the adopted distance bins with the relative number of the separations. The 20 separations lying below the spatial resolution are found where the beam is strongly elongated, and are larger than the beam minor axis. In addition, a limited number of separations are not included in the figure as they exceed the selected intervals: 180 in the near sample (162 of which have $d_{cl}\,\ge\,$9\,kpc), and six in the far sample.  {\rm \textit{Right panels}}: Overall distribution of the projected linear separations in the near ({\rm \textit{top}}) and far ({\rm \textit{bottom}}) sample, respectively. The thick colored lines (blue and green) show the probability distribution function estimated with the KDE method. The gray shaded area indicates the interval spanned by the circularized beams in the corresponding ALMAGAL images.}
    \label{fig:PhysSeparationDistances}
\end{figure*}

The right panels of Fig.\,\ref{fig:PhysSeparationDistances} show the distributions of the projected linear separations between the cores given by the MST edges, $l_{i}$, obtained with the revised heliocentric distances for the ALMAGAL clumps. Most of these separations ($\sim70$\%) exceed the spatial resolution of the data by a factor of $\sim3$. In general, they range from $\lesssim1000$ to $\sim100000$ au (corresponding to $\lesssim0.005-0.5$\,pc), with the median of the two distributions equal to $\sim6600$ and $\sim7400$ au for near and far sample, respectively.  
Although the two distributions have a similar median and first quartile, they do not derive from the same underlying distribution, as indicated by the Kolmogorov-Smirnov (KS) test rejecting this null hypothesis with a $p\,\lesssim10^{-5}$. The major difference between the two distributions is given by the large tail present in the far sample due to an increase in the number of separations with $l_{i}\,>\,15000$\,au, caused by the wider physical area covered by the FoV in those clumps. Those separations produce a secondary bump in the distribution at $\sim30000$\,au. 

We present the projected linear separations as a function of the clump distance in the left panels of Fig.\,\ref{fig:PhysSeparationDistances} to compare them with the spatial resolution and the FoV of the observations. Clumps with heliocentric distances greater than 9\,kpc are omitted in the plot for better visual clarity, but are included in the analysis. We found separations between cores ranging from the beam up to half the FoV's size, indicating that the enlargement of the observed area includes additional cores located at the outskirts of the system that are more widely separated. The inclusion of these cores introduces a weak trend in the median and mode of the distribution, calculated in bins of width 500\,pc, as also shown in Fig.\,\ref{fig:PhysSeparationDistances}. The mode increases steadily with the distance for both samples and varies from $\sim3000-3500$\,au at a distance of 2\,kpc up to $\sim7000-7500$\,au for $d_{cl}\approx\,4$\,kpc. A similar trend, but with slightly larger values, is found for the median of $l_{i}$, which varies from $\sim4000$ to $\sim9000$\,au and $\sim7000$ to $\sim8000$\,au for the near and far group, respectively. The median and mode of the distributions of $l_{i}$ have the largest variations for distance $d_{cl}\,<\,3.7$\,kpc, where we most likely observed only the inner region of the cluster, as discussed in Sect.\,\ref{sect:CenterField}. This suggests that typical separations in the central portion of these clusters may be shorter, with differences in the estimated characteristic lengths in the densest regions of the clump. The gradual shift of the median towards larger values is ascribed to the inclusion of the cores that are more widely separated. The variations of the mode are explained by the fact that the inclusion of cores with larger separations extends the interval of the data analyzed with the KDE method. As the interval increases, the KDE computation adopts a wider bandwidth, with the effect of shifting the position of the peak of the continuous distribution to larger values. Finally, the distributions are affected by the sensitivity limit of the observations, since in the clumps located closer to the Sun it is possible to identify fainter cores. The ALMAGAL catalog is complete for objects M$\,\gtrsim\,0.2$\,M$_{\odot}$ \citep{Coletta2025}, but less massive cores are regularly detected in clumps with $d_{cl}\,\leq\,4$\,kpc. To deal with this effect, we discuss separately the near and far groups. Moreover, we also checked whether our results holds when selecting only the clumps located in narrow distance intervals.

\subsubsection{Deprojection of the measured separations}
\label{sect:deprojection}

The measured separations correspond to the projection in the plane of the sky of the intrinsic 3D separations between cores hosted in the clump. The distribution of 3D separations can be statistically derived by evaluating the effect of the projection on any generic vector randomly oriented in the 3D space. The details of the calculations are reported in Appendix \ref{AppendixA}. On average, the projection along the line of sight of a generic vector oriented in the 3D space reduces its intrinsic length by a factor equal to $\pi/4$ (see Eq.\,\ref{Eq:ExpValue}). As a first approximation, we corrected the measured separation between each core pair by this factor to scale to an average deprojected value, which we consider as the intrinsic separation $l^{depr}_{i}$. In such a case, the distributions of $l^{depr}_{i}$ retain the shape of the projected separations as we simply applied a scaling factor between the two quantities.

\begin{table*}
\caption{Clump-averaged thermal Jeans parameters and reference values of the distribution of projected linear core separations from MST analysis.
}
\begin{tabular}{lllllllllllll}

\hline
ALMAGAL ID       & $N_{cores}$ & $M^{th}_{J}$\tablefootmark{a} & $\lambda^{th}_{J}$\tablefootmark{a} & $l_{min}$\tablefootmark{b} & $l_{max}$\tablefootmark{b} & $l_{mean}$\tablefootmark{b} & $l_{median}$\tablefootmark{b} & $l_{25\%}$\tablefootmark{b}  & $l_{75\%}$\tablefootmark{b}   & $l_{mode}$\tablefootmark{c}  & $l_{-68\%}$\tablefootmark{c} & $l_{+68\%}$\tablefootmark{c} \\

& - & M$_{\odot}$ & au & au & au &  au & au & au & au & au &  au & au \\
\hline
AG010.2283-0.2064 & 9     & 0.5              & 5251          & 744    & 23835  & 6519    & 3723      & 2443   & 13387  & 7503     & 2773       & 21050      \\
AG010.3226-0.1599 & 11    & 1.3              & 8512          & 2964   & 16956  & 5861    & 4336      & 3283   & 7756   & 8776     & 6160       & 13596      \\
AG010.6185-0.0314 & 6     & 1.0              & 13997         & 1982   & 21710  & 9671    & 2419      & 2007   & 20236  & 2708     & 1162       & 26768      \\
AG010.6691-0.2196 & 5     & 0.8              & 10478         & 6226   & 37678  & 15394   & 10598     & 7072   & 37678  & 12291    & 6788       & 24793      \\
AG010.8851+0.1225 & 4     & 2.9              & 18062         & 3518   & 5452   & 4219    & 3687      & 3518   & 5452   &   -       &   -         &  -          \\
AG011.0829-0.5311 & 6     & 0.6              & 7802          & 2991   & 23130  & 12189   & 10585     & 7676   & 16563  & 13387    & 5816       & 26683      \\
AG011.3614+0.8017 & 5     & 1.0              & 11826         & 9575   & 66471  & 26125   & 17459     & 10997  & 66471  & 24358    & 12821      & 52485      \\
AG011.9039-0.1403 & 11    & 2.0              & 12787         & 4624   & 25002  & 12247   & 13000     & 5659   & 16270  & 21127    & 6760       & 32537      \\
AG011.9183-0.6122 & 13    & 1.7              & 10855         & 1589   & 30588  & 8677    & 6800      & 4456   & 11055  & 9616     & 4054       & 22133      \\
AG011.9370-0.6160 & 22    & 2.1              & 11578         & 1872   & 41345  & 9146    & 6556      & 3477   & 9418   & 8530     & 2779       & 17002      \\
AG012.2087-0.1017 & 23    & 0.8              & 5044          & 1012   & 7181   & 3037    & 2962      & 2135   & 3813   & 5615     & 3780       & 8545       \\
AG012.4040-0.4681 & 6     & 1.2              & 12773         & 14294  & 41567  & 27021   & 28396     & 21264  & 29583  & 38511    & 24560      & 55522      \\
AG012.4977-0.2233 & 6     & 0.8              & 10009         & 15213  & 34785  & 26925   & 29814     & 20368  & 34443  & 51966    & 33141      & 67659      \\
AG012.6794-0.1830 & 8     & 0.8              & 6641          & 1898   & 22572  & 8803    & 8355      & 3595   & 11824  & 19404    & 6398       & 43866      \\
AG012.7207-0.2175 & 7     & 0.9              & 8644          & 4273   & 10152  & 7353    & 8588      & 4495   & 10017  & 15317    & 7458       & 20427      \\
AG012.7360-0.1031 & 11    & 1.3              & 15477         & 3035   & 20060  & 11353   & 12359     & 8505   & 13506  & 15213    & 10302      & 21802      \\
AG012.8535-0.2265 & 11    & 1.5              & 12384         & 2709   & 31920  & 13985   & 10561     & 4667   & 23744  & 28203    & 5135       & 57232      \\
AG012.8887+0.4890 & 15    & 2.0              & 9109          & 1221   & 20909  & 7090    & 4771      & 2083   & 9136   & 5304     & 2375       & 28271      \\
AG012.9008-0.2404 & 10    & 0.7              & 6696          & 1905   & 9649   & 4610    & 4311      & 3565   & 5014   & 9111     & 6670       & 12296      \\
AG012.9048-0.0306 & 14    & 1.7              & 12699         & 651    & 52167  & 11452   & 4036      & 3047   & 7341   & 4533     & 2104       & 10310      \\
AG012.9084-0.2604 & 23    & 2.3              & 15929         & 1408   & 16930  & 5937    & 6124      & 3398   & 7657   & 7850     & 5006       & 11802      \\
\hline
\end{tabular}
\tablefoot{
\tablefoottext{a}{Clump-averaged quantities computed from properties reported in \citet{Molinari2025}}\tablefoottext{b}{ Minimum, $l_{min}$; maximum, $l_{max}$; mean, $l_{mean}$; median, $l_{median}$; first quartile, $l_{25\%}$; and third quartile, $l_{75\%}$, of the distribution of  projected separations.}\tablefoottext{c}{The modal, $l_{mode}$, and the values at $\pm68\%$ peak intensity, $l_{\pm68\%}$, are determined from empirical density distributions computed with the Gaussian kernel density estimate (KDE) method \citep{Silverman1986}. The kernel bandwidth is derived from the data applying the Scott's rule \citep{Scott1992}.} The complete version of this table is available at the CDS.
}
\label{Tab:Clumps}
   
\end{table*}

\subsubsection{Characteristic separations between cores}

The variety of observed patterns is reflected in different distributions of the separations between the hosted cores in each ALMAGAL field. To provide a representative characterization of these distributions, we determined the minimum, maximum and median of the separation $l_{i}$, which we report in Table\,\ref{Tab:Clumps}. We also included the mode of the distribution to indicate the most frequently observed separation, which is particularly relevant for systems where one or more groups of cores are surrounded by a more widespread population. In these systems the median would provide a bias estimator, which may not be representative of the typical observed separation. In each clump we estimated the continuous probability density function $P^{k}(l^{k}_{i})$ of the separations of the hosted cores using the Gaussian KDE method implemented in the $scipy$ library. We identified the peak position of $P^{k}(l^{k}_{i})$  to determine the mode, $l_{mode}$, and the spread of the distribution, the latter defined as the width of $P^{k}(l_{i})$ at $68\%$ of the intensity peak, avoiding any assumption of the shape of the PDFs, which we observed to be often quite asymmetric. These quantities are measured for the 437 clumps hosting more than 5 cores, which provide at least four measurements for the core separations, and are included in Table\,\ref{Tab:Clumps}. Statistically deprojected values in the $k$-\textit{th} clump, $l^{depr,k}_{i}$, can be obtained by scaling by the constant factor $4/\pi$ (see the discussion in Sect\,\ref{sect:deprojection}). \\

\begin{figure}
    \centering
    \includegraphics[width=1\linewidth]{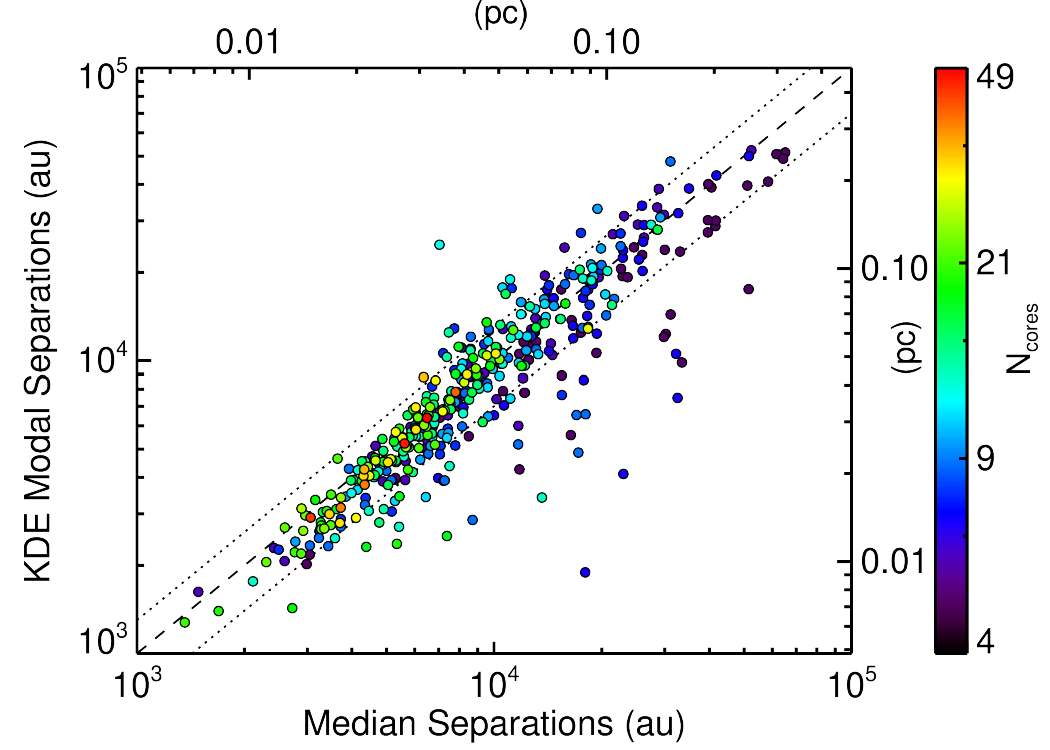}
    \caption{Relation between the median of the core separations in each clump and the mode determined with the KDE method. The colors indicate fields with different number of cores, ranging from 4 to 50 in logarithmic scale. The dashed line indicates the 1:1 relation, while the dotted lines indicate the range of variation of $\pm25$\%. }
    \label{fig:RelationEstimators}
\end{figure}

Our estimates of the characteristic, clump-averaged, separation are shown in Fig.\,\ref{fig:RelationEstimators}. They vary from $\sim1500$ to $\sim60000$\,au, and values larger than $\sim15000$\,au are found mainly in clumps with a low level of fragmentation. A decrease in the average separations in highly fragmented systems is expected for simple geometrical reasons, but we verified that the observed trend is not caused by the reduction of the available area per source. In fact, we compared with the expected decrease in separations when sources are uniformly distributed over a field with a fixed angular size and found that the observed averages are sensibly smaller than these expectations. \\
However, we note that the clump-averaged estimates provide limited information about the distribution of separations in a clump. Individual distributions $P^{k}(l^{k}_{i})$ have a diversity of shapes, often presenting multi-modal patterns, features that are not easily captured by a single averaged quantity. A thorough statistical analysis and classification of the distribution $P^{k}(l^{k}_{i})$ will be conducted in a future study. Here, we compared $l^{median,k}$ and $l^{mode,k}$ in Fig.\,\ref{fig:RelationEstimators} to report the occurrence of asymmetric distributions. These two quantities agree within 25\% for 316 systems ($\sim72\%$), and represent cases where $P^{k}(l^{k})$ is rather symmetric around a central value equal to the measurements provided. However, noticeable differences between the two estimators are not rare. In systems where they differ, the distributions of core separations $P^{k}(l^{depr,k})$ are sensitively skewed and asymmetric. Quantitatively, the median separation is larger than the mode in 96 systems ($\sim22\%$ of the sample), with such a pattern found in both near and far systems, with 59 and 37 cases, respectively. The opposite case, with a median smaller than the mode, is rarer and is found in only 25 systems ($\leq5\%$ of the sample).

Asymmetric distributions can be caused by the complex morphology of the 3D clump structure and the spatial distribution of cores. However, it is important to note that it is extremely unlikely to obtain multi-modal distributions for the projected separations starting from a single peaked distribution for the intrinsic 3D separations. One possibility is that these patterns are due to the limited degree of fragmentation observed, which could not be sufficient to properly sample the true distribution. Alternatively, they may also be the signature of different fragmentation conditions in groups of ALMAGAL clumps, with diverse characteristic separation lengths. We observed cases where a fraction of the core population is located farther away than the main group of more tightly packed cores. Indications that multiple characteristic lengths may be present in the fragmentation of massive clumps were already found in the literature \citep{Zhang2021}.
 
\subsection{Relation between the core separations and the clump properties}
\label{sec:SeparationvsClumpProperties}

The core positions directly probe the fragmentation of the clump as they indicate where the dense material collapsed. The distribution of patterns and core separations reflects the diversity of initial conditions or evolutionary paths that are observed in the ALMAGAL systems. This wide observed diversity has also been highlighted by \citet{Elia2025arXiv}, who related the properties of the core population to the ones of the hosting clumps finding a significant scatter in all their relations, interpreted as a result of different fragmentation conditions \citep{Elia2025arXiv}. Nonetheless, they also identified statistical trends between some fragmentation metrics. For example, they found that the number of cores, $N_{cores}$, increases with $\Sigma_{cl}$ and the evolutionary indicators $L/M$ ratio and T$_{bol}$. Similarly, here we discuss how the separations between cores relate to the clump-averaged properties, using the values derived from the {\it Herschel} Hi-GAL data \citep{Molinari2025} (see Sect.\,\ref{Sect:ClumpProperties}).

\subsubsection{Correlation with the clump evolution }
\label{sect:SeparationEvolutionLM}

\begin{table}
    \caption{Size of subsamples defined by different intervals of $L/M$ ratio and split into the near and far groups.}
    \centering
    \begin{tabular}{cccc}
      \hline
    $L/M$  &  N$_{clump}$ & N$_{clump}$ \\
     \hline
     (L$_{\odot}$/M$_{\odot})$& (C2+C5)  & (C3+C6)  & Total\\
     & near sample & far sample & \\
     \hline
       $L/M\leq\,1$  & 89  & 40 & 129 \\
       $1\,\leq\,L/M\,\leq\,10$  & 111 & 65 & 176 \\
        $L/M\geq\,10$   &  130 &  79 & 209\\
     \hline
    \end{tabular}
    \label{tab:ClumpSampleLovM}
    
\end{table}

The formation of cores is a consequence of the collapse of the dense material, which proceeds over timescales that are a few times the free-fall time of the gas. The degree of fragmentation changes with time, as more clump material may become gravitationally unstable, depending on the local conditions, and cores evolve into stars. 
The local conditions vary over the entire clump evolution because of the internal dynamics of the clump or the feedback of the newly formed stars. Changes in the separations between cores during the clump evolution, if present, are useful to trace the variations of these conditions, determining how fragmentation proceeds.

\begin{figure*}
    \centering
    \includegraphics[width=0.97\linewidth]{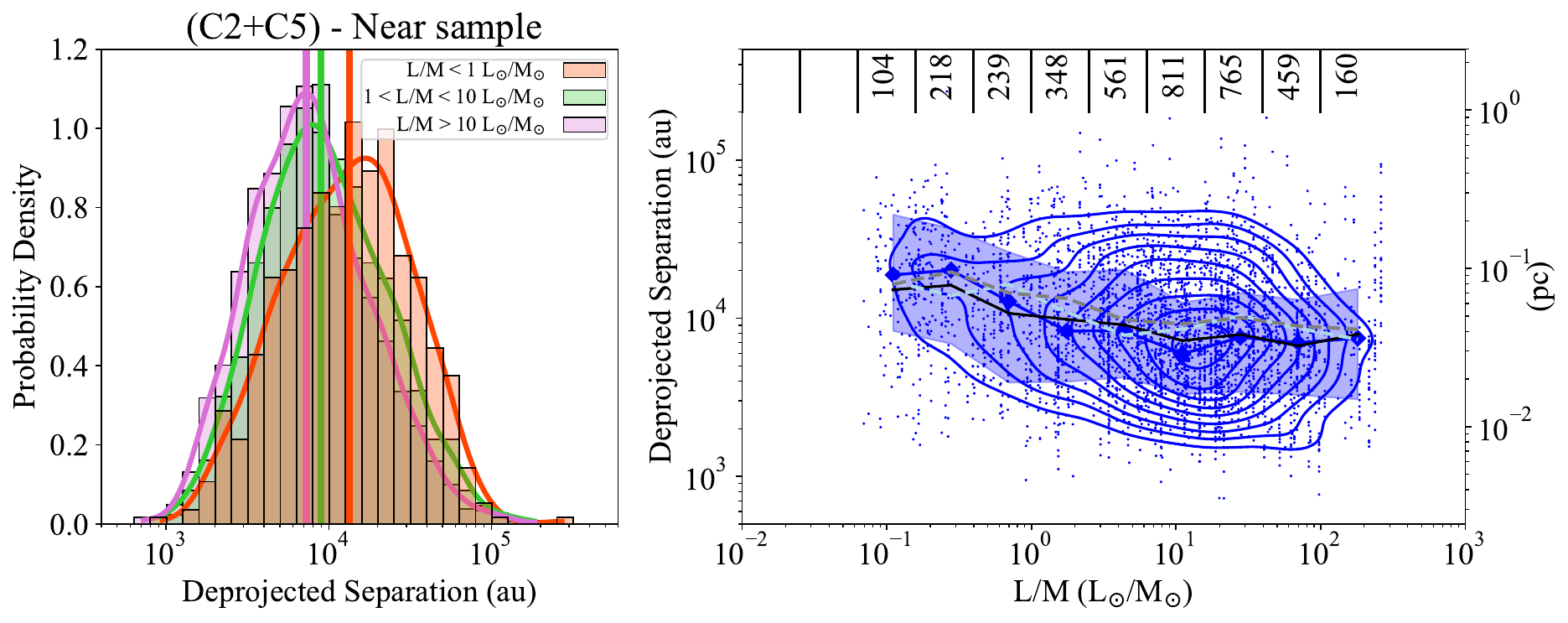}
    \includegraphics[width=0.97\linewidth]{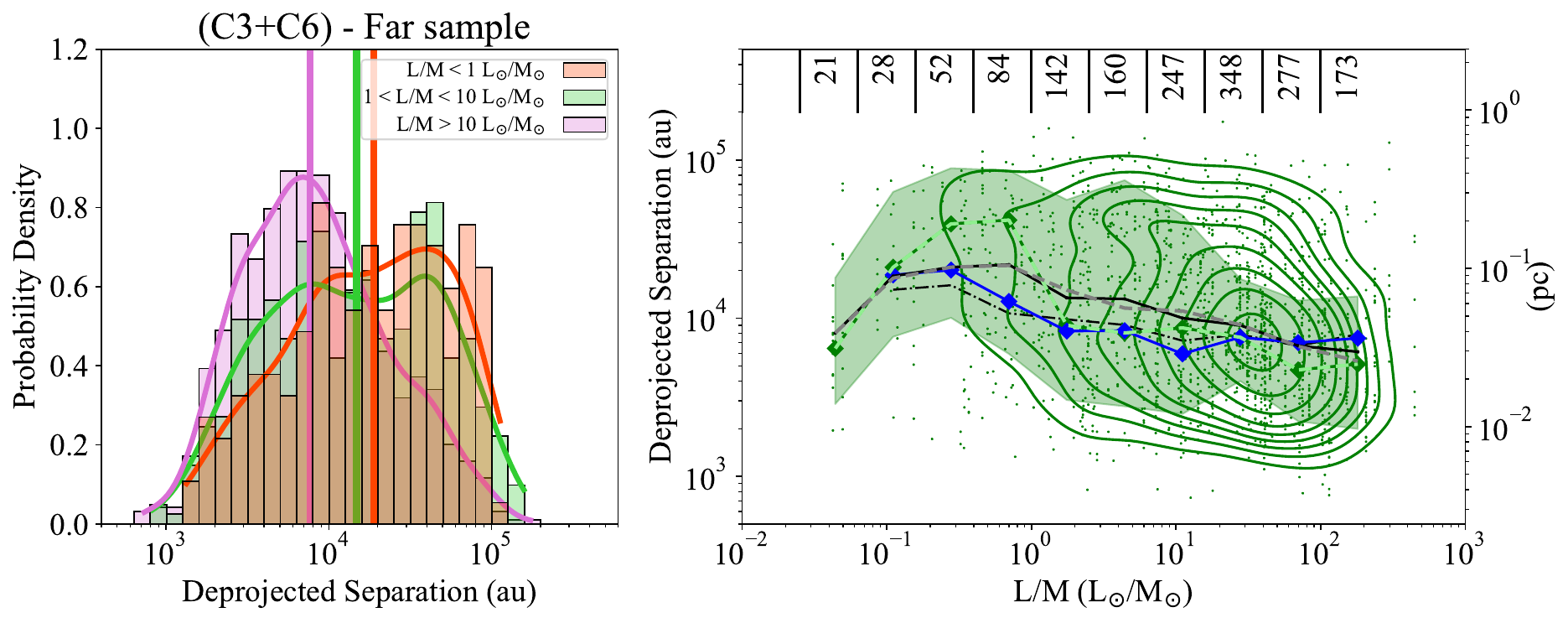}
    \caption{
    \textit{Left panels}: Density distribution of the deprojected core separations measured in the all clumps with $L/M\,\leq\,1$ (orange), $1\,\leq\,L/M\,\leq\,10$ ( green), and  $L/M\,\geq\,10$\,L$_{\odot}$/M$_{\odot}$ (purple) with the continuous distribution computed with the KDE method overlapping with the colored continuous line. Clumps of the near and the far are shown in the top and bottom panel, respectively. The colored vertical thick lines show the median of these distributions.  
    Right panels: Deprojected core separations as function of the clump $L/M$ for the systems in the near (\textit{top panel}) and the far (\textit{bottom panel}) subsample, respectively. The thick black lines indicate the median of the distributions of the core separations computed in bins of $L/M$ of width 0.3\,dex.  The thick colored lines, in blue and in green for the near and far subsample, respectively, indicate the peak of these distributions, while the colored area traces their spread defined as the width at 68\% of peak intensity. The dashed lines show the variation of the mode (in light blue and green) and of the median computed over subsamples having $3.7\,\leq\,d_{cl}\,\leq\,4.7$\,kpc and $4.7\,\leq\,d_{cl}\,\leq\,6.5$\,kpc.  
    The numbers at the top give the size of the subsamples from which the distribution in each bin is computed.
    The colored contours trace, from the outside inward, fractions of the sample lying outside that rise from 15\% to 95\%, with a step of 10\% for the near ({\rm in blue}) and far ({\rm in green}) subsample, respectively. The median and the peak derived for the near sample are also indicated in the bottom panel for comparison purposes.}
    \label{fig:SeparationEvolutionaryLM}
\end{figure*}

We show in {the left panels of} Fig.\,\ref{fig:SeparationEvolutionaryLM} the normalized density distributions of the deprojected separations measured in clumps divided into three main classes according to the value of the ratio $L/M$,\footnote{Two clumps, AG014.9963-0.6733 and AG305.2021+0.2073, both of the near sample do not have estimates for luminosity and mass, as they are saturated in {\it Herschel} data. Those systems have 4 and 15 cores respectively, so a total of 17 separations are excluded from the analysis.} which is usually adopted as a diagnostic for the evolutionary state of the clump \citep{Molinari2016} (see Sect.\,\ref{Sect:ClumpProperties}). The classes were chosen to separate the clumps in the youngest phases of their evolution, which we considered as the systems with a ratio $L/M\,<\,1$\,L$_{\odot}/$M$_{\odot}$, from those that are highly evolved and host multiple protostars, which we selected here as those with a ratio $L/M\,>\,10$\,L$_{\odot}/$M$_{\odot}$. These values match the intervals identified by \citet{Molinari2016b} for different phases of clump evolution, and allow us to also compare our results with those of the surveys ASHES \citep{Sanhueza2019}, CORE \citep{Beuther2018}, and ASSEMBLE \citep{Xu2024}, each of which observed systems in a specific evolutionary phase. The two classes of young and evolved systems are composed of 129 and 209 clumps, respectively, leaving 176 clumps in the intermediate one (see Table \ref{tab:ClumpSampleLovM}). 
Figure\,\ref{fig:SeparationEvolutionaryLM} shows that the distribution of the young class, with $L/M\,<\,1$\,L$_{\odot}/$M$_{\odot}$, and the evolved one, with $L/M\,>\,10$\,L$_{\odot}/$M$_{\odot}$ are substantially different. The peak and median decrease to smaller values in the population of evolved clumps, which is also characterized by a narrower distribution. The Kolmogorov-Smirnov and Mann-Whitney U tests highlight the differences of these distributions, since they both return a $p$-value $\ll\,10^{-7}$, rejecting the null hypothesis that these two datasets derive from the same underlying population. Specifically, in young systems with $L/M\,<\,1$\,L$_{\odot}/$M$_{\odot}$, the near and far distributions have median values equal to $\sim12000$ and $\sim18000$\,au ($\sim0.06-0.085$\,pc), respectively. However, these values describe only partially the distributions presented in Fig.\,\ref{fig:SeparationEvolutionaryLM}, which have complex shapes. The distribution peaks estimated with the KDE method correspond to $\sim14000$ and $\sim34000$\,au, respectively. These values overlap with the separations equal to $\sim14000-35000$\,au measured in the 39 protoclusters observed by the ASHES survey \citep{Sanhueza2019,Morii2024}. We note that while the ASHES survey maps a larger angular area than ALMAGAL's single pointings for near clumps, the two surveys cover a similar physical area in the more distant far sample, for which the comparison is less biased by the observed area. The distribution of the latter sample is broad and negatively skewed, with a peak that matches the upper values reported by \citet{Morii2024}. An additional comparison is provided by the 12 young starless clumps observed by \citet{Svoboda2019} where the median of the measured projected separations is 0.083\,pc, corresponding to a deprojected value of $\sim22000$\,au, which is marginally larger than the median values reported above, but still compatible with the ones measured in the youngest systems of our ALMAGAL sample.

In the evolved clumps with $L/M\,>\,10$\,L$_{\odot}/$M$_{\odot}$, the separations decrease to median values of $\sim7300-7700$\,au ($\sim0.035-0.038$\,pc). These values match the results reported by the CORE \citep{Beuther2018} and ASSEMBLE \citep{Xu2024} surveys, which targeted the high-mass and high-luminosity clumps characterized by $10\,\leq\,L/M\,\leq\,1000$\,L$_{\odot}/$M$_{\odot}$ and that are in the later phases of their evolution. The median of the separations measured in the 20 clumps of the CORE survey is equal to $\sim6000$\,au, with a interquartile interval varing from $\sim4000$\,au to $\sim15000$\,au, once their measurements are corrected for the projection effect (see Sect.\,\ref{sect:deprojection}). Even if this value is marginally smaller than our measurements, it is considered compatible given the widths of the distributions presented in Fig.\,\ref{fig:SeparationEvolutionaryLM}. The typical separation measured in the 11 high-mass clumps of the ASSEMBLE survey is equal $\sim0.035$\,pc ($\sim7200$\,au) which is in agreement with our results.

A considerable fraction of the ALMAGAL clumps have a $L/M$ ratio between the two values discussed above. In these systems, we find a different distribution of the separations for the clumps of the near and far sample. In the near sample, the distribution is similar to that of the evolved systems, the opposite holds for the far sample, where it resembles that of the young clumps. We measured median values between $\sim9000$ and $\sim15000$\,au, with the spread influenced by the wide area probed in the far sample. However, the KS test rejects the null hypothesis that each of the two sets of distributions derives from a unique underlying distribution with $p\lesssim10^{-7}$ and $p\lesssim{0.05}$ for the near and far sample, respectively. We mainly attribute the different shapes of the two distributions to the different degrees of fragmentation of their clump population. The near sample is dominated by highly fragmented clumps with N$_{cores}\,>\,8$, which strongly impact the shape of the distribution since their separations contribute for more than $\sim80\%$ of the total. Conversely, the far sample consists mainly of poorly fragmented clumps ( N$_{cores}\,\leq\,8$), whose separations are a relevant ($\sim65\%$) fraction of the distribution presented in Fig.\,\ref{fig:SeparationEvolutionaryLM}. In summary, the different shapes are produced by  the different relative contributions of poorly and highly fragmented systems combined with the evidence that separations are typically larger in clumps with N$_{cores}\,\leq\,8$ than in those with a higher degree of fragmentation. A minor, secondary effect may be caused by the larger physical area observed in the far group, which favors the measurements of wider separations. 

Only a few clumps with $1\,<\,L/M\,<\,10$\,L$_{\odot}/$M$_{\odot}$ were observed before ALMAGAL, a large fraction of them were obtained by the SQUALO \citep{Traficante2023} and the DIHCA survey \citep{Ishihara2024}. A typical separation of $\sim7000$\,au is reported from the 30 clumps of the DIHCA survey, although this includes systems with higher $L/M$, since their sample covers $\sim1$ to $50$\,L$_{\odot}/$M$_{\odot}$. The average separation and, in general, the entire distribution closely match the results from the more evolved clump populations, similar to what is found for the near sample of the ALMAGAL survey. In contrast, the comparison with the SQUALO survey is severely limited since only 3 clumps (out of 13 surveyed in total) fall within this interval of $L/M$ \citep{Traficante2023}. \citet{Traficante2023} reported a low level of fragmentation, measuring minimum separations between 13000 and 29000\,au. Although these values should be considered as a lower limit because of the projection effect, these wide separations are nonetheless comparable with our results from the far sample systems.\\

In summary, our statistical analysis reveals an evolutionary trend in core separation. When ALMAGAL clumps are grouped by their $L/M$ ratio, we find that the cores in more evolved systems are generally closer to one another than those in younger systems.
Evidence of a relationship between core separations and the evolutionary diagnostic $L/M$ was already proposed by \citet{Traficante2023}, who compared the separations in the 13 young and moderately evolved targets of their SQUALO survey to those of the 20 high-mass and high-luminosity clumps of the CORE survey \citep{Beuther2018}. Similar indications were also found by including the 30 moderately evolved systems ($\sim1\,<\,L/M\,<\,50$\,L$_{\odot}/$M$_{\odot}$) of the DIHCA survey \citep{Ishihara2024} and comparing the 11 massive, luminous, and evolved clumps of the ASSEMBLE survey \citep{Xu2024} with the 39 high-mass young clumps observed by the ASHES survey \citep{Sanhueza2019,Morii2024}.\\

The sample size and the uniform coverage across the $L/M$ parameter space of the ALMAGAL survey provide robust statistical significance to this trend. It allows us to move beyond a coarse three-group comparison and analyze in detail the variation as a continuous function of $L/M$. The right panels of Fig.\,\ref{fig:SeparationEvolutionaryLM} show the entire ensemble of separations as a function of the clump $L/M$ ratio.  Although the separations $l^{depr}_{i}$ have a wide scatter for any given ratio $L/M$, we find a steady decreasing trend traced by the median of the distributions computed in equally spaced bins in logarithm of $L/M$. This behavior is not caused by the weak correlation found with distance. In fact, it is still present when selecting systems in the two distance intervals $3.7\,\leq\,d_{cl}\,\leq\,4.7$\,kpc and $4.7\,\leq\,d_{cl}\,\leq\,6.5$\,kpc, for which the statistical values of the separations do not show any dependence on distance (see Sect.\,\ref{sect:distributionSeparations}), despite the minor quantitative differences emerging due to the smaller sample sizes. Apparently, the two samples near and far show different trends. In the near sample, we found that the median has a major decrease from $\sim15000$ to $\sim10000$\,au occurring until $L/M\approx1$\,L$_{\odot}/$M$_{\odot}$, followed by a minor decrease to $7000-7800$\,au for increasing values up $L/M\approx100$\,L$_{\odot}/$M$_{\odot}$. In the far sample, the median is typically larger in the early phases, with a value of $\sim20000$\,au, and shows a steeper and more steady decrease starting at $L/M\approx1$\,L$_{\odot}/$M$_{\odot}$. These differences are due to the smaller sample size, lower level of fragmentation, and wider observed area in clumps of the far sample, as discussed above. The shift of the median to larger values is caused by the increase in the statistical weight of clumps with a limited degree of fragmentation whose cores are more distantly separated, as discussed above. It is interesting to compare the trend traced by the mode of the distributions. The right panel of Fig.\,\ref{fig:SeparationEvolutionaryLM} shows that in the two samples the modes are coincident for $L/M\gtrsim1$\,L$_{\odot}/$M$_{\odot}$. Instead, we observe a match between the values of the near sample and the shapes of the 2D density contours for $L/M\,<1$\,L$_{\odot}/$M$_{\odot}$. These shapes suggest a potential bimodality in the distribution, with one of the peaks that follows the trend found in the near sample. More generally, the distributions of the separations in the far sample are all broad for clumps with $L/M\,\leq\,10$\,L$_{\odot}/$M$_{\odot}$. Moreover, the profiles of the KDE distributions shown in the left panel of Fig.\,\ref{fig:SeparationEvolutionaryLM} present two local maxima, one approximately at $\sim40000$\,au and another that shifts from $\sim10000$ down to $\sim7000$\,au.
The maximum at large separations is built mainly, but not exclusively, by systems with a low degree of fragmentation, whose cores are more widely distributed. In fact, about $\sim72$\% of the separations greater than $\sim20000$\,au of the far sample are measured in clumps with $N_{cores}\le8$. To verify that these maxima correspond to a statistically significant bimodality, we ran the \citet{Silverman1981} test, following the implementation described in \citet[see their Appendix\,A]{Beraldo2021}, applied to the distributions in different intervals of $L/M$. In a few intervals of $L/M$, we were able to assess that the corresponding distribution of separations has more than one significant peak. However, we were unable to extend this to the entire $L/M$ range due to the limited sample size in a few intervals of $L/M$. In general, we found that the null hypothesis that at least two peaks are present can be assessed only with a probability $Prob\,>\,0.8$, although it becomes significant ($Prob\,>\,0.95$) over limited intervals of $L/M$. 

\begin{figure}
    \centering
    \includegraphics[width=1\linewidth]{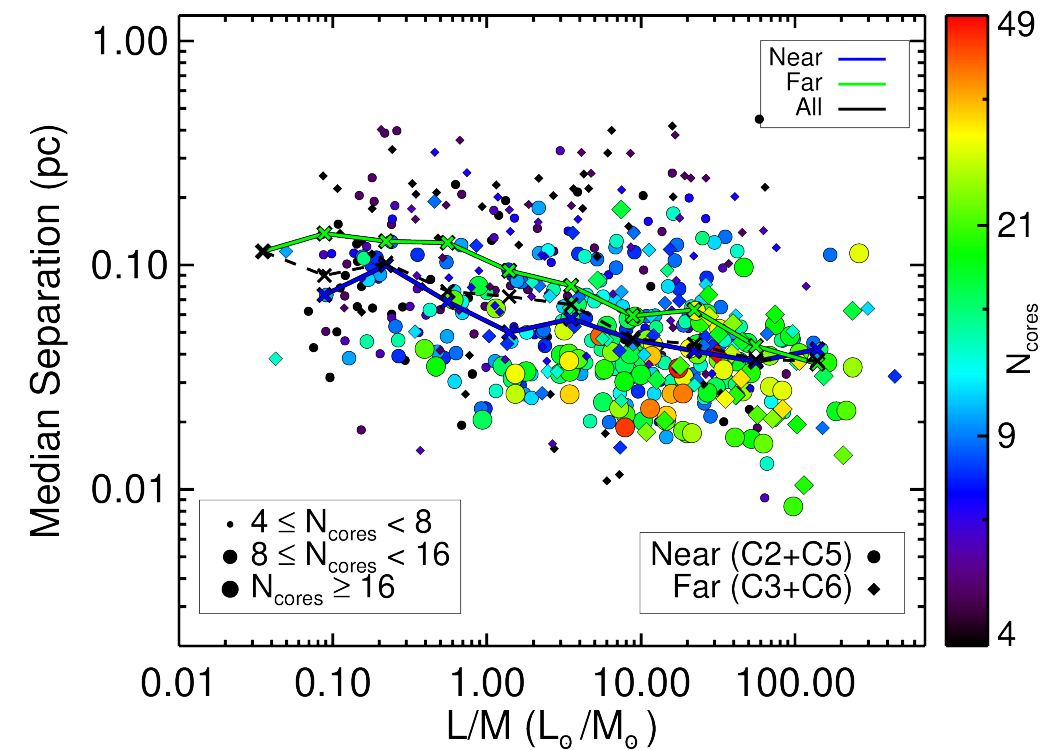}
    \caption{Relation between the clump-averaged separations between the cores provided by the median and the $L/M$ ratio,  color-coded to indicate the number of cores detected in the clump. To support the visualization, different sizes are adopted to identify systems  $4\,<\,N_{cores}\,<\,8$ (low), $8\,\leq\,N_{cores}\,<\,16$ (intermediate), and $\,N_{cores}\,\geq\,16$ (high). The different symbols refer to the clumps of the near (circle) and far (diamond) sample. The solid lines show the median of the distribution of the separations computed in bins of $L/M$ of width 0.4 dex for the near (blue) and far (green) sample, respectively, while the black dashed line corresponds to the median of all the clumps with $N_{cores}\geq4$. 
    }
    \label{fig:SeparationEvolutionaryLMAverageQuantity}
\end{figure}

Finally, we present in Fig.\,\ref{fig:SeparationEvolutionaryLMAverageQuantity} the same analysis using clump-averaged separations, showing only the median separation as a function of the  $L/M$ ratio since the mode has a similar trend. Similarly to the case of the entire ensemble, the averaged quantities also show a gradual decrease in the separation between the cores with $L/M$, as we determined a Spearman correlation coefficient equal to $\rho_{s} = -0.37\,[-0.38]$ for the median [mode] with a $p$-value $\ll10^{-9}$. These values indicate that a correlation is indeed present; however, the Pearson correlation coefficient of $\rho_{p} = -0.26\,[-0.28]$ suggests that it is not a single-linear relation over the entire interval of $L/M$. Figure\,\ref{fig:SeparationEvolutionaryLMAverageQuantity} also indicates that the observed relationship is connected to the degree of fragmentation, which also statistically increases with the $L/M$ ratio (see the discussion in \citealt{Elia2025arXiv}). As discussed for the case of the entire ensemble, the relationship changes for $L/M\sim1$\,L$_{\odot}/$M$_{\odot}$, above which it becomes rather flat. This value sets a boundary above which $\sim90\%$ of the fields are intermediate or clustered, according to the classification introduced in Sect.\,\ref{Sect:SpatialDistributionALMAGAL}, or, alternatively, have a moderate to high degree of fragmentation, as they mostly host more than eight cores.

\subsubsection{Relation with clump surface and volume density }
\label{Sect:SeparationClumpSurfaceDens}

\begin{figure}
    \centering
    \includegraphics[width=1\linewidth]{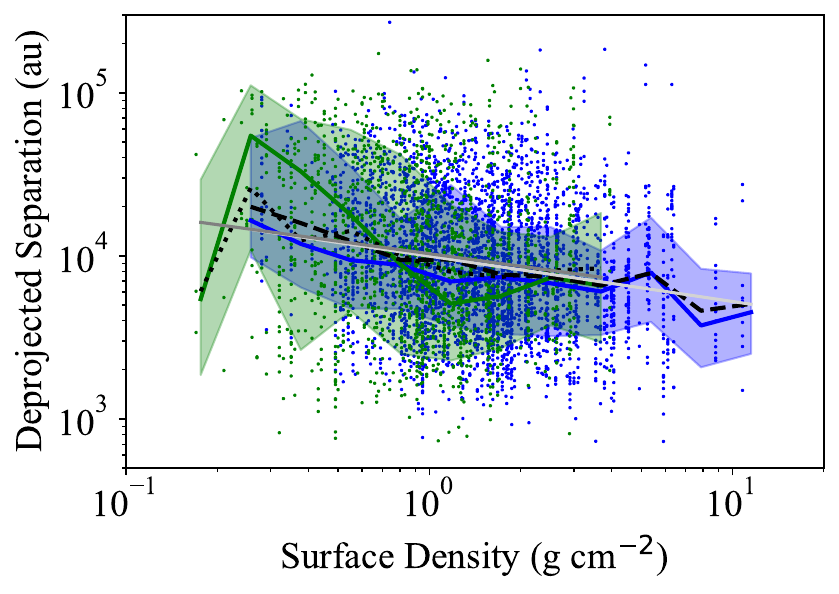}
    \includegraphics[width=1\linewidth]{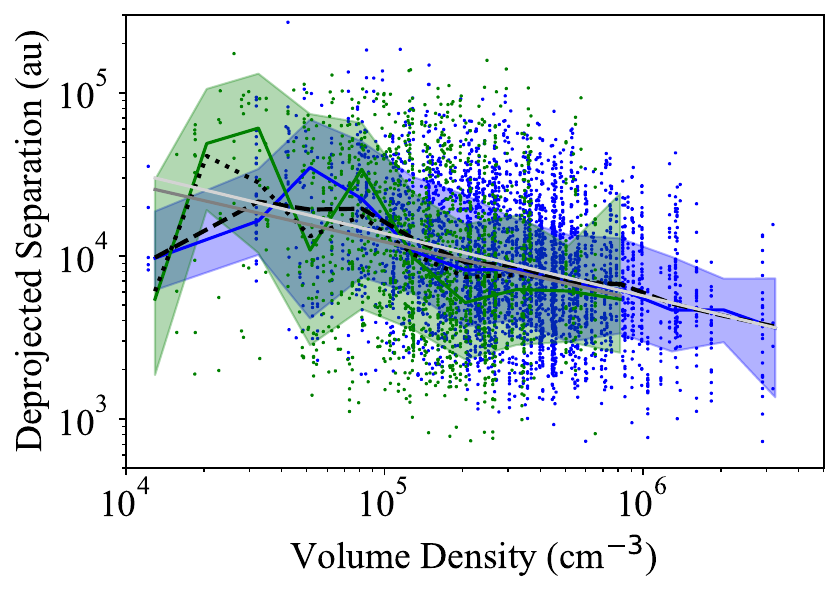} 
    \caption{Deprojected core separations against the clump surface ({\rm \textit{top panel}}) and volume ({\rm \textit{bottom panel}}) density. The distributions of the core separations are computed in bins of $\Sigma_{cl}$ and $n_{cl}$ of width 0.15 and 0.2 dex, respectively. The peaks and spreads of these distributions, defined as the width at 68\% peak intensity, are indicated by lines and area for the near (blue) and the far (green) samples, respectively. The medians of the distributions for these two sets are indicated as dashed and dotted black lines, respectively. The best fitting power-law relation computed over the entire set of separations and whose results are reported in Table\,\ref{Tab:CoeffFitDensity} are indicated as light gray and dark gray lines for the near and far sample, respectively.
    }
    \label{fig:SeparationDensityColumnandVolume}
\end{figure}

The ALMAGAL clumps cover systems with surface densities, $\Sigma_{cl}$, varying from $\sim0.1$ to $\sim10$\,g\,cm$^{-2}$, which allow us to determine whether there is a connection between the places where the material collapsed and the averaged density of the clump. Density is a key quantity in the fragmentation theory and, in the case of Jeans fragmentation, sets the scales that ultimately shape the separation between cores.  Figure \ref{fig:SeparationDensityColumnandVolume} shows the distribution of all the separations as a function of the clump surface, $\Sigma_{cl}$, and volume number densities, $n_{cl}$ for the near and far samples. The clump-averaged $\Sigma_{cl}$ and $n_{cl}$ are computed with the assumption of spherical geometry for the clump as

\begin{equation}
\label{Eq:Densities}
\Sigma_{cl} = \frac{M_{cl}}{\pi R_{cl}^2}\quad\quad \text{and}\quad\quad n_{cl} = \frac{3M_{cl}}{4\pi\, \mu m_{H}\, R_{cl}^3},
\end{equation}

\noindent where $M_{cl}$ and $R_{cl}$ are the clump mass and radius, respectively, $m_{H}$ is the atomic mass and $\mu$ is the mean molecular weight per free particle adopted to be equal to 2.37, which corresponds to a molecular gas with solar composition\footnote{This value of $\mu$ is derived for a mixture composed only by H$_{2}$ and He with mass ratio equal to the solar one, i.e., $X\approx0.71$ and $Y\approx0.27$ (see Appendix A in \citealt{Kauffmann2008}).}\citep{Kauffmann2008}. 

Quantitatively, the typical separation between the cores steadily decreases from $\sim25000$\,au to $\sim8000$\,au in clumps with an increasing $\Sigma_{cl}$ from $\sim0.3$\,g\,cm$^{-2}$ to $\sim5$\,g\,cm$^{-2}$, and a drop to $\sim6000$\,au for $\Sigma_{cl}\,\geq\,5$\,g\,cm$^{-2}$, where the relation appears to be flattening. A decreasing trend is also found when the separations are compared to the clump volume density $n_{cl}$, with a substantial decrease between $\sim10^{4}$ and $\sim10^{6}$\,cm$^{-3}$ (see the lower panel of Fig.\,\ref{fig:SeparationDensityColumnandVolume}). Both relations show a wide scatter of the measured separations, and weak linear relations appear only from the statistical estimators, i.e., the median and the mode. We checked for possible power-law relations between $l^{depr}$ versus $\Sigma_{cl}$ and $n_{cl}$ by fitting to the data with the functional forms $l^{depr}\propto\Sigma_{cl}^{\alpha}$ and $l^{depr}\propto n_{cl}^{\beta}$. The derived exponents of these fits are reported in Table\,\ref{Tab:CoeffFitDensity} and indicate that a weak correlation exists between the separations and these quantities. The empirical trends traced by the medians closely match the result of the fits derived using all the separations in each of the two samples and show a tight power-law behavior for $n_{cl}\,\geq\,10^{5}$\,cm$^{-3}$.

\begin{table*}
    \caption{Results of the fits of a power-law relationship between separations and surface and volume density}
   \begin{center}
    \begin{tabular}{ccccccccc}
        \hline
         Sample     & \multicolumn{2}{c}{$A$\tablefootmark{a}} &\multicolumn{2}{c}{$B$\tablefootmark{a}} & \multicolumn{2}{c}{$C$\tablefootmark{b}} & \multicolumn{2}{c}{$D$\tablefootmark{b}}\\
        \hline
        {} &  Median\tablefootmark{c} & All data\tablefootmark{d} &  Median\tablefootmark{c} & All data\tablefootmark{d} & Median\tablefootmark{c}  & All data\tablefootmark{d}  & Median\tablefootmark{c}  & All data\tablefootmark{d}  \\
        \hline
        near & 4.02$\pm$0.02 & 4.00$\pm$0.01 & -0.34$\pm$0.04 & -0.28$\pm$0.02 & 5.54$\pm$0.02 &  6.05$\pm$0.10& -0.29$\pm$0.06 & -0.38$\pm$0.02 \\
        far  &  4.01$\pm$0.01 & 3.99$\pm$0.06 & -0.17$\pm$0.15 & -0.25$\pm$0.05 & 5.40$\pm$0.70 & 5.90$\pm$0.21 &-0.27$\pm$0.14 & -0.36$\pm$0.04\\
        \hline
    \end{tabular}
    \end{center}
    \tablefoot{
\tablefoottext{a}{Fitted relationship $\log{l^{depr}}\,=\,A + B \log{\Sigma_{cl}}$.}\tablefoottext{b}{Fitted relationship $\log{l^{depr}}\,=\,C + D \log{ n_{cl}}$.}\tablefoottext{c}{Fit performed to the median values of separations calculated in bins of width 0.3 dex.}\tablefoottext{d}{Fit performed using the entire set of separations}
}
\label{Tab:CoeffFitDensity}
\end{table*}

\subsection{Comparison with the Jeans fragmentation}
\label{sect:SectJeans}

In this section, we compare the observed distribution of core separations with the expectations of the Jeans fragmentation theory for a non-magnetized, homogeneous, isothermal and self-gravitating clump.  Under these conditions, there is a critical mass (and a corresponding length) above which the medium is not able to remain stable and therefore collapses \citep{Jeans1902}. These critical values depend only on the density, $n$, and temperature, $T$ (see Appendix\,\ref{AppendixC}). 
Although these conditions refer to an idealized case which may not be representative of real star-forming clumps, the analytical prescription given by Jeans theory offers a simplified model against which  the observed fragmentation in the ALMAGAL clumps can be compared. Specifically, we are looking for whether thermal fragmentation is sufficient to statistically describe the separations between cores or whether additional contributions, such as turbulent support, are needed. Since core masses are affected by higher uncertainties than core separations, we focused our analysis on the comparison between the latter and the thermal Jeans lengths, while we report the results for the masses in Appendix\,\ref{AppendixD}.

\subsubsection{The Jeans length and mass for ALMAGAL clumps}
\label{Sect:HiGAL_JeansLengthMass}

We computed the thermal Jeans length, $\lambda_{J}^{th}$, and mass, $M_{J}^{th}$, for the ALMAGAL clumps with Eq.\,\ref{Eq:JeansLenght} and Eq.\,\ref{Eq:JeansMass}, adopting as temperature the one of the dust, $T_{d}$, derived from {\it Herschel}/Hi-GAL data and reported in Table\,A.1 in \citet{Molinari2025}. At the typical densities of ALMAGAL clumps ($n\,>\,10^{4}$\,cm$^{-3}$), gas and dust are expected to be thermally coupled \citep{Goldsmith2001}. This has been observationally confirmed, with relatively small uncertainties discussed below, in a sample of 1086 Hi-GAL clumps \citep{Merello2019}. In these clumps, on average dust temperatures were found to be consistent with temperatures measured from NH$_{3}$ emission \citep{Merello2019}, considered a reliable temperature tracer for the gas phase \citep{Juvela2012, Guzmman2015}. The values of $\lambda_{J}^{th}$ and $M_{J}^{th}$ for each ALMAGAL clump are reported in Table\,\ref{Tab:Clumps}. The thermal Jeans length ranges from $\sim3500$ to $\sim60000$\,au ($\sim0.017-0.3$\,pc), with a median value of $\sim14500$\,au ($\sim0.07$\,pc), and Jean masses  from $\sim0.2$ to 14\,M$_{\odot}$, with a median equal to 0.89\,M$_{\odot}$. 

The uncertainties of these quantities are determined by those on the temperature and density, which depend on both their direct measurements and on the assumptions adopted for their estimates. 
The fit to the {\it Herschel} data typically provides small uncertainties for $T_{d}$, quantified as $\sim2$\,K \citep{Elia2017, Svoboda2019}. This value translates into a different contribution for the young and evolved systems of the ALMAGAL sample, characterized by a typical temperature of $\approx10$ and $\approx40$\,K, respectively.  The uncertainties in $\lambda^{th}_{J}$ and $M^{th}_{J}$ vary from 2\% to 10\% and 6\% to 30\%, with the highest values for younger systems. 
Although the dust and gas temperature are found to be in agreement on average, their relationship presents a measurable scatter whose amplitude quantifies how well we can assess the thermal coupling between these two phases. Typical discrepancies between $T_{g}$ and $T_{d}$ are of the order $\sim25$\% (see Fig.\,2 in \citet{Merello2019}), where the distribution of $\log({T_{g}/T_{d}})$ has a width of $\sim$0.2 dex. These discrepancies introduce an additional random contribution of $\sim13$\% and $\sim38$\% to the uncertainties in $\lambda^{th}_{J}$ and $M^{th}_{J}$, respectively. In addition to the random scatter, $T_{d}$ and $T_{g}$ show a systematic offset in different stages of the clump evolution, which may lead to an underestimation or overestimation of $\lambda^{th}_{J}$ and $M^{th}_{J}$. Ammonia-derived gas temperatures are found to be systematically $\sim20\%$ higher than $T_{d}$ in young systems such as prestellar clumps \citep{Merello2019}, where star formation activity is absent, while no significant offset is found in evolved systems. However, those evolved phases also have other temperature diagnostics as additional molecular tracers become available. The gas temperatures $T_{g}$ estimated from these diagnostics are generally higher than $T_{d}$ derived from the fitting of the spectral energy distribution of the {\it Herschel} data \citep{Molinari2016,Lin2022}.
Specifically, \citet{Molinari2016} estimated gas temperatures from CH$_{3}$C$_{2}$H(2-1) emission finding that they are about $\sim40$\% higher than dust ones. Similar discrepancies were also found by \cite{Lin2022}, who instead analyzed the CH$_{3}$CN and CH$_{3}$OH emission lines in a few clumps with $L/M\,\geq\,10$\,L$_{\odot}/$M$_{\odot}$, although they also report extreme cases associated with \textsc{Hii} regions, where estimated $T_{g}$ are as high as $\sim130$\,K which is a few times higher than $T_{d}$. 

The discrepancy between gas and dust temperature may be explained by different reasons. Although fitting the dust continuum provides an accurate determination of $T_{d}$, this emission may include contributions from the foreground or background. In addition, it is not always clear whether the molecular line emission probes the entire clump or only a fraction of it \citep{Mininni2025}, in particular in more evolved systems where temperature gradients are certainly set by the hosted protostars. However, the systematic difference between $T_{d}$ and $T_{g}$ indicates that the clump-averaged $\lambda^{th}_{J}$ and $M^{th}_{J}$ may be different from our estimates in different evolutionary stages. Nonetheless, the expected average increase of these quantities does not exceed $\sim20$\% and $\sim40$\% for $\lambda^{th}_{J}$ and $M^{th}_{J}$, respectively. At the same time, we consider  an additional random contribution of $\sim10$\% and $\sim30$\% to the uncertainties in $\lambda^{th}_{J}$ and $M^{th}_{J}$ induced by the typical scatter of $\sim20$\% between the two estimates $T_{g}$ and $T_{d}$.

Unlike the case of temperature, the average volume density $n_{c}$ is not directly measured, but is typically inferred from measurements of the clump mass and radius. Assuming an uncertainty of 50\% in $M_{cl}$, 20\% in the measured source size and 15\% in the clump distance, the uncertainty in $n_{cl}$ is about $\sim60$\%, which results in $\sim30$\% for $\lambda^{th}_{J}$ and $M^{th}_{J}$.
These estimates are obtained by assuming a uniform distribution of the material, which may not be realistic due to the density stratification present in clumps \citep{Beuther2002, Williams2005, Palau2014,Beuther2024}. In these systems, the density profile has a typical power-law shape, i.e., $\rho = \rho_{0} \left ( \frac{R}{R_{0}} \right )^{-p}$ with $\rho_{0}$ and $R_{0}$ the density at the reference radius $R_{0}$. The local density in the inner region of size $R'$ is therefore enhanced by a factor
$\left ( \frac{R'}{R_{0}} \right )^{-p}$, which depends on the value of $p$. 
Observations of high-mass clumps indicate that $p$ varies in the interval $1\,<\,p\,<\,2$, with an average value of $p\approx1.6$ \citep[][see also the discussion in \citealt{Beuther2025} for a summary of the recent measurements]{Beuther2002, Williams2005, Palau2014, Beuther2024}. The observed variation appears to be connected to the clump evolutionary state. In fact, \citet{Lin2022} reported an increase from $p\approx1$ to $\sim1.5$ in their sample, and \citet{Gieser2022} found slightly shallower profiles with $p\approx1.6$ in the younger regions, whereas they also measured $p\,=\,2.0\pm0.2$ in the evolved massive clumps of the CORE surveys \citep{Gieser2021,Beuther2024}.

In principle, the density profile and the exponent $p$ could be constrained for each clump by fitting the Hi-GAL observations. However, this would require a careful and detailed radiative transfer modeling to properly account for the temperature structure, which is beyond the scope of this work. Our aim here is to quantify the amplitude of the uncertainty that the assumed density structure introduces into our estimates for Jeans properties. In the case of ALMAGAL clumps, fragmentation is effectively taking place in a region defined by the cluster radius, $R^{ph}_{cluster}$ (see Sect. \ref{Sect:SpatialDistributionALMAGAL} and Table\,\ref{Tab:FieldDistribution}). Due to density stratification, the average density is higher in the central regions, which results in a decrease for $\lambda'^{th}_{J}$ and $M'^{th}_{J}$ by a factor $C^{\frac{p}{2}}$, where $C= \left ( R^{ph}_{cluster} / R_{c} \right )$. The median value of $C$ is $\sim0.46$, with quartiles of $\sim0.36$ and $\sim0.59$, which translate into an expected $\lambda_{J}^{th^{'}} = 0.68_{-0.1}^{+0.09} \times \lambda_{J}^{th}$, for the young systems with $p\approx1$, and $\lambda_{J}^{th^{'}} = 0.46_{-0.1}^{+0.13} \times \lambda_{J}^{th}$ for evolved systems where $p\approx2$. 

In general, our assumptions in $T_{g}$ and $n_{c}$ introduce systematic shifts in the average estimates of $\lambda_{J}^{th}$ and $M_{J}^{th}$, that go in opposite directions. The amplitudes of these shifts are similar, and, on average, compensate; therefore, we did not apply any correction. However, since their contribution leads to an increase in the spread of the uncertainty, in our analysis we considered a typical random error on $\lambda_{J}^{th}$ of at least $\sim50$\%.

\subsubsection{ALMAGAL core separations compared to Jeans length}
\label{sect:ComparisonSeparationJeansLength}

\begin{figure}
    \centering
    \includegraphics[width=1\linewidth]{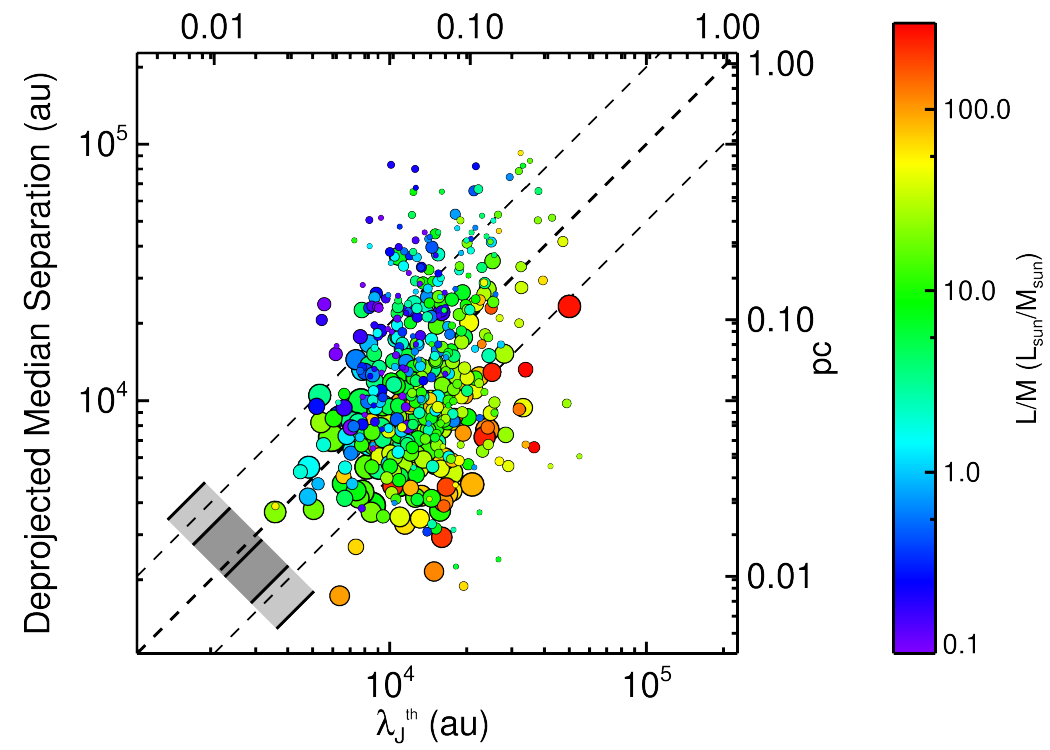}
    \includegraphics[width=1\linewidth]{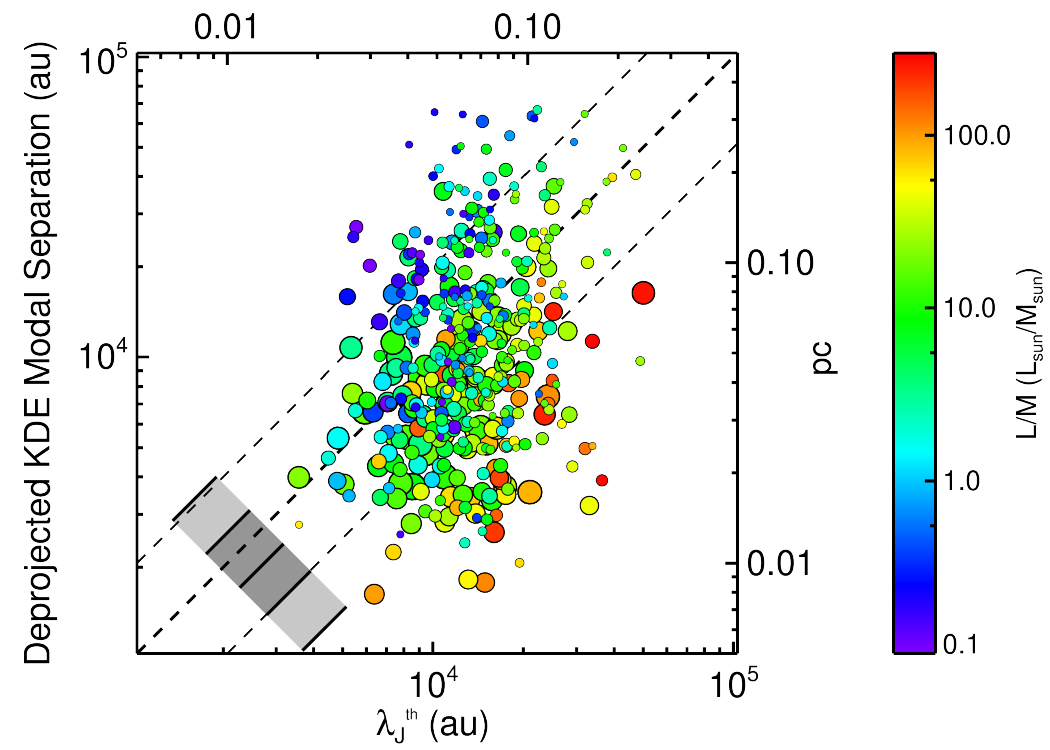} 
    \caption{Comparison between the clump-averaged estimators for the core separations, specifically the median (\textit{top panel}) and mode (\textit{bottom panel}) of the distribution in each clump, and the computed Jeans length. The fragmentation level in the clump is indicated by the symbol sizes, with increasing size going from low ($4\,\leq\,N_\mathrm{cores}\,<\,8$) and intermediate ($8\leq N_\mathrm{cores}<16$), up to high  ($N_\mathrm{cores}\geq16)$ core numbers. 
    The dashed lines are drawn as a reference and correspond to the separation equal to $\lambda_{J}^{th}$ (thick line) and 2$\times$ and 0.5$\times$ $\lambda_{J}^{th}$ (thin lines), respectively. The gray shaded boxes indicate the amplitude of the distributions of the ratio $l^{depr}_{i}$/$\lambda_{J}^{th}$ (see text), and correspond to the interquartile range (dark gray) and the 10th to 90th percentile range (light gray), respectively. }
    \label{fig:AveragedSeparationsJeansLengh}
\end{figure}

Figure\,\ref{fig:AveragedSeparationsJeansLengh} shows the comparison of the two clump-averaged estimators for the core separation, the median $l^{median}$ and the mode $l^{mode}$  (see Sect.\ref{sect:distributionSeparations}), with the estimated thermal Jeans length, $\lambda_{J}^{th}$. 
In these plots, we mark the level of fragmentation and $L/M$ of the clump with the symbol size and the color, respectively. A large fraction of ALMAGAL clumps have core separations that are consistent with the thermal Jeans length $\lambda_{J}^{th}$. The mode [median] separations range from $\sim0.07\,[0.1]\times$ to $\sim6.5\,[11.0]\times$ $\lambda_{J}^{th}$. 
The distributions of the ratio $l^{median}/ \lambda^{th}_{J}$ and $l^{mode} / \lambda^{th}_{J}$ are centered on values close to 1 and cover the narrow intervals included in Fig.\,\ref{fig:AveragedSeparationsJeansLengh} with dark and light gray shaded areas representing the interquartile and the 10th to 90th percentile interval, respectively. Quantitatively, the median and the mode are centered around 0.91$\times$ and 0.76$\times$ $\lambda_{J}^{th}$, respectively. The first and third quartiles are equal to $0.56\times-1.60\times$ and $0.5\times-1.26\times$ $\lambda_{J}^{th}$ for the median and the mode, respectively. The 10th-90th percentile ranges extend from 0.29$\times$ to $2.10\times$ for the mode and $0.36\times$ to $2.53\times$ $\lambda_{J}^{th}$ for the median. Both the clump-averaged median and mode reveal a similar behaviour with the thermal Jeans length.

\begin{figure*}
    \centering
    \includegraphics[width=0.97\linewidth]{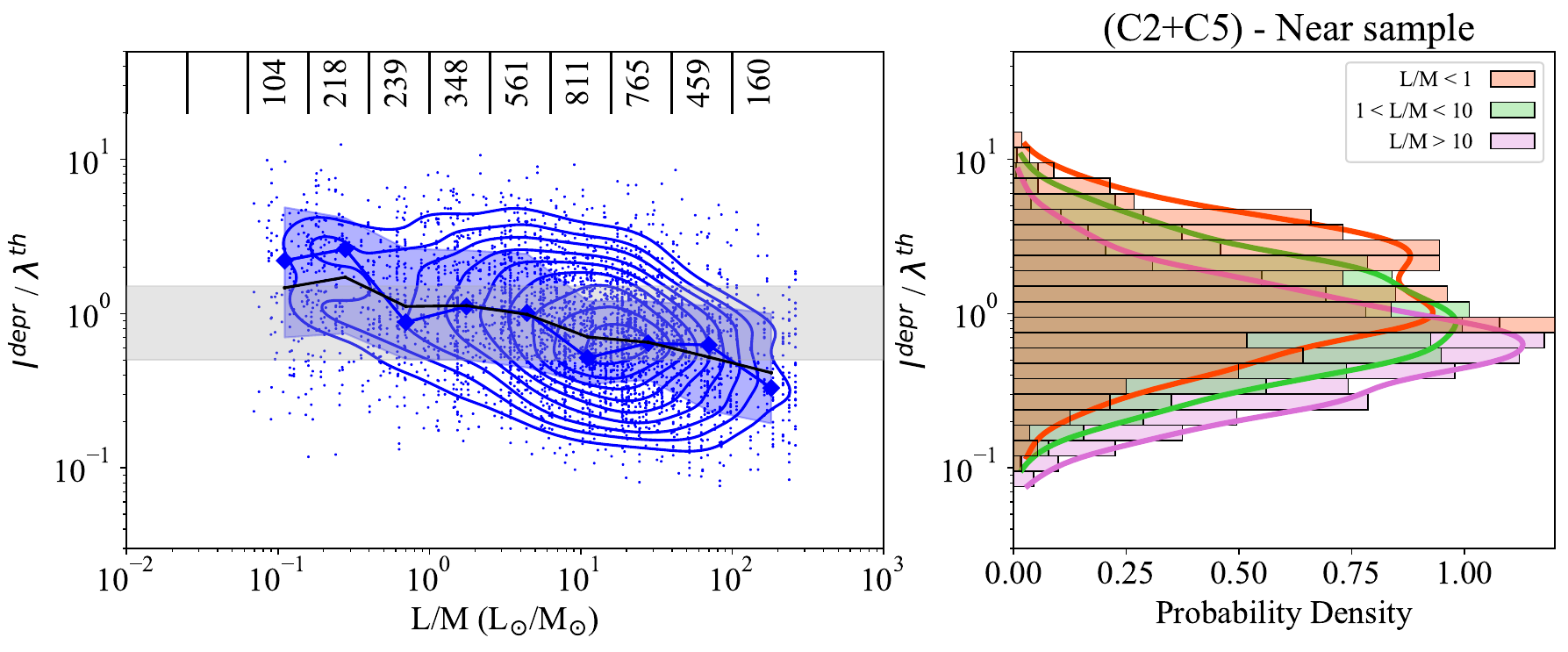}
    \includegraphics[width=0.97\linewidth]{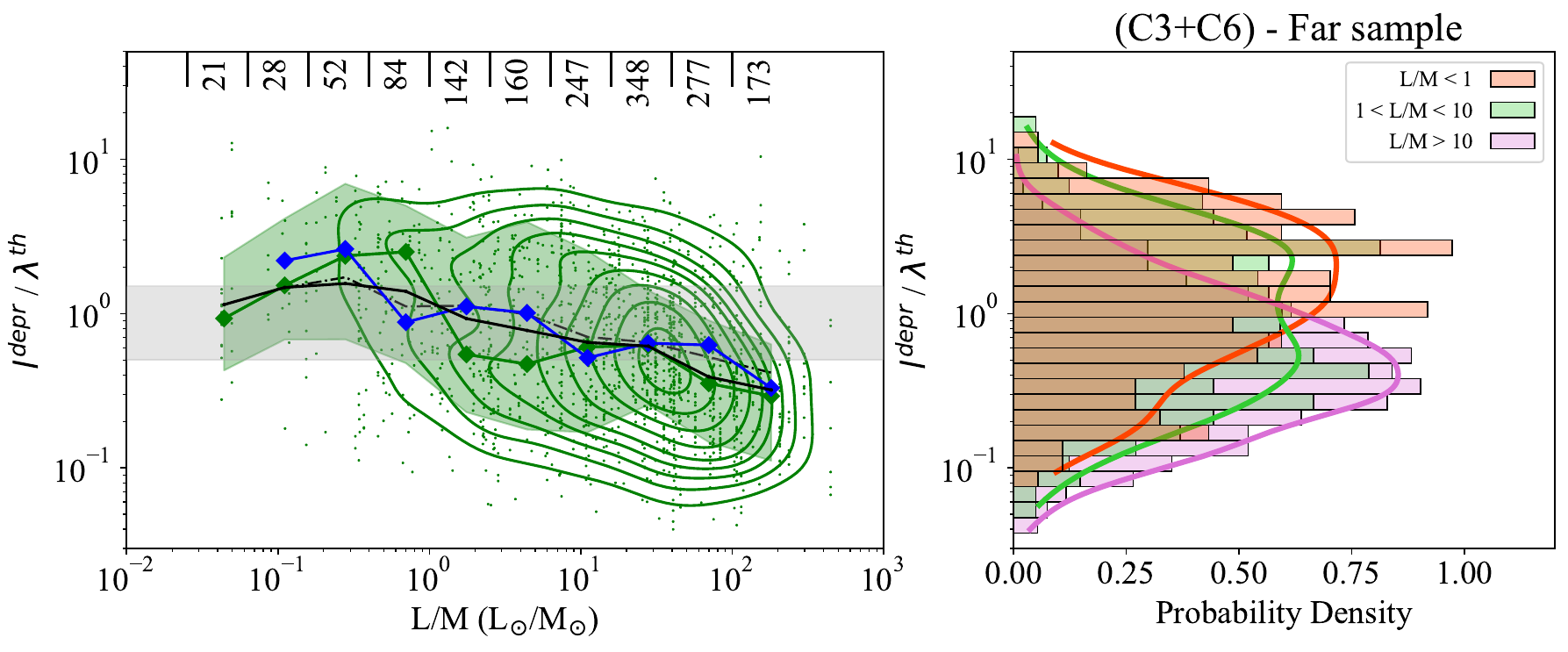}
    \caption{\textit{Left panel}: Comparison of the ratio $R$ of the deprojected separations to $\lambda^{th}_{J}$ and the clump $L/M$ ratio for the near (\textit{top panel}) and far sample (\textit{bottom panel}), respectively. The black dashed line shows the median of the distributions of $R$ computed in bins of $L/M$ equally spaced in logarithm of width 0.4 dex.  The thick colored lines and the colored area trace the mode and the spread (defined as the width at 68\% of the peak) of these distributions, adopting the blue and green colors for the near and far sample, respectively. The numbers at the top give the size of the subsamples adopted to compute the distribution in each bin. The gray shaded area shows the interval $0.5\,\leq\,R\,\leq1.5$, corresponding to an uncertainty on the $\lambda_{J}^{th}$ of a factor of 2. The colored contours trace, from the outside inward, fractions of the sample lying outside that rise from 15\% to 95\% with a step of 10\% for the near ({\rm in blue}) and the far ({\rm in green}) subsample, respectively. Right panel: Normalized density histogram and estimated probability density distribution estimated with the KDE method (colored continuous lines) of the ratio $R\,=\,l_{i}/\lambda^{th}_{J}$ measured in the three subsamples selected by different intervals of $L/M$ tracing clumps in different evolutionary phases: young  ($L/M\,\leq\,1$\,L$_{\odot}$/M$_{\odot}$ in orange), intermediate ($1\,\leq\,L/M\,\leq\,10$\,L$_{\odot}$/M$_{\odot}$ in green), and evolved ($L/M\,\geq\,10$\,L$_{\odot}$/M$_{\odot}$ in purple).    
    }
    \label{fig:SeparationsJeansLengh}
\end{figure*}

Although on average the core separations are in agreement with the thermal Jeans length, we observe that the systems with different $L/M$ are distributed in different regions of Fig.\,\ref{fig:AveragedSeparationsJeansLengh}, indicating that the ratio $R =l^{depr}/\lambda^{th}_{J}$ varies with $L/M$ and, therefore, changes during the clump evolution. The trend is shown in Fig.\ref{fig:SeparationsJeansLengh}, where we present a panoramic view of how the ratio $R$ varies with $L/M$ for all separations of the ensemble. We found a decreasing trend which is the result of two effects: the decrease in the core separations $l^{depr}$ discussed in Sect.\,\ref{sect:SeparationEvolutionLM}, and a steady increase in $\lambda_{J}^{th}$ with $L/M$. We note that the increase in $\lambda_{J}^{th}$ is mainly driven by the general warming of the clump material by the radiation of newly born stars. Although it also depends on density $n_{cl}$, this has a marginal impact when averaging over bins of $L/M$ ratio. In fact, our sample includes clumps with density spanning over the entire range of values in each of those bins (see Fig.\ref{Fig:ClumpProperties}). More generally, the selected ALMAGAL clumps do not show an increase in their density with evolution (see Sect.\,\ref{Sect:ClumpProperties}).

Given the uncertainties in $\lambda_{J}^{th}$, our result is that the observed separations are consistent with the thermal Jeans length computed with the average properties derived from {\it Herschel} observations for all clumps with the $L/M$ ratio between 1 and 100\,L$_{\odot}/$M$_{\odot}$. Figure\,\ref{fig:SeparationsJeansLengh} shows an average systematic decrease in $R$ with $L/M$ and that the core separations become significantly smaller than the thermal Jeans lengths in the evolved systems with $L/M\,\gg\,50$\,L$_{\odot}/$M$_{\odot}$ ($L/M\,\gg\,100$\,L$_{\odot}/$M$_{\odot}$ in the case of near sample). This trend for $L/M\gtrsim1$\,L$_{\odot}/$M$_{\odot}$ is mainly driven by an increase in $\lambda_{J}^{th}$ and not by a decrease in the core separations. The observation of sub-Jeans separations in these evolved stages does not rule out gravity-dominated thermal fragmentation. In these phases, protostellar feedback could create local density enhancements that exceed the mean clump density used for our $\lambda_{J}^{th}$ calculation. Therefore, Jeans-like fragmentation of these locally denser regions would naturally produce cores with separations smaller than our clump averaged estimates. 

It is interesting to look in detail at the distribution of $R$ for young and poorly fragmented clumps with $L/M\,<\,0.5-1$\,L$_{\odot}/$M$_{\odot}$. The distributions have a median $R\sim1.5$, but the shape of the density contours suggests a bimodality, which we confirmed with the statistical tests described in Sect.\,\ref{sect:SeparationEvolutionLM}. Although the two peaks for the far sample systems are connected to the bimodal distribution of $l_{i}$, they are also present in the near systems. One of the two peaks of the distributions is centered at $R\sim1$, indicating a population of cores which are separated by a thermal Jeans length, while the second peak is shifted at higher values, which are centered at about $R\sim3$.

\subsection{How are cores distributed in the fields?}
\label{Sect:DiagnosticsFragmentation}

In previous sections, we analyzed where cores are located in star-forming clumps and their relative separations to discuss how fragmentation takes place in potentially cluster-forming systems.  
How cores are distributed in the systems and the identification of potential different mass-dependent clustering of the cores offer additional information on the cluster formation process. 
To such aim, we analyzed two additional diagnostics, the $Q$ parameter and the mass segregation ratio, $\Lambda_{MSR}$, which are typically used to characterize young proto-clusters \citep{Allison2009, Cartwright2004, Parker2018, Dib2018}, and were also recently estimated for groups of cores \citep{Dib2019, Sanhueza2019, Sadaghiani2020, Avison2023, Morii2024, Xu2024}.

\subsubsection{The $Q$ Parameter}
\label{Sect:Qparameter}

\begin{table*}[]

\begin{center}
\caption{Diagnostics of spatial statistics in ALMAGAL clumps.}
\begin{tabular}{cccccccccccc}
\hline
ALMAGAL ID       & N & $Q$-par\tablefootmark{a} & $\overline{m}$\tablefootmark{a} & $\overline{s}$\tablefootmark{a} & $N^{crit}_{MSR}$\tablefootmark{b} & $\Lambda^{max}_{MST}$\tablefootmark{b} & $N^{max}_{MST}$\tablefootmark{b} & MSI\tablefootmark{c}   & MS Flag\tablefootmark{d} & Prob1\tablefootmark{e} & Prob2\tablefootmark{e} \\
\hline
AG010.2283-0.2064 & 9      & 0.78 & 0.38     & 0.49    & -        &         & -        & 1.154 &            &       &       \\
AG010.3226-0.1599 & 11     & 0.71 & 0.42     & 0.59    & -        & -       & -        & 1.011 &            &       &       \\
AG011.9039-0.1403 & 11     & 0.65 & 0.46     & 0.70    & 5        & 3.50    & 2        & 1.854 & Rob MS     & 1.000 &       \\
AG011.9183-0.6122 & 13     & 0.72 & 0.36     & 0.51    & -        & -       & -        & 1.223 &            &       &       \\
AG011.9370-0.6160 & 22     & 0.85 & 0.39     & 0.46    & 3        & 2.71    & 3        & 1.368 & Pos MS     &       &       \\
AG012.2087-0.1017 & 23     & 0.80 & 0.50     & 0.63    & -        & -       & -        & 1.574 &            &       &       \\
AG012.6794-0.1830 & 8      & 0.61 & 0.46     & 0.76    & 2        & 0.57    & 2        & 0.799 & Pos Inv MS &       &       \\
AG012.7360-0.1031 & 11     & 0.67 & 0.47     & 0.70    & -        & -       & -        & 1.183 &            &       &       \\
AG012.8535-0.2265 & 11     & 0.70 & 0.57     & 0.82    & -        & -       & -        & 1.170 &            &       &       \\
AG012.8887+0.4890 & 15     & 0.70 & 0.38     & 0.54    & 4        & 6.37    & 2        & 2.209 & Rob MS     & 1.000 &       \\
AG012.9008-0.2404 & 10     & 0.62 & 0.60     & 0.95    & 2        & 2.58    & 2        & 1.218 & Rob MS     & 0.964 &       \\
AG012.9048-0.0306 & 14     & 0.60 & 0.36     & 0.61    & 2        & 0.50    & 2        & 0.862 & Rob Inv MS &       & 0.992 \\
AG012.9084-0.2604 & 23     & 0.89 & 0.68     & 0.77    & 4        & 2.91    & 4        & 1.484 & Rob MS     & 0.948 &       \\
AG012.9156-0.3341 & 8      & 0.75 & 0.49     & 0.66    & 4        & 3.02    & 4        & 2.060 & Rob MS     & 0.930 &       \\
\hline
\end{tabular}
\tablefoot{
\tablefoottext{a}{Mean edge length of cluster MST, $\overline{m}$; the cluster correlation length, $\overline{s}$; and the $Q$ parameter defined as $Q\,=\,\overline{m}/\overline{s}$, following \citet{Cartwright2004}.}\tablefoottext{b}{Quantities of the mass segregation ratio (MSR) profile: maximum number of cores for which MSR profile is different from unity, $N^{crit}_{MST}$; the MSR maximum value, $\Lambda^{max}_{MSR}$; and its corresponding number of cores, $N^{max}_{MST}$. These quantities are measured only in the 110 systems where the mass segregation (MS) is assessed.}\tablefoottext{c}{Mass segregation integral, MSI, defined by \citet{Xu2024}.}\tablefoottext{d}{ 
Flags reporting whether a signature of direct (Pos MS) or inverse (Pos Inv MS) mass segregation is found and if it is robust (additional flag Rob) against the Monte Carlo simulations described in the text.}\tablefoottext{e}{Probability of presence of of direct (Prob1) or inverse (Prob2) mass segregation evaluated with Monte Carlo simulations.}
The complete version of this table is available at the CDS.
}
\label{Tab:Q_ClumpParameters}
\end{center}
\end{table*}

Statistical analysis provides several diagnostics capable to characterize the spatial distribution of young stellar groups. \citet{Cartwright2004} introduced the $Q$ parameter as a tool to quantify the degree of substructures in the distribution of the members of young stellar clusters and, if present, to identify cases where their radial density profiles are centrally concentrated \citep{Cartwright2004, Parker2015, Jaffa2017}. The $Q$ parameter is defined as the ratio of the mean edge length of the cluster MST, $\overline{m}$, and the correlation length of the cluster, $\overline{s}$, both with normalization factors defined to remove any \text{dependence} on the number of members and the size of the cluster. In formula these are 

\begin{equation}
\overline{m} = \sum_{i=1}^{Nc-1} \frac{l_{i}}{\sqrt{N_{cores}\pi R_{cluster}^2}}, 
\end{equation}

\noindent where $l_{i}$ are the MST edges and $R_{cluster}$ is the size of the cluster defined as in Sect.\,\ref{sect:CenterField} and

\begin{equation}
\overline{s} = \frac{ \overline{l_{i}} } {R_{cluster}},
\end{equation}

\noindent with $\overline{l_{i}}$ defined as the average separation between the cluster members. The ratio $Q\,=\,\overline{m}/\overline{s}$ allows us to distinguish between two specific spatial configurations, one where the members are radially distributed with a central concentration and the other where they have fractal type of subclustering. \citet{Cartwright2004} showed that systems with members distributed in smooth and centrally concentrated patterns have $Q\,>\,0.8$, with values that increase for higher concentrations represented by higher exponents $p$ in the number density profile $N(r)\propto r^{-p}$. In contrast, systems whose members are organized in a fractal type of subclustering, described with a fractal dimension $D$, have $Q\,<\,0.8$, and smaller values of $Q$ correspond to systems with a higher level of subclustering (lower values of $D$). 

Although this diagnostic is typically used in the statistical analysis of young star clusters \citep{Parker2014,GregorioHetem2015, Dib2018a}, it has only recently been applied to the distribution of star-forming cores, finding, for example, that $Q$ correlates with star formation activity \citep{Dib2019}. Specifically, $Q$ increases with the density of star formation rate in low-mass star-forming clouds \citep{Dib2019}. Moreover, 
the comparison between the core population observed by the ASHES \citep{Sanhueza2019, Morii2024} and ASSEMBLE surveys \citep{Xu2024} shows that different spatial arrangements are found depending on the evolutionary stage of the clump. Most of the ASHES clumps ($\sim70$\%), that are young and with $L/M\,\lesssim\,2$\,L$_{\odot}/$M$_{\odot}$, have $Q\lesssim0.8$, which characterizes subclustered systems. In contrast, higher values of $Q$ are typically measured in the more evolved clumps of the ASSEMBLE survey that have $L/M\gtrsim10$\,L$_{\odot}/$M$_{\odot}$ \citep{Xu2024}. The presence of a weak correlation between the $Q$ parameter and the ratio $L/M$ has been proposed to reflect a time evolution effect, where the signature of a gradual transition from a subclustered pattern to a centrally condensed configuration is a consequence of gravitational collapse \citep{Xu2024}. We tested the statistical robustness of the correlation $Q$ versus $L/M$
with the ALMAGAL sample, which largely extends the sample size of the ASHES and ASSEMBLE surveys from which it was initially derived. 

The $Q$ parameter measured in the ALMAGAL clumps with $N_{cores}\geq4$ shows values ranging from 0.4 to 0.95, with a median equal to 0.65, and half of the systems having 0.6\,<\,$Q$\,<\,0.75. These values would suggest that the ALMAGAL sample is composed mainly of clumps where the cores are distributed in subclusters, if the classification criteria introduced by \citet{Cartwright2004} were applied. However, recently \cite{Avison2023} questioned the diagnostic usefulness of the $Q$ parameter when applied to a small number of sources ($N_{cores}\,<\,10$). They simulated clusters composed of small numbers of members ($N_{cores}\,=\,5$ and 10) to characterize the observations of the TEMPO survey, where they found between 2 and 15 fragments \citep{Avison2023}. Most of the ALMAGAL clumps with these conditions ($N_{cores}\,<\,10$) would be classified as highly subclustered as $\sim80$\% of them have $Q\leq0.6$. To quantitatively assess whether the $Q$ parameter can be used as a useful diagnostic in the ALMAGAL fields that have a wider interval of the number of cores than TEMPO survey ($N_{cores}$ from 2 to 49), we extended the analysis of \citet{Avison2023}. This analysis is presented in Appendix\,\ref{AppendixE} and is also applicable to the ASHES and ASSEMBLE fields, which host 8 to 37 and 13 to 37 fragments, respectively \citep{Morii2024,Xu2024}.

Our result is that the $Q$ parameter retains a diagnostic power in probabilistic terms only for systems with $N_{cores}\,\geq\,10$. We defined two intervals of $Q$ where it is most likely that the cores follow a centrally concentrated ($Q\gtrsim0.8$ for high power-law exponent $p$) or subclustered distribution ($Q\lesssim0.65$, lower fractal dimension ), with a value of $Q\,=\,0.75$ that approximately separates these two regimes. These results revise the initial impression on the ALMAGAL systems and indicate that both spatial distribution patterns are present in the 226 ALMAGAL systems with $N_{cores}\,\geq\,10$. In fact, adopting the tentative limit of $Q=0.75$, there are 155 fields ($\sim68$\%) with $Q\,<\,0.75$ and 71 fields ($\sim32$\%) with $Q\,\geq\,0.75$, and similar proportions are found if higher cuts on $N_{cores}$ are applied. However, our simulations show that the statistical distributions of $Q$ for the two classes are not strongly separated, but they instead overlap. This led us to assess the classification with sufficient reliability only for the 60 fields where $Q\,<\,0.65$ and for the 38 where $Q\,>\,0.8$. Although we reduce the sample size with this choice, we were still able to confirm that both patterns are observed in massive dense clumps. 

\begin{figure}
    \centering
    \includegraphics[width=1\linewidth]{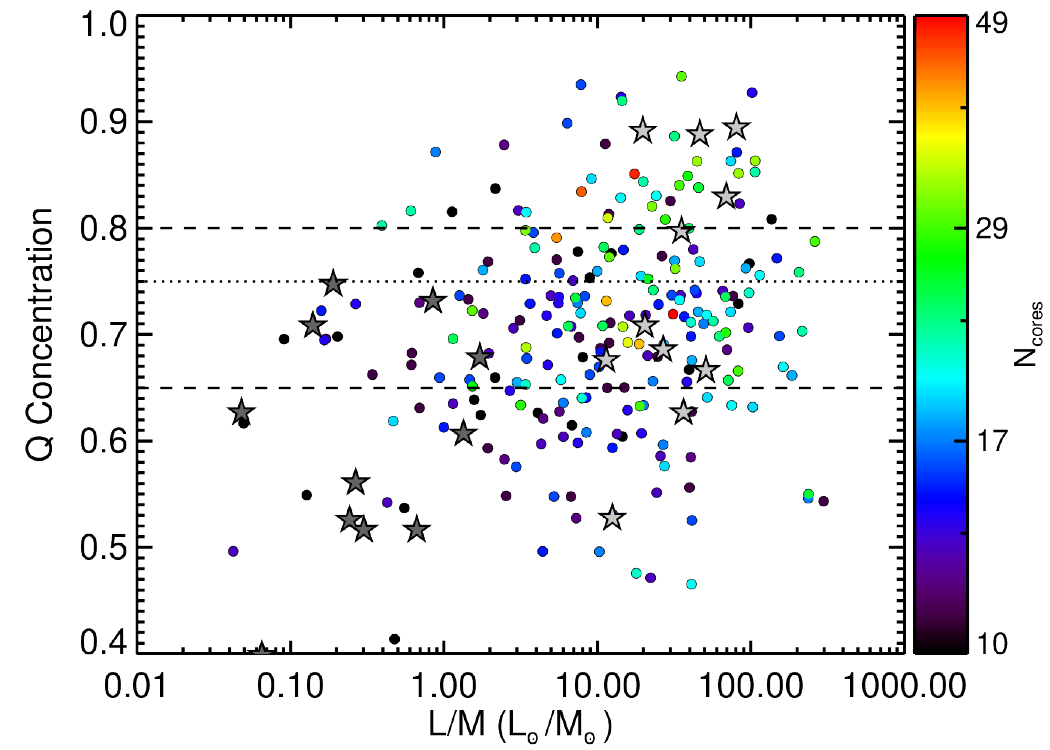}
    \caption{Relation between the $Q$ parameter and the $L/M$ ratio of ALMAGAL clumps which host at least ten cores. The two dashed lines values $Q\,=\,0.65$ and $Q\,=\,0.8$ defining the intervals for which it is most likely that the hosted cores follow a fractal type of subclustering ($Q\,\leq\,0.65$) and radial distribution with a central concentration ($Q\,\geq\,0.8$), as indicated by the Monte Carlo simulations in Appendix\,\ref{AppendixE}. The dotted line points to the $Q\,=\,0.75$, which separates the two regimes with a lower statistical significance. Estimates from the ASHES  \citep[dark gray;][]{Sanhueza2019} and ASSEMBLE surveys \citep[light gray;][]{Xu2024} are included, and are shown as star markers. 
    }
    \label{Fig:QparLoverM}
\end{figure}

We show in Fig.\,\ref{Fig:QparLoverM} the values of the $Q$ parameter as a function of the clump $L/M$ ratio for the 226 ALMAGAL fields with $N_{cores}\,\geq\,10$, and we included  measurements from the ASHES pilot \citep{Sanhueza2019} and the ASSEMBLE survey \citep{Xu2024}. The 39 fields with $L/M\leq2$\,L$_{\odot}$/M$_{\odot}$ expand the sample provided by the ASHES survey for the early evolutionary phases \citep{Sanhueza2019, Morii2024}, and indicate that these early phases have cores that are mainly distributed in fractal-type subclusters, with 13 fields showing $Q\,<\,0.65$. Fields with a clear signature of a concentrated distribution are rare in these young clumps, and only four of them can be robustly classified as centrally concentrated ($Q\,>\,0.8$). Although the occurrence of systems with a more pronounced centrally concentrated distributions (i.e., $Q\,>\,0.8$) is certainly higher in the later stages of their evolution, mild or high subclustered distributions of cores are still  present at later stages, as $Q\,<\,0.65$ is found in 27 systems out of 124 with $L/M\,>\,10$\,L$_{\odot}/$M$_{\odot}$.

We do not find a significant relationship between $Q$ and $L/M$, on the contrary,  we observe a large scatter of $Q$ for any interval of $L/M$, which is not connected to the level of fragmentation or to other clump-averaged properties, for example the clump surface density. It is unclear what the reasons are behind this large scatter, which may not be only caused by statistics but also by additional factors not considered in our analysis, as, for example, the surrounding local environment. A weak dependence of $Q$ with the $L/M$ ratio may be identified, but the Pearson and Spearman correlation coefficients equal to $\rho_{p}$ = 0.18 and $\rho_{s}$ = 0.185 with $p$-values of 0.0067 and 0.0053 are significantly lower than the values reported by \cite{Xu2024}: $\rho_{p}$ = 0.60 and $\rho_{s}$ = 0.57. This indicates that if a transition between spatial distribution patterns exists in massive clumps, it may require additional factors than simply the internal clump evolution.

\subsubsection{Mass segregation in the ALMAGAL fields}

A further clue on the formation and early evolution process can be extracted by also taking into account the mass of the cores.
Observations of evolved star clusters reveal that they are often mass segregated, with massive objects that are more concentrated towards the cluster centers than the low-mass ones \citep{Hillenbrand1998}. This observational signature is often interpreted as a consequence of the dynamical evolution caused by close encounters between pairs of cluster members that lead the system to a global state of energy equipartition. As a result, the less massive cluster components rise to higher orbits around the gravitational center, whereas the more massive cluster components move more slowly, falling into lower orbits. However, young clusters may have been born with a primordial MS set by the formation process itself. In fact, primordial MS with respect to the center of the gravitational potential is a natural outcome of the competitive accretion model for high-mass stars \citep{Bonnell2006}. 

Evidence of MS has already been found in observations of young cluster systems in both high-mass and intermediate- to low-mass environments \citep{Chen2007,Gennaro2011,Parker2018,Plunkett2018,Dib2018a,Dib2019,Sadaghiani2020,Nony2021}. {This indicates that MS is not produced merely by long-term dynamical evolution, but it can be already established in the early phases. However, it remains unclear whether it is primordial, i.e., set in the initial formation stages, or the result of a rapid, early} dynamical evolution of the system of cores. For instance, weak or no MS signature is found in the 39 young clumps of the ASHES survey \citep{Sanhueza2019,Morii2023}. In contrast, supporting the idea that it develops over time, clear segregation is observed in the more evolved systems, with 8 of 11 clumps in the ASSEMBLE survey presenting clear signatures of MS \citep{Xu2024}. 

We searched for MS signature in ALMAGAL systems using the mass segregation ratio $\Lambda_{MSR}$ \citep{Allison2009,Parker2018}, which compares the total length of the MST of the $N_{MST}$ most massive cores with the typical one of the entire protocluster. The value of $\Lambda_{MSR}$ for $N_{MST}$ cores is given by

\begin{equation}
\Lambda_{MSR}(N_{MST}) = \frac{\left < l_{random} \right >}{l_{M}} \pm \frac{\sigma_{l,random}}{l_{M}},
\end{equation}

\noindent where $\left < l_{random} \right >$ is the average of the total length of the MST derived by randomly selecting a total of $N_{MST}$ cores from the cluster, while $l_{M}$ is the total length of the MST of the $N_{MST}$ most massive cores. We computed the average $\left < l_{random} \right >$ and the dispersion $\sigma_{l,random}$ by randomly selecting the cores 1000 times as in \cite{Morii2023} and \cite{Xu2024}. If MS is present, then the $N_{MST}$ most massive cores are more concentrated, and the total length of their MST would be shorter than the one built from randomly selected cluster members, which will give $\Lambda_{MSR}\geq1$. In contrast, $\Lambda_{MSR}$ would be $\approx1$ if the cores had a similar spatial distribution regardless of their mass. 

We adopted the core masses estimated from \citet{Coletta2025} and computed $\Lambda_{MSR}(N_{MST})$ in the 296 ALMAGAL fields hosting at least eight cores to limit the uncertainties due to their small number. This threshold corresponds to the minimum number of detections in the ASHES field for which the ratio $\Lambda_{MSR}$ was also estimated \citep{Morii2024}. 
If we do not take into account any uncertainty in the core mass, we find that a signature of mass segregation is revealed in 171 fields, which have a MSR profile that differs from unity for more than the random uncertainty $\sigma_{l,random}/l_{M}$ (it exceeds unity in 131 fields). 
We adopted a MonteCarlo approach to take into account the uncertainty on the core masses. For each system, we generated 500 unique realizations where every core mass was drawn from a Gaussian distribution centered on its value, M$_{core}$, with a standard deviation of 20\%. We computed the $\Lambda^{j}_{MSR}(N_{MST})$ with the relative error for each of these 500 synthetic realizations, fully mapping the range of possible outcomes given the assumed mass uncertainties. The fraction of realizations where $\Lambda^{j}_{MSR}(N_{MST})$ exceeds unity over the dispersion $\sigma_{l,random}$ defines the probability that mass segregation is present in the system. We classified a system as \text{robustly mass segregated} if this fraction is higher than $70\%$, a condition found in 110 fields ($\sim37$\% of the total).  This considerable fraction suggests that cores are already segregated in mass prior to dispersal of the gas present in the clump, and raises the question of whether segregation is set at the initial formation stage or is a consequence of the early evolution, which requires a comparative analysis between systems at different evolutionary stages. 

Given their large numbers, we do not show all MSR profiles. Instead, we characterized each profile with metrics indicating average value, peak, and extent, reporting those quantities in Table\,\ref{Tab:Q_ClumpParameters}. We included the mass segregation integral (MSI), $I^{MSR}_{\Lambda}$ introduced by \cite{Xu2024} to quantify the fractional area of the MRS profile that exceeds unity. This quantity allows the comparison of the MSR profiles of different systems and is defined as

\begin{equation}
I^{MSR}_{\Lambda} = \frac{\sum_{i=2}^{N^{max}_{MST}}\Lambda_{MSR,i} }{N^{max}_{MST}-1},
\end{equation}

\noindent where $\Lambda_{MSR,i}$ is the value of the profile $\Lambda_{MSR}(N_{i})$ for the most massive cores $N_{i}$. Morever, we determined the maximum of the profile, $\Lambda^{max}_{MSR}(N^{max}_{MST})$, and the largest number of cores, $N^{crit}_{MST}$, where it still remains above unity. These two quantities extend the information provided by the MSI, as they indicate how much more concentrated the most massive cores are than any other randomly selected group and the largest number of cores where the spatial distribution is found different.  

\begin{figure}
    \centering
    \includegraphics[width=1\linewidth]{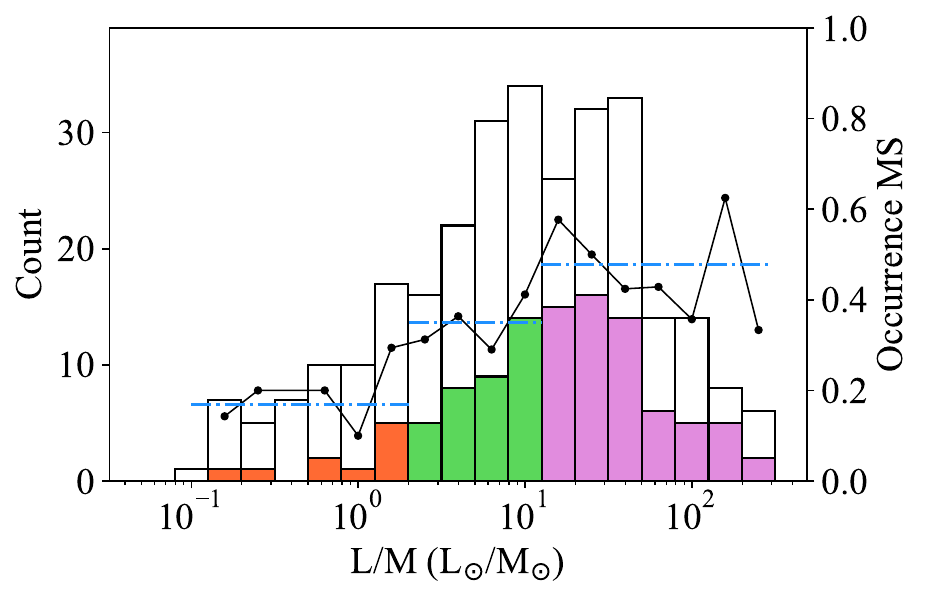}
    \caption{Distribution of the ALMAGAL clumps hosting at least 10 cores in different bins of the ratio $L/M$. The colored histograms show the systems were a signature of MS is found, with different colors referring to the different intervals of $L/M$ ratio, i.e., orange $L/M\,\leq\,2$\,L$_{\odot}$/M$_{\odot}$, green $2\,<\,L/M\,\leq\,10$\,L$_{\odot}$/M$_{\odot}$ and purple $L/M\,>\,10$\,L$_{\odot}$/M$_{\odot}$. The solid black line, which refers to the right y-axis scale, indicate the relative fraction of systems with MS with respect to the observed ones in bins of $L/M$, while the dot-dashed line shows the average occurrence fraction of MS systems in the three intervals of $L/M$.}
    \label{Fig:FractionMassSegregation}
\end{figure}

Figure\,\ref{Fig:FractionMassSegregation} shows the distribution of clumps with $N_{cores},\geq,8$ as a function of the ratio $L/M$ for systems with and without MS signature, indicating how frequently MS is found during the entire evolutionary sequence of massive clumps. We divided the sample of 296 clumps into three different groups based on the $L/M$ ratio: $L/M\,<\,2$\,L$_{\odot}/$M$_{\odot}$, $2\,<\,L/M\,\leq\,10$\,L$_{\odot}/$M$_{\odot}$, and $L/M\,>\,10$\,L$_{\odot}/$M$_{\odot}$. The first and latter intervals are also sampled by the targets of the ASHES and ASSEMBLE surveys, allowing us to extend their results with larger sample sizes. These groups consist of 59, 83, and 154 clumps and have an average number of cores equal to 12, 15 and 17, respectively. The 110 systems where mass segregated is robustly confirmed divide in 10, 25, and 75 systems in each of these three groups. All of those systems have a different MSI distribution, and we rejected the null hypothesis that they derive from the same underlying one with the Kolmogorov-Smirnov and Mann-Whitney U tests with confidence levels set by $p_{KS}\,<\,0.029$ in the two cases with $L/M\,<\,2$\,L$_{\odot}$/M$_{\odot}$ and $2\,<\,L/M\,<\,10$\,L$_{\odot}$/M$_{\odot}$, and with a more robust assessment of $p_{KS},p_{U}\lesssim10^{-5}$ in the more numerous group with $L/M\,>\,10$\,L$_{\odot}$/M$_{\odot}$.

The occurrence of MS in the young clumps of the ALMAGAL sample ($\sim17\%$ with 10 out of 59) is similar (and compatible within the uncertainty of the counting statistic) to the one measured in the ASHES sample (13\% with 5 out of 39). These systems have a $\Lambda_{MST}^{max}$ between $\sim2.5$ and $\sim7$ with an average value of $\sim5$, indicating that when MS is present, even in these early phases, it usually has a moderate to high impact on the distribution of the cores. One peculiar system shows a very high value of $\Lambda_{MST}^{max}\sim20$, but visual inspection of the system indicates that it is closely associated with multiple mid-infrared bright sources and, thus potentially affected by the strong feedback of nearby massive young stellar objects. Since there is only one case, we consider it an outlier. The fraction of clumps with MS steadily increases in more advanced stages of evolution. It first rises to 30\% (25 out of 83) and then reaches $\sim49$\% (75 out of 154) for $2\,<\,L/M\,<\,10$\,L$_{\odot}$/M$_{\odot}$ and $L/M\,>\,10$\,L$_{\odot}$/M$_{\odot}$, respectively. An indication of this effect was already suggested by \citet{Xu2024} who found that 8 out of 11 clumps  with $L/M\,>\,10$\,L$_{\odot}$/M$_{\odot}$ are mass segregated, but the large sample size provided by the ALMAGAL large program allows us to robustly and clearly assess this trend, also measuring the increase for intermediate evolutionary stages. 

\begin{figure}
    \includegraphics[width=0.99\linewidth]{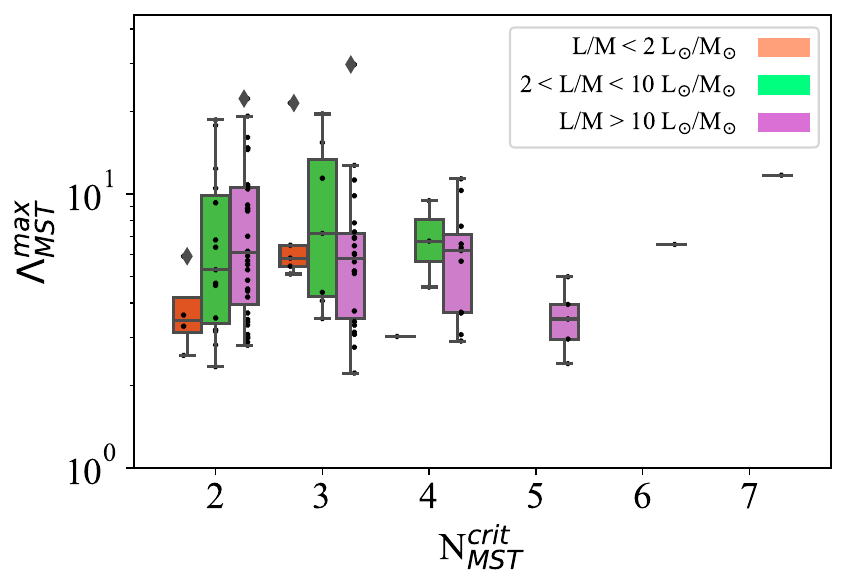}
    \caption{Boxplots showing the distribution of maximum value of the mass segregation ratio, $\Lambda_{MSR}^{max}$, divided for the different values $N^{crit}_{MST}$ that identify the interval where the MSR profile is statistically higher than unity, in the 110 ALMAGAL clumps where signatures of MS are found. Those clumps are divided into three groups according to the $L/M$ ratio composed of 59, 83, and 154 systems. The rectangular area of the boxplots shows the interquartile intervals, while the whiskers represent the 10th to 90th percentile intervals; the diamond symbols indicate the cases outside this interval.
    }
    \label{fig:MSRparameters}
\end{figure}

Furthermore, we also found that its signature affects a larger number of cores and becomes stronger in more evolved systems. Figure\,\ref{fig:MSRparameters} shows the distributions of the maximum value of the mass segregation ratio, $\Lambda_{MST}^{max}$, and of $N_{MST}^{crit}$ in the three aforementioned samples. The intensity of segregation is always greater than $\sim2.5$, with most systems having $\Lambda_{MSR}\,>\,3$, which is considered a high level of MS \citep{Pouteau2023}. 
The median value of $\Lambda_{MST}^{max}$ is $\sim5$ in less evolved clumps, and increases to $\sim6.4$ and $\sim6.2$ for the intermediate and more evolved subsamples, respectively. This suggests that the cores with higher mass become more concentrated as the clump evolves.

Segregation generally affects only two or three most massive cores of the clusters, as $\sim80$\% of the system where segregation is present have $N^{crit}_{MST}\,\leq\,3$, with 54 and 34 systems where $N^{crit}_{MST}\,=\,2$ and $3$, respectively. While the evolution leads to an average increase in the number of cores \citep{Elia2025arXiv}, at the same time we found that the number of cores showing a tighter spatial distribution also increases and may involve up to the five most massive cores in the system (only two systems show $N^{crit}_{MST}\,\ge\,6$ ).

In conclusion, although MS may be present in the earliest stages, it is about three times more common in evolved clumps ($L/M\,>\,10$\,L$_{\odot}$/M$_{\odot}$). Segregation is more pronounced in systems in more evolved phases, when cores with masses of a few M$_{\odot}$ in $\sim1000-2000$\,au are commonly found in the system (see Fig.\,18 in  \citealt{Coletta2025}). 
The gravitational concentration of the entire clump or the gas accretion toward its center has been suggested to be responsible for the observed increase in MS \citep{Xu2024}. We do not find any connection between the signature of MS and the clump surface density $\Sigma_{cl}$ or the cluster radius $R_{c}$, as it would be expected if gravitational collapse was present. However, it is important to note that {\it Herschel} derived surface densities might not be sufficient to highlight the increase in density of the clump central region because of their poor resolution. Further analysis of the dynamics and the gas distribution in clumps with high signatures of MS is necessary using molecular line tracers at high angular resolution to verify whether clump material is efficiently transferred toward central regions and the most massive cores.

\section{Discussion}
\label{sect:Discussion}

ALMAGAL observations reveal that fragmentation in high-mass clumps is widely diversified, with star-forming cores spatially arranged across a broad spectrum of configurations, from sparse, mildly concentrated, up to highly clustered. This observed heterogeneity, which mirrors the wide scatter found for core masses and fragmentation degrees discussed by \citet{Coletta2025} and \citet{Elia2025arXiv}, does not preclude a deeper analysis. Rather,  systematic analysis of  spatial distributions and core-to-core separations provides a powerful diagnostic tool for the physical mechanisms that govern where and how fragmentation takes place. Our data show that most of the small-scale ($\sim1400$\,au) cores are confined within the inner clump region ($\leq\,0.35$\,pc). Additional cores forming at larger distances are rare or, if present, are likely low-mass objects undetectable in ALMAGAL fields, which would be compatible with their formation in low-density environment.
This result might favor an in situ formation scenario for clusters \citep{Dib2010}. However, it does not contradict the more dynamic conveyor belt-type models \citep{Krumholz2020, VazquesSemadeni2019} that involve anisotropic rather than monolithic contraction, with material being fed into the clump along large-scale flows. The Global Hierarchical Collapse (GHC) model, an example of these dynamic models, predicts gravity-driven accretion along extended filaments to supply material to cores and hubs \citep{VazquesSemadeni2019, VazquesSemadeni2024}. GHC simulations show that low-mass cores may form along these peripheral filaments \citep{Gomez2014,VazquesSemadeni2017}, a scenario consistent with observed cores found along filaments connected to high-density clumps \citep{Myers2009, Kirk2013, Peretto2014, Hacar2017, Chen2019, AlvarezGutierrez2024}. The ALMAGAL cores found on the outskirts of the clumps may be formed in similar flows whose underlying filaments remain undetected because of the limited sensitivity and coverage. In this context, the presence of aligned cores in several ALMAGAL systems can be an additional footprint of the presence of these flows. In fact, numerical simulations show that the cluster structure retains indications of the native distribution and organization in structures of the star-forming material \citep{VazquesSemadeni2017}. \\

Two significant results are found from the analysis of the separations between cores in the ALMAGAL high-mass clumps. The separation is on average of the order, or smaller, of the pure thermal Jeans length over almost the entire early evolution of the clump sampled by the ALMAGAL survey. This evidence confirms and extends the results of previous studies based on a limited number of clumps  \citep{Palau2015,Palau2018, Beuther2018, Liu2020, Zhang2021, Morii2024, Ishihara2024}. Separations close to the thermal Jeans length are predicted in the case that the fragmentation process is driven mainly by gravitational contraction \citep{Palau2014,VazquesSemadeni2024}. 
Nonetheless, we found indications of systems in which the separations between the cores are significantly larger than the thermal Jeans length, $\lambda_{J}^{th}$, by an average factor of three. One possible explanation is that there is a contribution due to turbulence, which provides additional support and increases the Jeans length \citep{MacLow2004}. The turbulent Jeans length can be computed by replacing in Eq.\,\ref{Eq:JeansLenght} the sound speed $c_{s}$ with the gas velocity dispersion $\sigma_{v}$ given by $\sigma_{v} = \sqrt{c_{s}^{2} + \sigma_{nth}^{2}}$, where $\sigma_{nth}^{2}$ are the non-thermal motions present in the gas. The typical properties of the ALMAGAL clumps in the earliest phases have a median value of $c_{s} \approx0.2$\,km\,s$^{-1}$; therefore, to explain the observed ratio $R\sim3$ it would be sufficient to have a moderate non-thermal velocity dispersion $\sigma_{nt} \approx0.6$\,km\,s$^{-1}$, which corresponds to Mach number $\mathcal{M}\approx\,5$, evaluated as $\mathcal{M}\,=\,\sqrt{3}\sigma_{nth}/c_{s}$. To compare it with the average conditions present in the entire hosting molecular cloud, we matched the position of the ALMAGAL clumps with the clouds identified in the SEDIGISM survey using the \,$^{13}$CO(2-1) emission \citep{DuarteCabral2021}. We found that the typical $\sigma_{v}$ of the matched clouds is $\sim1.8$\,km\,s$^{-1}$, with a distribution with interquartile interval from $\sim1.3$ to $\sim2.6$\,km\,s$^{-1}$. These values indicate a remarkable contribution of non-thermal velocity dispersion at the cloud scale, which would produce significantly larger separations between cores than the observed ones.  

Another possibility is that different fragmentation regimes, each with its own characteristic length, are present in a fraction of ALMAGAL clumps, which would also explain the asymmetries found in the distribution of separations $P(l_{i})$. These regimes may be due to different contributions of the magnetic field, local turbulence, and density distribution \citep{Fontani2018}, be the result of the assembly of several subclusters \citep{Morii2023}, or the fragmentation in the filaments connected to the accretion flow expected in the GHC model \citep{VazquesSemadeni2019}. A more robust interpretation of our observations requires a one-to-one comparison with dedicated simulations such as these developed within the Rosetta Stone project which model the formation and temporal evolution of massive clumps \citep{Lebreuilly2025}, produce post-processed maps at multiple wavelengths \citep{Tung2025}, and create mock observations that closely resemble ALMA data with high fidelity \citep{Nucara2025}.

In addition to the comparison with the Jeans fragmentation theory, an important result emerging from the ALMAGAL clusters of cores is the progressive statistical decrease exhibited by the separations with the $L/M$ ratio, which can be interpreted as evidence that cores are more tightly packed in more evolved systems compared to the younger ones. This evidence has been advocated in support of the GHC model for high-mass and cluster formation as it naturally derives from the global contraction of the systems \citep{Traficante2023, Xu2024, VazquesSemadeni2024}. Our analysis shows that typical separations decrease by about a factor of three, with a major decrease from $\sim20000$ down to $\sim10000$\,au taking place at $L/M\approx1$\,L$_{\odot}$/M$_{\odot}$. Above this value, our data show only marginal variations, and the observed trend does not clearly reflect a decrease in core separations due to the ongoing global collapse of the clump. If a dynamical shrinking of the cluster is occurring, it is not the dominant effect traced by the median of the population of separation measured in ALMAGAL fields. The value $L/M\approx1$\,L$_{\odot}$/M$_{\odot}$ is reached when the luminosity of already formed protostars becomes relevant for the entire clump \citep{Molinari2008}, and corresponds to an increase in the degree of fragmentation observed in these systems. The relation of the degree of fragmentation with evolution is discussed by \citet{Elia2025arXiv} who reported a steady increase on average in the number of cores over the entire interval 0.1$\,\leq\,L/M\,\leq100$\,L$_{\odot}$/M$_{\odot}$. The increase in fragmentation is also presented in Fig.\,\ref{fig:SeparationEvolutionaryLMAverageQuantity}, which also shows the average variation of the core separations. We note that the observed decrease in these separations follows the transition from clumps characterized mainly by a limited number of cores ($N_{cores}\,\leq\,8$ ) to  more crowded systems. 

We attempted to draw an overall scenario from ALMAGAL observations. Our results indicate that fragmentation occurs in multiple substructures of the clump, which are generally widely separated by $\sim15000-30000$\,au ($\sim0.07-0.14$\,pc), measured in systems where fragmentation is at the onset. These substructures host a limited number of already collapsed cores, but also fragments that did not yet have enough time to grow sufficiently to reach a high contrast over the parent structure to be detectable, as proposed by \citet{Louvet2024} to justify the observed small fragmentation present in young systems. As the clump evolves, additional cores collapse and become detectable in the ALMAGAL fields with a sufficiently high significance. These newly collapsed cores form in the midst of the original population, but a minor fraction of these objects also form in the outskirts of the clump, even though less material is available. Those effects produce a decrease in the observed average separation while the additional cores form by collapsing material.  \\
Nonetheless, it is notable that systems with cores distributed following a smooth and radially concentrated profile start to be observed at the same $L/M$ value where the typical core separations decrease. In addition to the increase in fragmentation level, the emergence of those systems may be the signature of a gradual merging and dispersion of the initial, fractal distribution in subclusters, that characterizes the young ALMAGAL clumps. This outcome is consistent with predictions from dynamical-type models where conveyor belt-like processes are active \citep{Longmore2014,Krumholz2020,VazquesSemadeni2024}.  
According to these predictions, time evolution causes an increase in the degree of concentration of cores as the system shrinks and dynamically relaxes, as proposed by \citet{Xu2024}. However, while our results indicate that smooth, centrally concentrated distributions occur more frequently at later stages, we do not observe a tight correlation with evolution that would robustly support this scenario over the entire sample.  Instead, we still observe that subclustering signatures are present in a considerable fraction of systems with high values of the $L/M$ ratio. As this pattern is often observed also in young embedded stellar clusters \citep{Kuhn2014}, this suggests the presence of mechanisms that work against the merging of these substructures or that are able to slow the process down for long timescales. The scatter of $Q$ values and, more generally, the observed heterogeneity of the fragmentation in the ALMAGAL sample questions whether a unique path of evolution, like the one described above, can be traced for all the massive dense clumps observed. Additional factors that are not taken into account in this analysis may influence the evolution of the spatial distribution of cores and lead to the scatter of $Q$ values for each given interval of the $L/M$ ratio. 
Low values of the $Q$ parameter are also measured when the main system increases its radial concentration, but additional substructures form in the outskirts of the field and become statistically significant. Moreover, it has been found that the intensity of the internal mechanical feedback affects the final structure and distribution of stars in clusters. Numerical simulations including protostellar jets have typically higher values of $Q$ over time than those without any feedback or with only the radiative one \citep{Verliat2022}. 
In addition to the evolutionary effect, the variations in measured $Q$ may reflect differences in the initial conditions with which the region had initially assembled and fragmented, as pointed out by \citet{Dib2019} from the analysis of nearby low-mass star-forming regions. As the ALMAGAL fields cover different regions in the Galaxy, the clumps may be formed with different physical conditions imprinted by the large scale environment. In summary, our results are compatible with a dynamical evolutionary scenario, which defines the evolutionary path of a subset of massive clumps, and we cannot exclude the possibility that dynamical effects such as gravitational collapse, system shrinking, and dynamical relaxation are active in massive clumps.

\section{Conclusion}

We  analyzed the spatial distribution of the 5728 dense cores extracted by \citet{Coletta2025} in the 514 ALMAGAL clumps with at least four detections. We provided a quantitative and statistically robust characterization of the fragmentation process in  massive and potentially cluster-forming clumps, determining where the cores are located, the typical length of their reciprocal separations, and how they are distributed. Our measurements compose a comprehensive set of observational constraints for the fragmentation process  against which to compare the predictions of different theories of cluster formation. We focused our analysis on how the properties of the statistical distribution change during the evolution of the systems before gas dispersal.

Our results are as follows:

\begin{itemize} 

\item  The ALMAGAL clumps show a wide variety of fragmentation patterns, with cores distributed over regions that extend from $\lesssim$0.1 up to $\sim0.8$\,pc,  which closely match the inner size of the hosting clump as determined from {\it Herschel} observations. Although cores are also observed in the outskirts of the clump, they are either rare or are low mass (and thus often undetected), as expected from formation in a low-density environment. We introduced a tentative classification based on the components and elongation of the convex hull polygon defined from the core positions, finding that several configuration patterns are present in the sample. We classified them into aligned, sparse, circular, elongated, and subclustered arrangements, with the most common pattern being moderately elongated, with aspect ratio $\sim2-3$, although we note that highly elongated systems are also present.

\item The projected separations between the cores measured with the MST method show a broad distribution with a median value of $\sim6600$\,au and $\sim7400$\,au for clumps that are located at heliocentric distances 2\,$\,<\,d_{cl}\,<\,4.5$\,kpc and 4.5\,$\,<\,d_{cl}\,<\,7.5$\,kpc, respectively. Smaller median separations are measured in systems with $d_{cl}\,\leq\,3.7$\,kpc, suggesting that there are different characteristic separations in the inner regions of the clumps, which are expected to be denser. Evidence that a single characteristic length for a clump may not completely describe the observed fragmentation is also found via a comparison of the medians and modes of the distribution of the core separations, which differ from each other in 121 systems ($\sim24$\% of the analyzed sample). These differences are due to a local subclustering or to the presence of a more widely spaced population of cores in the outskirts of the system that followed a different fragmentation pattern.   

\item We identified a statistically significant shift in the distribution of core separations with the $L/M$ ratio, which can be interpreted as a variation during clump evolution. The observed shift varies significantly in the early phases ($L/M\,\lesssim\,2$\,L$_{\odot}$/M$_{\odot}$) when it drops from $\sim20000$\,au to $\sim10000$\,au for $2\,\leq\,L/M\,\leq\,10$\,L$_{\odot}/$M$_{\odot}$, and displays smaller changes for more evolved phases dropping to $\sim7000$\,au for systems with $L/M\,\geq\,10$\,L$_{\odot}$/M$_{\odot}$. Although the observed trend can be statistically described by a steady decrease in these two intervals, the larger variation appears to be produced by an increase in the degree of fragmentation in more evolved systems, as more cores are detected. 

\item We found a good agreement between the core separations and the thermal Jeans lengths $\lambda_{J}^{th}$ computed from the clump-averaged physical parameters, with a median ranging between $\sim0.5$ and $\sim2$ times $\lambda_{J}^{th}$. This agreement remains when the separations are divided according to the clump $L/M$. Nevertheless, we identified a small population of cores whose separations are centered at $\sim3\times\lambda_{J}^{th}$, which are still smaller than the prediction of turbulent fragmentation of the hosting molecular cloud. These cores are mostly hosted in clumps with low levels of fragmentation and in early phases of evolution, with $L/M\,\lesssim\,0.5-1$\,L$_{\odot}$/M$_{\odot}$. However, our results are compatible with thermal Jeans fragmentation being active in massive clumps. Although the masses are affected by large uncertainties and systematics, we note that the typical mass of cores is compatible with the thermal Jeans mass only in young and pristine clumps, where we observe that the core mass distribution is centered at $M_{core}\sim0.6$\,M$^{th}_{J}$. 

\item We did not find robust evidence of a progressive transition of the structure of the spatial distribution of the cores with evolution. The analysis using the $Q$ parameter indicates that the cores are often distributed in a fractal type of subclustering in the early stages characterized by $L/M\,\leq\,1$\,L$_{\odot}$/M$_{\odot}$. Although we found several clusters that are most likely distributed with radially concentrated profiles in the evolved clumps with $L/M\,\geq\,10$\,L$_{\odot}$/M$_{\odot}$, subclustering is still often detected  in those phases. The existence of these systems suggests that, in addition to the clump average properties, there are other factors that play a role in shaping the structure of the cluster over the evolution.

\item Signatures of mass segregation are found in 110 fields out of the 296 ALMAGAL clumps hosting more than eight cores. The occurrence of mass segregation is lower ($\sim17$\%) in young systems, but steadily increases with the evolutionary stage to $\sim30$\% in intermediate-evolved clumps with  2$\,\lesssim\,L/M\,<10$\,L$_{\odot}$/M$_{\odot}$ and up to $\sim48$\% in the more evolved systems in the ALMAGAL sample with $L/M\,\geq\,10$\,L$_{\odot}$/M$_{\odot}$.

\end{itemize}

\section*{Data availability}

Tables\,\ref{Tab:FieldDistribution}, \ref{Tab:Clumps}, and \ref{Tab:Q_ClumpParameters} are only available in electronic form at the CDS via anonymous ftp to \url{cdsarc.u-strasbg.fr} (\url{130.79.128.5}) or via \url{http://cdsweb.u-strasbg.fr/cgi-bin/qcat?J/A+A/}.

\begin{acknowledgements}
We are very grateful to the anonymous referee for constructive comments that helped clarify a number of points in the paper. The Teams at INAF-IAPS, Heidelberg, Paris-Saclay and University of Bologna acknowledge financial support from the European Research Council via the ERC Synergy Grant ``ECOGAL'' (project ID 855130). R.S.K. also acknowledges financial support from the Heidelberg Cluster of Excellence (EXC 2181 - 390900948) ``STRUCTURES'', funded by the German Excellence Strategy, and from the German Ministry for Economic Affairs and Climate Action in project ``MAINN'' (funding ID 50OO2206), and is also grateful to the Harvard Radcliffe Institute for Advanced Studies and Harvard-Smithsonian Center for Astrophysics for their hospitality and support during his sabbatical. 

A.S-M. acknowledges support from the RyC2021-032892-I grant funded by MCIN/AEI/10.13039/501100011033 and by the European Union `Next Generation EU'/PRTR, as well as the program Unidad de Excelencia Mar\'ia de Maezto CEZ2020-001058-M, and support from PID2023-144675NB-I00 (MCI-AEI-FEDER, UE).

Part of this research was carried out at the Jet Propulsion Laboratory, California Institute of Technology, under a contract with the National Aeronautics and Space Administration (80NM0018D0004).

R.K. acknowledges financial support via the Heisenberg Research Grant funded by the Deutsche Forschungsgemeinschaft (DFG, German Research Foundation) under grant no.~KU 2849/9, project no.~445783058.

P.S. was partially supported by a Grant-in-Aid for Scientific Research (KAKENHI Number JP22H01271 and JP23H01221) of JSPS. P.S. was supported by Yoshinori Ohsumi Fund (Yoshinori Ohsumi Award for Fundamental Research).

S.W. has been supported by the Collaborative Research Centre 1601 (SFB 1601) funded by the Deutsche Forschungsgemeinschaft (DFG).

L.B. gratefully acknowledges support from the ANID BASAL project FB210003.\\

This paper makes use of the following ALMA data: ADS/JAO.ALMA\#2019.1.00195.L. ALMA is a partnership of ESO (representing its member states), NSF (USA) and NINS (Japan), together with NRC (Canada), MOST and ASIAA (Taiwan), and KASI (Republic of Korea), in cooperation with the Republic of Chile. The Joint ALMA Observatory is operated by ESO, AUI/NRAO and NAOJ.
\end{acknowledgements}

\bibliographystyle{aa}
\bibliography{DistributionSourcesALMAGAL}

@ARTICLE{Adamo2020,
       author = {{Adamo}, Angela and {Zeidler}, Peter and {Kruijssen}, J.~M. Diederik and {Chevance}, M{\'e}lanie and {Gieles}, Mark and {Calzetti}, Daniela and {Charbonnel}, Corinne and {Zinnecker}, Hans and {Krause}, Martin G.~H.},
        title = "{Star Clusters Near and Far; Tracing Star Formation Across Cosmic Time}",
      journal = {\ssr},
         year = 2020,
        month = jun,
       volume = {216},
       number = {4},
          eid = {69},
        pages = {69},
          doi = {10.1007/s11214-020-00690-x},
archivePrefix = {arXiv},
       eprint = {2005.06188},
 primaryClass = {astro-ph.GA},
       adsurl = {https://ui.adsabs.harvard.edu/abs/2020SSRv..216...69A}
}

@ARTICLE{Adams2010,
       author = {{Adams}, Fred C.},
        title = "{The Birth Environment of the Solar System}",
      journal = {\araa},
         year = 2010,
        month = sep,
       volume = {48},
        pages = {47-85},
          doi = {10.1146/annurev-astro-081309-130830},
archivePrefix = {arXiv},
       eprint = {1001.5444},
 primaryClass = {astro-ph.SR},
       adsurl = {https://ui.adsabs.harvard.edu/abs/2010ARA&A..48...47A}
}

@ARTICLE{Allison2009,
       author = {{Allison}, Richard J. and {Goodwin}, Simon P. and {Parker}, Richard J. and {de Grijs}, Richard and {Portegies Zwart}, Simon F. and {Kouwenhoven}, M.~B.~N.},
        title = "{Dynamical Mass Segregation on a Very Short Timescale}",
      journal = {\apjl},
         year = 2009,
        month = aug,
       volume = {700},
       number = {2},
        pages = {L99-L103},
          doi = {10.1088/0004-637X/700/2/L99},
archivePrefix = {arXiv},
       eprint = {0906.4806},
 primaryClass = {astro-ph.GA},
       adsurl = {https://ui.adsabs.harvard.edu/abs/2009ApJ...700L..99A}
}

@ARTICLE{AlvarezGutierrez2024,
       author = {{{\'A}lvarez-Guti{\'e}rrez}, R.~H. and {Stutz}, A.~M. and {Sandoval-Garrido}, N. and {Louvet}, F. and {Motte}, F. and {Galv{\'a}n-Madrid}, R. and {Cunningham}, N. and {Sanhueza}, P. and {Bonfand}, M. and {Bontemps}, S. and {Gusdorf}, A. and {Ginsburg}, A. and {Csengeri}, T. and {Reyes}, S.~D. and {Salinas}, J. and {Baug}, T. and {Bronfman}, L. and {Busquet}, G. and {D{\'\i}az-Gonz{\'a}lez}, D.~J. and {Fernandez-Lopez}, M. and {Guzm{\'a}n}, A. and {Koley}, A. and {Liu}, H. -L. and {Olguin}, F.~A. and {Valeille-Manet}, M. and {Wyrowski}, F.},
        title = "{ALMA-IMF: XIII. N$_{2}$H$^{+}$ kinematic analysis of the intermediate protocluster G353.41}",
      journal = {\aap},
         year = 2024,
        month = sep,
       volume = {689},
          eid = {A74},
        pages = {A74},
          doi = {10.1051/0004-6361/202450321},
archivePrefix = {arXiv},
       eprint = {2404.07363},
 primaryClass = {astro-ph.GA},
       adsurl = {https://ui.adsabs.harvard.edu/abs/2024A&A...689A..74A}
}

@ARTICLE{Anderson2021,
       author = {{Anderson}, Michael and {Peretto}, Nicolas and {Ragan}, Sarah E. and {Rigby}, Andrew J. and {Avison}, Adam and {Duarte-Cabral}, Ana and {Fuller}, Gary A. and {Shirley}, Yancy L. and {Traficante}, Alessio and {Williams}, Gwenllian M.},
        title = "{An ALMA study of hub-filament systems - I. On the clump mass concentration within the most massive cores}",
      journal = {\mnras},
         year = 2021,
        month = dec,
       volume = {508},
       number = {2},
        pages = {2964-2978},
          doi = {10.1093/mnras/stab2674},
archivePrefix = {arXiv},
       eprint = {2109.07489},
 primaryClass = {astro-ph.GA},
       adsurl = {https://ui.adsabs.harvard.edu/abs/2021MNRAS.508.2964A}
}

@INPROCEEDINGS{Andre2014,
       author = {{Andr{\'e}}, P. and {Di Francesco}, J. and {Ward-Thompson}, D. and {Inutsuka}, S. -I. and {Pudritz}, R.~E. and {Pineda}, J.~E.},
        title = "{From Filamentary Networks to Dense Cores in Molecular Clouds: Toward a New Paradigm for Star Formation}",
    booktitle = {Protostars and Planets VI},
         year = 2014,
       editor = {{Beuther}, Henrik and {Klessen}, Ralf S. and {Dullemond}, Cornelis P. and {Henning}, Thomas},
        month = jan,
        pages = {27-51},
          doi = {10.2458/azu_uapress_9780816531240-ch002},
archivePrefix = {arXiv},
       eprint = {1312.6232},
 primaryClass = {astro-ph.GA},
       adsurl = {https://ui.adsabs.harvard.edu/abs/2014prpl.conf...27A}
}

@ARTICLE{Avison2023,
       author = {{Avison}, A. and {Fuller}, G.~A. and {Frimpong}, N. Asabre and {Etoka}, S. and {Hoare}, M. and {Jones}, B.~M. and {Peretto}, N. and {Traficante}, A. and {van der Tak}, F. and {Pineda}, J.~E. and {Beltr{\'a}n}, M. and {Wyrowski}, F. and {Thompson}, M. and {Lumsden}, S. and {Nagy}, Z. and {Hill}, T. and {Viti}, S. and {Fontani}, F. and {Schilke}, P.},
        title = "{Tracing Evolution in Massive Protostellar Objects - I. Fragmentation and emission properties of massive star-forming clumps in a luminosity-limited ALMA sample}",
      journal = {\mnras},
         year = 2023,
        month = dec,
       volume = {526},
       number = {2},
        pages = {2278-2300},
          doi = {10.1093/mnras/stad2824},
archivePrefix = {arXiv},
       eprint = {2309.05772},
 primaryClass = {astro-ph.GA},
       adsurl = {https://ui.adsabs.harvard.edu/abs/2023MNRAS.526.2278A}
}

@ARTICLE{Baldeschi2017,
       author = {{Baldeschi}, Adriano and {Elia}, D. and {Molinari}, S. and {Pezzuto}, S. and {Schisano}, E. and {Gatti}, M. and {Serra}, A. and {Merello}, M. and {Benedettini}, M. and {Di Giorgio}, A.~M. and {Liu}, J.~S.},
        title = "{Distance biases in the estimation of the physical properties of Hi-GAL compact sources - I. Clump properties and the identification of high-mass star-forming candidates}",
      journal = {\mnras},
         year = 2017,
        month = apr,
       volume = {466},
       number = {3},
        pages = {3682-3705},
          doi = {10.1093/mnras/stw3353},
archivePrefix = {arXiv},
       eprint = {1701.08035},
 primaryClass = {astro-ph.GA},
       adsurl = {https://ui.adsabs.harvard.edu/abs/2017MNRAS.466.3682B}
}

@ARTICLE{Ballesteros2018,
       author = {{Ballesteros-Paredes}, Javier and {V{\'a}zquez-Semadeni}, Enrique and {Palau}, Aina and {Klessen}, Ralf S.},
        title = "{Gravity or turbulence? - IV. Collapsing cores in out-of-virial disguise}",
      journal = {\mnras},
         year = 2018,
        month = sep,
       volume = {479},
       number = {2},
        pages = {2112-2125},
          doi = {10.1093/mnras/sty1515},
archivePrefix = {arXiv},
       eprint = {1710.07384},
 primaryClass = {astro-ph.GA},
       adsurl = {https://ui.adsabs.harvard.edu/abs/2018MNRAS.479.2112B}
}

@ARTICLE{Beraldo2021,
       author = {{Beraldo e Silva}, Leandro and {Debattista}, Victor P. and {Nidever}, David and {Amarante}, Jo{\~a}o A.~S. and {Garver}, Bethany},
        title = "{Co-formation of the thin and thick discs revealed by APOGEE-DR16 and Gaia-DR2}",
      journal = {\mnras},
         year = 2021,
        month = mar,
       volume = {502},
       number = {1},
        pages = {260-272},
          doi = {10.1093/mnras/staa3966},
archivePrefix = {arXiv},
       eprint = {2009.03346},
 primaryClass = {astro-ph.GA},
       adsurl = {https://ui.adsabs.harvard.edu/abs/2021MNRAS.502..260B}
}

@ARTICLE{Beuther2002,
       author = {{Beuther}, H. and {Schilke}, P. and {Menten}, K.~M. and {Motte}, F. and {Sridharan}, T.~K. and {Wyrowski}, F.},
        title = "{High-Mass Protostellar Candidates. II. Density Structure from Dust Continuum and CS Emission}",
      journal = {\apj},
         year = 2002,
        month = feb,
       volume = {566},
       number = {2},
        pages = {945-965},
          doi = {10.1086/338334},
archivePrefix = {arXiv},
       eprint = {astro-ph/0110370},
 primaryClass = {astro-ph},
       adsurl = {https://ui.adsabs.harvard.edu/abs/2002ApJ...566..945B}
}

@ARTICLE{Beuther2018,
       author = {{Beuther}, H. and {Mottram}, J.~C. and {Ahmadi}, A. and {Bosco}, F. and {Linz}, H. and {Henning}, Th. and {Klaassen}, P. and {Winters}, J.~M. and {Maud}, L.~T. and {Kuiper}, R. and {Semenov}, D. and {Gieser}, C. and {Peters}, T. and {Urquhart}, J.~S. and {Pudritz}, R. and {Ragan}, S.~E. and {Feng}, S. and {Keto}, E. and {Leurini}, S. and {Cesaroni}, R. and {Beltran}, M. and {Palau}, A. and {S{\'a}nchez-Monge}, {\'A}. and {Galvan-Madrid}, R. and {Zhang}, Q. and {Schilke}, P. and {Wyrowski}, F. and {Johnston}, K.~G. and {Longmore}, S.~N. and {Lumsden}, S. and {Hoare}, M. and {Menten}, K.~M. and {Csengeri}, T.},
        title = "{Fragmentation and disk formation during high-mass star formation. IRAM NOEMA (Northern Extended Millimeter Array) large program CORE}",
      journal = {\aap},
         year = 2018,
        month = sep,
       volume = {617},
          eid = {A100},
        pages = {A100},
          doi = {10.1051/0004-6361/201833021},
archivePrefix = {arXiv},
       eprint = {1805.01191},
 primaryClass = {astro-ph.GA},
       adsurl = {https://ui.adsabs.harvard.edu/abs/2018A&A...617A.100B}
}

@ARTICLE{Beuther2021,
       author = {{Beuther}, H. and {Gieser}, C. and {Suri}, S. and {Linz}, H. and {Klaassen}, P. and {Semenov}, D. and {Winters}, J.~M. and {Henning}, Th. and {Soler}, J.~D. and {Urquhart}, J.~S. and {Syed}, J. and {Feng}, S. and {M{\"o}ller}, T. and {Beltr{\'a}n}, M.~T. and {S{\'a}nchez-Monge}, {\'A}. and {Longmore}, S.~N. and {Peters}, T. and {Ballesteros-Paredes}, J. and {Schilke}, P. and {Moscadelli}, L. and {Palau}, A. and {Cesaroni}, R. and {Lumsden}, S. and {Pudritz}, R. and {Wyrowski}, F. and {Kuiper}, R. and {Ahmadi}, A.},
        title = "{Fragmentation and kinematics in high-mass star formation. CORE-extension targeting two very young high-mass star-forming regions}",
      journal = {\aap},
         year = 2021,
        month = may,
       volume = {649},
          eid = {A113},
        pages = {A113},
          doi = {10.1051/0004-6361/202040106},
archivePrefix = {arXiv},
       eprint = {2104.02420},
 primaryClass = {astro-ph.GA},
       adsurl = {https://ui.adsabs.harvard.edu/abs/2021A&A...649A.113B}
}

@ARTICLE{Beuther2024,
       author = {{Beuther}, H. and {Gieser}, C. and {Soler}, J.~D. and {Zhang}, Q. and {Rao}, R. and {Semenov}, D. and {Henning}, Th. and {Pudritz}, R. and {Peters}, T. and {Klaassen}, P. and {Beltr{\'a}n}, M.~T. and {Palau}, A. and {M{\"o}ller}, T. and {Johnston}, K.~G. and {Zinnecker}, H. and {Urquhart}, J. and {Kuiper}, R. and {Ahmadi}, A. and {S{\'a}nchez-Monge}, {\'A}. and {Feng}, S. and {Leurini}, S. and {Ragan}, S.~E.},
        title = "{Density distributions, magnetic field structures, and fragmentation in high-mass star formation}",
      journal = {\aap},
         year = 2024,
        month = feb,
       volume = {682},
          eid = {A81},
        pages = {A81},
          doi = {10.1051/0004-6361/202348117},
archivePrefix = {arXiv},
       eprint = {2311.11874},
 primaryClass = {astro-ph.GA},
       adsurl = {https://ui.adsabs.harvard.edu/abs/2024A&A...682A..81B}
}

@ARTICLE{Beuther2025,
       author = {{Beuther}, H. and {Kuiper}, R. and {Tafalla}, M.},
        title = "{Star Formation from Low to High Mass: A Comparative View}",
      journal = {\araa},
         year = 2025,
        month = aug,
       volume = {63},
       number = {1},
        pages = {1-44},
          doi = {10.1146/annurev-astro-013125-122023},
archivePrefix = {arXiv},
       eprint = {2501.16866},
 primaryClass = {astro-ph.GA},
       adsurl = {https://ui.adsabs.harvard.edu/abs/2025ARA&A..63....1B}
}

@ARTICLE{Bonnell2001,
       author = {{Bonnell}, I.~A. and {Bate}, M.~R. and {Clarke}, C.~J. and {Pringle}, J.~E.},
        title = "{Competitive accretion in embedded stellar clusters}",
      journal = {\mnras},
         year = 2001,
        month = may,
       volume = {323},
       number = {4},
        pages = {785-794},
          doi = {10.1046/j.1365-8711.2001.04270.x},
archivePrefix = {arXiv},
       eprint = {astro-ph/0102074},
 primaryClass = {astro-ph},
       adsurl = {https://ui.adsabs.harvard.edu/abs/2001MNRAS.323..785B}
}

@ARTICLE{Bonnell2003,
       author = {{Bonnell}, Ian A. and {Bate}, Matthew R. and {Vine}, Stephen G.},
        title = "{The hierarchical formation of a stellar cluster}",
      journal = {\mnras},
         year = 2003,
        month = aug,
       volume = {343},
       number = {2},
        pages = {413-418},
          doi = {10.1046/j.1365-8711.2003.06687.x},
archivePrefix = {arXiv},
       eprint = {astro-ph/0305082},
 primaryClass = {astro-ph},
       adsurl = {https://ui.adsabs.harvard.edu/abs/2003MNRAS.343..413B}
}

@ARTICLE{Bonnell2006,
       author = {{Bonnell}, Ian A. and {Bate}, Matthew R.},
        title = "{Star formation through gravitational collapse and competitive accretion}",
      journal = {\mnras},
         year = 2006,
        month = jul,
       volume = {370},
       number = {1},
        pages = {488-494},
          doi = {10.1111/j.1365-2966.2006.10495.x},
archivePrefix = {arXiv},
       eprint = {astro-ph/0604615},
 primaryClass = {astro-ph},
       adsurl = {https://ui.adsabs.harvard.edu/abs/2006MNRAS.370..488B}
}

@ARTICLE{Bressert2010,
       author = {{Bressert}, E. and {Bastian}, N. and {Gutermuth}, R. and {Megeath}, S.~T. and {Allen}, L. and {Evans}, Neal J., II and {Rebull}, L.~M. and {Hatchell}, J. and {Johnstone}, D. and {Bourke}, T.~L. and {Cieza}, L.~A. and {Harvey}, P.~M. and {Merin}, B. and {Ray}, T.~P. and {Tothill}, N.~F.~H.},
        title = "{The spatial distribution of star formation in the solar neighbourhood: do all stars form in dense clusters?}",
      journal = {\mnras},
         year = 2010,
        month = nov,
       volume = {409},
       number = {1},
        pages = {L54-L58},
          doi = {10.1111/j.1745-3933.2010.00946.x},
archivePrefix = {arXiv},
       eprint = {1009.1150},
 primaryClass = {astro-ph.SR},
       adsurl = {https://ui.adsabs.harvard.edu/abs/2010MNRAS.409L..54B}
}

@ARTICLE{Cartwright2004,
       author = {{Cartwright}, Annabel and {Whitworth}, Anthony P.},
        title = "{The statistical analysis of star clusters}",
      journal = {\mnras},
         year = 2004,
        month = feb,
       volume = {348},
       number = {2},
        pages = {589-598},
          doi = {10.1111/j.1365-2966.2004.07360.x},
archivePrefix = {arXiv},
       eprint = {astro-ph/0403474},
 primaryClass = {astro-ph},
       adsurl = {https://ui.adsabs.harvard.edu/abs/2004MNRAS.348..589C}
}

@ARTICLE{Cartwright2006,
       author = {{Cartwright}, Annabel and {Whitworth}, Anthony P. and {Nutter}, David},
        title = "{Methods for analysing structure in molecular clouds}",
      journal = {\mnras},
         year = 2006,
        month = jul,
       volume = {369},
       number = {3},
        pages = {1411-1418},
          doi = {10.1111/j.1365-2966.2006.10389.x},
       adsurl = {https://ui.adsabs.harvard.edu/abs/2006MNRAS.369.1411C}
}

@ARTICLE{Chen2007,
       author = {{Chen}, L. and {de Grijs}, R. and {Zhao}, J.~L.},
        title = "{Mass Segregation in Very Young Open Clusters: A Case Study of NGC 2244 and NGC 6530}",
      journal = {\aj},
         year = 2007,
        month = oct,
       volume = {134},
       number = {4},
        pages = {1368-1379},
          doi = {10.1086/521022},
archivePrefix = {arXiv},
       eprint = {0706.2723},
 primaryClass = {astro-ph},
       adsurl = {https://ui.adsabs.harvard.edu/abs/2007AJ....134.1368C}
}

@ARTICLE{Chen2019,
       author = {{Chen}, Huei-Ru Vivien and {Zhang}, Qizhou and {Wright}, M.~C.~H. and {Busquet}, Gemma and {Lin}, Yuxin and {Liu}, Hauyu Baobab and {Olguin}, F.~A. and {Sanhueza}, Patricio and {Nakamura}, Fumitaka and {Palau}, Aina and {Ohashi}, Satoshi and {Tatematsu}, Ken'ichi and {Liao}, Li-Wen},
        title = "{Filamentary Accretion Flows in the Infrared Dark Cloud G14.225-0.506 Revealed by ALMA}",
      journal = {\apj},
         year = 2019,
        month = apr,
       volume = {875},
       number = {1},
          eid = {24},
        pages = {24},
          doi = {10.3847/1538-4357/ab0f3e},
archivePrefix = {arXiv},
       eprint = {1903.04376},
 primaryClass = {astro-ph.GA},
       adsurl = {https://ui.adsabs.harvard.edu/abs/2019ApJ...875...24C}
}

@ARTICLE{Chen2021,
       author = {{Chen}, Yingtian and {Li}, Hui and {Vogelsberger}, Mark},
        title = "{Effects of initial density profiles on massive star cluster formation in giant molecular clouds}",
      journal = {\mnras},
         year = 2021,
        month = apr,
       volume = {502},
       number = {4},
        pages = {6157-6169},
          doi = {10.1093/mnras/stab491},
archivePrefix = {arXiv},
       eprint = {2006.00004},
 primaryClass = {astro-ph.GA},
       adsurl = {https://ui.adsabs.harvard.edu/abs/2021MNRAS.502.6157C}
}

@ARTICLE{Coletta2025,
       author = {{Coletta}, A. and {Molinari}, S. and {Schisano}, E. and {Traficante}, A. and {Elia}, D. and {Benedettini}, M. and {Mininni}, C. and {Soler}, J.~D. and {S{\'a}nchez-Monge}, {\'A}. and {Schilke}, P. and {Battersby}, C. and {Fuller}, G.~A. and {Beuther}, H. and {Zhang}, Q. and {Beltr{\'a}n}, M.~T. and {Jones}, B. and {Klessen}, R.~S. and {Walch}, S. and {Fontani}, F. and {Avison}, A. and {Brogan}, C.~L. and {Clarke}, S.~D. and {Hatchfield}, P. and {Hennebelle}, P. and {Ho}, P.~T.~P. and {Hunter}, T.~R. and {Johnston}, K.~G. and {Klaassen}, P.~D. and {Koch}, P.~M. and {Kuiper}, R. and {Lis}, D.~C. and {Liu}, T. and {Lumsden}, S.~L. and {Maruccia}, Y. and {M{\"o}ller}, T. and {Moscadelli}, L. and {Nucara}, A. and {Rigby}, A.~J. and {Rygl}, K.~L.~J. and {Sanhueza}, P. and {van der Tak}, F. and {Wells}, M.~R.~A. and {Wyrowski}, F. and {De Angelis}, F. and {Liu}, S. and {Ahmadi}, A. and {Bronfman}, L. and {Liu}, S. -Y. and {Su}, Y. -N. and {Tang}, Y. and {Testi}, L. and {Zinnecker}, H.},
        title = "{ALMAGAL: III. Compact source catalog: Fragmentation statistics and physical evolution of the core population}",
      journal = {\aap},
         year = 2025,
        month = apr,
       volume = {696},
          eid = {A151},
        pages = {A151},
          doi = {10.1051/0004-6361/202452706},
archivePrefix = {arXiv},
       eprint = {2503.05663},
 primaryClass = {astro-ph.GA},
       adsurl = {https://ui.adsabs.harvard.edu/abs/2025A&A...696A.151C}
}

@ARTICLE{Csengeri2017,
       author = {{Csengeri}, T. and {Bontemps}, S. and {Wyrowski}, F. and {Motte}, F. and {Menten}, K.~M. and {Beuther}, H. and {Bronfman}, L. and {Commer{\c{c}}on}, B. and {Chapillon}, E. and {Duarte-Cabral}, A. and {Fuller}, G.~A. and {Henning}, Th. and {Leurini}, S. and {Longmore}, S. and {Palau}, A. and {Peretto}, N. and {Schuller}, F. and {Tan}, J.~C. and {Testi}, L. and {Traficante}, A. and {Urquhart}, J.~S.},
        title = "{ALMA survey of massive cluster progenitors from ATLASGAL. Limited fragmentation at the early evolutionary stage of massive clumps}",
      journal = {\aap},
         year = 2017,
        month = apr,
       volume = {600},
          eid = {L10},
        pages = {L10},
          doi = {10.1051/0004-6361/201629754},
archivePrefix = {arXiv},
       eprint = {1703.03273},
 primaryClass = {astro-ph.GA},
       adsurl = {https://ui.adsabs.harvard.edu/abs/2017A&A...600L..10C}
}

@ARTICLE{Dale2011,
       author = {{Dale}, James E. and {Bonnell}, Ian},
        title = "{Ionizing feedback from massive stars in massive clusters: fake bubbles and untriggered star formation}",
      journal = {\mnras},
         year = 2011,
        month = jun,
       volume = {414},
       number = {1},
        pages = {321-328},
          doi = {10.1111/j.1365-2966.2011.18392.x},
archivePrefix = {arXiv},
       eprint = {1103.1532},
 primaryClass = {astro-ph.SR},
       adsurl = {https://ui.adsabs.harvard.edu/abs/2011MNRAS.414..321D}
}

@ARTICLE{Dib2010,
       author = {{Dib}, Sami and {Shadmehri}, Mohsen and {Padoan}, Paolo and {Maheswar}, G. and {Ojha}, D.~K. and {Khajenabi}, Fazeleh},
        title = "{The IMF of stellar clusters: effects of accretion and feedback}",
      journal = {\mnras},
         year = 2010,
        month = jun,
       volume = {405},
       number = {1},
        pages = {401-420},
          doi = {10.1111/j.1365-2966.2010.16451.x},
archivePrefix = {arXiv},
       eprint = {0908.4522},
 primaryClass = {astro-ph.GA},
       adsurl = {https://ui.adsabs.harvard.edu/abs/2010MNRAS.405..401D}
}

@ARTICLE{Dib2018a,
       author = {{Dib}, Sami and {Schmeja}, Stefan and {Parker}, Richard J.},
        title = "{Structure and mass segregation in Galactic stellar clusters}",
      journal = {\mnras},
         year = 2018,
        month = jan,
       volume = {473},
       number = {1},
        pages = {849-859},
          doi = {10.1093/mnras/stx2413},
archivePrefix = {arXiv},
       eprint = {1707.00744},
 primaryClass = {astro-ph.GA},
       adsurl = {https://ui.adsabs.harvard.edu/abs/2018MNRAS.473..849D}
}

@ARTICLE{Dib2018,
       author = {{Dib}, Sami and {Basu}, Shantanu},
        title = "{The emergence of the galactic stellar mass function from a non-universal IMF in clusters}",
      journal = {\aap},
         year = 2018,
        month = jun,
       volume = {614},
          eid = {A43},
        pages = {A43},
          doi = {10.1051/0004-6361/201732490},
archivePrefix = {arXiv},
       eprint = {1711.07487},
 primaryClass = {astro-ph.GA},
       adsurl = {https://ui.adsabs.harvard.edu/abs/2018A&A...614A..43D}
}

@ARTICLE{Dib2019,
       author = {{Dib}, Sami and {Henning}, Thomas},
        title = "{Star formation activity and the spatial distribution and mass segregation of dense cores in the early phases of star formation}",
      journal = {\aap},
         year = 2019,
        month = sep,
       volume = {629},
          eid = {A135},
        pages = {A135},
          doi = {10.1051/0004-6361/201834080},
archivePrefix = {arXiv},
       eprint = {1808.07057},
 primaryClass = {astro-ph.GA},
       adsurl = {https://ui.adsabs.harvard.edu/abs/2019A&A...629A.135D}
}

@ARTICLE{Dib2020,
       author = {{Dib}, Sami and {Bontemps}, Sylvain and {Schneider}, Nicola and {Elia}, Davide and {Ossenkopf-Okada}, Volker and {Shadmehri}, Mohsen and {Arzoumanian}, Doris and {Motte}, Fr{\'e}d{\'e}rique and {Heyer}, Mark and {Nordlund}, {\r{A}}ke and {Ladjelate}, Bilal},
        title = "{The structure and characteristic scales of molecular clouds}",
      journal = {\aap},
         year = 2020,
        month = oct,
       volume = {642},
          eid = {A177},
        pages = {A177},
          doi = {10.1051/0004-6361/202038849},
archivePrefix = {arXiv},
       eprint = {2007.08533},
 primaryClass = {astro-ph.GA},
       adsurl = {https://ui.adsabs.harvard.edu/abs/2020A&A...642A.177D}
}

@ARTICLE{Dib2023,
       author = {{Dib}, Sami},
        title = "{Variation of the High-mass Slope of the Stellar Initial Mass Function: Theory Meets Observations}",
      journal = {\apj},
         year = 2023,
        month = dec,
       volume = {959},
       number = {2},
          eid = {88},
        pages = {88},
          doi = {10.3847/1538-4357/ad09bc},
archivePrefix = {arXiv},
       eprint = {2309.10842},
 primaryClass = {astro-ph.GA},
       adsurl = {https://ui.adsabs.harvard.edu/abs/2023ApJ...959...88D}
}

@ARTICLE{Dobbs2022,
       author = {{Dobbs}, C.~L. and {Bending}, T.~J.~R. and {Pettitt}, A.~R. and {Bate}, M.~R.},
        title = "{The formation of massive stellar clusters in converging galactic flows with photoionization}",
      journal = {\mnras},
         year = 2022,
        month = jan,
       volume = {509},
       number = {1},
        pages = {954-973},
          doi = {10.1093/mnras/stab3036},
archivePrefix = {arXiv},
       eprint = {2110.09201},
 primaryClass = {astro-ph.GA},
       adsurl = {https://ui.adsabs.harvard.edu/abs/2022MNRAS.509..954D}
}

@MISC{DuarteCabral2021,
       author = {{Duarte-Cabral}, A. and {Colombo}, D. and {Urquhart}, J.~S. and {Ginsburg}, A. and {Russeil}, D. and {Schuller}, F. and {Anderson}, L.~D. and {Barnes}, P.~J. and {Beltr{\'a}n}, M.~T. and {Beuther}, H. and {Bontemps}, S. and {Bronfman}, L. and {Csengeri}, T. and {Dobbs}, C.~L. and {Eden}, D. and {Giannetti}, A. and {Kauffmann}, J. and {Mattern}, M. and {Medina}, S. -N.~X. and {Menten}, K.~M. and {Lee}, M. -Y. and {Pettitt}, A.~R. and {Riener}, M. and {Rigby}, A.~J. and {Traficante}, A. and {Veena}, V.~S. and {Wienen}, M. and {Wyrowski}, F. and {Agurto}, C. and {Azagra}, F. and {Cesaroni}, R. and {Finger}, R. and {Gonzalez}, E. and {Henning}, T. and {Hernandez}, A.~K. and {Kainulainen}, J. and {Leurini}, S. and {Lopez}, S. and {Mac-Auliffe}, F. and {Mazumdar}, P. and {Molinari}, S. and {Motte}, F. and {Muller}, E. and {Nguyen-Luong}, Q. and {Parra}, R. and {Perez-Beaupuits}, J. -P. and {Montenegro-Montes}, F.~M. and {Moore}, T.~J.~T. and {Ragan}, S.~E. and {S{\'a}nchez-Monge}, A. and {Sanna}, A. and {Schilke}, P. and {Schisano}, E. and {Schneider}, N. and {Suri}, S. and {Testi}, L. and {Torstensson}, K. and {Venegas}, P. and {Wang}, K. and {Zavagno}, A.},
        title = "{The SEDIGISM survey: molecular clouds in the inner Galaxy}",
         year = 2021,
        month = jan,
        pages = {3027-3049},
          doi = {10.1093/mnras/staa2480},
archivePrefix = {arXiv},
       eprint = {2012.01502},
 primaryClass = {astro-ph.GA},
    publisher = {OUP},
       adsurl = {https://ui.adsabs.harvard.edu/abs/2021MNRAS.500.3027D}
}

@ARTICLE{Elia2017,
       author = {{Elia}, Davide and {Molinari}, S. and {Schisano}, E. and {Pestalozzi}, M. and {Pezzuto}, S. and {Merello}, M. and {Noriega-Crespo}, A. and {Moore}, T.~J.~T. and {Russeil}, D. and {Mottram}, J.~C. and {Paladini}, R. and {Strafella}, F. and {Benedettini}, M. and {Bernard}, J.~P. and {Di Giorgio}, A. and {Eden}, D.~J. and {Fukui}, Y. and {Plume}, R. and {Bally}, J. and {Martin}, P.~G. and {Ragan}, S.~E. and {Jaffa}, S.~E. and {Motte}, F. and {Olmi}, L. and {Schneider}, N. and {Testi}, L. and {Wyrowski}, F. and {Zavagno}, A. and {Calzoletti}, L. and {Faustini}, F. and {Natoli}, P. and {Palmeirim}, P. and {Piacentini}, F. and {Piazzo}, L. and {Pilbratt}, G.~L. and {Polychroni}, D. and {Baldeschi}, A. and {Beltr{\'a}n}, M.~T. and {Billot}, N. and {Cambr{\'e}sy}, L. and {Cesaroni}, R. and {Garc{\'\i}a-Lario}, P. and {Hoare}, M.~G. and {Huang}, M. and {Joncas}, G. and {Liu}, S.~J. and {Maiolo}, B.~M.~T. and {Marsh}, K.~A. and {Maruccia}, Y. and {M{\`e}ge}, P. and {Peretto}, N. and {Rygl}, K.~L.~J. and {Schilke}, P. and {Thompson}, M.~A. and {Traficante}, A. and {Umana}, G. and {Veneziani}, M. and {Ward-Thompson}, D. and {Whitworth}, A.~P. and {Arab}, H. and {Bandieramonte}, M. and {Becciani}, U. and {Brescia}, M. and {Buemi}, C. and {Bufano}, F. and {Butora}, R. and {Cavuoti}, S. and {Costa}, A. and {Fiorellino}, E. and {Hajnal}, A. and {Hayakawa}, T. and {Kacsuk}, P. and {Leto}, P. and {Li Causi}, G. and {Marchili}, N. and {Martinavarro-Armengol}, S. and {Mercurio}, A. and {Molinaro}, M. and {Riccio}, G. and {Sano}, H. and {Sciacca}, E. and {Tachihara}, K. and {Torii}, K. and {Trigilio}, C. and {Vitello}, F. and {Yamamoto}, H.},
        title = "{The Hi-GAL compact source catalogue - I. The physical properties of the clumps in the inner Galaxy (-71.0{\textdegree} < {\ensuremath{\ell}} < 67.0{\textdegree})}",
      journal = {\mnras},
         year = 2017,
        month = oct,
       volume = {471},
       number = {1},
        pages = {100-143},
          doi = {10.1093/mnras/stx1357},
archivePrefix = {arXiv},
       eprint = {1706.01046},
 primaryClass = {astro-ph.GA},
       adsurl = {https://ui.adsabs.harvard.edu/abs/2017MNRAS.471..100E}
}

@ARTICLE{Elia2021,
       author = {{Elia}, Davide and {Merello}, M. and {Molinari}, S. and {Schisano}, E. and {Zavagno}, A. and {Russeil}, D. and {M{\`e}ge}, P. and {Martin}, P.~G. and {Olmi}, L. and {Pestalozzi}, M. and {Plume}, R. and {Ragan}, S.~E. and {Benedettini}, M. and {Eden}, D.~J. and {Moore}, T.~J.~T. and {Noriega-Crespo}, A. and {Paladini}, R. and {Palmeirim}, P. and {Pezzuto}, S. and {Pilbratt}, G.~L. and {Rygl}, K.~L.~J. and {Schilke}, P. and {Strafella}, F. and {Tan}, J.~C. and {Traficante}, A. and {Baldeschi}, A. and {Bally}, J. and {di Giorgio}, A.~M. and {Fiorellino}, E. and {Liu}, S.~J. and {Piazzo}, L. and {Polychroni}, D.},
        title = "{The Hi-GAL compact source catalogue - II. The 360{\textdegree} catalogue of clump physical properties}",
      journal = {\mnras},
         year = 2021,
        month = jun,
       volume = {504},
       number = {2},
        pages = {2742-2766},
          doi = {10.1093/mnras/stab1038},
archivePrefix = {arXiv},
       eprint = {2104.04807},
 primaryClass = {astro-ph.GA},
       adsurl = {https://ui.adsabs.harvard.edu/abs/2021MNRAS.504.2742E}
}

@ARTICLE{Elia2025arXiv,
       author = {{Elia}, D. and {Coletta}, A. and {Molinari}, S. and {Schisano}, E. and {Benedettini}, M. and {S{\'a}nchez-Monge}, {\'A}. and {Traficante}, A. and {Mininni}, C. and {Nucara}, A. and {Pezzuto}, S. and {Schilke}, P. and {Soler}, J.~D. and {Avison}, A. and {Beltr{\'a}n}, M.~T. and {Beuther}, H. and {Clarke}, S. and {Fuller}, G.~A. and {Klessen}, R.~S. and {Kuiper}, R. and {Lebreuilly}, U. and {Lis}, D.~C. and {M{\"o}ller}, T. and {Moscadelli}, L. and {Rigby}, A.~J. and {Sanhueza}, P. and {van der Tak}, F. and {Zhang}, Q. and {Rygl}, K.~L.~J. and {Merello}, M. and {Battersby}, C. and {Ho}, P.~T.~P. and {Klaassen}, P.~D. and {Koch}, P.~M. and {Allande}, J. and {Bronfman}, L. and {Fontani}, F. and {Hennebelle}, P. and {Jones}, B. and {Liu}, T. and {Stroud}, G. and {Wells}, M.~R.~A. and {Ahmadi}, A. and {Brogan}, C.~L. and {De Angelis}, F. and {Hunter}, T.~R. and {Johnston}, K.~G. and {Law}, C.~Y. and {Liu}, S.~J. and {Liu}, S.-Y. and {Maruccia}, Y. and {Pelkonen}, V.-M. and {Su}, Y.-N. and {Tang}, Y. and {Testi}, L. and {Walch}, S. and {Zhang}, T. and {Zinnecker}, H.},
        title = "{ALMAGAL: V. Relations between the core populations and the parent clump physical properties}",
      journal = {\aap},
         year = 2026,
        month = jan,
       volume = {705},
          eid = {A100},
        pages = {A100},
          doi = {10.1051/0004-6361/202554764},
archivePrefix = {arXiv},
       eprint = {2511.10825},
 primaryClass = {astro-ph.GA},
       adsurl = {https://ui.adsabs.harvard.edu/abs/2026A&A...705A.100E}
}

@ARTICLE{Elmegreen1996,
       author = {{Elmegreen}, Bruce G. and {Falgarone}, Edith},
        title = "{A Fractal Origin for the Mass Spectrum of Interstellar Clouds}",
      journal = {\apj},
         year = 1996,
        month = nov,
       volume = {471},
        pages = {816},
          doi = {10.1086/178009},
       adsurl = {https://ui.adsabs.harvard.edu/abs/1996ApJ...471..816E}
}

@INPROCEEDINGS{Elmegreen2010,
       author = {{Elmegreen}, Bruce G.},
        title = "{The nature and nurture of star clusters}",
    booktitle = {Star Clusters: Basic Galactic Building Blocks Throughout Time and Space},
         year = 2010,
       editor = {{de Grijs}, Richard and {L{\'e}pine}, Jacques R.~D.},
       series = {IAU Symposium},
       volume = {266},
        month = jan,
        pages = {3-13},
          doi = {10.1017/S1743921309990809},
archivePrefix = {arXiv},
       eprint = {0910.4638},
 primaryClass = {astro-ph.GA},
       adsurl = {https://ui.adsabs.harvard.edu/abs/2010IAUS..266....3E}
}

@ARTICLE{Fontani2018,
       author = {{Fontani}, F. and {Commer{\c{c}}on}, B. and {Giannetti}, A. and {Beltr{\'a}n}, M.~T. and {S{\'a}nchez-Monge}, {\'A}. and {Testi}, L. and {Brand}, J. and {Tan}, J.~C.},
        title = "{Fragmentation properties of massive protocluster gas clumps: an ALMA study}",
      journal = {\aap},
         year = 2018,
        month = jul,
       volume = {615},
          eid = {A94},
        pages = {A94},
          doi = {10.1051/0004-6361/201832672},
archivePrefix = {arXiv},
       eprint = {1804.02429},
 primaryClass = {astro-ph.GA},
       adsurl = {https://ui.adsabs.harvard.edu/abs/2018A&A...615A..94F}
}

@ARTICLE{Gennaro2011,
       author = {{Gennaro}, M. and {Brandner}, W. and {Stolte}, A. and {Henning}, Th.},
        title = "{Mass segregation and elongation of the starburst cluster Westerlund 1}",
      journal = {\mnras},
         year = 2011,
        month = apr,
       volume = {412},
       number = {4},
        pages = {2469-2488},
          doi = {10.1111/j.1365-2966.2010.18068.x},
archivePrefix = {arXiv},
       eprint = {1011.5223},
 primaryClass = {astro-ph.GA},
       adsurl = {https://ui.adsabs.harvard.edu/abs/2011MNRAS.412.2469G}
}

@ARTICLE{Gieser2021,
       author = {{Gieser}, C. and {Beuther}, H. and {Semenov}, D. and {Ahmadi}, A. and {Suri}, S. and {M{\"o}ller}, T. and {Beltr{\'a}n}, M.~T. and {Klaassen}, P. and {Zhang}, Q. and {Urquhart}, J.~S. and {Henning}, Th. and {Feng}, S. and {Galv{\'a}n-Madrid}, R. and {de Souza Magalh{\~a}es}, V. and {Moscadelli}, L. and {Longmore}, S. and {Leurini}, S. and {Kuiper}, R. and {Peters}, T. and {Menten}, K.~M. and {Csengeri}, T. and {Fuller}, G. and {Wyrowski}, F. and {Lumsden}, S. and {S{\'a}nchez-Monge}, {\'A}. and {Maud}, L. and {Linz}, H. and {Palau}, A. and {Schilke}, P. and {Pety}, J. and {Pudritz}, R. and {Winters}, J.~M. and {Pi{\'e}tu}, V.},
        title = "{Physical and chemical structure of high-mass star-forming regions. Unraveling chemical complexity with CORE: the NOEMA large program}",
      journal = {\aap},
         year = 2021,
        month = apr,
       volume = {648},
          eid = {A66},
        pages = {A66},
          doi = {10.1051/0004-6361/202039670},
archivePrefix = {arXiv},
       eprint = {2102.11676},
 primaryClass = {astro-ph.GA},
       adsurl = {https://ui.adsabs.harvard.edu/abs/2021A&A...648A..66G}
}

@ARTICLE{Gieser2022,
       author = {{Gieser}, C. and {Beuther}, H. and {Semenov}, D. and {Suri}, S. and {Soler}, J.~D. and {Linz}, H. and {Syed}, J. and {Henning}, Th. and {Feng}, S. and {M{\"o}ller}, T. and {Palau}, A. and {Winters}, J.~M. and {Beltr{\'a}n}, M.~T. and {Kuiper}, R. and {Moscadelli}, L. and {Klaassen}, P. and {Urquhart}, J.~S. and {Peters}, T. and {Longmore}, S.~N. and {S{\'a}nchez-Monge}, {\'A}. and {Galv{\'a}n-Madrid}, R. and {Pudritz}, R.~E. and {Johnston}, K.~G.},
        title = "{Clustered star formation at early evolutionary stages. Physical and chemical analysis of the young star-forming regions ISOSS J22478+6357 and ISOSS J23053+5953}",
      journal = {\aap},
         year = 2022,
        month = jan,
       volume = {657},
          eid = {A3},
        pages = {A3},
          doi = {10.1051/0004-6361/202141857},
archivePrefix = {arXiv},
       eprint = {2110.01896},
 primaryClass = {astro-ph.SR},
       adsurl = {https://ui.adsabs.harvard.edu/abs/2022A&A...657A...3G}
}

@ARTICLE{Girichidis2011,
       author = {{Girichidis}, Philipp and {Federrath}, Christoph and {Banerjee}, Robi and {Klessen}, Ralf S.},
        title = "{Importance of the initial conditions for star formation - I. Cloud evolution and morphology}",
      journal = {\mnras},
         year = 2011,
        month = jun,
       volume = {413},
       number = {4},
        pages = {2741-2759},
          doi = {10.1111/j.1365-2966.2011.18348.x},
archivePrefix = {arXiv},
       eprint = {1008.5255},
 primaryClass = {astro-ph.SR},
       adsurl = {https://ui.adsabs.harvard.edu/abs/2011MNRAS.413.2741G}
}

@ARTICLE{Girichidis2020,
       author = {{Girichidis}, Philipp and {Offner}, Stella S.~R. and {Kritsuk}, Alexei G. and {Klessen}, Ralf S. and {Hennebelle}, Patrick and {Kruijssen}, J.~M. Diederik and {Krause}, Martin G.~H. and {Glover}, Simon C.~O. and {Padovani}, Marco},
        title = "{Physical Processes in Star Formation}",
      journal = {\ssr},
         year = 2020,
        month = jun,
       volume = {216},
       number = {4},
          eid = {68},
        pages = {68},
          doi = {10.1007/s11214-020-00693-8},
archivePrefix = {arXiv},
       eprint = {2005.06472},
 primaryClass = {astro-ph.GA},
       adsurl = {https://ui.adsabs.harvard.edu/abs/2020SSRv..216...68G}
}

@ARTICLE{Goldsmith2001,
       author = {{Goldsmith}, Paul F.},
        title = "{Molecular Depletion and Thermal Balance in Dark Cloud Cores}",
      journal = {\apj},
         year = 2001,
        month = aug,
       volume = {557},
       number = {2},
        pages = {736-746},
          doi = {10.1086/322255},
       adsurl = {https://ui.adsabs.harvard.edu/abs/2001ApJ...557..736G}
}

@ARTICLE{Gomez2014,
       author = {{G{\'o}mez}, Gilberto C. and {V{\'a}zquez-Semadeni}, Enrique},
        title = "{Filaments in Simulations of Molecular Cloud Formation}",
      journal = {\apj},
         year = 2014,
        month = aug,
       volume = {791},
       number = {2},
          eid = {124},
        pages = {124},
          doi = {10.1088/0004-637X/791/2/124},
archivePrefix = {arXiv},
       eprint = {1308.6298},
 primaryClass = {astro-ph.GA},
       adsurl = {https://ui.adsabs.harvard.edu/abs/2014ApJ...791..124G}
}

@ARTICLE{Goodwin2004,
       author = {{Goodwin}, S.~P. and {Whitworth}, A.~P.},
        title = "{The dynamical evolution of fractal star clusters: The survival of substructure}",
      journal = {\aap},
         year = 2004,
        month = jan,
       volume = {413},
        pages = {929-937},
          doi = {10.1051/0004-6361:20031529},
archivePrefix = {arXiv},
       eprint = {astro-ph/0310333},
 primaryClass = {astro-ph},
       adsurl = {https://ui.adsabs.harvard.edu/abs/2004A&A...413..929G}
}

@ARTICLE{GregorioHetem2015,
       author = {{Gregorio-Hetem}, J. and {Hetem}, A. and {Santos-Silva}, T. and {Fernandes}, B.},
        title = "{Statistical fractal analysis of 25 young star clusters}",
      journal = {\mnras},
         year = 2015,
        month = apr,
       volume = {448},
       number = {3},
        pages = {2504-2513},
          doi = {10.1093/mnras/stv111},
archivePrefix = {arXiv},
       eprint = {1501.04230},
 primaryClass = {astro-ph.SR},
       adsurl = {https://ui.adsabs.harvard.edu/abs/2015MNRAS.448.2504G}
}

@ARTICLE{Grudic2018,
       author = {{Grudi{\'c}}, Michael Y. and {Guszejnov}, D{\'a}vid and {Hopkins}, Philip F. and {Lamberts}, Astrid and {Boylan-Kolchin}, Michael and {Murray}, Norman and {Schmitz}, Denise},
        title = "{From the top down and back up again: star cluster structure from hierarchical star formation}",
      journal = {\mnras},
         year = 2018,
        month = nov,
       volume = {481},
       number = {1},
        pages = {688-702},
          doi = {10.1093/mnras/sty2303},
archivePrefix = {arXiv},
       eprint = {1708.09065},
 primaryClass = {astro-ph.GA},
       adsurl = {https://ui.adsabs.harvard.edu/abs/2018MNRAS.481..688G}
}

@ARTICLE{Grudic2021,
       author = {{Grudi{\'c}}, Michael Y. and {Guszejnov}, D{\'a}vid and {Hopkins}, Philip F. and {Offner}, Stella S.~R. and {Faucher-Gigu{\`e}re}, Claude-Andr{\'e}},
        title = "{STARFORGE: Towards a comprehensive numerical model of star cluster formation and feedback}",
      journal = {\mnras},
         year = 2021,
        month = sep,
       volume = {506},
       number = {2},
        pages = {2199-2231},
          doi = {10.1093/mnras/stab1347},
archivePrefix = {arXiv},
       eprint = {2010.11254},
 primaryClass = {astro-ph.IM},
       adsurl = {https://ui.adsabs.harvard.edu/abs/2021MNRAS.506.2199G}
}

@ARTICLE{Gutermuth2009,
       author = {{Gutermuth}, R.~A. and {Megeath}, S.~T. and {Myers}, P.~C. and {Allen}, L.~E. and {Pipher}, J.~L. and {Fazio}, G.~G.},
        title = "{A Spitzer Survey of Young Stellar Clusters Within One Kiloparsec of the Sun: Cluster Core Extraction and Basic Structural Analysis}",
      journal = {\apjs},
         year = 2009,
        month = sep,
       volume = {184},
       number = {1},
        pages = {18-83},
          doi = {10.1088/0067-0049/184/1/18},
archivePrefix = {arXiv},
       eprint = {0906.0201},
 primaryClass = {astro-ph.SR},
       adsurl = {https://ui.adsabs.harvard.edu/abs/2009ApJS..184...18G}
}

@ARTICLE{Guszejnov2021,
       author = {{Guszejnov}, D{\'a}vid and {Grudi{\'c}}, Michael Y. and {Hopkins}, Philip F. and {Offner}, Stella S.~R. and {Faucher-Gigu{\`e}re}, Claude-Andr{\'e}},
        title = "{STARFORGE: the effects of protostellar outflows on the IMF}",
      journal = {\mnras},
         year = 2021,
        month = apr,
       volume = {502},
       number = {3},
        pages = {3646-3663},
          doi = {10.1093/mnras/stab278},
archivePrefix = {arXiv},
       eprint = {2010.11249},
 primaryClass = {astro-ph.GA},
       adsurl = {https://ui.adsabs.harvard.edu/abs/2021MNRAS.502.3646G}
}

@ARTICLE{Guzmman2015,
       author = {{Guzm{\'a}n}, Andr{\'e}s E. and {Sanhueza}, Patricio and {Contreras}, Yanett and {Smith}, Howard A. and {Jackson}, James M. and {Hoq}, Sadia and {Rathborne}, Jill M.},
        title = "{Far-infrared Dust Temperatures and Column Densities of the MALT90 Molecular Clump Sample}",
      journal = {\apj},
         year = 2015,
        month = dec,
       volume = {815},
       number = {2},
          eid = {130},
        pages = {130},
          doi = {10.1088/0004-637X/815/2/130},
archivePrefix = {arXiv},
       eprint = {1511.00762},
 primaryClass = {astro-ph.SR},
       adsurl = {https://ui.adsabs.harvard.edu/abs/2015ApJ...815..130G}
}

@ARTICLE{Jia2025,
       author = {{Jia}, Bo-Sheng and {Zhang}, Guo-Yin and {Men'shchikov}, Alexander and {Dib}, Sami and {Li}, Jin-Zeng and {Wang}, Ke and {Li}, Di and {Li}, Xue-Mei and {Ren}, Zhi-Yuan and {Zhang}, Chang and {Pervaiz}, Nageen and {Xiao}, Lin},
        title = "{Properties of the Cores and Filaments in the Ophiuchus Molecular Cloud and its L1688 Hub-filament System}",
      journal = {Research in Astronomy and Astrophysics},
         year = 2025,
        month = aug,
       volume = {25},
       number = {8},
          eid = {085018},
        pages = {085018},
          doi = {10.1088/1674-4527/ade383},
archivePrefix = {arXiv},
       eprint = {2506.20619},
 primaryClass = {astro-ph.GA},
       adsurl = {https://ui.adsabs.harvard.edu/abs/2025RAA....25h5018J}
}

@ARTICLE{Hacar2017,
       author = {{Hacar}, A. and {Alves}, J. and {Tafalla}, M. and {Goicoechea}, J.~R.},
        title = "{Gravitational collapse of the OMC-1 region}",
      journal = {\aap},
         year = 2017,
        month = jun,
       volume = {602},
          eid = {L2},
        pages = {L2},
          doi = {10.1051/0004-6361/201730732},
archivePrefix = {arXiv},
       eprint = {1703.03464},
 primaryClass = {astro-ph.GA},
       adsurl = {https://ui.adsabs.harvard.edu/abs/2017A&A...602L...2H}
}

@ARTICLE{Hacar2018,
       author = {{Hacar}, A. and {Tafalla}, M. and {Forbrich}, J. and {Alves}, J. and {Meingast}, S. and {Grossschedl}, J. and {Teixeira}, P.~S.},
        title = "{An ALMA study of the Orion Integral Filament. I. Evidence for narrow fibers in a massive cloud}",
      journal = {\aap},
         year = 2018,
        month = mar,
       volume = {610},
          eid = {A77},
        pages = {A77},
          doi = {10.1051/0004-6361/201731894},
archivePrefix = {arXiv},
       eprint = {1801.01500},
 primaryClass = {astro-ph.GA},
       adsurl = {https://ui.adsabs.harvard.edu/abs/2018A&A...610A..77H}
}

@INPROCEEDINGS{Hacar2023,
       author = {{Hacar}, A. and {Clark}, S.~E. and {Heitsch}, F. and {Kainulainen}, J. and {Panopoulou}, G.~V. and {Seifried}, D. and {Smith}, R.},
        title = "{Initial Conditions for Star Formation: a Physical Description of the Filamentary ISM}",
    booktitle = {Protostars and Planets VII},
         year = 2023,
       editor = {{Inutsuka}, S. and {Aikawa}, Y. and {Muto}, T. and {Tomida}, K. and {Tamura}, M.},
       series = {Astronomical Society of the Pacific Conference Series},
       volume = {534},
        month = jul,
        pages = {153},
          doi = {10.48550/arXiv.2203.09562},
archivePrefix = {arXiv},
       eprint = {2203.09562},
 primaryClass = {astro-ph.GA},
       adsurl = {https://ui.adsabs.harvard.edu/abs/2023ASPC..534..153H}
}

@ARTICLE{Hillenbrand1998,
       author = {{Hillenbrand}, Lynne A. and {Hartmann}, Lee W.},
        title = "{A Preliminary Study of the Orion Nebula Cluster Structure and Dynamics}",
      journal = {\apj},
         year = 1998,
        month = jan,
       volume = {492},
       number = {2},
        pages = {540-553},
          doi = {10.1086/305076},
       adsurl = {https://ui.adsabs.harvard.edu/abs/1998ApJ...492..540H}
}

@article{Hoffman1983,
    author = {Richard Hoffman and Anil K Jain},
    title = {A test of randomness based on the minimal spanning tree},
    journal = {Pattern Recognition Letters},
    volume = {1},
    number = {3},
    pages = {175-180},
    year = {1983},
    issn = {0167-8655},
    doi = {https://doi.org/10.1016/0167-8655(83)90059-4},
    url = {https://www.sciencedirect.com/science/article/pii/0167865583900594},
}

@ARTICLE{Ishihara2024,
       author = {{Ishihara}, Kousuke and {Sanhueza}, Patricio and {Nakamura}, Fumitaka and {Saito}, Masao and {Chen}, Huei-Ru Vivien and {Li}, Shanghuo and {Olguin}, Fernando and {Taniguchi}, Kotomi and {Morii}, Kaho and {Lu}, Xing and {Luo}, Qiu-yi and {Sakai}, Takeshi and {Zhang}, Qizhou},
        title = "{Digging into the Interior of Hot Cores with ALMA (DIHCA). IV. Fragmentation in High-mass Star-forming Clumps}",
      journal = {\apj},
         year = 2024,
        month = oct,
       volume = {974},
       number = {1},
          eid = {95},
        pages = {95},
          doi = {10.3847/1538-4357/ad630f},
archivePrefix = {arXiv},
       eprint = {2407.06845},
 primaryClass = {astro-ph.GA},
       adsurl = {https://ui.adsabs.harvard.edu/abs/2024ApJ...974...95I}
}

@ARTICLE{Ishihara2025,
       author = {{Ishihara}, Kousuke and {Nakamura}, Fumitaka and {Sanhueza}, Patricio and {Saito}, Masao},
        title = "{Turbulent fragmentation as the primary driver of core formation in Polaris Flare and Lupus I}",
      journal = {\aap},
         year = 2025,
        month = mar,
       volume = {695},
          eid = {L25},
        pages = {L25},
          doi = {10.1051/0004-6361/202452427},
archivePrefix = {arXiv},
       eprint = {2503.06613},
 primaryClass = {astro-ph.GA},
       adsurl = {https://ui.adsabs.harvard.edu/abs/2025A&A...695L..25I}
}

@ARTICLE{Jaffa2017,
       author = {{Jaffa}, S.~E. and {Whitworth}, A.~P. and {Lomax}, O.},
        title = "{Q$^{+}$: characterizing the structure of young star clusters}",
      journal = {\mnras},
         year = 2017,
        month = apr,
       volume = {466},
       number = {1},
        pages = {1082-1092},
          doi = {10.1093/mnras/stw3140},
archivePrefix = {arXiv},
       eprint = {1611.10149},
 primaryClass = {astro-ph.GA},
       adsurl = {https://ui.adsabs.harvard.edu/abs/2017MNRAS.466.1082J}
}

@ARTICLE{Jeans1902,
       author = {{Jeans}, J.~H.},
        title = "{The Stability of a Spherical Nebula}",
      journal = {Philosophical Transactions of the Royal Society of London Series A},
         year = 1902,
        month = jan,
       volume = {199},
        pages = {1-53},
          doi = {10.1098/rsta.1902.0012},
       adsurl = {https://ui.adsabs.harvard.edu/abs/1902RSPTA.199....1J}
}

@ARTICLE{Juvela2012,
       author = {{Juvela}, M. and {Harju}, J. and {Ysard}, N. and {Lunttila}, T.},
        title = "{Reliability of NH$_{3}$ as the temperature probe of cold cloud cores}",
      journal = {\aap},
         year = 2012,
        month = feb,
       volume = {538},
          eid = {A133},
        pages = {A133},
          doi = {10.1051/0004-6361/201118257},
archivePrefix = {arXiv},
       eprint = {1201.3062},
 primaryClass = {astro-ph.GA},
       adsurl = {https://ui.adsabs.harvard.edu/abs/2012A&A...538A.133J}
}

@ARTICLE{Lin2022,
       author = {{Lin}, Y. and {Wyrowski}, F. and {Liu}, H.~B. and {Izquierdo}, A.~F. and {Csengeri}, T. and {Leurini}, S. and {Menten}, K.~M.},
        title = "{The evolution of temperature and density structures of OB cluster-forming molecular clumps}",
      journal = {\aap},
         year = 2022,
        month = feb,
       volume = {658},
          eid = {A128},
        pages = {A128},
          doi = {10.1051/0004-6361/202142023},
archivePrefix = {arXiv},
       eprint = {2112.01115},
 primaryClass = {astro-ph.GA},
       adsurl = {https://ui.adsabs.harvard.edu/abs/2022A&A...658A.128L}
}

@ARTICLE{Kainulainen2017,
       author = {{Kainulainen}, J. and {Stutz}, A.~M. and {Stanke}, T. and {Abreu-Vicente}, J. and {Beuther}, H. and {Henning}, T. and {Johnston}, K.~G. and {Megeath}, S.~T.},
        title = "{Resolving the fragmentation of high line-mass filaments with ALMA: the integral shaped filament in Orion A}",
      journal = {\aap},
         year = 2017,
        month = apr,
       volume = {600},
          eid = {A141},
        pages = {A141},
          doi = {10.1051/0004-6361/201628481},
archivePrefix = {arXiv},
       eprint = {1603.05688},
 primaryClass = {astro-ph.GA},
       adsurl = {https://ui.adsabs.harvard.edu/abs/2017A&A...600A.141K}
}

@ARTICLE{Kauffmann2008,
       author = {{Kauffmann}, J. and {Bertoldi}, F. and {Bourke}, T.~L. and {Evans}, II, N.~J. and {Lee}, C.~W.},
        title = "{MAMBO mapping of Spitzer c2d small clouds and cores}",
      journal = {\aap},
         year = 2008,
        month = sep,
       volume = {487},
       number = {3},
        pages = {993-1017},
          doi = {10.1051/0004-6361:200809481},
archivePrefix = {arXiv},
       eprint = {0805.4205},
 primaryClass = {astro-ph},
       adsurl = {https://ui.adsabs.harvard.edu/abs/2008A&A...487..993K}
}

@ARTICLE{Kauffmann2010,
       author = {{Kauffmann}, Jens and {Pillai}, Thushara},
        title = "{How Many Infrared Dark Clouds Can form Massive Stars and Clusters?}",
      journal = {\apjl},
         year = 2010,
        month = nov,
       volume = {723},
       number = {1},
        pages = {L7-L12},
          doi = {10.1088/2041-8205/723/1/L7},
archivePrefix = {arXiv},
       eprint = {1009.1617},
 primaryClass = {astro-ph.GA},
       adsurl = {https://ui.adsabs.harvard.edu/abs/2010ApJ...723L...7K}
}

@ARTICLE{Kirk2013,
       author = {{Kirk}, Helen and {Myers}, Philip C. and {Bourke}, Tyler L. and {Gutermuth}, Robert A. and {Hedden}, Abigail and {Wilson}, Grant W.},
        title = "{Filamentary Accretion Flows in the Embedded Serpens South Protocluster}",
      journal = {\apj},
         year = 2013,
        month = apr,
       volume = {766},
       number = {2},
          eid = {115},
        pages = {115},
          doi = {10.1088/0004-637X/766/2/115},
archivePrefix = {arXiv},
       eprint = {1301.6792},
 primaryClass = {astro-ph.GA},
       adsurl = {https://ui.adsabs.harvard.edu/abs/2013ApJ...766..115K}
}

@ARTICLE{Konyves2015,
       author = {{K{\"o}nyves}, V. and {Andr{\'e}}, Ph. and {Men'shchikov}, A. and {Palmeirim}, P. and {Arzoumanian}, D. and {Schneider}, N. and {Roy}, A. and {Didelon}, P. and {Maury}, A. and {Shimajiri}, Y. and {Di Francesco}, J. and {Bontemps}, S. and {Peretto}, N. and {Benedettini}, M. and {Bernard}, J. -Ph. and {Elia}, D. and {Griffin}, M.~J. and {Hill}, T. and {Kirk}, J. and {Ladjelate}, B. and {Marsh}, K. and {Martin}, P.~G. and {Motte}, F. and {Nguy{\^e}n Luong}, Q. and {Pezzuto}, S. and {Roussel}, H. and {Rygl}, K.~L.~J. and {Sadavoy}, S.~I. and {Schisano}, E. and {Spinoglio}, L. and {Ward-Thompson}, D. and {White}, G.~J.},
        title = "{A census of dense cores in the Aquila cloud complex: SPIRE/PACS observations from the Herschel Gould Belt survey}",
      journal = {\aap},
         year = 2015,
        month = dec,
       volume = {584},
          eid = {A91},
        pages = {A91},
          doi = {10.1051/0004-6361/201525861},
archivePrefix = {arXiv},
       eprint = {1507.05926},
 primaryClass = {astro-ph.GA},
       adsurl = {https://ui.adsabs.harvard.edu/abs/2015A&A...584A..91K}
}

@ARTICLE{Krause2020,
       author = {{Krause}, Martin G.~H. and {Offner}, Stella S.~R. and {Charbonnel}, Corinne and {Gieles}, Mark and {Klessen}, Ralf S. and {V{\'a}zquez-Semadeni}, Enrique and {Ballesteros-Paredes}, Javier and {Girichidis}, Philipp and {Kruijssen}, J.~M. Diederik and {Ward}, Jacob L. and {Zinnecker}, Hans},
        title = "{The Physics of Star Cluster Formation and Evolution}",
      journal = {\ssr},
         year = 2020,
        month = jun,
       volume = {216},
       number = {4},
          eid = {64},
        pages = {64},
          doi = {10.1007/s11214-020-00689-4},
archivePrefix = {arXiv},
       eprint = {2005.00801},
 primaryClass = {astro-ph.GA},
       adsurl = {https://ui.adsabs.harvard.edu/abs/2020SSRv..216...64K}
}

@ARTICLE{Kruijssen2012,
       author = {{Kruijssen}, J.~M. Diederik},
        title = "{On the fraction of star formation occurring in bound stellar clusters}",
      journal = {\mnras},
         year = 2012,
        month = nov,
       volume = {426},
       number = {4},
        pages = {3008-3040},
          doi = {10.1111/j.1365-2966.2012.21923.x},
archivePrefix = {arXiv},
       eprint = {1208.2963},
 primaryClass = {astro-ph.CO},
       adsurl = {https://ui.adsabs.harvard.edu/abs/2012MNRAS.426.3008K}
}

@ARTICLE{Krumholz2006,
       author = {{Krumholz}, Mark R.},
        title = "{Radiation Feedback and Fragmentation in Massive Protostellar Cores}",
      journal = {\apjl},
         year = 2006,
        month = apr,
       volume = {641},
       number = {1},
        pages = {L45-L48},
          doi = {10.1086/503771},
archivePrefix = {arXiv},
       eprint = {astro-ph/0603026},
 primaryClass = {astro-ph},
       adsurl = {https://ui.adsabs.harvard.edu/abs/2006ApJ...641L..45K}
}

@ARTICLE{Krumholz2008,
       author = {{Krumholz}, Mark R. and {McKee}, Christopher F.},
        title = "{A minimum column density of 1gcm$^{-2}$ for massive star formation}",
      journal = {\nat},
         year = 2008,
        month = feb,
       volume = {451},
       number = {7182},
        pages = {1082-1084},
          doi = {10.1038/nature06620},
archivePrefix = {arXiv},
       eprint = {0801.0442},
 primaryClass = {astro-ph},
       adsurl = {https://ui.adsabs.harvard.edu/abs/2008Natur.451.1082K}
}

@INPROCEEDINGS{Krumholz2014,
       author = {{Krumholz}, M.~R. and {Bate}, M.~R. and {Arce}, H.~G. and {Dale}, J.~E. and {Gutermuth}, R. and {Klein}, R.~I. and {Li}, Z. -Y. and {Nakamura}, F. and {Zhang}, Q.},
        title = "{Star Cluster Formation and Feedback}",
    booktitle = {Protostars and Planets VI},
         year = 2014,
       editor = {{Beuther}, Henrik and {Klessen}, Ralf S. and {Dullemond}, Cornelis P. and {Henning}, Thomas},
        month = jan,
        pages = {243-266},
          doi = {10.2458/azu_uapress_9780816531240-ch011},
archivePrefix = {arXiv},
       eprint = {1401.2473},
 primaryClass = {astro-ph.GA},
       adsurl = {https://ui.adsabs.harvard.edu/abs/2014prpl.conf..243K}
}

@ARTICLE{Krumholz2020,
       author = {{Krumholz}, Mark R. and {McKee}, Christopher F.},
        title = "{How do bound star clusters form?}",
      journal = {\mnras},
         year = 2020,
        month = may,
       volume = {494},
       number = {1},
        pages = {624-641},
          doi = {10.1093/mnras/staa659},
archivePrefix = {arXiv},
       eprint = {1909.01565},
 primaryClass = {astro-ph.GA},
       adsurl = {https://ui.adsabs.harvard.edu/abs/2020MNRAS.494..624K}
}

@ARTICLE{Kuhn2014,
       author = {{Kuhn}, Michael A. and {Feigelson}, Eric D. and {Getman}, Konstantin V. and {Baddeley}, Adrian J. and {Broos}, Patrick S. and {Sills}, Alison and {Bate}, Matthew R. and {Povich}, Matthew S. and {Luhman}, Kevin L. and {Busk}, Heather A. and {Naylor}, Tim and {King}, Robert R.},
        title = "{The Spatial Structure of Young Stellar Clusters. I. Subclusters}",
      journal = {\apj},
         year = 2014,
        month = jun,
       volume = {787},
       number = {2},
          eid = {107},
        pages = {107},
          doi = {10.1088/0004-637X/787/2/107},
archivePrefix = {arXiv},
       eprint = {1403.4252},
 primaryClass = {astro-ph.GA},
       adsurl = {https://ui.adsabs.harvard.edu/abs/2014ApJ...787..107K}
}

@ARTICLE{Lada2003,
       author = {{Lada}, Charles J. and {Lada}, Elizabeth A.},
        title = "{Embedded Clusters in Molecular Clouds}",
      journal = {\araa},
         year = 2003,
        month = jan,
       volume = {41},
        pages = {57-115},
          doi = {10.1146/annurev.astro.41.011802.094844},
archivePrefix = {arXiv},
       eprint = {astro-ph/0301540},
 primaryClass = {astro-ph},
       adsurl = {https://ui.adsabs.harvard.edu/abs/2003ARA&A..41...57L}
}

@ARTICLE{Lebreuilly2025,
       author = {{Lebreuilly}, Ugo and {Traficante}, Alessio and {Nucara}, Alice and {Tung}, Ngo-Duy and {Hennebelle}, Patrick and {Molinari}, Sergio and {Klessen}, Ralf S. and {Testi}, Leonardo and {Pelkonen}, Veli-Matti and {Benedettini}, Milena and {Coletta}, Alessandro and {Elia}, Davide and {Mininni}, Chiara and {Pezzuto}, Stefania and {Soler}, Juan D. and {Suin}, Paolo and {Toci}, Claudia},
        title = "{The Rosetta Stone Project: I. A suite of radiative magnetohydrodynamics simulations of high-mass star-forming clumps}",
      journal = {\aap},
         year = 2025,
        month = sep,
       volume = {701},
          eid = {A217},
        pages = {A217},
          doi = {10.1051/0004-6361/202554774},
archivePrefix = {arXiv},
       eprint = {2507.08436},
 primaryClass = {astro-ph.SR},
       adsurl = {https://ui.adsabs.harvard.edu/abs/2025A&A...701A.217L}
}

@ARTICLE{Li2019,
       author = {{Li}, Hui and {Vogelsberger}, Mark and {Marinacci}, Federico and {Gnedin}, Oleg Y.},
        title = "{Disruption of giant molecular clouds and formation of bound star clusters under the influence of momentum stellar feedback}",
      journal = {\mnras},
         year = 2019,
        month = jul,
       volume = {487},
       number = {1},
        pages = {364-380},
          doi = {10.1093/mnras/stz1271},
archivePrefix = {arXiv},
       eprint = {1904.11987},
 primaryClass = {astro-ph.GA},
       adsurl = {https://ui.adsabs.harvard.edu/abs/2019MNRAS.487..364L}
}

@ARTICLE{Li2024G,
       author = {{Li}, Guang-Xing},
        title = "{Tides in clouds: control of star formation by long-range gravitational force}",
      journal = {\mnras},
         year = 2024,
        month = feb,
       volume = {528},
       number = {1},
        pages = {L52-L58},
          doi = {10.1093/mnrasl/slad149},
archivePrefix = {arXiv},
       eprint = {2309.16125},
 primaryClass = {astro-ph.GA},
       adsurl = {https://ui.adsabs.harvard.edu/abs/2024MNRAS.528L..52L}
}

@ARTICLE{Liu2020,
       author = {{Liu}, Tie and {Evans}, Neal J. and {Kim}, Kee-Tae and {Goldsmith}, Paul F. and {Liu}, Sheng-Yuan and {Zhang}, Qizhou and {Tatematsu}, Ken'ichi and {Wang}, Ke and {Juvela}, Mika and {Bronfman}, Leonardo and {Cunningham}, Maria R. and {Garay}, Guido and {Hirota}, Tomoya and {Lee}, Jeong-Eun and {Kang}, Sung-Ju and {Li}, Di and {Li}, Pak-Shing and {Mardones}, Diego and {Qin}, Sheng-Li and {Ristorcelli}, Isabelle and {Tej}, Anandmayee and {Toth}, L. Viktor and {Wu}, Jing-Wen and {Wu}, Yue-Fang and {Yi}, Hee-weon and {Yun}, Hyeong-Sik and {Liu}, Hong-Li and {Peng}, Ya-Ping and {Li}, Juan and {Li}, Shang-Huo and {Lee}, Chang Won and {Shen}, Zhi-Qiang and {Baug}, Tapas and {Wang}, Jun-Zhi and {Zhang}, Yong and {Issac}, Namitha and {Zhu}, Feng-Yao and {Luo}, Qiu-Yi and {Soam}, Archana and {Liu}, Xun-Chuan and {Xu}, Feng-Wei and {Wang}, Yu and {Zhang}, Chao and {Ren}, Zhiyuan and {Zhang}, Chao},
        title = "{ATOMS: ALMA Three-millimeter Observations of Massive Star-forming regions - I. Survey description and a first look at G9.62+0.19}",
      journal = {\mnras},
         year = 2020,
        month = aug,
       volume = {496},
       number = {3},
        pages = {2790-2820},
          doi = {10.1093/mnras/staa1577},
archivePrefix = {arXiv},
       eprint = {2006.01549},
 primaryClass = {astro-ph.GA},
       adsurl = {https://ui.adsabs.harvard.edu/abs/2020MNRAS.496.2790L}
}

@ARTICLE{Liu2022,
       author = {{Liu}, Hong-Li and {Tej}, Anandmayee and {Liu}, Tie and {Goldsmith}, Paul F. and {Stutz}, Amelia and {Juvela}, Mika and {Qin}, Sheng-Li and {Xu}, Feng-Wei and {Bronfman}, Leonardo and {Evans}, Neal J. and {Saha}, Anindya and {Issac}, Namitha and {Tatematsu}, Ken'ichi and {Wang}, Ke and {Li}, Shanghuo and {Zhang}, Siju and {Baug}, Tapas and {Dewangan}, Lokesh and {Wu}, Yue-Fang and {Zhang}, Yong and {Lee}, Chang Won and {Liu}, Xun-Chuan and {Zhou}, Jianwen and {Soam}, Archana},
        title = "{ATOMS: ALMA Three-millimeter Observations of Massive Star-forming regions - IX. A pilot study towards IRDC G034.43+00.24 on multi-scale structures and gas kinematics}",
      journal = {\mnras},
         year = 2022,
        month = apr,
       volume = {511},
       number = {3},
        pages = {4480-4489},
          doi = {10.1093/mnras/stac378},
archivePrefix = {arXiv},
       eprint = {2202.11307},
 primaryClass = {astro-ph.GA},
       adsurl = {https://ui.adsabs.harvard.edu/abs/2022MNRAS.511.4480L}
}

@ARTICLE{Liu2024,
       author = {{Liu}, Xunchuan and {Liu}, Tie and {Zhu}, Lei and {Garay}, Guido and {Liu}, Hong-Li and {Goldsmith}, Paul and {Evans}, Neal and {Kim}, Kee-Tae and {Liu}, Sheng-Yuan and {Xu}, Fengwei and {Lu}, Xing and {Tej}, Anandmayee and {Mai}, Xiaofeng and {Bronfman}, Leonardo and {Li}, Shanghuo and {Mardones}, Diego and {Stutz}, Amelia and {Tatematsu}, Ken'ichi and {Wang}, Ke and {Zhang}, Qizhou and {Qin}, Sheng-Li and {Zhou}, Jianwen and {Luo}, Qiuyi and {Zhang}, Siju and {Cheng}, Yu and {He}, Jinhua and {Gu}, Qilao and {Li}, Ziyang and {Zhang}, Zhenying and {Zhang}, Suinan and {Saha}, Anindya and {Dewangan}, Lokesh and {Sanhueza}, Patricio and {Shen}, Zhiqiang},
        title = "{The ALMA-QUARKS Survey. I. Survey Description and Data Reduction}",
      journal = {Research in Astronomy and Astrophysics},
         year = 2024,
        month = feb,
       volume = {24},
       number = {2},
          eid = {025009},
        pages = {025009},
          doi = {10.1088/1674-4527/ad0d5c},
archivePrefix = {arXiv},
       eprint = {2311.08651},
 primaryClass = {astro-ph.GA},
       adsurl = {https://ui.adsabs.harvard.edu/abs/2024RAA....24b5009L}
}

@INPROCEEDINGS{Longmore2014,
       author = {{Longmore}, S.~N. and {Kruijssen}, J.~M.~D. and {Bastian}, N. and {Bally}, J. and {Rathborne}, J. and {Testi}, L. and {Stolte}, A. and {Dale}, J. and {Bressert}, E. and {Alves}, J.},
        title = "{The Formation and Early Evolution of Young Massive Clusters}",
    booktitle = {Protostars and Planets VI},
         year = 2014,
       editor = {{Beuther}, Henrik and {Klessen}, Ralf S. and {Dullemond}, Cornelis P. and {Henning}, Thomas},
        month = jan,
        pages = {291-314},
          doi = {10.2458/azu_uapress_9780816531240-ch013},
archivePrefix = {arXiv},
       eprint = {1401.4175},
 primaryClass = {astro-ph.GA},
       adsurl = {https://ui.adsabs.harvard.edu/abs/2014prpl.conf..291L}
}

@ARTICLE{Louvet2024,
       author = {{Louvet}, F. and {Sanhueza}, P. and {Stutz}, A. and {Men'shchikov}, A. and {Motte}, F. and {Galv{\'a}n-Madrid}, R. and {Bontemps}, S. and {Pouteau}, Y. and {Ginsburg}, A. and {Csengeri}, T. and {Di Francesco}, J. and {Dell'Ova}, P. and {Gonz{\'a}lez}, M. and {Didelon}, P. and {Braine}, J. and {Cunningham}, N. and {Thomasson}, B. and {Lesaffre}, P. and {Hennebelle}, P. and {Bonfand}, M. and {Gusdorf}, A. and {{\'A}lvarez-Guti{\'e}rrez}, R.~H. and {Nony}, T. and {Busquet}, G. and {Olguin}, F. and {Bronfman}, L. and {Salinas}, J. and {Fernandez-Lopez}, M. and {Moraux}, E. and {Liu}, H.~L. and {Lu}, X. and {Huei-Ru}, V. and {Towner}, A. and {Valeille-Manet}, M. and {Brouillet}, N. and {Herpin}, F. and {Lefloch}, B. and {Baug}, T. and {Maud}, L. and {L{\'o}pez-Sepulcre}, A. and {Svoboda}, B.},
        title = "{ALMA-IMF: XV. Core mass function in the high-mass star formation regime}",
      journal = {\aap},
         year = 2024,
        month = oct,
       volume = {690},
          eid = {A33},
        pages = {A33},
          doi = {10.1051/0004-6361/202345986},
archivePrefix = {arXiv},
       eprint = {2407.18719},
 primaryClass = {astro-ph.GA},
       adsurl = {https://ui.adsabs.harvard.edu/abs/2024A&A...690A..33L}
}

@ARTICLE{Lumsden2013,
       author = {{Lumsden}, S.~L. and {Hoare}, M.~G. and {Urquhart}, J.~S. and {Oudmaijer}, R.~D. and {Davies}, B. and {Mottram}, J.~C. and {Cooper}, H.~D.~B. and {Moore}, T.~J.~T.},
        title = "{The Red MSX Source Survey: The Massive Young Stellar Population of Our Galaxy}",
      journal = {\apjs},
         year = 2013,
        month = sep,
       volume = {208},
       number = {1},
          eid = {11},
        pages = {11},
          doi = {10.1088/0067-0049/208/1/11},
archivePrefix = {arXiv},
       eprint = {1308.0134},
 primaryClass = {astro-ph.GA},
       adsurl = {https://ui.adsabs.harvard.edu/abs/2013ApJS..208...11L}
}

@ARTICLE{MacLow2004,
       author = {{Mac Low}, Mordecai-Mark and {Klessen}, Ralf S.},
        title = "{Control of star formation by supersonic turbulence}",
      journal = {Reviews of Modern Physics},
         year = 2004,
        month = jan,
       volume = {76},
       number = {1},
        pages = {125-194},
          doi = {10.1103/RevModPhys.76.125},
archivePrefix = {arXiv},
       eprint = {astro-ph/0301093},
 primaryClass = {astro-ph},
       adsurl = {https://ui.adsabs.harvard.edu/abs/2004RvMP...76..125M}
}

@ARTICLE{Maschberger2010,
       author = {{Maschberger}, Th. and {Clarke}, C.~J. and {Bonnell}, I.~A. and {Kroupa}, P.},
        title = "{Properties of hierarchically forming star clusters}",
      journal = {\mnras},
         year = 2010,
        month = may,
       volume = {404},
       number = {2},
        pages = {1061-1080},
          doi = {10.1111/j.1365-2966.2010.16346.x},
archivePrefix = {arXiv},
       eprint = {1002.4401},
 primaryClass = {astro-ph.GA},
       adsurl = {https://ui.adsabs.harvard.edu/abs/2010MNRAS.404.1061M}
}

@ARTICLE{McKee2003,
       author = {{McKee}, Christopher F. and {Tan}, Jonathan C.},
        title = "{The Formation of Massive Stars from Turbulent Cores}",
      journal = {\apj},
         year = 2003,
        month = mar,
       volume = {585},
       number = {2},
        pages = {850-871},
          doi = {10.1086/346149},
archivePrefix = {arXiv},
       eprint = {astro-ph/0206037},
 primaryClass = {astro-ph},
       adsurl = {https://ui.adsabs.harvard.edu/abs/2003ApJ...585..850M}
}

@ARTICLE{Megeath2016,
       author = {{Megeath}, S.~T. and {Gutermuth}, R. and {Muzerolle}, J. and {Kryukova}, E. and {Hora}, J.~L. and {Allen}, L.~E. and {Flaherty}, K. and {Hartmann}, L. and {Myers}, P.~C. and {Pipher}, J.~L. and {Stauffer}, J. and {Young}, E.~T. and {Fazio}, G.~G.},
        title = "{The Spitzer Space Telescope Survey of the Orion A and B Molecular Clouds. II. The Spatial Distribution and Demographics of Dusty Young Stellar Objects}",
      journal = {\aj},
         year = 2016,
        month = jan,
       volume = {151},
       number = {1},
          eid = {5},
        pages = {5},
          doi = {10.3847/0004-6256/151/1/5},
archivePrefix = {arXiv},
       eprint = {1511.01202},
 primaryClass = {astro-ph.GA},
       adsurl = {https://ui.adsabs.harvard.edu/abs/2016AJ....151....5M}
}

@ARTICLE{Megeath2022,
       author = {{Megeath}, S.~T. and {Gutermuth}, R.~A. and {Kounkel}, M.~A.},
        title = "{Low Mass Stars as Tracers of Star and Cluster Formation}",
      journal = {\pasp},
         year = 2022,
        month = apr,
       volume = {134},
       number = {1034},
          eid = {042001},
        pages = {042001},
          doi = {10.1088/1538-3873/ac4c9c},
archivePrefix = {arXiv},
       eprint = {2203.03655},
 primaryClass = {astro-ph.GA},
       adsurl = {https://ui.adsabs.harvard.edu/abs/2022PASP..134d2001M}
}

@ARTICLE{Merello2019,
       author = {{Merello}, M. and {Molinari}, S. and {Rygl}, K.~L.~J. and {Evans}, N.~J. and {Elia}, D. and {Schisano}, E. and {Traficante}, A. and {Shirley}, Y. and {Svoboda}, B. and {Goldsmith}, P.~F.},
        title = "{Thermal balance and comparison of gas and dust properties of dense clumps in the Hi-GAL survey}",
      journal = {\mnras},
         year = 2019,
        month = mar,
       volume = {483},
       number = {4},
        pages = {5355-5379},
          doi = {10.1093/mnras/sty3453},
archivePrefix = {arXiv},
       eprint = {1812.06134},
 primaryClass = {astro-ph.GA},
       adsurl = {https://ui.adsabs.harvard.edu/abs/2019MNRAS.483.5355M}
}

@ARTICLE{Myers2009,
       author = {{Myers}, Philip C.},
        title = "{Filamentary Structure of Star-forming Complexes}",
      journal = {\apj},
         year = 2009,
        month = aug,
       volume = {700},
       number = {2},
        pages = {1609-1625},
          doi = {10.1088/0004-637X/700/2/1609},
archivePrefix = {arXiv},
       eprint = {0906.2005},
 primaryClass = {astro-ph.GA},
       adsurl = {https://ui.adsabs.harvard.edu/abs/2009ApJ...700.1609M}
}

@ARTICLE{Mininni2025,
       author = {{Mininni}, C. and {Molinari}, S. and {Soler}, J.~D. and {S{\'a}nchez-Monge}, {\'A}. and {Coletta}, A. and {Benedettini}, M. and {Traficante}, A. and {Schisano}, E. and {Elia}, D. and {Pezzuto}, S. and {Nucara}, A. and {Schilke}, P. and {Battersby}, C. and {Ho}, P.~T.~P. and {Beltr{\'a}n}, M.~T. and {Beuther}, H. and {Fuller}, G.~A. and {Jones}, B. and {Klessen}, R.~S. and {Zhang}, Q. and {Walch}, S. and {Tang}, Y. and {Ahmadi}, A. and {Allande}, J. and {Avison}, A. and {Brogan}, C.~L. and {De Angelis}, F. and {Fontani}, F. and {Hennebelle}, P. and {Hunter}, T.~R. and {Johnston}, K.~G. and {Koch}, P. and {Kuiper}, R. and {Law}, C.-Y. and {Lis}, D.~C. and {Liu}, S. and {Liu}, T. and {Liu}, S.-Y. and {Moscadelli}, L. and {M{\"o}ller}, T. and {Rigby}, A.~J. and {Rygl}, K.~L.~J. and {Sanhueza}, P. and {Testi}, L. and {Su}, Y.-N. and {van der Tak}, F.~F.~S. and {Wells}, M.~R.~A. and {Bronfman}, L. and {Zhang}, T. and {Zinnecker}, H.},
        title = "{ALMAGAL: IV. Morphological comparison of molecular and thermal dust emission using the histogram of oriented gradients method}",
      journal = {\aap},
         year = 2025,
        month = jun,
       volume = {699},
          eid = {A34},
        pages = {A34},
          doi = {10.1051/0004-6361/202452700},
archivePrefix = {arXiv},
       eprint = {2504.12963},
 primaryClass = {astro-ph.GA},
       adsurl = {https://ui.adsabs.harvard.edu/abs/2025A&A...699A..34M}
}

@ARTICLE{Molinari2008,
       author = {{Molinari}, S. and {Pezzuto}, S. and {Cesaroni}, R. and {Brand}, J. and {Faustini}, F. and {Testi}, L.},
        title = "{The evolution of the spectral energy distribution in massive young stellar objects}",
      journal = {\aap},
         year = 2008,
        month = apr,
       volume = {481},
       number = {2},
        pages = {345-365},
          doi = {10.1051/0004-6361:20078661},
       adsurl = {https://ui.adsabs.harvard.edu/abs/2008A&A...481..345M}
}

@ARTICLE{Molinari2010,
       author = {{Molinari}, S. and {Swinyard}, B. and {Bally}, J. and {Barlow}, M. and {Bernard}, J. -P. and {Martin}, P. and {Moore}, T. and {Noriega-Crespo}, A. and {Plume}, R. and {Testi}, L. and {Zavagno}, A. and {Abergel}, A. and {Ali}, B. and {Anderson}, L. and {Andr{\'e}}, P. and {Baluteau}, J. -P. and {Battersby}, C. and {Beltr{\'a}n}, M.~T. and {Benedettini}, M. and {Billot}, N. and {Blommaert}, J. and {Bontemps}, S. and {Boulanger}, F. and {Brand}, J. and {Brunt}, C. and {Burton}, M. and {Calzoletti}, L. and {Carey}, S. and {Caselli}, P. and {Cesaroni}, R. and {Cernicharo}, J. and {Chakrabarti}, S. and {Chrysostomou}, A. and {Cohen}, M. and {Compiegne}, M. and {de Bernardis}, P. and {de Gasperis}, G. and {di Giorgio}, A.~M. and {Elia}, D. and {Faustini}, F. and {Flagey}, N. and {Fukui}, Y. and {Fuller}, G.~A. and {Ganga}, K. and {Garcia-Lario}, P. and {Glenn}, J. and {Goldsmith}, P.~F. and {Griffin}, M. and {Hoare}, M. and {Huang}, M. and {Ikhenaode}, D. and {Joblin}, C. and {Joncas}, G. and {Juvela}, M. and {Kirk}, J.~M. and {Lagache}, G. and {Li}, J.~Z. and {Lim}, T.~L. and {Lord}, S.~D. and {Marengo}, M. and {Marshall}, D.~J. and {Masi}, S. and {Massi}, F. and {Matsuura}, M. and {Minier}, V. and {Miville-Desch{\^e}nes}, M. -A. and {Montier}, L.~A. and {Morgan}, L. and {Motte}, F. and {Mottram}, J.~C. and {M{\"u}ller}, T.~G. and {Natoli}, P. and {Neves}, J. and {Olmi}, L. and {Paladini}, R. and {Paradis}, D. and {Parsons}, H. and {Peretto}, N. and {Pestalozzi}, M. and {Pezzuto}, S. and {Piacentini}, F. and {Piazzo}, L. and {Polychroni}, D. and {Pomar{\`e}s}, M. and {Popescu}, C.~C. and {Reach}, W.~T. and {Ristorcelli}, I. and {Robitaille}, J. -F. and {Robitaille}, T. and {Rod{\'o}n}, J.~A. and {Roy}, A. and {Royer}, P. and {Russeil}, D. and {Saraceno}, P. and {Sauvage}, M. and {Schilke}, P. and {Schisano}, E. and {Schneider}, N. and {Schuller}, F. and {Schulz}, B. and {Sibthorpe}, B. and {Smith}, H.~A. and {Smith}, M.~D. and {Spinoglio}, L. and {Stamatellos}, D. and {Strafella}, F. and {Stringfellow}, G.~S. and {Sturm}, E. and {Taylor}, R. and {Thompson}, M.~A. and {Traficante}, A. and {Tuffs}, R.~J. and {Umana}, G. and {Valenziano}, L. and {Vavrek}, R. and {Veneziani}, M. and {Viti}, S. and {Waelkens}, C. and {Ward-Thompson}, D. and {White}, G. and {Wilcock}, L.~A. and {Wyrowski}, F. and {Yorke}, H.~W. and {Zhang}, Q.},
        title = "{Clouds, filaments, and protostars: The Herschel Hi-GAL Milky Way}",
      journal = {\aap},
         year = 2010,
        month = jul,
       volume = {518},
          eid = {L100},
        pages = {L100},
          doi = {10.1051/0004-6361/201014659},
archivePrefix = {arXiv},
       eprint = {1005.3317},
 primaryClass = {astro-ph.GA},
       adsurl = {https://ui.adsabs.harvard.edu/abs/2010A&A...518L.100M}
}

@ARTICLE{Molinari2011,
       author = {{Molinari}, S. and {Schisano}, E. and {Faustini}, F. and {Pestalozzi}, M. and {di Giorgio}, A.~M. and {Liu}, S.},
        title = "{Source extraction and photometry for the far-infrared and sub-millimeter continuum in the presence of complex backgrounds}",
      journal = {\aap},
         year = 2011,
        month = jun,
       volume = {530},
          eid = {A133},
        pages = {A133},
          doi = {10.1051/0004-6361/201014752},
archivePrefix = {arXiv},
       eprint = {1011.3946},
 primaryClass = {astro-ph.GA},
       adsurl = {https://ui.adsabs.harvard.edu/abs/2011A&A...530A.133M}
}

@ARTICLE{Molinari2016,
       author = {{Molinari}, S. and {Schisano}, E. and {Elia}, D. and {Pestalozzi}, M. and {Traficante}, A. and {Pezzuto}, S. and {Swinyard}, B.~M. and {Noriega-Crespo}, A. and {Bally}, J. and {Moore}, T.~J.~T. and {Plume}, R. and {Zavagno}, A. and {di Giorgio}, A.~M. and {Liu}, S.~J. and {Pilbratt}, G.~L. and {Mottram}, J.~C. and {Russeil}, D. and {Piazzo}, L. and {Veneziani}, M. and {Benedettini}, M. and {Calzoletti}, L. and {Faustini}, F. and {Natoli}, P. and {Piacentini}, F. and {Merello}, M. and {Palmese}, A. and {Del Grande}, R. and {Polychroni}, D. and {Rygl}, K.~L.~J. and {Polenta}, G. and {Barlow}, M.~J. and {Bernard}, J. -P. and {Martin}, P.~G. and {Testi}, L. and {Ali}, B. and {Andr{\'e}}, P. and {Beltr{\'a}n}, M.~T. and {Billot}, N. and {Carey}, S. and {Cesaroni}, R. and {Compi{\`e}gne}, M. and {Eden}, D. and {Fukui}, Y. and {Garcia-Lario}, P. and {Hoare}, M.~G. and {Huang}, M. and {Joncas}, G. and {Lim}, T.~L. and {Lord}, S.~D. and {Martinavarro-Armengol}, S. and {Motte}, F. and {Paladini}, R. and {Paradis}, D. and {Peretto}, N. and {Robitaille}, T. and {Schilke}, P. and {Schneider}, N. and {Schulz}, B. and {Sibthorpe}, B. and {Strafella}, F. and {Thompson}, M.~A. and {Umana}, G. and {Ward-Thompson}, D. and {Wyrowski}, F.},
        title = "{Hi-GAL, the Herschel infrared Galactic Plane Survey: photometric maps and compact source catalogues. First data release for the inner Milky Way: +68{\textdegree} {\ensuremath{\geq}} l {\ensuremath{\geq}} -70{\textdegree}}",
      journal = {\aap},
         year = 2016,
        month = jul,
       volume = {591},
          eid = {A149},
        pages = {A149},
          doi = {10.1051/0004-6361/201526380},
archivePrefix = {arXiv},
       eprint = {1604.05911},
 primaryClass = {astro-ph.GA},
       adsurl = {https://ui.adsabs.harvard.edu/abs/2016A&A...591A.149M}
}

@ARTICLE{Molinari2016b,
       author = {{Molinari}, S. and {Merello}, M. and {Elia}, D. and {Cesaroni}, R. and {Testi}, L. and {Robitaille}, T.},
        title = "{Calibration of Evolutionary Diagnostics in High-mass Star Formation}",
      journal = {\apjl},
         year = 2016,
        month = jul,
       volume = {826},
       number = {1},
          eid = {L8},
        pages = {L8},
          doi = {10.3847/2041-8205/826/1/L8},
archivePrefix = {arXiv},
       eprint = {1604.06192},
 primaryClass = {astro-ph.GA},
       adsurl = {https://ui.adsabs.harvard.edu/abs/2016ApJ...826L...8M}
}

@ARTICLE{Molinari2025,
       author = {{Molinari}, S. and {Schilke}, P. and {Battersby}, C. and {Ho}, P.~T.~P. and {S{\'a}nchez-Monge}, {\'A}. and {Traficante}, A. and {Jones}, B. and {Beltr{\'a}n}, M.~T. and {Beuther}, H. and {Fuller}, G.~A. and {Zhang}, Q. and {Klessen}, R.~S. and {Walch}, S. and {Tang}, Y. -W. and {Benedettini}, M. and {Elia}, D. and {Coletta}, A. and {Mininni}, C. and {Schisano}, E. and {Avison}, A. and {Law}, C.~Y. and {Nucara}, A. and {Soler}, J.~D. and {Stroud}, G. and {Wallace}, J. and {Wells}, M.~R.~A. and {Ahmadi}, A. and {Brogan}, C.~L. and {Hunter}, T.~R. and {Liu}, S. -Y. and {Pezzuto}, S. and {Su}, Y. -N. and {Zimmermann}, B. and {Zhang}, T. and {Wyrowski}, F. and {De Angelis}, F. and {Liu}, S. and {Clarke}, S.~D. and {Fontani}, F. and {Klaassen}, P.~D. and {Koch}, P. and {Johnston}, K.~G. and {Lebreuilly}, U. and {Liu}, T. and {Lumsden}, S.~L. and {Moeller}, T. and {Moscadelli}, L. and {Kuiper}, R. and {Lis}, D. and {Peretto}, N. and {Pfalzner}, S. and {Rigby}, A.~J. and {Sanhueza}, P. and {Rygl}, K.~L.~J. and {van der Tak}, F. and {Zinnecker}, H. and {Amaral}, F. and {Bally}, J. and {Bronfman}, L. and {Cesaroni}, R. and {Goh}, K. and {Hoare}, M.~G. and {Hatchfield}, P. and {Hennebelle}, P. and {Henning}, T. and {Kim}, K. -T. and {Kim}, W. -J. and {Maud}, L. and {Merello}, M. and {Nakamura}, F. and {Plume}, R. and {Qin}, S. -L. and {Svoboda}, B. and {Testi}, L. and {Veena}, V.~S. and {Walker}, D.},
        title = "{ALMAGAL: I. The ALMA evolutionary study of high-mass protocluster formation in the Galaxy: Presentation of the survey and early results}",
      journal = {\aap},
         year = 2025,
        month = apr,
       volume = {696},
          eid = {A149},
        pages = {A149},
          doi = {10.1051/0004-6361/202452702},
archivePrefix = {arXiv},
       eprint = {2503.05555},
 primaryClass = {astro-ph.GA},
       adsurl = {https://ui.adsabs.harvard.edu/abs/2025A&A...696A.149M}
}

@ARTICLE{Morii2021,
       author = {{Morii}, Kaho and {Sanhueza}, Patricio and {Nakamura}, Fumitaka and {Jackson}, James M. and {Li}, Shanghuo and {Beuther}, Henrik and {Zhang}, Qizhou and {Feng}, Siyi and {Tafoya}, Daniel and {Guzm{\'a}n}, Andr{\'e}s E. and {Izumi}, Natsuko and {Sakai}, Takeshi and {Lu}, Xing and {Tatematsu}, Ken'ichi and {Ohashi}, Satoshi and {Silva}, Andrea and {Olguin}, Fernando A. and {Contreras}, Yanett},
        title = "{The ALMA Survey of 70 {\ensuremath{\mu}}m Dark High-mass Clumps in Early Stages (ASHES). IV. Star Formation Signatures in G023.477}",
      journal = {\apj},
         year = 2021,
        month = dec,
       volume = {923},
       number = {2},
          eid = {147},
        pages = {147},
          doi = {10.3847/1538-4357/ac2365},
archivePrefix = {arXiv},
       eprint = {2109.01231},
 primaryClass = {astro-ph.GA},
       adsurl = {https://ui.adsabs.harvard.edu/abs/2021ApJ...923..147M}
}

@ARTICLE{Morii2023,
       author = {{Morii}, Kaho and {Sanhueza}, Patricio and {Nakamura}, Fumitaka and {Zhang}, Qizhou and {Sabatini}, Giovanni and {Beuther}, Henrik and {Lu}, Xing and {Li}, Shanghuo and {Garay}, Guido and {Jackson}, James M. and {Olguin}, Fernando A. and {Tafoya}, Daniel and {Tatematsu}, Ken'ichi and {Izumi}, Natsuko and {Sakai}, Takeshi and {Silva}, Andrea},
        title = "{The ALMA Survey of 70 {\ensuremath{\mu}}m Dark High-mass Clumps in Early Stages (ASHES). IX. Physical Properties and Spatial Distribution of Cores in IRDCs}",
      journal = {\apj},
         year = 2023,
        month = jun,
       volume = {950},
       number = {2},
          eid = {148},
        pages = {148},
          doi = {10.3847/1538-4357/acccea},
archivePrefix = {arXiv},
       eprint = {2304.01757},
 primaryClass = {astro-ph.GA},
       adsurl = {https://ui.adsabs.harvard.edu/abs/2023ApJ...950..148M}
}

@ARTICLE{Morii2024,
       author = {{Morii}, Kaho and {Sanhueza}, Patricio and {Zhang}, Qizhou and {Nakamura}, Fumitaka and {Li}, Shanghuo and {Sabatini}, Giovanni and {Olguin}, Fernando A. and {Beuther}, Henrik and {Tafoya}, Daniel and {Izumi}, Natsuko and {Tatematsu}, Ken'ichi and {Sakai}, Takeshi},
        title = "{The ALMA Survey of 70 {\ensuremath{\mu}}m Dark High-mass Clumps in Early Stages (ASHES). XI. Statistical Study of Early Fragmentation}",
      journal = {\apj},
         year = 2024,
        month = may,
       volume = {966},
       number = {2},
          eid = {171},
        pages = {171},
          doi = {10.3847/1538-4357/ad32d0},
archivePrefix = {arXiv},
       eprint = {2403.07058},
 primaryClass = {astro-ph.GA},
       adsurl = {https://ui.adsabs.harvard.edu/abs/2024ApJ...966..171M}
}

@ARTICLE{Moscadelli2021,
       author = {{Moscadelli}, L. and {Beuther}, H. and {Ahmadi}, A. and {Gieser}, C. and {Massi}, F. and {Cesaroni}, R. and {S{\'a}nchez-Monge}, {\'A}. and {Bacciotti}, F. and {Beltr{\'a}n}, M.~T. and {Csengeri}, T. and {Galv{\'a}n-Madrid}, R. and {Henning}, Th. and {Klaassen}, P.~D. and {Kuiper}, R. and {Leurini}, S. and {Longmore}, S.~N. and {Maud}, L.~T. and {M{\"o}ller}, T. and {Palau}, A. and {Peters}, T. and {Pudritz}, R.~E. and {Sanna}, A. and {Semenov}, D. and {Urquhart}, J.~S. and {Winters}, J.~M. and {Zinnecker}, H.},
        title = "{Multi-scale view of star formation in IRAS 21078+5211: from clump fragmentation to disk wind}",
      journal = {\aap},
         year = 2021,
        month = mar,
       volume = {647},
          eid = {A114},
        pages = {A114},
          doi = {10.1051/0004-6361/202039837},
archivePrefix = {arXiv},
       eprint = {2102.04872},
 primaryClass = {astro-ph.GA},
       adsurl = {https://ui.adsabs.harvard.edu/abs/2021A&A...647A.114M}
}

@ARTICLE{Motte2018,
       author = {{Motte}, Fr{\'e}d{\'e}rique and {Bontemps}, Sylvain and {Louvet}, Fabien},
        title = "{High-Mass Star and Massive Cluster Formation in the Milky Way}",
      journal = {\araa},
         year = 2018,
        month = sep,
       volume = {56},
        pages = {41-82},
          doi = {10.1146/annurev-astro-091916-055235},
archivePrefix = {arXiv},
       eprint = {1706.00118},
 primaryClass = {astro-ph.GA},
       adsurl = {https://ui.adsabs.harvard.edu/abs/2018ARA&A..56...41M}
}

@ARTICLE{Motte2022,
       author = {{Motte}, F. and {Bontemps}, S. and {Csengeri}, T. and {Pouteau}, Y. and {Louvet}, F. and {Stutz}, A.~M. and {Cunningham}, N. and {L{\'o}pez-Sepulcre}, A. and {Brouillet}, N. and {Galv{\'a}n-Madrid}, R. and {Ginsburg}, A. and {Maud}, L. and {Men'shchikov}, A. and {Nakamura}, F. and {Nony}, T. and {Sanhueza}, P. and {{\'A}lvarez-Guti{\'e}rrez}, R.~H. and {Armante}, M. and {Baug}, T. and {Bonfand}, M. and {Busquet}, G. and {Chapillon}, E. and {D{\'\i}az-Gonz{\'a}lez}, D. and {Fern{\'a}ndez-L{\'o}pez}, M. and {Guzm{\'a}n}, A.~E. and {Herpin}, F. and {Liu}, H. -L. and {Olguin}, F. and {Towner}, A.~P.~M. and {Bally}, J. and {Battersby}, C. and {Braine}, J. and {Bronfman}, L. and {Chen}, H. -R.~V. and {Dell'Ova}, P. and {Di Francesco}, J. and {Gonz{\'a}lez}, M. and {Gusdorf}, A. and {Hennebelle}, P. and {Izumi}, N. and {Joncour}, I. and {Lee}, Y. -N. and {Lefloch}, B. and {Lesaffre}, P. and {Lu}, X. and {Menten}, K.~M. and {Mignon-Risse}, R. and {Molet}, J. and {Moraux}, E. and {Mundy}, L. and {Nguyen Luong}, Q. and {Reyes}, N. and {Reyes Reyes}, S.~D. and {Robitaille}, J. -F. and {Rosolowsky}, E. and {Sandoval-Garrido}, N.~A. and {Schuller}, F. and {Svoboda}, B. and {Tatematsu}, K. and {Thomasson}, B. and {Walker}, D. and {Wu}, B. and {Whitworth}, A.~P. and {Wyrowski}, F.},
        title = "{ALMA-IMF. I. Investigating the origin of stellar masses: Introduction to the Large Program and first results}",
      journal = {\aap},
         year = 2022,
        month = jun,
       volume = {662},
          eid = {A8},
        pages = {A8},
          doi = {10.1051/0004-6361/202141677},
archivePrefix = {arXiv},
       eprint = {2112.08182},
 primaryClass = {astro-ph.GA},
       adsurl = {https://ui.adsabs.harvard.edu/abs/2022A&A...662A...8M}
}

@ARTICLE{Nony2021,
       author = {{Nony}, T. and {Robitaille}, J. -F. and {Motte}, F. and {Gonzalez}, M. and {Joncour}, I. and {Moraux}, E. and {Men'shchikov}, A. and {Didelon}, P. and {Louvet}, F. and {Buckner}, A.~S.~M. and {Schneider}, N. and {Lumsden}, S.~L. and {Bontemps}, S. and {Pouteau}, Y. and {Cunningham}, N. and {Fiorellino}, E. and {Oudmaijer}, R. and {Andr{\'e}}, P. and {Thomasson}, B.},
        title = "{Mass segregation and sequential star formation in NGC 2264 revealed by Herschel}",
      journal = {\aap},
         year = 2021,
        month = jan,
       volume = {645},
          eid = {A94},
        pages = {A94},
          doi = {10.1051/0004-6361/202039353},
archivePrefix = {arXiv},
       eprint = {2011.05939},
 primaryClass = {astro-ph.GA},
       adsurl = {https://ui.adsabs.harvard.edu/abs/2021A&A...645A..94N}
}

@ARTICLE{Nucara2025,
       author = {{Nucara}, Alice and {Traficante}, Alessio and {Lebreuilly}, Ugo and {Tung}, Ngo-Duy and {Molinari}, Sergio and {Hennebelle}, Patrick and {Testi}, Leonardo and {Klessen}, Ralf S. and {Pelkonen}, Veli-Matti and {Avison}, Adam and {Benedettini}, Milena and {Coletta}, Alessandro and {De Angelis}, Fabrizio and {Elia}, Davide and {Fuller}, Gary A. and {Jones}, Bethany M. and {Mercimek}, Seyma and {Mininni}, Chiara and {Pezzuto}, Stefania and {Pillai}, Thushara and {Roccatagliata}, Veronica and {Schisano}, Eugenio and {Soler}, Juan D. and {Suin}, Paolo and {Toci}, Claudia and {Walker}, Daniel},
        title = "{The Rosetta Stone project: III. ALMA synthetic observations of fragmentation in high-mass star-forming clumps}",
      journal = {\aap},
         year = 2025,
        month = sep,
       volume = {701},
          eid = {A219},
        pages = {A219},
          doi = {10.1051/0004-6361/202554775},
archivePrefix = {arXiv},
       eprint = {2507.11032},
 primaryClass = {astro-ph.SR},
       adsurl = {https://ui.adsabs.harvard.edu/abs/2025A&A...701A.219N}
}

@INPROCEEDINGS{Li2014,
       author = {{Li}, H. -B. and {Goodman}, A. and {Sridharan}, T.~K. and {Houde}, M. and {Li}, Z. -Y. and {Novak}, G. and {Tang}, K.~S.},
        title = "{The Link Between Magnetic Fields and Cloud/Star Formation}",
    booktitle = {Protostars and Planets VI},
         year = 2014,
       editor = {{Beuther}, Henrik and {Klessen}, Ralf S. and {Dullemond}, Cornelis P. and {Henning}, Thomas},
        month = jan,
        pages = {101-123},
          doi = {10.2458/azu_uapress_9780816531240-ch005},
archivePrefix = {arXiv},
       eprint = {1404.2024},
 primaryClass = {astro-ph.GA},
       adsurl = {https://ui.adsabs.harvard.edu/abs/2014prpl.conf..101L}
}

@ARTICLE{Padoan2020,
       author = {{Padoan}, Paolo and {Pan}, Liubin and {Juvela}, Mika and {Haugb{\o}lle}, Troels and {Nordlund}, {\r{A}}ke},
        title = "{The Origin of Massive Stars: The Inertial-inflow Model}",
      journal = {\apj},
         year = 2020,
        month = sep,
       volume = {900},
       number = {1},
          eid = {82},
        pages = {82},
          doi = {10.3847/1538-4357/abaa47},
archivePrefix = {arXiv},
       eprint = {1911.04465},
 primaryClass = {astro-ph.GA},
       adsurl = {https://ui.adsabs.harvard.edu/abs/2020ApJ...900...82P}
}

@ARTICLE{Palau2014,
       author = {{Palau}, Aina and {Estalella}, Robert and {Girart}, Josep M. and {Fuente}, Asunci{\'o}n and {Fontani}, Francesco and {Commer{\c{c}}on}, Benoit and {Busquet}, Gemma and {Bontemps}, Sylvain and {S{\'a}nchez-Monge}, {\'A}lvaro and {Zapata}, Luis A. and {Zhang}, Qizhou and {Hennebelle}, Patrick and {di Francesco}, James},
        title = "{Fragmentation of Massive Dense Cores Down to <\raisebox{-0.5ex}\textasciitilde 1000 AU: Relation between Fragmentation and Density Structure}",
      journal = {\apj},
         year = 2014,
        month = apr,
       volume = {785},
       number = {1},
          eid = {42},
        pages = {42},
          doi = {10.1088/0004-637X/785/1/42},
archivePrefix = {arXiv},
       eprint = {1401.8292},
 primaryClass = {astro-ph.GA},
       adsurl = {https://ui.adsabs.harvard.edu/abs/2014ApJ...785...42P}
}

@ARTICLE{Palau2015,
       author = {{Palau}, Aina and {Ballesteros-Paredes}, Javier and {V{\'a}zquez-Semadeni}, Enrique and {S{\'a}nchez-Monge}, {\'A}lvaro and {Estalella}, Robert and {Fall}, S. Michael and {Zapata}, Luis A. and {Camacho}, Vianey and {G{\'o}mez}, Laura and {Naranjo-Romero}, Ra{\'u}l and {Busquet}, Gemma and {Fontani}, Francesco},
        title = "{Gravity or turbulence? - III. Evidence of pure thermal Jeans fragmentation at {\ensuremath{\sim}}0.1 pc scale}",
      journal = {\mnras},
         year = 2015,
        month = nov,
       volume = {453},
       number = {4},
        pages = {3785-3797},
          doi = {10.1093/mnras/stv1834},
archivePrefix = {arXiv},
       eprint = {1504.07644},
 primaryClass = {astro-ph.GA},
       adsurl = {https://ui.adsabs.harvard.edu/abs/2015MNRAS.453.3785P}
}

@ARTICLE{Palau2018,
       author = {{Palau}, Aina and {Zapata}, Luis A. and {Rom{\'a}n-Z{\'u}{\~n}iga}, Carlos G. and {S{\'a}nchez-Monge}, {\'A}lvaro and {Estalella}, Robert and {Busquet}, Gemma and {Girart}, Josep M. and {Fuente}, Asunci{\'o}n and {Commer{\c{c}}on}, Benoit},
        title = "{Thermal Jeans Fragmentation within {\ensuremath{\sim}}1000 au in OMC-1S}",
      journal = {\apj},
         year = 2018,
        month = mar,
       volume = {855},
       number = {1},
          eid = {24},
        pages = {24},
          doi = {10.3847/1538-4357/aaad03},
archivePrefix = {arXiv},
       eprint = {1706.04623},
 primaryClass = {astro-ph.GA},
       adsurl = {https://ui.adsabs.harvard.edu/abs/2018ApJ...855...24P}
}

@ARTICLE{Palau2021,
       author = {{Palau}, Aina and {Zhang}, Qizhou and {Girart}, Josep M. and {Liu}, Junhao and {Rao}, Ramprasad and {Koch}, Patrick M. and {Estalella}, Robert and {Chen}, Huei-Ru Vivien and {Liu}, Hauyu Baobab and {Qiu}, Keping and {Li}, Zhi-Yun and {Zapata}, Luis A. and {Bontemps}, Sylvain and {Ho}, Paul T.~P. and {Beuther}, Henrik and {Ching}, Tao-Chung and {Shinnaga}, Hiroko and {Ahmadi}, Aida},
        title = "{Does the Magnetic Field Suppress Fragmentation in Massive Dense Cores?}",
      journal = {\apj},
         year = 2021,
        month = may,
       volume = {912},
       number = {2},
          eid = {159},
        pages = {159},
          doi = {10.3847/1538-4357/abee1e},
archivePrefix = {arXiv},
       eprint = {2010.12099},
 primaryClass = {astro-ph.GA},
       adsurl = {https://ui.adsabs.harvard.edu/abs/2021ApJ...912..159P}
}

@ARTICLE{Parker2014,
       author = {{Parker}, Richard J.},
        title = "{Dynamics versus structure: breaking the density degeneracy in star formation}",
      journal = {\mnras},
         year = 2014,
        month = dec,
       volume = {445},
       number = {4},
        pages = {4037-4044},
          doi = {10.1093/mnras/stu2054},
archivePrefix = {arXiv},
       eprint = {1410.0004},
 primaryClass = {astro-ph.GA},
       adsurl = {https://ui.adsabs.harvard.edu/abs/2014MNRAS.445.4037P}
}

@ARTICLE{Parker2015,
       author = {{Parker}, Richard J. and {Goodwin}, Simon P.},
        title = "{Comparisons between different techniques for measuring mass segregation}",
      journal = {\mnras},
         year = 2015,
        month = jun,
       volume = {449},
       number = {4},
        pages = {3381-3392},
          doi = {10.1093/mnras/stv539},
archivePrefix = {arXiv},
       eprint = {1503.02692},
 primaryClass = {astro-ph.GA},
       adsurl = {https://ui.adsabs.harvard.edu/abs/2015MNRAS.449.3381P}
}

@ARTICLE{Parker2017,
       author = {{Parker}, Richard J. and {Alves de Oliveira}, Catarina},
        title = "{Dynamical histories of the IC 348 and NGC 1333 star-forming regions in Perseus}",
      journal = {\mnras},
         year = 2017,
        month = jul,
       volume = {468},
       number = {4},
        pages = {4340-4350},
          doi = {10.1093/mnras/stx739},
archivePrefix = {arXiv},
       eprint = {1703.08547},
 primaryClass = {astro-ph.GA},
       adsurl = {https://ui.adsabs.harvard.edu/abs/2017MNRAS.468.4340P}
}

@ARTICLE{Parker2018,
       author = {{Parker}, Richard J.},
        title = "{On the spatial distributions of dense cores in Orion B}",
      journal = {\mnras},
         year = 2018,
        month = may,
       volume = {476},
       number = {1},
        pages = {617-629},
          doi = {10.1093/mnras/sty249},
archivePrefix = {arXiv},
       eprint = {1801.09680},
 primaryClass = {astro-ph.GA},
       adsurl = {https://ui.adsabs.harvard.edu/abs/2018MNRAS.476..617P}
}

@ARTICLE{Peretto2014,
       author = {{Peretto}, N. and {Fuller}, G.~A. and {Andr{\'e}}, Ph. and {Arzoumanian}, D. and {Rivilla}, V.~M. and {Bardeau}, S. and {Duarte Puertas}, S. and {Guzman Fernandez}, J.~P. and {Lenfestey}, C. and {Li}, G. -X. and {Olguin}, F.~A. and {R{\"o}ck}, B.~R. and {de Villiers}, H. and {Williams}, J.},
        title = "{SDC13 infrared dark clouds: Longitudinally collapsing filaments?}",
      journal = {\aap},
         year = 2014,
        month = jan,
       volume = {561},
          eid = {A83},
        pages = {A83},
          doi = {10.1051/0004-6361/201322172},
archivePrefix = {arXiv},
       eprint = {1311.0203},
 primaryClass = {astro-ph.GA},
       adsurl = {https://ui.adsabs.harvard.edu/abs/2014A&A...561A..83P}
}

@ARTICLE{Plunkett2018,
       author = {{Plunkett}, Adele L. and {Fern{\'a}ndez-L{\'o}pez}, Manuel and {Arce}, H{\'e}ctor G. and {Busquet}, Gemma and {Mardones}, Diego and {Dunham}, Michael M.},
        title = "{Distribution of Serpens South protostars revealed with ALMA}",
      journal = {\aap},
         year = 2018,
        month = jul,
       volume = {615},
          eid = {A9},
        pages = {A9},
          doi = {10.1051/0004-6361/201732372},
archivePrefix = {arXiv},
       eprint = {1804.02405},
 primaryClass = {astro-ph.SR},
       adsurl = {https://ui.adsabs.harvard.edu/abs/2018A&A...615A...9P}
}

@ARTICLE{Portegies2010,
       author = {{Portegies Zwart}, Simon F. and {McMillan}, Stephen L.~W. and {Gieles}, Mark},
        title = "{Young Massive Star Clusters}",
      journal = {\araa},
         year = 2010,
        month = sep,
       volume = {48},
        pages = {431-493},
          doi = {10.1146/annurev-astro-081309-130834},
archivePrefix = {arXiv},
       eprint = {1002.1961},
 primaryClass = {astro-ph.GA},
       adsurl = {https://ui.adsabs.harvard.edu/abs/2010ARA&A..48..431P}
}

@ARTICLE{Pouteau2023,
       author = {{Pouteau}, Y. and {Motte}, F. and {Nony}, T. and {Gonz{\'a}lez}, M. and {Joncour}, I. and {Robitaille}, J. -F. and {Busquet}, G. and {Galv{\'a}n-Madrid}, R. and {Gusdorf}, A. and {Hennebelle}, P. and {Ginsburg}, A. and {Csengeri}, T. and {Sanhueza}, P. and {Dell'Ova}, P. and {Stutz}, A.~M. and {Towner}, A.~P.~M. and {Cunningham}, N. and {Louvet}, F. and {Men'shchikov}, A. and {Fern{\'a}ndez-L{\'o}pez}, M. and {Schneider}, N. and {Armante}, M. and {Bally}, J. and {Baug}, T. and {Bonfand}, M. and {Bontemps}, S. and {Bronfman}, L. and {Brouillet}, N. and {D{\'\i}az-Gonz{\'a}lez}, D. and {Herpin}, F. and {Lefloch}, B. and {Liu}, H. -L. and {Lu}, X. and {Nakamura}, F. and {Nguyen-Luong}, Q. and {Olguin}, F. and {Tatematsu}, K. and {Valeille-Manet}, M.},
        title = "{ALMA-IMF. VI. Investigating the origin of stellar masses: Core mass function evolution in the W43-MM2\&MM3 mini-starburst}",
      journal = {\aap},
         year = 2023,
        month = jun,
       volume = {674},
          eid = {A76},
        pages = {A76},
          doi = {10.1051/0004-6361/202244776},
archivePrefix = {arXiv},
       eprint = {2212.09307},
 primaryClass = {astro-ph.GA},
       adsurl = {https://ui.adsabs.harvard.edu/abs/2023A&A...674A..76P}
}

@ARTICLE{Prim1957,
       author = {{Prim}, R.~C.},
        title = "{Shortest Connection Networks And Some Generalizations}",
      journal = {Bell System Technical Journal},
         year = 1957,
        month = nov,
       volume = {36},
       number = {6},
        pages = {1389-1401},
          doi = {10.1002/j.1538-7305.1957.tb01515.x},
       adsurl = {https://ui.adsabs.harvard.edu/abs/1957BSTJ...36.1389P}
}

@ARTICLE{Rebolledo2020,
       author = {{Rebolledo}, David and {Guzm{\'a}n}, Andr{\'e}s E. and {Contreras}, Yanett and {Garay}, Guido and {Medina}, S. -N.~X. and {Sanhueza}, Patricio and {Green}, Anne J. and {Castro}, Camila and {Guzm{\'a}n}, Viviana and {Burton}, Michael G.},
        title = "{Effect of Feedback of Massive Stars in the Fragmentation, Distribution, and Kinematics of the Gas in Two Star-forming Regions in the Carina Nebula}",
      journal = {\apj},
         year = 2020,
        month = mar,
       volume = {891},
       number = {2},
          eid = {113},
        pages = {113},
          doi = {10.3847/1538-4357/ab6d76},
archivePrefix = {arXiv},
       eprint = {2001.06969},
 primaryClass = {astro-ph.GA},
       adsurl = {https://ui.adsabs.harvard.edu/abs/2020ApJ...891..113R}
}

@ARTICLE{Rigby2024,
       author = {{Rigby}, Andrew J. and {Peretto}, Nicolas and {Anderson}, Michael and {Ragan}, Sarah E. and {Priestley}, Felix D. and {Fuller}, Gary A. and {Thompson}, Mark A. and {Traficante}, Alessio and {Watkins}, Elizabeth J. and {Williams}, Gwenllian M.},
        title = "{The dynamic centres of infrared-dark clouds and the formation of cores}",
      journal = {\mnras},
         year = 2024,
        month = feb,
       volume = {528},
       number = {2},
        pages = {1172-1197},
          doi = {10.1093/mnras/stae030},
archivePrefix = {arXiv},
       eprint = {2401.04238},
 primaryClass = {astro-ph.GA},
       adsurl = {https://ui.adsabs.harvard.edu/abs/2024MNRAS.528.1172R}
}

@article{Ripley1977,
    author={Ripley, B. D. and Rasson, J.-P.},
    title="{Finding the edge of a Poisson forest}", 
    volume={14}, 
    DOI={10.2307/3213451}, 
    number={3}, 
    journal={Journal of Applied Probability},
    year={1977}, 
    pages={483–491}
}

@ARTICLE{Sanchez2009,
       author = {{S{\'a}nchez}, N{\'e}stor and {Alfaro}, Emilio J.},
        title = "{The Spatial Distribution of Stars in Open Clusters}",
      journal = {\apj},
         year = 2009,
        month = may,
       volume = {696},
       number = {2},
        pages = {2086-2093},
          doi = {10.1088/0004-637X/696/2/2086},
archivePrefix = {arXiv},
       eprint = {0902.1071},
 primaryClass = {astro-ph.SR},
       adsurl = {https://ui.adsabs.harvard.edu/abs/2009ApJ...696.2086S}
}

@ARTICLE{Sadaghiani2020,
       author = {{Sadaghiani}, M. and {S{\'a}nchez-Monge}, {\'A}. and {Schilke}, P. and {Liu}, H.~B. and {Clarke}, S.~D. and {Zhang}, Q. and {Girart}, J.~M. and {Seifried}, D. and {Aghababaei}, A. and {Li}, H. and {Ju{\'a}rez}, C. and {Tang}, K.~S.},
        title = "{Physical properties of the star-forming clusters in NGC 6334. A study of the continuum dust emission with ALMA}",
      journal = {\aap},
         year = 2020,
        month = mar,
       volume = {635},
          eid = {A2},
        pages = {A2},
          doi = {10.1051/0004-6361/201935699},
archivePrefix = {arXiv},
       eprint = {1911.06579},
 primaryClass = {astro-ph.SR},
       adsurl = {https://ui.adsabs.harvard.edu/abs/2020A&A...635A...2S}
}

@ARTICLE{SanchezMonge2025,
       author = {{S{\'a}nchez-Monge}, {\'A}. and {Brogan}, C.~L. and {Hunter}, T.~R. and {Ahmadi}, A. and {Avison}, A. and {Beltr{\'a}n}, M.~T. and {Beuther}, H. and {Coletta}, A. and {Fuller}, G.~A. and {Johnston}, K.~G. and {Jones}, B. and {Liu}, S. -Y. and {Mininni}, C. and {Molinari}, S. and {Schilke}, P. and {Schisano}, E. and {Su}, Y. -N. and {Traficante}, A. and {Zhang}, Q. and {Battersby}, C. and {Benedettini}, M. and {Elia}, D. and {Ho}, P.~T.~P. and {Klaassen}, P.~D. and {Klessen}, R.~S. and {Law}, C.~Y. and {Lis}, D.~C. and {Liu}, T. and {Maud}, L. and {M{\"o}ller}, T. and {Moscadelli}, L. and {Pezzuto}, S. and {Rygl}, K.~L.~J. and {Sanhueza}, P. and {Soler}, J.~D. and {Stroud}, G. and {Tang}, Y. and {van der Tak}, F.~F.~S. and {Walker}, D.~L. and {Wallace}, J. and {Walch}, S. and {Wells}, M.~R.~A. and {Wyrowski}, F. and {Zhang}, T. and {Allande}, J. and {Bronfman}, L. and {Dann}, E. and {De Angelis}, F. and {Fontani}, F. and {Henning}, Th. and {Kim}, W. -J. and {Kuiper}, R. and {Merello}, M. and {Nakamura}, F. and {Nucara}, A. and {Rigby}, A.~J.},
        title = "{ALMAGAL: II. The ALMA evolutionary study of high-mass protocluster formation in the Galaxy: ALMA data processing and pipeline}",
      journal = {\aap},
         year = 2025,
        month = apr,
       volume = {696},
          eid = {A150},
        pages = {A150},
          doi = {10.1051/0004-6361/202452703},
archivePrefix = {arXiv},
       eprint = {2503.05559},
 primaryClass = {astro-ph.GA},
       adsurl = {https://ui.adsabs.harvard.edu/abs/2025A&A...696A.150S}
}

@ARTICLE{Sanhueza2017,
       author = {{Sanhueza}, Patricio and {Jackson}, James M. and {Zhang}, Qizhou and {Guzm{\'a}n}, Andr{\'e}s E. and {Lu}, Xing and {Stephens}, Ian W. and {Wang}, Ke and {Tatematsu}, Ken'ichi},
        title = "{A Massive Prestellar Clump Hosting No High-mass Cores}",
      journal = {\apj},
         year = 2017,
        month = jun,
       volume = {841},
       number = {2},
          eid = {97},
        pages = {97},
          doi = {10.3847/1538-4357/aa6ff8},
archivePrefix = {arXiv},
       eprint = {1704.08264},
 primaryClass = {astro-ph.GA},
       adsurl = {https://ui.adsabs.harvard.edu/abs/2017ApJ...841...97S}
}

@ARTICLE{Sanhueza2019,
       author = {{Sanhueza}, Patricio and {Contreras}, Yanett and {Wu}, Benjamin and {Jackson}, James M. and {Guzm{\'a}n}, Andr{\'e}s E. and {Zhang}, Qizhou and {Li}, Shanghuo and {Lu}, Xing and {Silva}, Andrea and {Izumi}, Natsuko and {Liu}, Tie and {Miura}, Rie E. and {Tatematsu}, Ken'ichi and {Sakai}, Takeshi and {Beuther}, Henrik and {Garay}, Guido and {Ohashi}, Satoshi and {Saito}, Masao and {Nakamura}, Fumitaka and {Saigo}, Kazuya and {Veena}, V.~S. and {Nguyen-Luong}, Quang and {Tafoya}, Daniel},
        title = "{The ALMA Survey of 70 {\ensuremath{\mu}}m Dark High-mass Clumps in Early Stages (ASHES). I. Pilot Survey: Clump Fragmentation}",
      journal = {\apj},
         year = 2019,
        month = dec,
       volume = {886},
       number = {2},
          eid = {102},
        pages = {102},
          doi = {10.3847/1538-4357/ab45e9},
archivePrefix = {arXiv},
       eprint = {1909.07985},
 primaryClass = {astro-ph.GA},
       adsurl = {https://ui.adsabs.harvard.edu/abs/2019ApJ...886..102S}
}

@ARTICLE{Sanhueza2025,
       author = {{Sanhueza}, Patricio and {Liu}, Junhao and {Morii}, Kaho and {Girart}, Josep Miquel and {Zhang}, Qizhou and {Stephens}, Ian W. and {Jackson}, James M. and {Cort{\'e}s}, Paulo C. and {Koch}, Patrick M. and {Cyganowski}, Claudia J. and {Saha}, Piyali and {Beuther}, Henrik and {Zhang}, Suinan and {Beltr{\'a}n}, Maria T. and {Cheng}, Yu and {Olguin}, Fernando A. and {Lu}, Xing and {Choudhury}, Spandan and {Pattle}, Kate and {Fern{\'a}ndez-L{\'o}pez}, Manuel and {Hwang}, Jihye and {Kang}, Ji-hyun and {Karoly}, Janik and {Ginsburg}, Adam and {Lyo}, A. -Ran and {Taniguchi}, Kotomi and {Jiao}, Wenyu and {Eswaraiah}, Chakali and {Luo}, Qiu-yi and {Wang}, Jia-Wei and {Commer{\c{c}}on}, Beno{\^\i}t and {Li}, Shanghuo and {Xu}, Fengwei and {Chen}, Huei-Ru Vivien and {Zapata}, Luis A. and {Chung}, Eun Jung and {Nakamura}, Fumitaka and {Panigrahy}, Sandhyarani and {Sakai}, Takeshi},
        title = "{Magnetic Fields in Massive Star-forming Regions (MagMaR). V. The Magnetic Field at the Onset of High-mass Star Formation}",
      journal = {\apj},
         year = 2025,
        month = feb,
       volume = {980},
       number = {1},
          eid = {87},
        pages = {87},
          doi = {10.3847/1538-4357/ad9d40},
archivePrefix = {arXiv},
       eprint = {2412.08790},
 primaryClass = {astro-ph.GA},
       adsurl = {https://ui.adsabs.harvard.edu/abs/2025ApJ...980...87S}
}

@INPROCEEDINGS{Scalo1985,
       author = {{Scalo}, J.~M.},
        title = "{Fragmentation and hierarchical structure in the interstellar medium.}",
    booktitle = {Protostars and Planets II},
         year = 1985,
       editor = {{Black}, D.~C. and {Matthews}, M.~S.},
        month = jan,
        pages = {201-296},
       adsurl = {https://ui.adsabs.harvard.edu/abs/1985prpl.conf..201S}
}

@ARTICLE{Schisano2014,
       author = {{Schisano}, E. and {Rygl}, K.~L.~J. and {Molinari}, S. and {Busquet}, G. and {Elia}, D. and {Pestalozzi}, M. and {Polychroni}, D. and {Billot}, N. and {Carey}, S. and {Paladini}, R. and {Noriega-Crespo}, A. and {Moore}, T.~J.~T. and {Plume}, R. and {Glover}, S.~C.~O. and {V{\'a}zquez-Semadeni}, E.},
        title = "{The Identification of Filaments on Far-infrared and Submillimiter Images: Morphology, Physical Conditions and Relation with Star Formation of Filamentary Structure}",
      journal = {\apj},
         year = 2014,
        month = aug,
       volume = {791},
       number = {1},
          eid = {27},
        pages = {27},
          doi = {10.1088/0004-637X/791/1/27},
archivePrefix = {arXiv},
       eprint = {1406.4443},
 primaryClass = {astro-ph.SR},
       adsurl = {https://ui.adsabs.harvard.edu/abs/2014ApJ...791...27S}
}

@ARTICLE{Schisano2020,
       author = {{Schisano}, Eugenio and {Molinari}, S. and {Elia}, D. and {Benedettini}, M. and {Olmi}, L. and {Pezzuto}, S. and {Traficante}, A. and {Brescia}, M. and {Cavuoti}, S. and {di Giorgio}, A.~M. and {Liu}, S.~J. and {Moore}, T.~J.~T. and {Noriega-Crespo}, A. and {Riccio}, G. and {Baldeschi}, A. and {Becciani}, U. and {Peretto}, N. and {Merello}, M. and {Vitello}, F. and {Zavagno}, A. and {Beltr{\'a}n}, M.~T. and {Cambr{\'e}sy}, L. and {Eden}, D.~J. and {Li Causi}, G. and {Molinaro}, M. and {Palmeirim}, P. and {Sciacca}, E. and {Testi}, L. and {Umana}, G. and {Whitworth}, A.~P.},
        title = "{The Hi-GAL catalogue of dusty filamentary structures in the Galactic plane}",
      journal = {\mnras},
         year = 2020,
        month = mar,
       volume = {492},
       number = {4},
        pages = {5420-5456},
          doi = {10.1093/mnras/stz3466},
archivePrefix = {arXiv},
       eprint = {1912.04020},
 primaryClass = {astro-ph.GA},
       adsurl = {https://ui.adsabs.harvard.edu/abs/2020MNRAS.492.5420S}
}

@ARTICLE{Schneider2012,
       author = {{Schneider}, N. and {Csengeri}, T. and {Hennemann}, M. and {Motte}, F. and {Didelon}, P. and {Federrath}, C. and {Bontemps}, S. and {Di Francesco}, J. and {Arzoumanian}, D. and {Minier}, V. and {Andr{\'e}}, Ph. and {Hill}, T. and {Zavagno}, A. and {Nguyen-Luong}, Q. and {Attard}, M. and {Bernard}, J. -Ph. and {Elia}, D. and {Fallscheer}, C. and {Griffin}, M. and {Kirk}, J. and {Klessen}, R. and {K{\"o}nyves}, V. and {Martin}, P. and {Men'shchikov}, A. and {Palmeirim}, P. and {Peretto}, N. and {Pestalozzi}, M. and {Russeil}, D. and {Sadavoy}, S. and {Sousbie}, T. and {Testi}, L. and {Tremblin}, P. and {Ward-Thompson}, D. and {White}, G.},
        title = "{Cluster-formation in the Rosette molecular cloud at the junctions of filaments}",
      journal = {\aap},
         year = 2012,
        month = apr,
       volume = {540},
          eid = {L11},
        pages = {L11},
          doi = {10.1051/0004-6361/201118566},
archivePrefix = {arXiv},
       eprint = {1203.6472},
 primaryClass = {astro-ph.GA},
       adsurl = {https://ui.adsabs.harvard.edu/abs/2012A&A...540L..11S}
}

@ARTICLE{Schneider2022,
       author = {{Schneider}, N. and {Ossenkopf-Okada}, V. and {Clarke}, S. and {Klessen}, R.~S. and {Kabanovic}, S. and {Veltchev}, T. and {Bontemps}, S. and {Dib}, S. and {Csengeri}, T. and {Federrath}, C. and {Di Francesco}, J. and {Motte}, F. and {Andr{\'e}}, Ph. and {Arzoumanian}, D. and {Beattie}, J.~R. and {Bonne}, L. and {Didelon}, P. and {Elia}, D. and {K{\"o}nyves}, V. and {Kritsuk}, A. and {Ladjelate}, B. and {Myers}, Ph. and {Pezzuto}, S. and {Robitaille}, J.~F. and {Roy}, A. and {Seifried}, D. and {Simon}, R. and {Soler}, J. and {Ward-Thompson}, D.},
        title = "{Understanding star formation in molecular clouds. IV. Column density PDFs from quiescent to massive molecular clouds}",
      journal = {\aap},
         year = 2022,
        month = oct,
       volume = {666},
          eid = {A165},
        pages = {A165},
          doi = {10.1051/0004-6361/202039610},
archivePrefix = {arXiv},
       eprint = {2207.14604},
 primaryClass = {astro-ph.GA},
       adsurl = {https://ui.adsabs.harvard.edu/abs/2022A&A...666A.165S}
}

@ARTICLE{Schmeja2006,
       author = {{Schmeja}, S. and {Klessen}, R.~S.},
        title = "{Evolving structures of star-forming clusters}",
      journal = {\aap},
         year = 2006,
        month = apr,
       volume = {449},
       number = {1},
        pages = {151-159},
          doi = {10.1051/0004-6361:20054464},
archivePrefix = {arXiv},
       eprint = {astro-ph/0511448},
 primaryClass = {astro-ph},
       adsurl = {https://ui.adsabs.harvard.edu/abs/2006A&A...449..151S}
}

@ARTICLE{Schmeja2008,
       author = {{Schmeja}, S. and {Kumar}, M.~S.~N. and {Ferreira}, B.},
        title = "{The structures of embedded clusters in the Perseus, Serpens and Ophiuchus molecular clouds}",
      journal = {\mnras},
         year = 2008,
        month = sep,
       volume = {389},
       number = {3},
        pages = {1209-1217},
          doi = {10.1111/j.1365-2966.2008.13442.x},
archivePrefix = {arXiv},
       eprint = {0805.2049},
 primaryClass = {astro-ph},
       adsurl = {https://ui.adsabs.harvard.edu/abs/2008MNRAS.389.1209S}
}

@BOOK{Scott1992,
       author = {{Scott}, D.~W.},
        title = "{Multivariate Density Estimation}",
    year = 1992,
    publisher={Wiley}
}

@ARTICLE{Shima2017,
       author = {{Shima}, Kazuhiro and {Tasker}, Elizabeth J. and {Habe}, Asao},
        title = "{Does feedback help or hinder star formation? The effect of photoionization on star formation in giant molecular clouds}",
      journal = {\mnras},
         year = 2017,
        month = may,
       volume = {467},
       number = {1},
        pages = {512-523},
          doi = {10.1093/mnras/stw3279},
archivePrefix = {arXiv},
       eprint = {1612.06381},
 primaryClass = {astro-ph.GA},
       adsurl = {https://ui.adsabs.harvard.edu/abs/2017MNRAS.467..512S}
}

@ARTICLE{Silverman1981,
       author = {{Silverman}, B.~W.},
        title = "{Using Kernel Density Estimates to Investigate Multimodality}",
      journal = {J. Roy. Stat. Soc},
         year = 1981,
        month = jan,
       volume = {43},
        pages = {97-99},
       adsurl = {https://ui.adsabs.harvard.edu/abs/1981JRSSB..43...97S}
}

@BOOK{Silverman1986,
       author = {{Silverman}, B.~W.},
        title = "{``Density estimation for statistics and data analysis'''}",
         year = 1986,
       publisher = {Chapman \& Hall},
       adsurl = {https://ui.adsabs.harvard.edu/abs/1986desd.book.....S}
}

@ARTICLE{Smith2020,
       author = {{Smith}, Rowan J. and {Tre{\ss}}, Robin G. and {Sormani}, Mattia C. and {Glover}, Simon C.~O. and {Klessen}, Ralf S. and {Clark}, Paul C. and {Izquierdo}, Andr{\'e}s F. and {Duarte-Cabral}, Ana and {Zucker}, Catherine},
        title = "{The Cloud Factory I: Generating resolved filamentary molecular clouds from galactic-scale forces}",
      journal = {\mnras},
         year = 2020,
        month = feb,
       volume = {492},
       number = {2},
        pages = {1594-1613},
          doi = {10.1093/mnras/stz3328},
archivePrefix = {arXiv},
       eprint = {1911.05753},
 primaryClass = {astro-ph.GA},
       adsurl = {https://ui.adsabs.harvard.edu/abs/2020MNRAS.492.1594S}
}

@ARTICLE{Svoboda2019,
       author = {{Svoboda}, Brian E. and {Shirley}, Yancy L. and {Traficante}, Alessio and {Battersby}, Cara and {Fuller}, Gary A. and {Zhang}, Qizhou and {Beuther}, Henrik and {Peretto}, Nicolas and {Brogan}, Crystal and {Hunter}, Todd},
        title = "{ALMA Observations of Fragmentation, Substructure, and Protostars in High-mass Starless Clump Candidates}",
      journal = {\apj},
         year = 2019,
        month = nov,
       volume = {886},
       number = {1},
          eid = {36},
        pages = {36},
          doi = {10.3847/1538-4357/ab40ca},
archivePrefix = {arXiv},
       eprint = {1908.10374},
 primaryClass = {astro-ph.GA},
       adsurl = {https://ui.adsabs.harvard.edu/abs/2019ApJ...886...36S}
}

@ARTICLE{Tang2019,
       author = {{Tang}, Ya-Wen and {Koch}, Patrick M. and {Peretto}, Nicolas and {Novak}, Giles and {Duarte-Cabral}, Ana and {Chapman}, Nicholas L. and {Hsieh}, Pei-Ying and {Yen}, Hsi-Wei},
        title = "{Gravity, Magnetic Field, and Turbulence: Relative Importance and Impact on Fragmentation in the Infrared Dark Cloud G34.43+00.24}",
      journal = {\apj},
         year = 2019,
        month = jun,
       volume = {878},
       number = {1},
          eid = {10},
        pages = {10},
          doi = {10.3847/1538-4357/ab1484},
archivePrefix = {arXiv},
       eprint = {1903.12397},
 primaryClass = {astro-ph.GA},
       adsurl = {https://ui.adsabs.harvard.edu/abs/2019ApJ...878...10T}
}

@ARTICLE{Traficante2023,
       author = {{Traficante}, A. and {Jones}, B.~M. and {Avison}, A. and {Fuller}, G.~A. and {Benedettini}, M. and {Elia}, D. and {Molinari}, S. and {Peretto}, N. and {Pezzuto}, S. and {Pillai}, T. and {Rygl}, K.~L.~J. and {Schisano}, E. and {Smith}, R.~J.},
        title = "{The SQUALO project (Star formation in QUiescent And Luminous Objects) I: clump-fed accretion mechanism in high-mass star-forming objects}",
      journal = {\mnras},
         year = 2023,
        month = apr,
       volume = {520},
       number = {2},
        pages = {2306-2327},
          doi = {10.1093/mnras/stad272},
archivePrefix = {arXiv},
       eprint = {2301.09917},
 primaryClass = {astro-ph.GA},
       adsurl = {https://ui.adsabs.harvard.edu/abs/2023MNRAS.520.2306T}
}

@ARTICLE{Tung2025,
       author = {{Tung}, Ngo-Duy and {Traficante}, Alessio and {Lebreuilly}, Ugo and {Nucara}, Alice and {Testi}, Leonardo and {Hennebelle}, Patrick and {Klessen}, Ralf S. and {Molinari}, Sergio and {Pelkonen}, Veli-Matti and {Benedettini}, Milena and {Coletta}, Alessandro and {Elia}, Davide and {Fuller}, Gary A. and {Pezzuto}, Stefania and {Soler}, Juan D. and {Toci}, Claudia},
        title = "{The Rosetta Stone Project: II. The correlation between star formation efficiency and L/M indicator for the evolutionary stages of star-forming clumps in post-processed radiative magnetohydrodynamics simulations}",
      journal = {\aap},
         year = 2025,
        month = sep,
       volume = {701},
          eid = {A218},
        pages = {A218},
          doi = {10.1051/0004-6361/202554773},
archivePrefix = {arXiv},
       eprint = {2507.09936},
 primaryClass = {astro-ph.GA},
       adsurl = {https://ui.adsabs.harvard.edu/abs/2025A&A...701A.218T}
}

@ARTICLE{Xu2024,
       author = {{Xu}, Fengwei and {Wang}, Ke and {Liu}, Tie and {Tang}, Mengyao and {Evans}, Neal J., II and {Palau}, Aina and {Morii}, Kaho and {He}, Jinhua and {Sanhueza}, Patricio and {Liu}, Hong-Li and {Stutz}, Amelia and {Zhang}, Qizhou and {Chen}, Xi and {Li}, Pak Shing and {G{\'o}mez}, Gilberto C. and {V{\'a}zquez-Semadeni}, Enrique and {Li}, Shanghuo and {Mai}, Xiaofeng and {Lu}, Xing and {Liu}, Meizhu and {Chen}, Li and {Li}, Chuanshou and {Shi}, Hongqiong and {Ren}, Zhiyuan and {Li}, Di and {Garay}, Guido and {Bronfman}, Leonardo and {Dewangan}, Lokesh and {Juvela}, Mika and {Lee}, Chang Won and {Zhang}, S. and {Yue}, Nannan and {Wang}, Chao and {Ge}, Yifei and {Jiao}, Wenyu and {Luo}, Qiuyi and {Zhou}, J. -W. and {Tatematsu}, Ken'ichi and {Chibueze}, James O. and {Su}, Keyun and {Sun}, Shenglan and {Ristorcelli}, I. and {Toth}, L. Viktor},
        title = "{The ALMA Survey of Star Formation and Evolution in Massive Protoclusters with Blue Profiles (ASSEMBLE): Core Growth, Cluster Contraction, and Primordial Mass Segregation}",
      journal = {\apjs},
         year = 2024,
        month = jan,
       volume = {270},
       number = {1},
          eid = {9},
        pages = {9},
          doi = {10.3847/1538-4365/acfee5},
archivePrefix = {arXiv},
       eprint = {2309.14684},
 primaryClass = {astro-ph.GA},
       adsurl = {https://ui.adsabs.harvard.edu/abs/2024ApJS..270....9X}
}

@ARTICLE{VazquesSemadeni2017,
       author = {{V{\'a}zquez-Semadeni}, Enrique and {Gonz{\'a}lez-Samaniego}, Alejandro and {Col{\'\i}n}, Pedro},
        title = "{Hierarchical star cluster assembly in globally collapsing molecular clouds}",
      journal = {\mnras},
         year = 2017,
        month = may,
       volume = {467},
       number = {2},
        pages = {1313-1328},
          doi = {10.1093/mnras/stw3229},
archivePrefix = {arXiv},
       eprint = {1611.00088},
 primaryClass = {astro-ph.GA},
       adsurl = {https://ui.adsabs.harvard.edu/abs/2017MNRAS.467.1313V}
}

@ARTICLE{VazquesSemadeni2019,
       author = {{V{\'a}zquez-Semadeni}, Enrique and {Palau}, Aina and {Ballesteros-Paredes}, Javier and {G{\'o}mez}, Gilberto C. and {Zamora-Avil{\'e}s}, Manuel},
        title = "{Global hierarchical collapse in molecular clouds. Towards a comprehensive scenario}",
      journal = {\mnras},
         year = 2019,
        month = dec,
       volume = {490},
       number = {3},
        pages = {3061-3097},
          doi = {10.1093/mnras/stz2736},
archivePrefix = {arXiv},
       eprint = {1903.11247},
 primaryClass = {astro-ph.GA},
       adsurl = {https://ui.adsabs.harvard.edu/abs/2019MNRAS.490.3061V}
}

@ARTICLE{VazquesSemadeni2024,
       author = {{V{\'a}zquez-Semadeni}, Enrique and {Palau}, Aina and {G{\'o}mez}, Gilberto C. and {Arroyo-Ch{\'a}vez}, Griselda and {Alig}, Christian and {Ballesteros-Paredes}, Javier and {Camacho}, Vianey and {Traficante}, Alessio and {Gonz{\'a}lez-Samaniego}, Alejandro and {Zamora-Avil{\'e}s}, Manuel and {Burkert}, Andreas},
        title = "{The Turbulent Support (TS) and Global Hierarchical Collapse (GHC) models for molecular clouds compared. Differences, convergence, and myths}",
      journal = {\mnras},
         year = 2025,
        month = nov,
          doi = {10.1093/mnras/staf2059},
archivePrefix = {arXiv},
       eprint = {2408.10406},
 primaryClass = {astro-ph.GA},
       adsurl = {https://ui.adsabs.harvard.edu/abs/2025MNRAS.tmp.1946V}
}

@INPROCEEDINGS{Williams2000,
       author = {{Williams}, J.~P. and {Blitz}, L. and {McKee}, C.~F.},
        title = "{The Structure and Evolution of Molecular Clouds: from Clumps to Cores to the IMF}",
    booktitle = {Protostars and Planets IV},
         year = 2000,
       editor = {{Mannings}, V. and {Boss}, A.~P. and {Russell}, S.~S.},
        month = may,
        pages = {97},
          doi = {10.48550/arXiv.astro-ph/9902246},
archivePrefix = {arXiv},
       eprint = {astro-ph/9902246},
 primaryClass = {astro-ph},
       adsurl = {https://ui.adsabs.harvard.edu/abs/2000prpl.conf...97W}
}

@ARTICLE{Williams2005,
       author = {{Williams}, S.~J. and {Fuller}, G.~A. and {Sridharan}, T.~K.},
        title = "{The circumstellar environments of high-mass protostellar objects. II. Dust continuum models}",
      journal = {\aap},
         year = 2005,
        month = apr,
       volume = {434},
       number = {1},
        pages = {257-274},
          doi = {10.1051/0004-6361:20034114},
       adsurl = {https://ui.adsabs.harvard.edu/abs/2005A&A...434..257W}
}

@ARTICLE{Verliat2022,
       author = {{Verliat}, Antoine and {Hennebelle}, Patrick and {Gonz{\'a}lez}, Marta and {Lee}, Yueh-Ning and {Geen}, Sam},
        title = "{Influence of protostellar jets and HII regions on the formation and evolution of stellar clusters}",
      journal = {\aap},
         year = 2022,
        month = jul,
       volume = {663},
          eid = {A6},
        pages = {A6},
          doi = {10.1051/0004-6361/202141765},
archivePrefix = {arXiv},
       eprint = {2202.02237},
 primaryClass = {astro-ph.GA},
       adsurl = {https://ui.adsabs.harvard.edu/abs/2022A&A...663A...6V}
}

@ARTICLE{Zavala-Molina2023,
       author = {{Zavala-Molina}, Rafael and {Ballesteros-Paredes}, Javier and {Gazol}, Adriana and {Palau}, Aina},
        title = "{The effect of tidal forces on the Jeans instability criterion in star-forming regions}",
      journal = {\mnras},
         year = 2023,
        month = sep,
       volume = {524},
       number = {3},
        pages = {4614-4630},
          doi = {10.1093/mnras/stad2091},
archivePrefix = {arXiv},
       eprint = {2306.11106},
 primaryClass = {astro-ph.GA},
       adsurl = {https://ui.adsabs.harvard.edu/abs/2023MNRAS.524.4614Z}
}

@ARTICLE{Zhang2014,
       author = {{Zhang}, Qizhou and {Qiu}, Keping and {Girart}, Josep M. and {Liu}, Hauyu Baobab and {Tang}, Ya-Wen and {Koch}, Patrick M. and {Li}, Zhi-Yun and {Keto}, Eric and {Ho}, Paul T.~P. and {Rao}, Ramprasad and {Lai}, Shih-Ping and {Ching}, Tao-Chung and {Frau}, Pau and {Chen}, How-Huan and {Li}, Hua-Bai and {Padovani}, Marco and {Bontemps}, Sylvain and {Csengeri}, Timea and {Ju{\'a}rez}, Carmen},
        title = "{Magnetic Fields and Massive Star Formation}",
      journal = {\apj},
         year = 2014,
        month = sep,
       volume = {792},
       number = {2},
          eid = {116},
        pages = {116},
          doi = {10.1088/0004-637X/792/2/116},
archivePrefix = {arXiv},
       eprint = {1407.3984},
 primaryClass = {astro-ph.GA},
       adsurl = {https://ui.adsabs.harvard.edu/abs/2014ApJ...792..116Z}
}

@ARTICLE{Zhang2015,
       author = {{Zhang}, Qizhou and {Wang}, Ke and {Lu}, Xing and {Jim{\'e}nez-Serra}, Izaskun},
        title = "{Fragmentation of Molecular Clumps and Formation of a Protocluster}",
      journal = {\apj},
         year = 2015,
        month = may,
       volume = {804},
       number = {2},
          eid = {141},
        pages = {141},
          doi = {10.1088/0004-637X/804/2/141},
archivePrefix = {arXiv},
       eprint = {1503.03017},
 primaryClass = {astro-ph.SR},
       adsurl = {https://ui.adsabs.harvard.edu/abs/2015ApJ...804..141Z}
}

@ARTICLE{Zhang2021,
       author = {{Zhang}, S. and {Zavagno}, A. and {L{\'o}pez-Sepulcre}, A. and {Liu}, H. and {Louvet}, F. and {Figueira}, M. and {Russeil}, D. and {Wu}, Y. and {Yuan}, J. and {Pillai}, T.~G.~S.},
        title = "{H II regions and high-mass starless clump candidates. II. Fragmentation and induced star formation at  0.025 pc scale: an ALMA continuum study}",
      journal = {\aap},
         year = 2021,
        month = feb,
       volume = {646},
          eid = {A25},
        pages = {A25},
          doi = {10.1051/0004-6361/202038421},
archivePrefix = {arXiv},
       eprint = {2012.07738},
 primaryClass = {astro-ph.GA},
       adsurl = {https://ui.adsabs.harvard.edu/abs/2021A&A...646A..25Z}
}

@ARTICLE{Zinnecker2007,
       author = {{Zinnecker}, Hans and {Yorke}, Harold W.},
        title = "{Toward Understanding Massive Star Formation}",
      journal = {\araa},
         year = 2007,
        month = sep,
       volume = {45},
       number = {1},
        pages = {481-563},
          doi = {10.1146/annurev.astro.44.051905.092549},
archivePrefix = {arXiv},
       eprint = {0707.1279},
 primaryClass = {astro-ph},
       adsurl = {https://ui.adsabs.harvard.edu/abs/2007ARA&A..45..481Z}
}

@ARTICLE{Zhou2022,
       author = {{Zhou}, Jian-Wen and {Liu}, Tie and {Evans}, Neal J. and {Garay}, Guido and {Goldsmith}, Paul F. and {G{\'o}mez}, Gilberto C. and {V{\'a}zquez-Semadeni}, Enrique and {Liu}, Hong-Li and {Stutz}, Amelia M. and {Wang}, Ke and {Juvela}, Mika and {He}, Jinhua and {Li}, Di and {Bronfman}, Leonardo and {Liu}, Xunchuan and {Xu}, Feng-Wei and {Tej}, Anandmayee and {Dewangan}, L.~K. and {Li}, Shanghuo and {Zhang}, Siju and {Zhang}, Chao and {Ren}, Zhiyuan and {Tatematsu}, Ken'ichi and {Shing Li}, Pak and {Won Lee}, Chang and {Baug}, Tapas and {Qin}, Sheng-Li and {Wu}, Yuefang and {Peng}, Yaping and {Zhang}, Yong and {Liu}, Rong and {Luo}, Qiu-Yi and {Ge}, Jixing and {Saha}, Anindya and {Chakali}, Eswaraiah and {Zhang}, Qizhou and {Kim}, Kee-Tae and {Ristorcelli}, Isabelle and {Shen}, Zhi-Qiang and {Li}, Jin-Zeng},
        title = "{ATOMS: ALMA Three-millimeter Observations of Massive Star-forming regions - XI. From inflow to infall in hub-filament systems}",
      journal = {\mnras},
         year = 2022,
        month = aug,
       volume = {514},
       number = {4},
        pages = {6038-6052},
          doi = {10.1093/mnras/stac1735},
archivePrefix = {arXiv},
       eprint = {2206.08505},
 primaryClass = {astro-ph.GA},
       adsurl = {https://ui.adsabs.harvard.edu/abs/2022MNRAS.514.6038Z}
}

@ARTICLE{Zhou2024a,
       author = {{Zhou}, J.~W. and {Dib}, S. and {Juvela}, M. and {Sanhueza}, P. and {Wyrowski}, F. and {Liu}, T. and {Menten}, K.~M.},
        title = "{Gas inflows from cloud to core scales in G332.83-0.55: Hierarchical hub-filament structures and tide-regulated gravitational collapse}",
      journal = {\aap},
         year = 2024,
        month = jun,
       volume = {686},
          eid = {A146},
        pages = {A146},
          doi = {10.1051/0004-6361/202449514},
archivePrefix = {arXiv},
       eprint = {2403.13442},
 primaryClass = {astro-ph.GA},
       adsurl = {https://ui.adsabs.harvard.edu/abs/2024A&A...686A.146Z}
}

@ARTICLE{Zhou2024,
       author = {{Zhou}, Jian-wen and {Kroupa}, Pavel and {Dib}, Sami},
        title = "{Self-similar cluster structures in massive star-forming regions: Isolated evolution from clumps to embedded clusters}",
      journal = {\aap},
         year = 2024,
        month = aug,
       volume = {688},
          eid = {L19},
        pages = {L19},
          doi = {10.1051/0004-6361/202450412},
archivePrefix = {arXiv},
       eprint = {2407.20150},
 primaryClass = {astro-ph.GA},
       adsurl = {https://ui.adsabs.harvard.edu/abs/2024A&A...688L..19Z}
}

@ARTICLE{Zhou2025,
       author = {{Zhou}, Jian-wen and {Dib}, Sami and {Kroupa}, Pavel},
        title = "{The post-gas expulsion coalescence of embedded clusters as an origin of open clusters}",
      journal = {\mnras},
         year = 2025,
        month = feb,
       volume = {537},
       number = {2},
        pages = {845-857},
          doi = {10.1093/mnras/staf076},
archivePrefix = {arXiv},
       eprint = {2501.08831},
 primaryClass = {astro-ph.GA},
       adsurl = {https://ui.adsabs.harvard.edu/abs/2025MNRAS.537..845Z}
}

\begin{appendix}

\section{Probability distribution of the projected separation}
\label{AppendixA}

\begin{figure}
    \centering
    \includegraphics[width=1\linewidth]{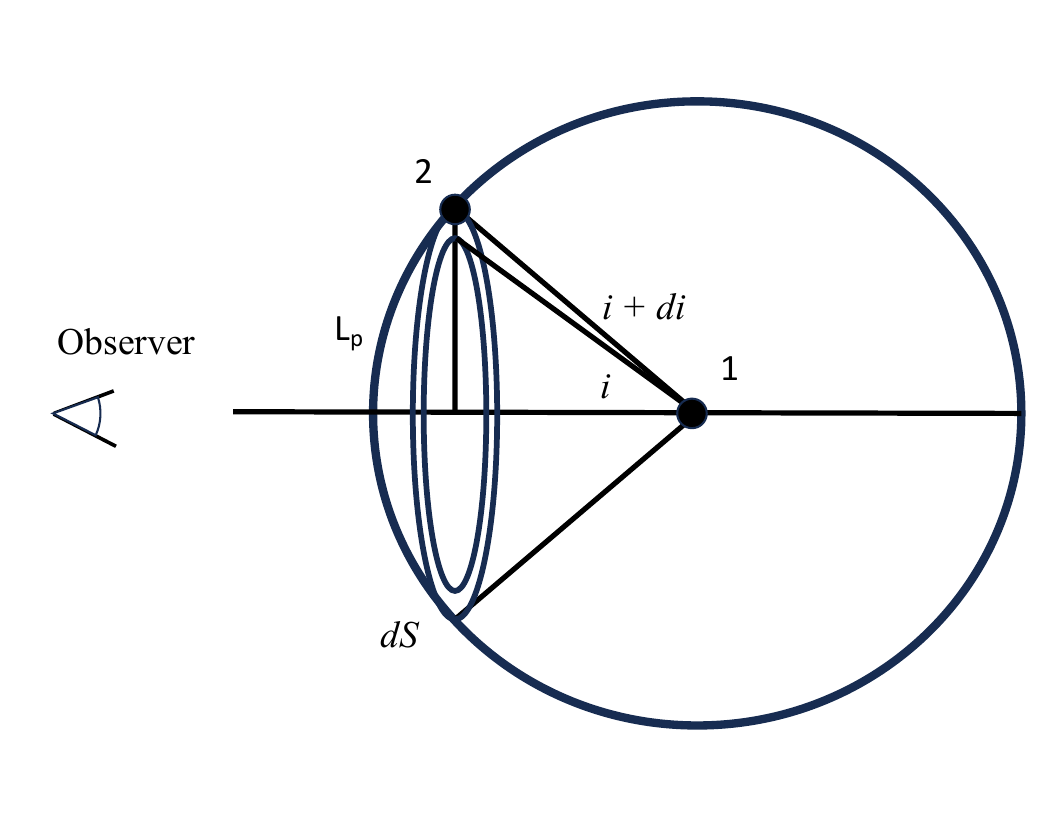}
    \caption{Scheme of projection of a pair of randomly oriented objects in 3D space }
    \label{fig:AppendixA}
\end{figure}

The measurable separations between cores are the projection on the plane of the sky of the real 3D distances between pair members. To derive these 3D distances, we can only rely on statistical analysis that, once applied to the distribution of projected separations, allows us to infer the 3D distance between cores. This approach is based on the estimation of the probability distribution density of the projected distances $L_{p}$, $p(L_{p})$ of a pair of objects that are randomly placed in the 3D space. As a first step, we consider only the effect of the projection on these two objects with fixed separation $L$, without considering any distribution of their separation. Assuming two points in the 3D space with a fixed separation $L$, the projected separation observed along a generic direction is equal to $L_{p} = L \sin{i}$, where $i$ is the inclination angle of the vector connecting the two points and the line of sight (see Fig.\,\ref{fig:AppendixA}). For a randomly oriented connecting vector between the two points, the probability of observing a separation $L_{p}$ is equal to the area of the annulus $dS = 2\pi L_{p}L\,di$ normalized to the surface of a sphere of radius $L$, in other words: 

\begin{equation}
\label{EqA1}
dP(L_{p}) = \frac{dS}{4 \pi L^{2}} =  \frac{L_{p}}{L} \frac{di}{2} = \frac{\sin{i}\,di}{2}.
\end{equation}

\noindent Differentiating $L_{p} = L \sin{i}$ and doing some algebra yields the following: 

\begin{equation}
\label{EqA2}
di = \frac{dL_{p}}{L\,\cos{i}} = \frac{1}{L}\frac{dL_{p}}{\sqrt{1 - \sin^2{i}}}.
\end{equation}

\noindent Replacing Eq.\,\ref{EqA2} in Eq.\,\ref{EqA1}, we obtain the probability density distribution function $p(L_{p})$,

\begin{equation}
p(L_{p}) = \frac{dP}{dL_{p}} = \frac{L_{p}}{L\sqrt{L^2 - L_{p}^2}},
\end{equation}

\noindent which is normalized to unity as $\int_{0}^{L} p(L_{p})\,dL_{p} = 1$.

The expected value $\left < L_{p} \right >$ of the projected separation $L_{p}$ is derived from the probability density distribution function $p(L_{p})$ as

\begin{equation}
\label{Eq:ExpValue}
\left < L_{p} \right > = \int_{0}^{L}L_{p}p(L_{p})\,dL_{p} = \frac{\pi}{4}L,
\end{equation}

\noindent providing the average factor that link the intrinsic 3D separation to the projected one in the case of random orientation for a pair of objects. A similar estimate has been provided by \cite{Ishihara2024}, against which this computation provides a statistical distribution. More generally, in the case that the 3D distance between the two points is not fixed, but has a probability distribution function $P'(L)$, then the projected separations will be distributed as 

\begin{equation}
\label{EqA4}
p(L_{p}) = \int \frac{1}{L} \frac{P'(L)\,L_{p}\, dL}{\sqrt{L^2 - L_{p}^2}},
\end{equation}whose formulation has the advantage of describing how to statistically deproject the projected probability distribution function $P'(L)$ that can be derived from the observed separations.  Equation\,\ref{EqA4} can be inverted with numerical methods, or by assuming a shape for $P'(L)$. An important aspect of Eq. \ref{EqA4} is that it is extremely unlikely for a unimodal 3D distribution $P'(L)$ to produce a multimodal 2D projected distribution $p(L_{p})$.

\section{Thermal Jeans analysis}
\label{AppendixC}

The linear analysis of the stability of a non-magnetized, homogeneous, isothermal, and self-gravitating clump shows that there is a critical length above which density perturbations are always unstable against gravitational contraction \citep{Jeans1902,MacLow2004}. This critical length, and the mass contained in a sphere of this diameter, are usually referred to as the thermal Jeans length, $\lambda_{J}^{th}$, and mass, $M_{J}^{th}$, which, in the absence of turbulent support, depend  on the density, $\rho_{cl}$, and the internal sound speed, $c_{s}$ as 

\begin{equation}
    \lambda_{J}^{th} = \left ( \frac{\pi c_{s}^{2}}{G \rho_{cl}} \right )^{\frac{1}{2}} \quad\quad \text{and}\quad\quad M_{J}^{th} = \frac{ \pi^{\frac{5}{2}}}{6} \frac{c_{s}^{3}}{\sqrt{G^{3}\rho_{cl}}},
    \label{EqLJ}
\end{equation}

\noindent where $G$ is the gravitational constant. 

Expressing the sound speed in terms of the temperature of the clump and assuming a mean molecular weight $\mu=2.37$, which corresponds to a molecular gas with solar composition (\citealt{Kauffmann2008}; see also our Sect.\,\ref{Sect:SeparationClumpSurfaceDens}), these relations become  

\begin{equation}
    \lambda_{J}^{th} \approx 4300\,\rm{\,au\,}\, \left ( \frac{T}{10\,K} \right )^{\frac{1}{2}}  \left ( \frac{n}{10^{6}\,cm^{-3}} \right )^{-\frac{1}{2}}\,and 
    \label{Eq:JeansLenght}
\end{equation}

\begin{equation}
    M_{J}^{th} \approx 0.28 \,\rm{\,M_{\odot}\,}\, \left ( \frac{T}{10\,K} \right )^{\frac{3}{2}}  \left ( \frac{n}{10^{6}\,cm^{-3}} \right )^{-\frac{1}{2}},   
    \label{Eq:JeansMass}
\end{equation}

\noindent where $n$ is the number (volume) density in the system.

\section{ALMAGAL core masses versus thermal Jeans mass}
\label{AppendixD}

The estimated thermal Jeans masses in the ALMAGAL clumps range from $\sim0.3$ to $\sim14$\,M$_{\odot}$, a range of masses that is fully observable by the sensitivity of the survey \citep{Molinari2025}, and that is above the completeness limit of $\sim0.2$\,M$_{\odot}$ of 
the ALMAGAL core catalog \citep{Coletta2025}.

Together, the distribution of the ratio $M_{core}/M_{J}^{th}$ presents a wide spread over about two orders of magnitude, but is heavily skewed towards low values of the ratio. The median value is $M_{core}/M_{J}^{th}\sim0.3$ and the first and third quartiles are equal to 0.1 and 0.75, respectively, indicating that most of the ALMAGAL cores have masses considerably lower than $M_{J}^{th}$. 
Specifically, 4987 of 6218, equal to $\sim80$\% of the entire sample, have $M_{core}\,<\,M_{J}^{th}$. Even taking into account the large uncertainties in the core mass and $M_{J}^{th}$, the result is that a large fraction of cores with sub-Jeans masses are present. Similar results were also found in other surveys, like, for example, the ASHES survey, where a large population of sub-Jeans mass cores has also been detected \citep{Sanhueza2019,Morii2023}. \\

Figure \ref{fig:MassMJcomparisonAverageParameters} shows the density distribution of the ratio $M_{core}/M_{J}^{th}$ as a function of $\Sigma_{cl}$ and $L/M$. We do not observe any strong correlation between the clump surface density $\Sigma_{cl}$ and the ratio M$_{core}$/M$_{J}^{th}$. The large majority ($\sim70$\%) of the detected cores are identified in clumps with 0.6$\lesssim\Sigma_{cl}\lesssim$2.5\,g\,cm$^{-2}$ whose distribution of M$_{core}$/M$_{J}^{th}$ does not depend on the clump surface density $\Sigma_{cl}$. They all have a wide distribution that extends from 0.02 to $\sim10$, with a typical peak around $\sim0.1-0.3$.

\begin{figure}
    \centering
    \includegraphics[width=1\linewidth]{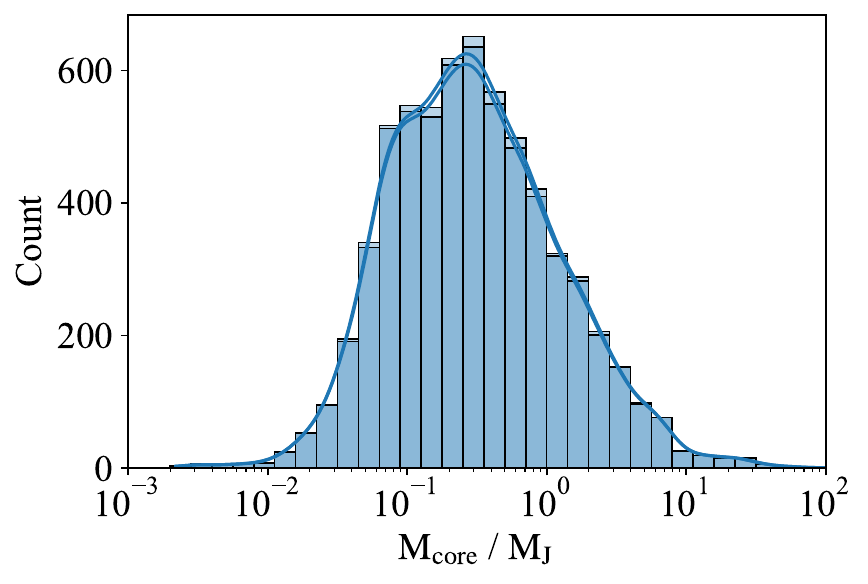}
    \caption{Distribution of the ratio $M_{core}/M_{J}^{th}$ observed in the ALMAGAL clumps. Light blue: Entire population of 6218 cores; dark blue: Population of cores in clumps with $N_{cores}\,\geq\,4$.}
    \label{fig:DistributionMassMJ}
\end{figure}

\begin{figure*}
    \centering
    \includegraphics[width=0.48\linewidth]{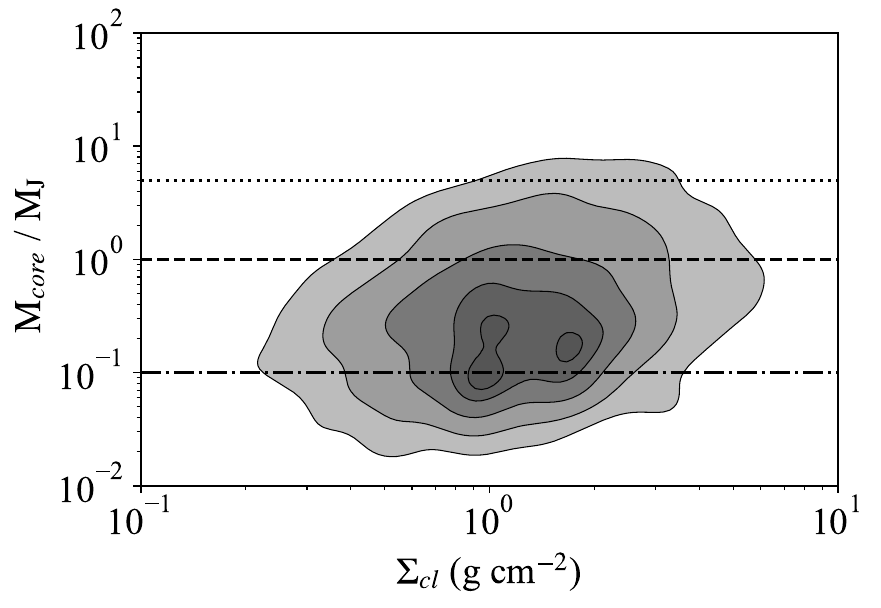}
    \includegraphics[width=0.48\linewidth]{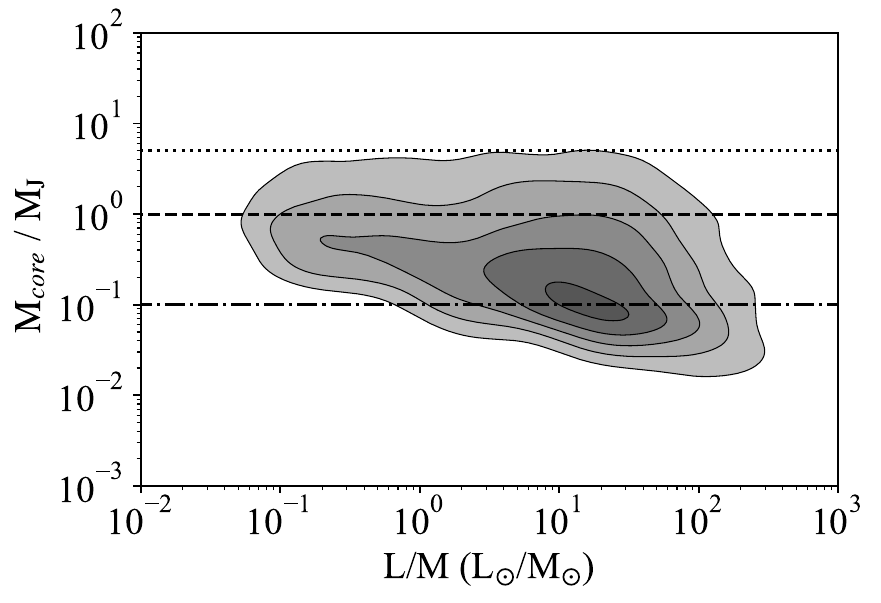}
    \caption{Density distribution plot of the core mass normalized to the thermal Jeans mass, $M_\mathrm{J}$, presented as function of the clump $L/M$ (\textit{right panel}) and surface density, $\Sigma_{c}$  (\textit{left panel}). The contours identify the different areas for which, going from the external to the internal, a fraction of 10\%, 25\%, 50\%, 75\%, 95\% of the entire distribution lies outside. The horizontal lines are reference values of the ratios: unity (dashed), 0.1 (dot-dashed), and 5 (dotted).
    }
    \label{fig:MassMJcomparisonAverageParameters}
\end{figure*}

It is more interesting to analyze the distribution of the ratio M$_{core}$/M$_{J}^{th}$ as a function of the $L/M$ ratio. While the ratio M$_{core}$/M$_{J}^{th}$ still shows a general wide spread, we find a broad general trend with the $L/M$ ratio. It is important to note that for clumps in the early stages of evolution, we identify a better agreement of the core masses with the computed thermal Jeans mass. In fact, the distribution of M$_{core}$/M$_{J}^{th}$ for cores in clumps with $L/M\lesssim1$\,L$_{\odot}/$M$_{\odot}$ peaks at $\sim0.6$, and shows a relatively narrow distribution with the first and third quartiles equal to 0.3 and 1.33. This agreement is not due to the sample size, as the distribution is derived from 1186 cores, equivalent to $\sim20$\% of the entire sample. 
On the other hand, for the core separations, we observe that the distribution M$_{core}$/M$_{J}^{th}$ shifts its peak towards lower values and broadens with increasing values of $L/M$. The peak gradually shifts to M$_{core}$/M$_{J}^{th}\sim0.1$ for clumps with $L/M\gtrsim50$\,L$_{\odot}/$M$_{\odot}$. Moreover, we note that while the peak shifts, the distribution becomes progressively more right skewed, presenting a considerable tail at high values of the ratio, and increase its width. 
This observed behavior is due to the combination of two effects. First, we measured an increase in the estimated M$_{J}^{th}$ with $L/M$ in the sample of ALMAGAL clumps, which shifts the distribution of the ratio M$_{core}$/M$_{J}^{th}$. The increase in M$_{J}^{th}$ is mainly driven by a general warm-up of the clump material in systems with higher $L/M$, due to the heating from the already formed protostars. The clump-averaged volume density, $n$, has a minor impact on the estimated M$_{J}^{th}$, since the observed ALMAGAL clumps have a distribution of $n$ that spans the same interval independently of $L/M$. Secondly, \citet{Coletta2025} found that the core mass distribution extends to higher masses in the clumps with $L/M$, which explains the broadening of the distribution of the ratio in these systems. 
Significant differences between M$_{core}$ and M$_{J}^{th}$ are measured in clumps with $L/M\,>\,2$\,L$_{\odot}/$M$_{\odot}$, where the average ratio M$_{core}$/M$_{J}^{th}$ is found to be $\lesssim0.3$, indicating that the newly formed cores have masses smaller than the thermal Jeans mass. One possible explanation for the abundance of sub-Jeans masses may reside in the increase in the number of cores with evolution. The formation of additional fragments increases the strength of the tidal forces present in the cluster. Those forces are suggested to also promote gravitational instabilities, with the effect of reducing the critical mass required for local collapse \citep{Zavala-Molina2023}

\section{Simulations of the $Q$ parameter in regimes with low number of fragments}
\label{AppendixE}

The $Q$ parameter introduced by \cite{Cartwright2004} is usually adopted to characterize the spatial distributions of the components of clusters, since its value distinguishes between systems where the elements are centrally concentrated and those where they are arranged in subclusters, which are well represented by a fractal distribution \citep{Cartwright2004,Parker2018,Dib2019}. \cite{Cartwright2004} used synthetic clusters to identify the intervals of the values of the $Q$ parameter that characterize these different spatial distributions. They found that $Q\simeq0.8$ separates between the two aforementioned distributions: centrally concentrated clusters have $Q\,>\,0.8$, while subclustered systems have $Q\,<\,0.8$. The threshold value of $Q\simeq0.8$ identifies systems whose elements are uniformly distributed. \\
However, as correctly pointed out by \cite{Avison2023}, this classification was defined for star clusters composed of a large number of components, from several tens to hundreds of elements. On the contrary, the typical cluster of cores observed with ALMA is generally composed of a smaller number of sources per field, up to a few tens in the case of the ALMAGAL survey. Poorly crowded clusters with 2 to 15 detected sources per field were found in the TEMPO survey by \citet{Avison2023}, who were the first to investigate the reliability of the $Q$ parameter applied to the case of small sample sizes. Their result was that $Q$ parameter can lead to inconclusive results when $N_{cores}\,<\,10$, as it typically returns values below the threshold value of 0.8 independently of the intrinsic spatial distribution of the cluster elements. Although their results suggest that the $Q$ parameter has a limited applicability to typical ALMA observations, they analyzed only two sets of synthetic clusters with $N_{core}=5$ and 10, as these values represent the degree of fragmentation observed in the TEMPO survey. Here, we extend their work to the case of ALMAGAL clumps, where \citet{Coletta2025} detected between 2 and 49 sources, investigating the reliability of the $Q$ parameter in these cases. 

We followed the prescription of \cite{Cartwright2004} to simulate synthetic clusters with centrally condensed and subclustered spatial distributions for their components. We simulated 3D clusters following the former type of distribution as spherical systems composed of $N$ objects radially distributed with a number density $N(r)\propto r^{-p}$. These are simply generated by creating $N$ triples of polar coordinates $(r, \theta, \phi)$ with the following formulae: 

\begin{equation}
r = \left ( \frac{(3 - p) R}{3} \right )^{\frac{1}{3-p}}, \\ 
\end{equation}
\begin{equation}
\theta = \arccos((2\Theta) -1),\\
\end{equation}
\begin{equation}
\phi  = 2\pi\Phi.
\end{equation}

\noindent Here $R$, $\Theta$, and $\Phi$ are numbers extracted randomly with a uniform distribution between 0 and 1. \\
Simulations of subclustered systems are generated by creating fractal distributions with the box method described in \cite{Goodwin2004} and also adopted by \cite{Cartwright2004} and \cite{Cartwright2006}. In short, these patterns are created by initially dividing the region of the cluster into eight cubes with each side equal to $R_{c}/2$, placing a first generation of parent seeds in the center of each of these cubes. Each cube is further split into 8 subcubes, whose centers are populated by a first generation of children, which may become fertile. Only fertile elements are kept in each cycle, and, in the next one, those children become parents, spawning a new generation. The iterative process creates the fractal distribution for the cluster that contains a larger or smaller number of substructures depending on the probability function adopted for the fertility. This probability is set by the fractal dimension $D$ as 2$^{(D-3)}$, allowing fewer children to mature for lower values of the fractal dimension $D$, and thus generating a system with more substructures. Moreover, to avoid extremely regular structures, we introduce a random displacement of amplitude 0.01$\times R_{c}$ to the position of the spawning children. Cycles are iterated until a number of cluster components larger than the target $N$ are generated, then we randomly select $N$ of these components to define a single synthetic cluster realization. \\
We ran seven sets of simulations: three for each of two types of distribution sampling with $p=1$, 2, and 2.9 (low, medium, and high concentrations) and with $D=$ 1.5, 2.0, and 2.5, (high, mild, and low level of subclustering), respectively, and one where the components are uniformly distributed in the cluster, which is simply generated adopting $p=0$ (an approximately similar result is obtained with $D=3$). Each set is composed of a group of synthetic clusters computed varying $N_{cores}$ from 4 to 30 with a step of 2. Our simulations are designed over a range of $N_{cores}$ smaller than that of the ALMAGAL survey, since higher number of fragments rarely occurs in our observations. We computed 10$^{5}$ synthetic realizations for each value of $N_{cores}$, projecting the 3D spatial distribution in two dimensions and then deriving the distribution of the $Q$ parameter. The large number of realizations is chosen to carefully determine the probability distribution function of the $Q$ parameter. To speed up the calculation of the 10$^{5}$ realizations, we initially simulated 10$^{4}$ systems with three times the intended number of sources $N_{cores}$, and then, for each system, we selected $N_{cores}$ elements at random ten times. 

\begin{figure*}
    \centering
    \includegraphics[width=0.32\linewidth]{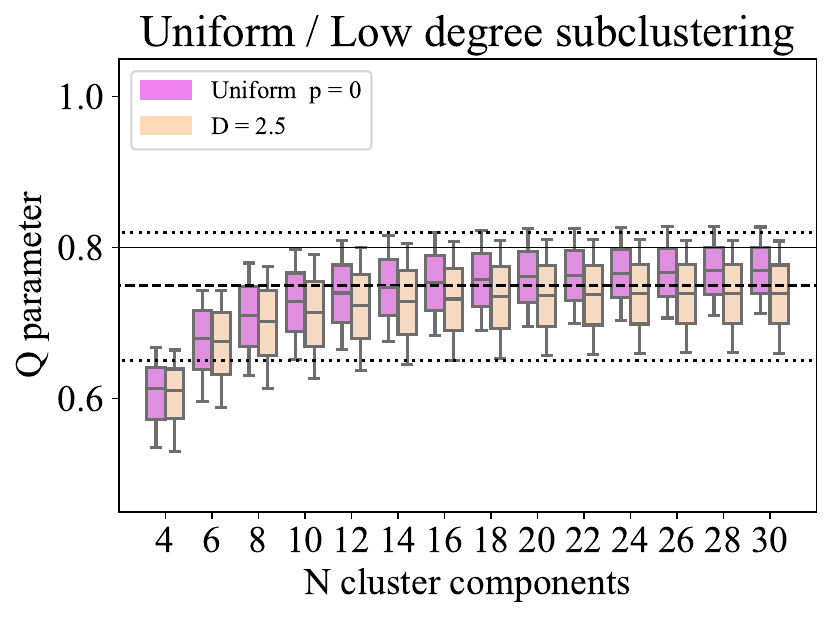}
    \includegraphics[width=0.32\linewidth]{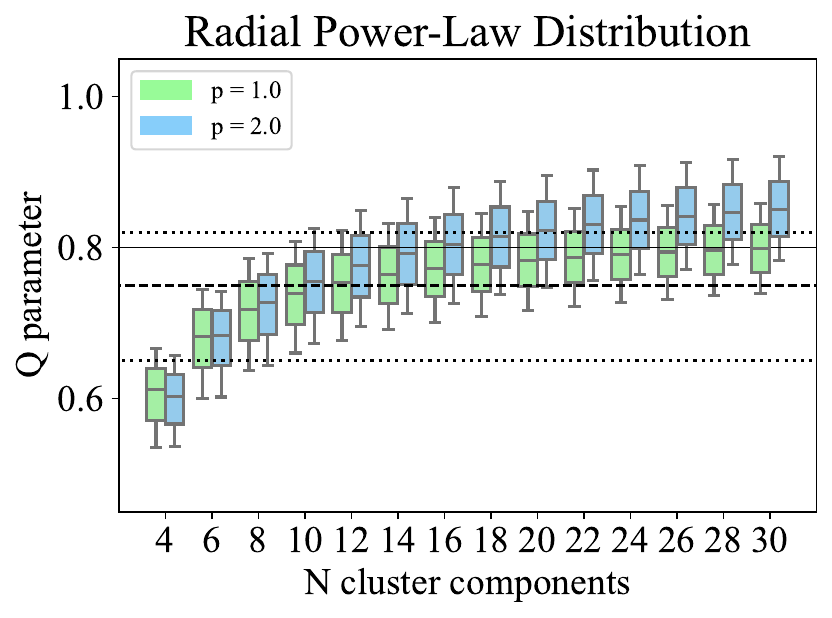}
    \includegraphics[width=0.32\linewidth]{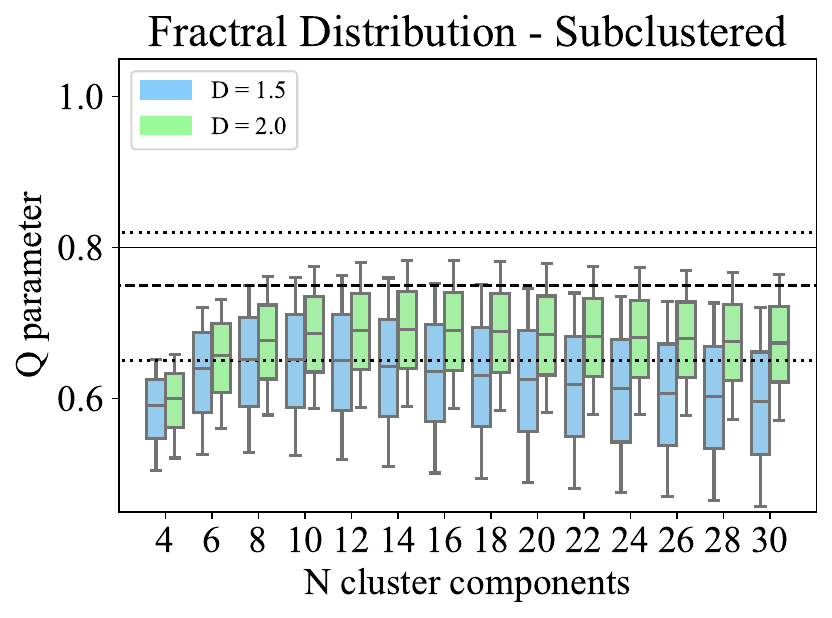}
    \caption{Boxplots representing the distributions of the  $Q$ parameter derived from the simulated clusters as a function of the number of cluster components. Each box indicates the median, the first, and the third quartile, while the whiskers extend from the 10th  to the 90th percentile of the distribution of measured $Q$. The left panel compares the results of the simulations with uniform distribution, obtained with $p$ = 0, and with the lowest level of subclustering in the simulation campaign, corresponding to a fractal dimension $D$ = 2.5. The central panel presents the results for clusters with a radial distribution of their components, for the two cases with exponents $p$ = 1 and $p$ = 2. The right panel presents the results for clusters with a fractal distribution of their components with two different levels of subclustering, determined by a fractal dimension $D$ = 2.0 (mild subclustering) and $D$ = 1.5 (high subclustering). The reference values for $Q$ discussed in the text are reported as horizontal lines, corresponding to $Q\,=\,0.65$ and $Q\,=\,0.83$ (dotted lines), 0.75 (dashed line), and the threshold $Q\,=\,0.8$ introduced by \citet[thin line]{Cartwright2004}.}
    \label{Fig:SimulationQpar}
\end{figure*}

\begin{figure*}
    \centering
    \includegraphics[width=0.32\linewidth]{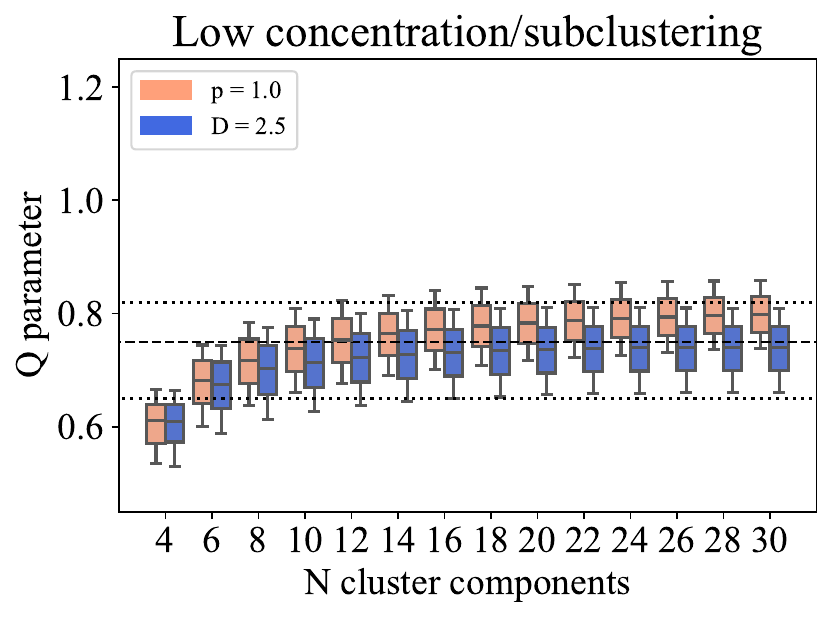}
    \includegraphics[width=0.32\linewidth]{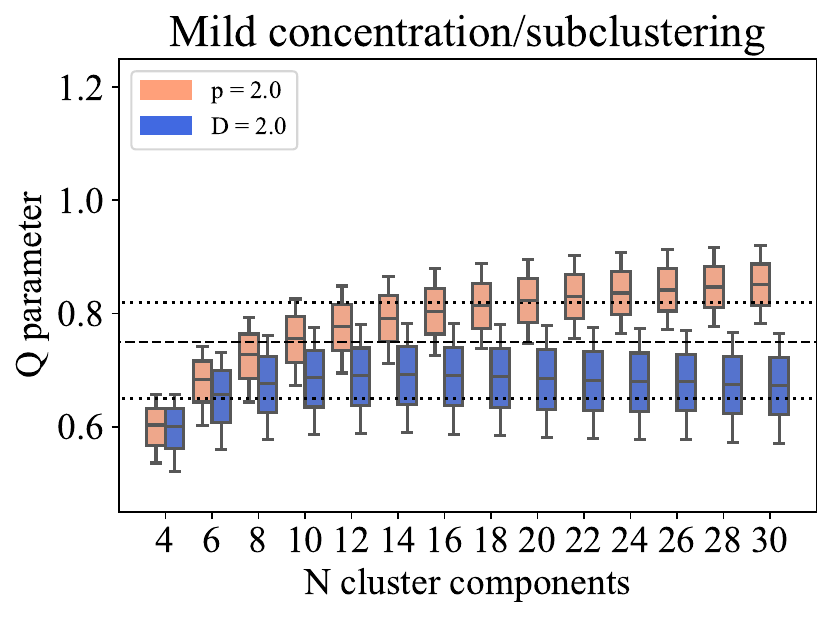}
    \includegraphics[width=0.32\linewidth]{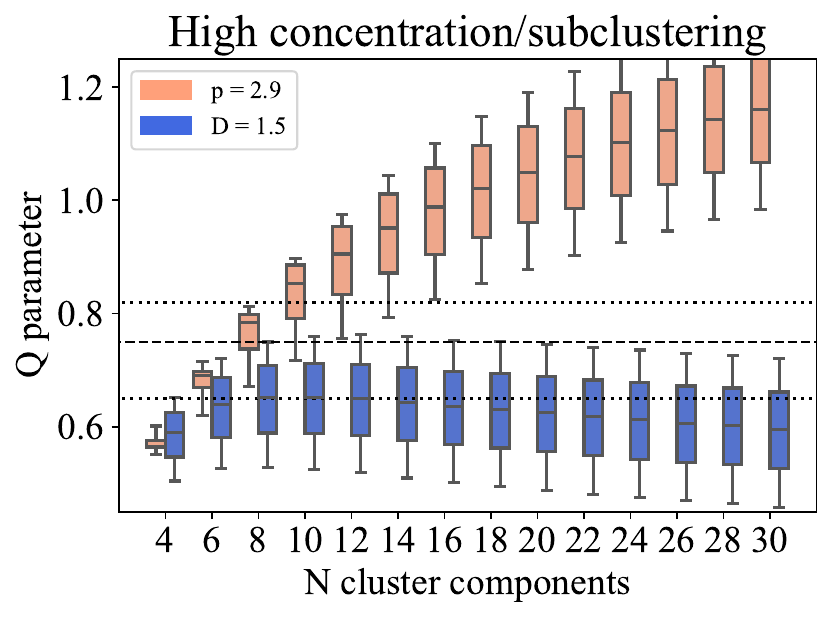}
    \caption{Same as Fig.\,\ref{Fig:SimulationQpar}, but showing the direct comparison of the distributions of the $Q$ parameter computed from the sets of synthetic clusters with the components distributed with a radial power law and with a fractal distribution. The comparison is shown for sets with low, mild, and high concentration degrees and level of subclustering. } 
    \label{Fig:SimulationQparComparison}
\end{figure*}

The simulation campaign is composed of 1.1M synthetic clusters, and the results are shown in Fig.\,\ref{Fig:SimulationQpar} where we report how the probability distribution functions (PDFs) of the $Q$ parameter, represented as boxplots at the 10, 25, 75, and 90th percentiles, vary as a function of the number of cluster elements $N$ for different types of spatial distribution. For each measured $Q$ parameter, these PDFs could be theoretically adopted to assess the likelihood of a system being  a concentrated cluster or a subclustered structure. However, we prefer to use these results to define the ranges of $Q$ that are most likely associated with one or the other type of spatial distribution. \\
The plots in Fig.\,\ref{Fig:SimulationQpar} show that in the case of very small populations ($N_{core}\,<\,8$) the measured PDFs of different spatial distributions strongly overlap, confirming the result of \citet{Avison2023} that the $Q$ parameter is highly unreliable in these systems. The situation improves for $N_{core}\geq\,10$, above which the PDFs of the two types of spatial distributions start to progressively shift one with respect to the other, although they still retain a certain degree of overlap. For this reason, we focus the discussion on the results of cases where $N_{core}\,\geq\,10$. \\
These plots show that the criteria introduced by \cite{Cartwright2004} cannot be directly applied to the case of small sample sizes. In particular, the threshold limit $Q\approx0.8$, which should also identify the case with uniform distribution, does not reflect the results obtained from our simulation campaign. Our simulations of the uniform distribution indicate that the median of the probability distribution functions slightly increases from $\sim0.73$ to 0.76 when the number of cluster components increases. More generally, the PDFs in this limit case are confined in an interval equal to $0.65\,\leq\,Q\,\lesssim\,0.82$. Values close to $\sim0.8$ are only found in the upper tails of the distribution, for values greater than the third quartile, while the average of the medians is equal to $Q\,=\,0.75$. The simulation with uniform distribution is a limit case for the two types of spatial distribution analyzed; therefore, its statistical properties can be adopted as a reference to discuss the results of the other simulations. If we look at the median of the PDFs, we notice that the clusters with centrally concentrated distribution have $Q\gtrsim$0.75, while those with mild and high levels of subclustering have $Q\,<\,$0.75. \\ 
However, adopting a defined value as a threshold limit does not take into account that PDFs have widths that extend over wide intervals of $Q$, and often overlap for different simulated spatial distributions. Figure\,\ref{Fig:SimulationQparComparison} shows the PDFs derived from the simulation campaign, comparing the sets of simulations for increasing degrees of concentration or subclustering. The two different spatial distributions may both produce systems with $Q$ values between $0.65\,\leq\,Q\,\lesssim\,0.8$, but outside this interval it can be determined with a reasonably high confidence whether the cluster components are distributed as a radial power-law or as subclusters. Simulations with subclustering have PDFs whose 90th percentile is always lower than 0.8, and systems with $Q\,>\,0.8$ are unlikely to have a fractal distribution. Similarly, since the 10th percentile of the probability distribution functions of the clusters with radial distribution is always higher than 0.65, we expect that when $Q\,<\,0.65$ is measured, these systems are most likely to present a level of subclustering.

\end{appendix}

\end{document}